\PassOptionsToPackage{table}{xcolor}
\documentclass[acmsmall]{acmart}
\usepackage{xcolor}
\usepackage{listings}

\usepackage{tabularx}
\usepackage{graphicx}
\usepackage{multirow}
\usepackage{colortbl}
\usepackage{tcolorbox}
\usepackage{enumitem}

\usepackage{xcolor}
\usepackage{adjustbox}
\usepackage{colortbl}
\usepackage{booktabs}

\definecolor{baselineCol}{HTML}{FFF2CC}
\definecolor{versionCol}{HTML}{CFE2F3}
\definecolor{callgraphCol}{HTML}{D9EAD3}
\definecolor{headerCol}{HTML}{D9D9D9}

\definecolor{darkgreen}{rgb}{0,0.5,0}

\usepackage{tikz}
\usetikzlibrary{calc}

\definecolor{darkgreen}{rgb}{0,0.5,0}

\usepackage{threeparttable}

\usepackage{amsmath}
\definecolor{darkgreen}{HTML}{006400}

\definecolor{revisioncolor}{HTML}{0000FF}
\newif\ifrevisedhighlight
\revisedhighlightfalse
\newcommand{\revised}[1]{\ifrevisedhighlight\textcolor{revisioncolor}{#1}\else#1\fi}

\newcommand{\barPercent}[1]{%
  \begin{tikzpicture}[baseline=(bar.base)]
    \def\maxVal{36}
    \def\halfWidth{0.3}
    \pgfmathsetmacro{\val}{#1}
    \pgfmathsetmacro{\absVal}{abs(\val)}
    \pgfmathsetmacro{\clampedVal}{min(\absVal,\maxVal)}
    \pgfmathsetmacro{\barFrac}{\clampedVal/\maxVal * \halfWidth}
    \draw[gray!30] (-\halfWidth,0) rectangle (\halfWidth,0.3);
    \ifdim \val pt < 0pt
      \fill[red!40] (0,0) rectangle ($(-\barFrac,0.3)$);
    \else
      \fill[green!60] (0,0) rectangle ($(\barFrac,0.3)$);
    \fi
    \coordinate (bar) at (0,0.15);
  \end{tikzpicture}%
}

\newcommand{\barPercentSeven}[1]{%
  \begin{tikzpicture}[baseline=(bar.base)]
    \def\maxVal{6}
    \def\halfWidth{0.3}
    \pgfmathsetmacro{\val}{#1}
    \pgfmathsetmacro{\absVal}{abs(\val)}
    \pgfmathsetmacro{\clampedVal}{min(\absVal,\maxVal)}
    \pgfmathsetmacro{\barFrac}{\clampedVal/\maxVal * \halfWidth}
    \draw[gray!30] (-\halfWidth,0) rectangle (\halfWidth,0.3);
    \ifdim \val pt < 0pt
      \fill[red!40] (0,0) rectangle ($(-\barFrac,0.3)$);
    \else
      \fill[green!60] (0,0) rectangle ($(\barFrac,0.3)$);
    \fi
    \coordinate (bar) at (0,0.15);
  \end{tikzpicture}%
}

\newcommand{\barPercentThirty}[1]{%
  \begin{tikzpicture}[baseline=(bar.base)]
    \def\maxVal{27}
    \def\halfWidth{0.3}
    \pgfmathsetmacro{\val}{#1}
    \pgfmathsetmacro{\absVal}{abs(\val)}
    \pgfmathsetmacro{\clampedVal}{min(\absVal,\maxVal)}
    \pgfmathsetmacro{\barFrac}{\clampedVal/\maxVal * \halfWidth}
    \draw[gray!30] (-\halfWidth,0) rectangle (\halfWidth,0.3);
    \ifdim \val pt < 0pt
      \fill[red!40] (0,0) rectangle ($(-\barFrac,0.3)$);
    \else
      \fill[green!60] (0,0) rectangle ($(\barFrac,0.3)$);
    \fi
    \coordinate (bar) at (0,0.15);
  \end{tikzpicture}%
}

\newcommand{\barPercentSeventy}[1]{%
  \begin{tikzpicture}[baseline=(bar.base)]
    \def\maxVal{35}
    \def\totalWidth{0.6}
    \pgfmathsetmacro{\val}{#1}
    \pgfmathsetmacro{\absVal}{abs(\val)}
    \pgfmathsetmacro{\clampedVal}{min(\absVal,\maxVal)}
    \pgfmathsetmacro{\barLength}{\clampedVal/\maxVal * \totalWidth}
    \draw[gray!30] (0,0) rectangle (\totalWidth,0.3);
    \fill[green!60] (0,0) rectangle (\barLength,0.3);
    \coordinate (bar) at (0,0.15);
  \end{tikzpicture}%
}

\newcommand{\barPercentSeventyLoss}[1]{%
  \begin{tikzpicture}[baseline=(bar.base)]
    \def\maxVal{35}
    \def\totalWidth{0.6}
    \pgfmathsetmacro{\val}{#1}
    \pgfmathsetmacro{\absVal}{abs(\val)}
    \pgfmathsetmacro{\clampedVal}{min(\absVal,\maxVal)}
    \pgfmathsetmacro{\barLength}{\clampedVal/\maxVal * \totalWidth}
    \draw[gray!30] (0,0) rectangle (\totalWidth,0.3);
    \fill[red!40] (0,0) rectangle (\barLength,0.3);
    \coordinate (bar) at (0,0.15);
  \end{tikzpicture}%
}

\AtBeginDocument{%
  }

\setcopyright{cc}
\setcctype{by}
\copyrightyear{2026}
\acmYear{2026}
\acmDOI{10.1145/3859926}

\acmJournal{TOSEM}

\begin{document}

\title{Enhancing Neural Code Representation With Additional Context}

\author{Huy Nguyen}
\email{huyxuann@student.unimelb.edu.au}
\orcid{0000-0002-8796-0762}
\affiliation{%
  \department{School of Computing and Information Systems}
  \institution{The University of Melbourne}
  \city{Melbourne}
  \state{Victoria}
  \country{Australia}
}

\author{Christoph Treude}
\email{ctreude@smu.edu.sg}
\orcid{0000-0002-6919-2149}
\affiliation{%
  \department{School of Computing and Information Systems}
  \institution{Singapore Management University}
  \city{Singapore}
  \country{Singapore}
}

\author{Patanamon Thongtanunam}
\email{patanamon.t@unimelb.edu.au}
\orcid{0000-0001-6328-8839}
\affiliation{%
  \department{School of Computing and Information Systems}
  \institution{The University of Melbourne}
  \city{Melbourne}
  \state{Victoria}
  \country{Australia}
}

\renewcommand{\shortauthors}{Nguyen et al.}

\begin{abstract}

Automated program comprehension underpins many software engineering tasks, from code summarisation to clone detection. While recent deep learning models have achieved strong results, they typically rely on source code alone and overlook other information that developers routinely use to understand programs, such as version history or structural relationships. This gap limits the ability of current models to capture the broader context in which code evolves and operates. To address this, we conduct an empirical study of how enriching code representations with these additional contextual signals affects the performance of neural models on key program comprehension tasks.
We evaluate four downstream software engineering tasks, code clone detection, code classification, code summarisation, and vulnerability detection, using SeSaMe (1,679 semantically similar Java methods from 11 open-source projects), CodeSearchNet (2,117 projects; 63,259 methods paired with natural-language documentation), and Vul4J (178 vulnerability-inducing methods). To ensure a robust comparison, we fine-tune five foundational code models (tree-based and pre-trained Transformer architectures), namely CodeBERT, GraphCodeBERT, CodeT5, PLBART, and ASTNN, under both code-only and context-augmented settings. We further expand the evaluation to seven code-specialised large language models (CodeLlama 7B/13B/34B, Qwen2.5-Coder 1.5B/7B/14B, and DeepSeek-Coder 6.7B), comparing prompt-based context injection against fine-tuned use of context.

\revised{Experimental results show that context improves classification performance for all five foundational models and, after fine-tuning, for most of the seven LLMs. For summarisation, the effect depends on the model family: version history improves all four foundational models on every metric, whereas context injected into the prompt of a decoder-only LLM helps two of the seven models (Qwen2.5-Coder 7B and 14B, via call-graph context) and lowers both lexical and semantic metrics for the other five.} In particular, encoding version history \revised{never hurts and usually improves clone detection across the five foundational models (up to +15.92\% F1 for CodeT5)} and summarisation (e.g., GraphCodeBERT +5.56\% METEOR; CodeT5 +3.05\% BLEU-4). By contrast, call-graph context delivers model- and task-dependent gains. Combining multiple types of context (version history, call graphs, and method age) can further enhance improvements, \revised{reaching up to +21.48\% macro-F1 in code classification and +15.04\% F1 in clone detection for specific models under concatenation-based aggregation.}

Finally, a human evaluation with two annotators on 100 Java snippets (rank-ordering with ties evaluation) showed that context-augmented summaries were ranked significantly higher than summaries from models that rely on source code only for Accuracy and Content Adequacy (Wilcoxon $p\le 0.026$; Cliff’s $\lvert\delta\rvert$ up to 0.55), while improvements in Conciseness were context-dependent; inter-rater agreement was high (Kendall’s $\tau_b$).

Our study highlights the potential benefit of utilising rich and diverse context to enhance the code comprehension capability of neural-based models. An exploratory evaluation on vulnerability detection provides initial evidence that context-enriched representations transfer to security-oriented tasks. It also opens new research directions for optimising contextual encoding strategies in software engineering, for instance automated bug localisation, where call-graph context helps trace fault propagation paths~\cite{wang2014version}.
\end{abstract}




\received{12 October 2025}
\received[revised]{4 June 2026}
\received[revised]{29 September 2026}
\received[accepted]{12 October 2026}

\maketitle

\section{Introduction}
\label{sec:introduction}

Automated program comprehension is a critical aspect of modern software engineering, which underpins crucial tasks such as maintenance and debugging, as well as code generation~\cite{gold2004understanding, sites2021understanding}. The recent proliferation of AI-powered code assistants has made this capability more critical than ever~\cite{bano2024large, gitclear2023CodingCopilot, pan2024assessing}. Although code can be generated rapidly, developers must still invest significant effort to understand, integrate, and validate it~\cite{gitclear2023CodingCopilot, vaithilingam2022expectation, treude2025generative}. However, the deep learning models designed for these tasks typically analyse source code in isolation. They often overlook the rich contextual information that human developers instinctively use, such as a method's version history or its structural relationships within a larger system (e.g., its callers and callees)~\cite{samoaa2022systematic}. This fundamental gap limits the ability of current models to capture the broader context in which code evolves and operates, hindering their full potential.

Recent advances in deep learning have improved source code representations for software comprehension tasks, yet notable gaps remain.
These methodological limitations are compounded by the datasets commonly used in the field, such as OnlineJudge and BigCloneBench~\cite{samoaa2022systematic}, which are widely accepted for their scale and annotation quality; however, they contain only source code and labels. They lack the richer contextual information necessary for a deeper investigation into software evolution or behaviour~\cite{krinke2022bigclonebench}. By contrast, modern code hosting platforms (e.g., GitHub) make it feasible to collect such types of context at scale~\cite{tian2022adding}.

Previous works demonstrated the value of incorporating contextual artefacts to improve software engineering tasks. For instance, Wang and Lo~\cite{wang2014version} improved bug localisation by combining structural information with version history, while other studies found that temporal metrics, such as method age, correlate with code stability~\cite{ruthruff2008predicting}. However, a key difference here is how this context was used. These earlier approaches typically treated contextual signals as derived metadata (such as commit logs) or engineered numerical features (such as file age) for statistical and information retrieval models.
While prior work has begun to explore encoding contextual source code into neural representations — Tian and Treude~\cite{tian2022adding} provided preliminary evidence for call-graph context using ASTNN on a small dataset, and our own prior conference work~\cite{nguyen2024encoding} extended this to version history context and initial context combinations using two models on the same dataset — both studies were limited in scale, model coverage, and task diversity, and neither systematically evaluated the joint encoding of multiple context types across aggregation strategies. Providing that large-scale empirical foundation, and combining version history, call graphs, and method age in a unified framework, is the primary contribution of this journal extension. This difference forms our primary hypothesis: that by enriching neural code representations with these raw contextual artifacts, deep learning models can achieve a more profound program comprehension.
Specifically, we investigate the impact of incorporating the full source code of a method's version history, its call graph, and its age, which we believe provides a richer semantic understanding beyond what static source code alone can offer.

\begin{figure*}[h]
  \centering
  \includegraphics[width=\linewidth]{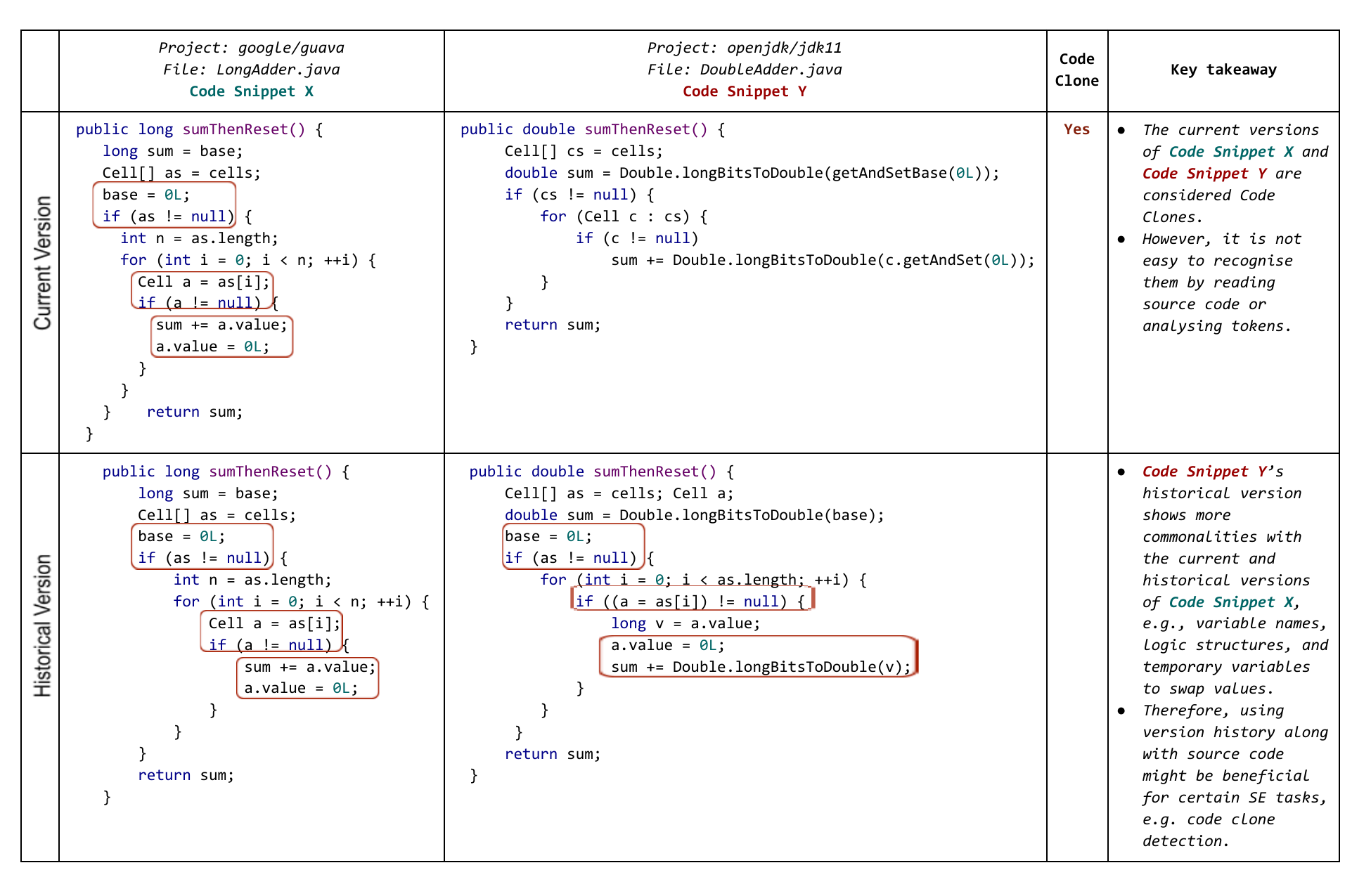}
  \caption{A motivating example of using Version History to detect code clones.}
  \label{fig:code_example}
  \Description{Example of using version history to support code clone detection.}
\end{figure*}

Figure~\ref{fig:code_example} presents a motivating example of a code clone pair, identified by the ground truth in the SeSaMe dataset~\cite{kamp2019sesame}. Although the current versions of these methods, which are originally mined from two different Java projects on GitHub, look substantially different, their historical versions reveal a greater degree of similarity than is apparent from a static analysis of the latest code. This illustrates how version history can provide critical evidence for recognising code clones, as evolutionary traces often expose recurring modifications and shared solutions that enrich program comprehension~\cite{wang2014version, hoang2020cc2vec}~\footnote{\href{https://github.com/openjdk/jdk11/blob/00d1900dc994bb745654c93bdca95ad3e579f720/src/java.base/share/classes/java/util/concurrent/atomic/DoubleAdder.java\#L155}{project: jdk11 | file: DoubleAdder.java | method: sumThenReset() | current version}}~\footnote{\href{https://github.com/openjdk/jdk11/blob/3f14786363e002a7f9f01713349a4ad687677f11/jdk/src/share/classes/java/util/concurrent/atomic/DoubleAdder.java\#L156}{project: jdk11 | file: DoubleAdder.java | method: sumThenReset() | historical version}}~\footnote{\href{https://github.com/google/guava/blob/fd919e54a55ba169dc7d9f54b7b3485aa7fa0970/android/guava/src/com/google/common/cache/LongAdder.java\#L123}{project: guava | file: LongAdder.java | method: sumThenReset() | current version}}~\footnote{\href{https://github.com/google/guava/blob/06cbdb1ed84f235b51fec6acc740d8261357545d/guava/src/com/google/common/cache/LongAdder.java\#L135}{project: guava | file: LongAdder.java | method: sumThenReset() | historical version}}.

A similar benefit comes from incorporating call-graph context in code summarisation. Consider the example in Figure~\ref{fig:sample-java} and Figure~\ref{fig:sample_callgraph}, where the benchmark’s ground-truth (reference) summary is \texttt{"Binds the given prefix to the given namespaces"} When fine-tuning pre-trained models such as CodeT5 or CodeBERT on source code alone, the generated summaries are often inadequate (e.g., \texttt{"Starts the namespace mapping"} or \texttt{"Start a prefix"}), as they over-rely on superficial cues from the method name and miss the key semantic action.

\begin{figure}[t]
  \centering
  \begin{minipage}{0.95\linewidth}
\begin{lstlisting}[language=Java]
@Override
public final void startPrefixMapping(String prefix, String uri) {
    namespaceContext.bindNamespaceUri(prefix, uri);
    namespaceContextChanged = true;
}
\end{lstlisting}
  \end{minipage}
  \caption{Sample Java code}
  \label{fig:sample-java}
  \Description[Sample Java code]{Sample Java code}
\end{figure}

\begin{figure}[t]
  \centering
  \begin{minipage}{0.95\linewidth}
\begin{lstlisting}[language=Java]
public void bindNamespaceUri(String prefix, String namespaceUri) {
    Assert.notNull(prefix, "No prefix given");
    Assert.notNull(namespaceUri, "No namespaceUri given");
    if (XMLConstants.DEFAULT_NS_PREFIX.equals(prefix)) {
        defaultNamespaceUri = namespaceUri;
    } else {
        prefixToNamespaceUri.put(prefix, namespaceUri);
        getPrefixesInternal(namespaceUri).add(prefix);
    }
}
\end{lstlisting}
  \end{minipage}
  \caption{Sample call hierarchy context}
  \label{fig:sample_callgraph}
\Description[Sample call hierarchy context]{Sample call hierarchy context}
\end{figure}

By including call hierarchy context (Figure \ref{fig:sample_callgraph}), such as the callee method \texttt{bindNamespaceUri} that explicitly checks for null values and manages namespace-to-prefix mappings, the models produce more accurate summaries. For example, the improved output \texttt{"Bind the given prefix to the given namespace URI"} captures the critical semantic element (\texttt{"URI"}) derived from the callee’s code. Compared with the original inadequate summaries, this enriched version reflects deeper semantic nuances and offers a more faithful description of the method’s functionality. This example highlights that fine-tuning deep learning models with additional context from call graphs can lead to higher-quality summaries, better aligning generated outputs with the underlying program semantics~\cite{tian2022adding, jiang2022hierarchical}.

In this work, we leverage three distinct datasets uniquely suited to our research objectives: (1) SeSaMe, which consists of 1,679 semantically similar Java methods from 11 open-source projects, tailored for classification tasks like code clone detection; (2) CodeSearchNet, a significantly larger dataset comprising 2,117 projects and 63,259 methods, primarily designed for code-to-text (code summarisation) tasks; and (3) Vul4J, a dataset of 178 vulnerability-inducing Java methods used for an exploratory vulnerability detection evaluation. We construct augmented datasets by mining version history and call-graph relations: in SeSaMe, 9,924 historical method versions and 10,308 call-graph items; in CodeSearchNet (Java), 193,766 versions and 333,573 call-graph items, with method lifetimes up to 7,368 days. While both datasets initially include only source code and corresponding labels, they also provide linkage information, enabling us to gather rich and diverse contextual data from external sources. We conduct a data mining process to extract additional context, gathering historical code versions from version control systems (e.g., GitHub) and call hierarchy using static analysis tools (e.g., Java-CallGraph). This process produces enriched datasets containing valuable contextual cues, significantly extending the original datasets in both scale and contextual depth. Moreover, we evaluate our approach using five foundational code models (tree-based and pre-trained Transformer architectures) for software engineering, including CodeBERT, GraphCodeBERT, CodeT5, PLBART, and ASTNN, covering prevalent neural architectures as experimental baselines. We further compare these against seven code-specialised large language models (LLMs) -- CodeLlama (7B, 13B, 34B), Qwen2.5-Coder (1.5B, 7B, 14B), and DeepSeek-Coder (6.7B) -- evaluating both prompt-based context injection and fine-tuned settings. To robustly validate our hypothesis that additional context encoding enhances code representations, we conduct both automated experimental evaluations and a human evaluation study, providing an objective analysis of the impact and applicability of our proposed approach across different downstream tasks.

We conducted an empirical study to address three research questions:

\begin{enumerate}[label=\textbf{RQ\arabic*},leftmargin=*]
    \item What is the impact of explicitly encoding additional context into learned vector representations on the performance of task-specific deep learning models?
    \item How does combining multiple types of encoded context into source code representations affect the performance of these models?
    \item What is the impact of different representation-level aggregation techniques on integrating source code with additional context for these models?
\end{enumerate}
\textbf{RQ1} empirically validates our hypothesis that additional context (i.e. version history or call graphs) enhances model performance by comparing models trained only on source code against those enriched with a single additional context. \textbf{RQ2} extends this investigation by combining multiple types of context simultaneously, such as version history, call graphs, and method age, to explore whether these combined context types bring complementary advantages over single-context scenarios. \textbf{RQ3} evaluates the effectiveness of various aggregation techniques, including concatenation, max-pooling, and concatenation of absolute differences, in selecting critical cues and preserving meaningful patterns when incorporating contextual information into code representations. Addressing these three research questions is essential, as it not only deepens our understanding of the role and importance of contextual information in neural code representations but also provides actionable insights for effectively leveraging such contextual information in practical applications.

Our results demonstrate that version history is the most reliable contextual signal, delivering substantial improvements of up to \textbf{+15.9\% F1} in clone detection and \textbf{+5.6\% METEOR} in summarisation. Call-graph information shows more mixed effects, proving especially useful for code classification, while its impact is less consistent for the other tasks. When multiple types of context are combined with effective aggregation strategies, performance gains are further amplified, \revised{reaching up to \textbf{+21.48\% macro-F1} in code classification (CodeT5) and \textbf{+15.04\% F1} in clone detection (PLBART).} \revised{A follow-up human evaluation of CodeT5 summaries is consistent with these findings, showing that context-augmented models produce summaries that are rated as significantly more accurate and adequate than those generated from source code alone.} Overall, our findings indicate that while context provides a clear benefit, the choice of which context to use is nuanced and depends on the specific program comprehension task.

This paper is an extended study building on our short technical paper~\cite{nguyen2024encoding} published at MSR 2024 (The 2024 International Conference on Mining Software Repositories), which offered preliminary evidence that encoding version history improves code representations using a small-scale setting (SeSaMe: 1,679 Java methods from 11 projects) and two models (ASTNN and CodeBERT) for two tasks (code clone detection and code classification). In this extension, we substantially broaden the empirical scope along several dimensions. First, we enlarge the data and task coverage by retaining SeSaMe and adding CodeSearchNet dataset, a much larger corpus spanning 2,117 projects and 63,259 methods, to study code-to-text generation (code summarisation) in addition to the prior classification tasks. Second, we expand the model spectrum from two to five representative architectures, covering a tree-based neural model (ASTNN), Transformer encoder-only models (CodeBERT, GraphCodeBERT), and encoder--decoder models (CodeT5, PLBART). Third, beyond version history, we mine and encode other types of additional context, including function call graphs and method age, then systematically compare representation-level aggregation strategies (Concatenation, Max-Pooling, and Concatenation of Absolute Differences) across code-only and context-augmented settings. Finally, we complement automated metrics with a human evaluation of 100 Java code snippets, enabling us to triangulate quantitative gains with human judgments of summary quality. We further extend the evaluation in two additional directions. We expand the model coverage to include seven code-specialised large language models (CodeLlama 7B/13B/34B, Qwen2.5-Coder 1.5B/7B/14B, and DeepSeek-Coder 6.7B), comparing prompt-based context injection against fine-tuned use of context to assess whether the benefits of additional context scale to modern decoder-only architectures. We also conduct an exploratory evaluation on vulnerability detection using the Vul4J dataset (178 vulnerability-inducing Java methods), providing initial evidence that context-enriched representations transfer beyond program comprehension tasks. Together, these extensions provide a more comprehensive assessment of when and how context enhances neural code representations across datasets, models, and tasks.

\textbf{The contributions of this work are:}
\begin{itemize}
    \item \textbf{Representation-level context encoding:} A scalable framework that integrates version history, call graphs, and number-of-days (derived from historical versions) into code representations during fine-tuning, using Concat, Max-Pooling, and Diff-Concat across five foundational code models (tree-based and pre-trained Transformer architectures): CodeBERT, GraphCodeBERT, CodeT5, PLBART, and ASTNN.
    \item \textbf{Context-augmented datasets:} Three datasets enriched with mined version history, call-graph relations, and number-of-days, suitable for four downstream tasks: Code Clone Detection, Code Classification, Code Summarisation, and Vulnerability Detection~\cite{dataset}.
    \item \textbf{Evaluation process across four SE tasks:} Automated experiments on all four SE tasks, including Code Clone Detection, Code Classification, Code Summarisation, and an exploratory evaluation on Vulnerability Detection (Vul4J), demonstrating where and when contextual information improves different models; plus a blinded, rank-order-with-ties human evaluation for code summarisation, showing where and when contextual information improves different models.
    \item \textbf{LLM comparison:} An expanded evaluation across seven code-specialised large language models (CodeLlama 7B/13B/34B, Qwen2.5-Coder 1.5B/7B/14B, and DeepSeek-Coder 6.7B), comparing prompt-based context injection against fine-tuned use of context, \revised{showing that, on the two classification tasks, every LLM reaches its highest performance with context after fine-tuning on context-augmented inputs, and that fine-tuned LLMs with prompt-level context are competitive with fine-tuned encoders using representation-level fusion.}
    \item \textbf{Aggregation insights:} An assessment of representation-level aggregation methods, confirming effective strategies for combining additional context in classification models.
\end{itemize}

\section{Related Work}

The effective representation of source code is fundamental to a wide range of automated software engineering tasks, from bug detection to code generation~\cite{ho2025ensesmells, ho2025empirical}. Research in this area has evolved from traditional approaches to sophisticated deep learning models that learn rich semantic and syntactic features directly from code artefacts. However, these models often analyse code in isolation, lacking the broader context that developers use to comprehend software. This section builds the framework for our study by first reviewing the foundational techniques for representing source code (Section~\ref{sec:neural_representations}). We then narrow our focus to the specific, under-explored potential of incorporating \emph{evolutionary} and \emph{structural} context (Section~\ref{sec:context_in_models}). Finally, we discuss how a suite of \emph{program comprehension tasks} can be used to evaluate the impact of such contextual enhancements (Section~\ref{sec:downstream_tasks}).

\subsection{Neural Representations of Source Code}
\label{sec:neural_representations}

To process source code, deep learning models must first convert it into a numerical format or a vector representation. The choice of this representation is critical, as it determines which features of the code, for instance, semantic, syntactic, or structural information, the model can effectively learn. Research in this area has progressed from representing code as a simple sequence of text to leveraging its complex and inherent structures through tree- and graph-based approaches.

\textbf{Token-based Representations.}
The most direct method for representing source code is to treat it as a sequence of tokens,  which is similar to words in a natural language. In this approach, the model learns semantic relationships based on the co-occurrence of tokens within a large corpus of code. To handle the vast and often out-of-vocabulary nature of identifiers in programming languages (e.g., \texttt{custom\_variable\_name}), these models employ subword tokenisation techniques like Byte-Pair Encoding (BPE). This approach has been popularised by large-scale Transformer-based models, most notably CodeBERT~\cite{feng2020codebert}, which learns representations from both source code and paired natural language comments. Other foundational models in this category include CuBERT~\cite{kanade2020learning} and CodeGPT~\cite{lu2021codexglue}.

\textbf{Tree-based Representations.}
To capture the grammatical and hierarchical structure of code, many methods leverage the Abstract Syntax Tree (AST). The AST provides a formal, tree-structured representation of the code's syntax, abstracting away superficial elements like punctuation. Models can learn from this structure by using architectures such as Tree-LSTMs~\cite{shido2019automatic} or other recursive neural networks. ASTNN~\cite{zhang2019novel, tian2022adding} is an influential work, which decomposes a large AST into a set of smaller statement sub-trees, encodes them into vectors, and aggregates them to form the final representation. Besides, another example, code2vec~\cite{alon2019code2vec}, represents code snippets by sampling and aggregating a bag of paths from the AST.

\textbf{Graph-based Representations.}
While the ASTs focus on capturing syntax, other methods leverage more complex graph structures to represent program semantics related to data and control flow. These methods utilise representations such as the Control-Flow Graph (CFG), which models the execution order of statements, and the Data-Flow Graph (DFG), which tracks the definition and use of data variables. Graph Neural Networks (GNNs)~\cite{leclair2020improved} are the primary architecture for learning from these structures by propagating information between related nodes in the graph. For instance, GraphCodeBERT~\cite{guo2020graphcodebert} enhances a token-based representation by explicitly incorporating the DFG, allowing it to understand variable relationships better. GNNs have also been applied directly to Program Dependence Graphs (PDGs) for specialised tasks such as vulnerability~\cite{zhou2019devign, nguyen2023multi} or code smell~\cite{ho2025ensesmells} detection.

\textbf{Hybrid Representations.}
The current trend is to not rely on a single representation, but to create hybrid representations that combine multiple views of the code for a more holistic understanding~\cite{samoaa2022systematic}. The reason is that no single representation is universally superior, where tokens excel at local semantics, trees capture syntax, and graphs can model a program's complexity and relationships~\cite{wang2023comparison}. Many modern techniques are inherently hybrid approaches, such as GraphCodeBERT (tokens + DFGs)~\cite{guo2020graphcodebert} and CodeT5~\cite{wang2021codet5}. The increasing adoption of these hybrid techniques has been noted in recent systematic literature reviews of the field~\cite{samoaa2022systematic, long2022multi}, which demonstrate their effectiveness across a range of software engineering tasks.

\textbf{Model's design choices that influence performance.}
Beyond the input structure, pretraining objectives and architecture families also impact downstream suitability~\cite{wang2021codet5}. Encoder-only models commonly use masked-language modelling (MLM) or replaced-token detection; graph-augmented variants add edge/flow prediction, and some models align code with NL via contrastive losses~\cite{le2022autopruner}. Encoder–decoder models (e.g., CodeT5, PLBART) are often preferred for generation (e.g., summarisation), whereas encoder-only architectures (e.g., CodeBERT) are strong for discrimination (e.g., classification/clone detection)~\cite{zeng2022extensive}. Our study selects representative models from these families in later sections and studies how additional context can be encoded at the representation level.

\textbf{Large Language Models for Code.}
The emergence of decoder-only large language models (LLMs) has introduced a new paradigm for code intelligence. Models such as CodeLlama~\cite{roziere2023code}, Qwen2.5-Coder~\cite{hui2024qwen2}, and DeepSeek-Coder~\cite{guo2024deepseek} extend general-purpose LLMs with code-specific pre-training, enabling strong performance on tasks such as code generation, completion, and summarisation across multiple programming languages. These models can be used in a prompt-based setting, where additional context is injected directly into the input prompt, or fine-tuned on task-specific data using parameter-efficient methods such as LoRA~\cite{hu2022lora}. A key distinction from the encoder-based models discussed above is that prompt-based context delivery relies on the model's self-attention mechanism to determine the relevance of each context source, without any task-specific supervision on the fusion process itself. Our study evaluates seven such models (CodeLlama 7B/13B/34B, Qwen2.5-Coder 1.5B/7B/14B, and DeepSeek-Coder 6.7B) under both prompt-based and fine-tuned settings, providing a direct comparison between prompt-level context injection and representation-level context encoding.

\subsection{Contextualising Code with Historical and Structural Information}
\label{sec:context_in_models}

While the representations discussed in Section~\ref{sec:neural_representations} effectively capture features from a static snapshot of code, they often analyse methods alone and miss the broader context that gives the code its full meaning. Research into human program comprehension has long established that developers rely on a rich and diverse variety of contextual information, such as a file's history or its relationship to other parts of the system, to understand a program's logic more accurately ~\cite{maletic2001supporting, kulkarni2014supporting}. This principle provides a strong motivation for moving beyond isolated and static representations. Therefore, this section reviews prior work that has begun to explore two specific, highly valuable forms of context: evolutionary and structural information~\cite{murphy2019beyond}.

\textbf{Evolutionary Context.}
Data generated during the software development process, particularly from version control systems like GitHub, provides a rich evolutionary context~\cite{widyasari2023topic}. This context reveals the \textit{intent} and \textit{history} behind the code through artefacts like commit messages, code changes (code diffs), and sequences of \textbf{historical versions}. A notable model leveraging this is CC2Vec~\cite{hoang2020cc2vec}, which demonstrated the value of this context by learning representations from commit messages paired with their associated \textit{code changes}.
Our work differs by investigating the potential of encoding the \textit{most recent historical versions within the model's token budget}, a richer but less explored signal that captures the complete state of a method at a point in time. Other studies have used simpler temporal features, such as file age or number of revisions, as signals for predicting code stability and maintenance effort~\cite{thongtanunam2013mining}.

\textbf{Structural Context.}
Beyond the source code in a single file, structural context defines a method's role and relationships within the broader program architecture. This context is typically derived from static analysis of the entire codebase and includes artefacts like the program's \textbf{call graph}, which identifies caller and callee relationships. While many models incorporate structure, they often focus on procedural information within a single method. For example, GraphCodeBERT~\cite{guo2020graphcodebert} uses the Data-Flow Graph (DFG) of a single function, and ASTNN~\cite{zhang2019novel} uses its Abstract Syntax Tree. In contrast, leveraging the inter-procedural context of callers and callees might provide a different level of understanding by situating a method within its operational environment, a technique shown to improve downstream task performance~\cite{tian2022adding, jiang2022hierarchical}.

While other types of context, such as dynamic execution traces or textual documentation, are undoubtedly valuable, this study focuses specifically on evolutionary and structural information. We select these for two key reasons. First, they are universally available in any version-controlled software project, making them a highly scalable source of information that does not require special effort or execution environments. Secondly, the potential of using the \textit{full source code} derived from these context types, instead of only metadata, diffs, or graph structures, remains a significant and under-explored research gap. Our work aims to address this by systematically investigating how these rich, code-related contexts can be effectively encoded into neural representations of source code.

\subsection{Evaluating Models on Program Comprehension Tasks}
\label{sec:downstream_tasks}

Our ultimate goal of enriching code representations is to enhance a model's ability to "understand" source code. In understanding software by humans, program comprehension is the cognitive process developers use to understand a program's function, structure, and behaviour~\cite{sites2021understanding}. For code models, this capability is not a single skill but a composite of several key abilities: the capacity to recognise semantic equivalence despite syntactic variation, to identify a method's high-level functional purpose, and to perform abstraction and explanation~\cite{maalej2014comprehension, ben2018neural}. To evaluate how well models achieve these abilities, the field relies on a suite of established downstream tasks, where each task serves as a proxy for one or more of these comprehension aspects~\cite{ben2018neural}. This section reviews the core tasks selected for our study, which are related to program comprehension and suitable for the multifaceted evaluation.

\textbf{Code Clone Detection and Similarity.}
This task evaluates a model's ability to recognise semantical equivalence within source code. By identifying code fragments that perform the same function despite syntactic differences (e.g., variable renamings, different loop structures), we test if the model has learned a deeper, more abstract representation of the code's meaning. Tree and graph-based models like ASTNN~\cite{zhang2019novel} and GraphCodeBERT~\cite{guo2020graphcodebert} are often used for this task, with performance benchmarked on datasets like BigCloneBench or POJ-104~\cite{svajlenko2015evaluating}.

\textbf{Code Classification and Tagging.}
\revised{In the general form of this task, a model assigns a method to one of a predefined set of categories, such as the algorithm it implements or the purpose it serves. In our instantiation on SeSaMe, the categories are the 11 source projects from which the methods were drawn, so the task is project-level classification. Project membership serves as a proxy for a method's architectural and domain family: methods from the same project share a domain, coding idioms, and design conventions, and recovering that family from the method and its context is a meaningful comprehension signal. It is not a direct measure of a method's abstract functional purpose independent of where it comes from, and we interpret the results accordingly (Section~\ref{sec:result_rq1}).} This task often serves as a strong baseline for evaluating the quality of learned representations from encoder-only models~\cite{yang2021learning, lecongetal2025llms}.

\textbf{Code Summarisation.}
The code-to-text tasks of summarisation assess a model's capacity for abstraction and explanation. The models need to generate a concise and accurate natural language description (e.g., a docstring or code comments) for a function. In this task, a model must not only understand what the code does but also be able to synthesise that understanding into a human-readable format. This task is dominated by encoder-decoder models like CodeT5~\cite{wang2021codet5} and PLBART~\cite{ahmad2021unified}, as it requires sequence generation of tokens.

By evaluating our approach across these representative tasks, where each of them targets a different aspect of comprehension, we provide an empirical assessment of the impact of historical and structural context. Strong performance across these distinct tasks provides more substantial evidence that a model has gained a generalisable and deeper understanding of the source code, rather than simply overfitting to a single task. We hypothesise that by enriching representations with the types of context explored in Section~\ref{sec:context_in_models}, we can provide models with a better understanding, leading to improved performance on these comprehension tasks and highlighting a promising direction for future research.

\section{Research Methodology}

\subsection{Research Questions}

\begin{figure*}[h]
  \centering
  \includegraphics[width=\linewidth]{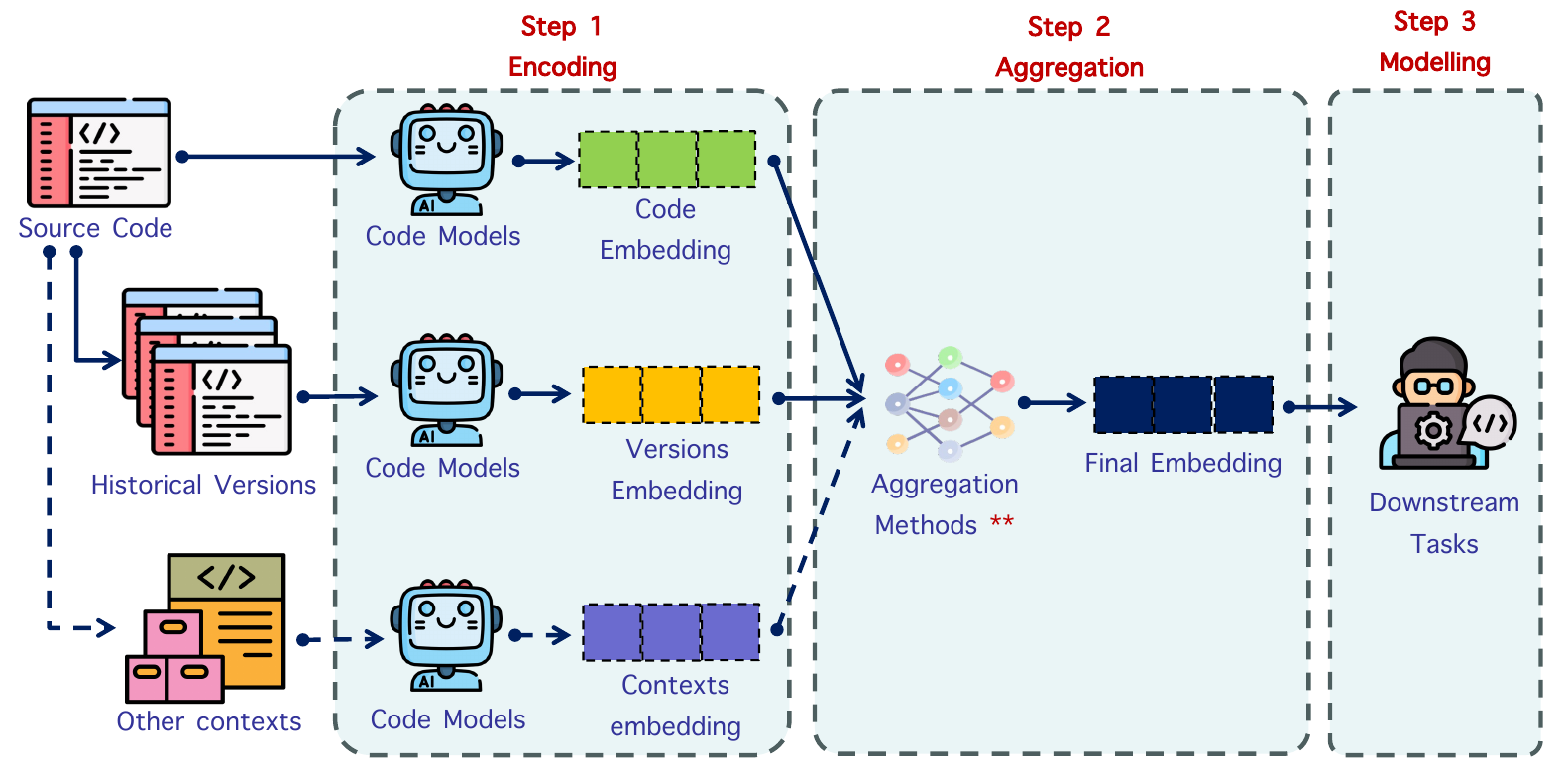}
  \captionsetup{justification=centering}
  \caption[Research Framework.]{Research Framework.\\
    {\footnotesize\textit{Icons designed by Magnific, Vectors Tank, and Afian Rochmah Afif from Flaticon (flaticon.com).}}}
  \label{fig:fig_framework}
  \Description{A three-step research framework showing the encoding of source code and additional context, the aggregation of their embeddings, and the final modelling for downstream tasks.}
  \begin{threeparttable}
     \begin{tablenotes}
       \item[**] \footnotesize The aggregation techniques are detailed in Section~\ref{sec:encoding_aggregation}.
     \end{tablenotes}
  \end{threeparttable}
\end{figure*}

Figure~\ref{fig:fig_framework} illustrates our approach to evaluate the feasibility of incorporating additional context, i.e., version history, into source code representations and its impact on the performance of software engineering tasks. Our study consists of three steps: (1) encoding contextual information into an embedding representation that allows it to be combined with source code representation, (2) exploring effective aggregation techniques to incorporate multiple embeddings into the final vector representation of a method, and (3) assessing downstream performance using representative deep learning models. We also investigate the benefits of combining more than one context, i.e. version history plus call graph.

Lastly, we explore combinations of multiple types of context and their effectiveness.

\paragraph{Task selection rationale.}
We evaluate three canonical program-comprehension tasks — code clone detection, code classification, and code summarisation — which represent the standard evaluation protocol for encoder-based code representation learning~\cite{lu2021codexglue,guo2022unixcoder,wang2021codet5}. These tasks are deliberately chosen because they provide clean, well-labelled supervision signals at the method level, enabling us to isolate the contribution of contextual encoding from confounding factors such as retrieval quality, prompt engineering, or multi-step planning that arise in more complex pipeline settings. Method-level evaluation remains the primary testbed for representation learning in the SE research community, with recent work at top venues (ICSE, FSE, ASE, TOSEM) continuing to use these benchmarks to evaluate code intelligence models~\cite{wan2022they,feng2020codebert,guo2020graphcodebert}. As a first step toward generalisation beyond program comprehension, we additionally conduct a prompt-based vulnerability detection experiment on the Vul4J dataset (Section~\ref{sec:vuln_generalisation}), providing initial evidence that context-enriched representations transfer to a security-oriented downstream task.

To evaluate the benefits of adding version history to source code representation, we set out the following three research questions.

\begin{enumerate}[label=\textbf{RQ\arabic*},leftmargin=*]

    \item \textbf{What is the impact of explicitly encoding additional context into learned vector representations on the performance of task-specific deep learning models?} \\
    This research question aims to empirically validate our hypothesis that combining additional context into code representations enhances model performance. In our experiments, we compare the performance of models under two scenarios: one that uses only the source code and another that encodes additional context (i.e., by integrating version history or call graph into the code representation). By evaluating downstream tasks (e.g., code clone detection and code summarisation) under these two conditions, we expect to demonstrate measurable improvements that justify the inclusion of encoding contextual information as a fundamental enhancement to code representation.

    \item \textbf{How does combining multiple types of encoded context into source code representations affect the performance of these models?} \\
    The second research question explores the feasibility and benefits of integrating multiple context sources simultaneously, rather than relying on a single context type. We aim to determine whether the combined richness of various contextual cues (e.g., version history and call graphs) results in further performance improvement, which may lead to a different conclusion than using a single context. We expect that combining various types of additional context will offer complementary advantages that enhance the overall model’s code comprehension capability.

    \item \textbf{What is the impact of different representation-level aggregation techniques on integrating source code with additional context for these models?} \\
    This question examines how various methods of combining additional context with the final code representation impact the downstream performance of deep learning models. By evaluating and comparing the results from different aggregation methods, including concatenation, max pooling, and concatenation of absolute differences, we aim to identify the most effective approach in enhancing the deep learning model's performance. The expectation is that some aggregation techniques might enable better preservation of meaningful patterns and relationships, leading to improved model performance in software engineering tasks.

\end{enumerate}

In general, these research questions aim to explore different aspects of encoding contextual data into code representation regarding the diversity of context types or methods used to combine them. In that way, the study proposes a foundational approach to utilising additional context to enhance neural-based code comprehension and generation in software engineering.

\subsection{Dataset Construction.}
\label{sec:dataset_construction}

To empirically evaluate the impact of additional context on neural code models, we constructed two large-scale, context-augmented datasets. The process began with two well-established, public datasets, which served as the foundation for our three program comprehension tasks. A third dataset, Vul4J, is introduced in Section~\ref{sec:vuln_generalisation} for an exploratory vulnerability detection evaluation. We then developed and applied a unified data mining pipeline to enrich these datasets with specific evolutionary and structural context. This section details the original datasets, our context augmentation process, and the final statistics of the resulting augmented datasets.

\subsubsection{Original Datasets for Program Comprehension Tasks}
\label{sec:original_datasets}

Our study utilises two primary datasets, each selected for its suitability for specific downstream tasks. A third dataset (Vul4J) is used for the exploratory vulnerability detection evaluation in Section~\ref{sec:vuln_generalisation}.

Firstly, we use \textbf{SeSaMe}~\cite{kamp2019sesame, nguyen2024encoding}, a curated dataset of semantically similar Java methods gathered from 11 well-known open-source projects. Its focus on semantic similarity makes it an ideal foundation for our \textit{Code Clone Detection} and \textit{Code Classification} tasks, which require a nuanced understanding of functional equivalence rather than just syntactic similarity.

Second, for the \textit{Code Summarisation} task, we use the \textbf{CodeSearchNet} corpus~\cite{husain2019codesearchnet, lu2021codexglue}. While CodeSearchNet is a large-scale multilingual benchmark, we specifically select its Java language partition for this study. This choice was driven by two practical requirements for our context augmentation process: the Java dataset provides reliable linkage to the original GitHub repositories, which is essential for mining version history, and is supported by available static analysis tools for extracting call hierarchy information. Its rich pairing of functions with high-quality docstrings provides the necessary data for training and evaluating models on the task of generating human-readable summaries from source code.

Finally, to probe whether context-enriched representations generalise beyond the three primary program-comprehension tasks, we incorporate a third benchmark ---
\textbf{Vul4J}~\cite{bui2022vul4j} --- exclusively for an exploratory zero-shot vulnerability detection experiment (Section~\ref{sec:vuln_generalisation}). Vul4J
is a curated collection of confirmed, CVE-linked Java vulnerabilities sourced from real-world open-source projects. Unlike SeSaMe and CodeSearchNet, Vul4J is not part
of the main context-augmented training pipeline; the full dataset construction process, method-level conversion procedure, and augmentation statistics are detailed in Section~\ref{sec:vuln_generalisation}.

\subsubsection{Context Mining Process}
\label{sec:context_augmentation}

To enrich the original datasets, we designed and executed a data mining pipeline to extract two key types of additional context: evolutionary and structural.

\textbf{Mining Evolutionary Context.} To gather the version history, we utilised \textit{PyDriller}~\cite{spadini2018pydriller}, a framework for analysing Git repositories. For each method in our original datasets, we traversed its full commit history to identify all previous versions. To ensure relevance, we used the static analysis tool \textit{Lizard} to parse the methods in each historical file and retained only those versions where the source code of a target method was actually modified. This initial mining step was highly successful, retaining 175,240 methods, or approximately 97\% of the original CodeSearchNet dataset. From this complete version history, we also derived the \textit{method age} (referred to as "number of days that the method existed"), providing a simple measure of its lifespan and stability.

\textbf{Mining Structural Context.} Capturing the call hierarchy is a more complex process as it requires a fully compilable project state. Our multi-step pipeline was as follows: (1) For each project, we first checked out the specific commit hash provided in the original dataset. (2) We then attempted to build the project into `.jar` files using its configured build system (\texttt{Maven} or \texttt{Gradle}). This step faced significant challenges, with many projects failing to compile due to issues such as missing dependencies, Java version incompatibilities, or broken build scripts. (3) For the projects that were built successfully, we ran \textit{java-callgraph}\footnote{\url{https://github.com/gousiosg/java-callgraph}}~\cite{tian2022adding} to generate the call hierarchy for our target methods. (4) We then filtered this data, removing methods that had no identifiable callers or callees within the project, as they provided no inter-procedural context. Finally, (5) we extracted the source code of the remaining \texttt{callers} and \texttt{callees} and mapped them to the corresponding methods in our dataset. This rigorous process ultimately provided call graph information for 63,259 methods, representing a final retention rate of approximately 35\% of the original dataset.

\subsubsection{Final Context-augmented Datasets and Statistics}
\label{sec:final_datasets}

This augmentation process resulted in two new, context-rich datasets. The descriptive statistics for these datasets are presented below, broken down by their standard training, validation, and testing partitions.

\textbf{Augmented SeSaMe Dataset.} \revised{The original SeSaMe dataset contains 1,679 unique Java methods. Not all of these could be augmented: some originating repositories were no longer available at mining time, and others could not be compiled from their pinned commit, which is a prerequisite for call-graph extraction. After applying the mining pipeline and retaining only methods for which the required context could be recovered, 1,587 distinct methods remain, forming the 847 annotated clone pairs that constitute our experimental splits.} As shown in Table~\ref{tab:sesame_stats}, the augmentation process yielded 9,924 historical code versions for \revised{these methods}. The data reveals a significant diversity in evolutionary history; while some methods have only a single version, others, particularly in projects like \textit{checkstyle}, have many. The method lifetimes are also highly variable, ranging from 34 days to over 17 years (6,334 days), providing a rich source of temporal information. \revised{Note that the per-partition \textit{Unique Methods} counts in Table~\ref{tab:sesame_stats} sum to 1,630 rather than 1,587, because a method may participate in clone pairs assigned to more than one partition; the clone pairs themselves are disjoint across partitions.}

\begin{table}[ht]
    \centering

    \caption{SeSaMe - Dataset Partitions and Additional Context Statistics}
    \label{tab:sesame_stats}

    \resizebox{\textwidth}{!}{%
    \begin{threeparttable}

    \begin{tabular}{l rr rr ccc ccc cc}
    \toprule
    & \multicolumn{2}{c}{\textbf{Dataset Size}} & \multicolumn{2}{c}{\textbf{Source Statistics}} & \multicolumn{3}{c}{\textbf{Version History}} & \multicolumn{3}{c}{\textbf{Call Graph}} & \multicolumn{2}{c}{\textbf{Method Age (days)}} \\
    \cmidrule(lr){2-3} \cmidrule(lr){4-5} \cmidrule(lr){6-8} \cmidrule(lr){9-11} \cmidrule(lr){12-13}

    \textbf{Partition}
    & \textbf{Clone Pairs}\tnote{a}
    & \textbf{Unique Methods}
    & \textbf{Projects}
    & \textbf{Files}
    & \textbf{Total} & \textbf{Max} & \textbf{Min}
    & \textbf{Total} & \textbf{Max} & \textbf{Min}
    & \textbf{Max} & \textbf{Min} \\
    \midrule

    \textbf{Train}
    & \revised{677}
    & \revised{1,296}
    & 11
    & 932
    & 8,074 & 727 & 1
    & 8,104 & 383 & 1
    & 6,334 & 34 \\

    \textbf{Valid}
    & \revised{84}
    & \revised{164}
    & 11
    & 151
    & 855 & 245 & 1
    & 1,341 & 205 & 1
    & 6,334 & 34 \\

    \textbf{Test}
    & \revised{86}
    & \revised{170}
    & 11
    & 156
    & 995 & 243 & 1
    & 863 & 166 & 1
    & 6,334 & 125 \\

    \bottomrule
    \end{tabular}%

    \begin{tablenotes}
        \small
        \item[a] Each clone pair consists of two code methods and a label indicating if they are clones.
    \end{tablenotes}

    \end{threeparttable}
    }
\end{table}

\textbf{Augmented CodeSearchNet Dataset.} Table~\ref{tab:csn_stats} details the statistics for the augmented CodeSearchNet dataset, which is significantly larger in scale. Our process successfully mined context for 63,259 unique Java methods across 2,117 projects, resulting in a total of 193,766 historical versions and 333,573 call graph entries across all partitions. The training set contains 179,471 historical versions, with some methods having up to 131 distinct historical versions. This large-scale dataset provides a robust foundation for evaluating our context-aware models on the code summarisation task.

\begin{table}[ht]
    \centering
    \caption{CodeSearchNet - Dataset Partitions and Additional Context Statistics}
    \label{tab:csn_stats}
    \resizebox{\textwidth}{!}{%
    \begin{threeparttable}
    \begin{tabular}{lcccccccccccc}
    \toprule
    & & & & \multicolumn{3}{c}{\textbf{Version History}} & \multicolumn{3}{c}{\textbf{Call Graph}} & \multicolumn{2}{c}{\textbf{Method Age (days)}} \\
    \cmidrule(lr){5-7} \cmidrule(lr){8-10} \cmidrule(lr){11-12}
    \textbf{Partition}
    & \textbf{Proj.}
    & \textbf{File}
    & \textbf{Method}
    & \textbf{Total} & \textbf{Max} & \textbf{Min}
    & \textbf{Total} & \textbf{Max} & \textbf{Min}
    & \textbf{Max} & \textbf{Min}\tnote{a} \\
    \midrule
    \textbf{Train}
    & 1,905 & 22,190 & 58,339
    & 179,471 & 131 & 1
    & 309,546 & 2,622 & 1
    & 7,368 & 0 \\
    \textbf{Valid}
    & 97 & 528 & 1,271
    & 3,918 & 29 & 1
    & 6,882 & 382 & 1
    & 4,742 & 0 \\
    \textbf{Test}
    & 115 & 1,412 & 3,649
    & 10,377 & 45 & 1
    & 17,145 & 1,189 & 1
    & 4,143 & 0 \\
    \bottomrule
    \end{tabular}%

    \begin{tablenotes}
        \item[a] \footnotesize A method age of 0 days indicates that the version sampled in the dataset corresponds to the method's initial creation commit.
    \end{tablenotes}
    \end{threeparttable}
    }
\end{table}

These newly constructed datasets, enriched with detailed evolutionary and structural information, i.e. version history and call graph, form the empirical backbone of our study, enabling a rigorous evaluation of how additional context impacts a model's program comprehension capabilities.

\subsection{Prompt-based Context Injection Protocols for LLM Evaluation}
\label{sec:llm_baselines}

To ensure a rigorous and fair comparison between our representation-level encoding and prompt-based context injection, we developed standardized prompting protocols tailored to the architectural requirements of each LLM family. These protocols ensure that the exact same contextual artifacts (Source Code, Version History, Call Graphs, and Method Age) used in our representation-level experiments are provided as text-based input for the prompt-based baselines. We provide the full templates below to ensure reproducibility.

\subsubsection{CodeLlama (Instruction Format)}
For the CodeLlama family (7B, 13B, 34B), we utilize the native \texttt{[INST]} and \texttt{<<SYS>>} instruction tags for prompt-based context injection. To ensure stable output formatting on binary classification tasks, we include one demonstration example in the prompt.

\begin{tcolorbox}[colback=gray!5,colframe=gray!50,title=CodeLlama Clone Detection Prompt]
\begin{small}
\begin{verbatim}
<s>[INST] <<SYS>>
You are an expert software engineering assistant specializing in code analysis. 
Your task is to analyze two Java methods (Method 1 and Method 2) and their 
surrounding context to determine if they are code clones (functionally 
identical or highly similar).

You may be provided with the following Context sections:
- '### Method 1 Source Code' and '### Method 2 Source Code'
- '### Method 1 Version History' and '### Method 2 Version History'
- '### Method 1 Caller/Callee Context'
- '### Method 1 Age' (in days)

Output exactly 'Yes' or 'No'. Do NOT provide explanations.
<</SYS>>

Example 1 (Clone):
### Method 1 Source Code
public int add(int a, int b) { return a + b; }
====================
### Method 2 Source Code
public int sum(int x, int y) { return x + y; }
Is Clone: Yes

Now analyze the following data:
{pair_data}

Are Method 1 and Method 2 code clones? Answer strictly with 'Yes' or 'No'.
### Is Clone: [/INST]
\end{verbatim}
\end{small}
\end{tcolorbox}

\subsubsection{Qwen2.5-Coder (ChatML Format)}
For the Qwen2.5-Coder family (1.5B, 7B, 14B), we utilize the \textbf{ChatML} format. This structure provides explicit boundaries for the system role, reducing the risk of the model from conflating the developer's instructions with the Java source code being analyzed.

\begin{tcolorbox}[colback=gray!5,colframe=gray!50,title=Qwen2.5-Coder Summarization Prompt]
\begin{small}
\begin{verbatim}
<|im_start|>system
You are an expert technical documentation assistant. Your task is to analyze 
Java code and its surrounding context to understand its exact purpose.
Evaluate the Source Code, Version History, and Call Context provided.<|im_end|>
<|im_start|>user
Here is the data:

{method_data}

Based on the information above, Write a SINGLE-sentence summary for the 
primary Java method.
Constraints:
1. Start directly with an action verb (e.g., 'Validates...', 'Parses...').
2. Do NOT start with 'The method...' or 'This code...'.

### Summary:<|im_end|>
<|im_start|>assistant
\end{verbatim}
\end{small}
\end{tcolorbox}

\subsubsection{DeepSeek-Coder (Instructional Format)}
DeepSeek-Coder-6.7B utilizes a distinct "Instruction/Response" structure. Given DeepSeek-Coder's strong pre-training on code, we found that a minimalist prompt structure worked best for project-level classification.

\begin{tcolorbox}[colback=gray!5,colframe=gray!50,title=DeepSeek-Coder Classification Prompt]
\begin{small}
\begin{verbatim}
You are an expert software engineering assistant specializing in code 
classification. Your task is to analyze a single Java method and its 
surrounding context to predict which open-source project it belongs to.

### Instruction:
Here is the data for the method:

{method_data}

Based on the information above, predict the exact project name from the 
allowed list: [openjdk11, deeplearning4j, eclipse.jdt.core, guava, 
commons-math, freemind, commons-collections, caffeine, checkstyle, 
commons-lang, trove].
Do not output any other text, explanation, or markdown formatting.

### Project Name:
### Response:
\end{verbatim}
\end{small}
\end{tcolorbox}

\subsubsection{On the Relationship Between Prompt-Level Aggregation and Representation-Level Fusion}

A key distinction motivates our dual evaluation of encoder models and LLMs: \emph{prompt-level context concatenation} and \emph{representation-level vector fusion} address the same underlying problem — how to combine multiple heterogeneous context sources — but through fundamentally different mechanisms. Larger context windows in modern LLMs solve the \emph{token length} problem (how much text a model can physically read), but they do not resolve the \emph{information fusion} problem (how to optimally weight and combine version history, call graphs, and method age into a useful representation). In our encoder-based framework, aggregation strategies (Concatenation, Max-Pooling, Diff-Concat) are \emph{learned} during fine-tuning, allowing the model to discover which fusion strategy produces the most discriminative combined representation for each task. By contrast, prompt-level concatenation relies on implicit self-attention to sort out relative importance across context sources — with no task-specific supervision on the fusion itself.

Our LLM experiments provide direct empirical evidence that this fusion problem does not disappear with larger context windows. Across our eight context orderings and combinations, we observe substantial sensitivity to the \emph{order} in which context sources appear in the prompt — even though all models have context windows far exceeding 512 tokens and receive the full context in every condition. Representative ordering effects on F1 include: CodeLlama-7B fine-tuned on Clone Detection (\texttt{code\_cg\_vh}: 0.5614 vs \texttt{code\_vh\_cg}: 0.7368, a 17.5-point gap); Qwen-1.5B zero-shot on Clone Detection (\texttt{code\_cg\_vh}: 0.0625 vs \texttt{code\_vh\_cg}: 0.2000, a 13.75-point gap); and Qwen-7B on Vulnerability Detection (\texttt{code\_cg\_vh}: 0.1818 vs \texttt{code\_vh\_cg}: 0.0526, a 12.9-point gap). These results confirm that ordering context in a prompt is a related context organisation problem with measurable consequences — and our representation-level aggregation study provides a principled, learnable counterpart to this implicit, order-sensitive process. \revised{Two variables are involved when LLM and encoder results are compared: whether the model is fine-tuned on the task, and how context is fused (prompt-level text concatenation or representation-level vector fusion). Our design separates them as follows. Each LLM is evaluated in a zero-shot mode and a fine-tuned mode; in the fine-tuned mode the model is trained on inputs that contain the contextual artifacts in the prompt, so the fine-tuned rows with context in Tables~\ref{tab:result_rq1_clone_class}, \ref{tab:result_rq1_summarisation}, \ref{tab:result_rq2_clone_detection}, \ref{tab:result_rq2_classification}, and \ref{tab:result_rq2_summarisation} are fine-tuned LLMs with prompt-level context fusion. Comparing the two modes within an LLM isolates the effect of fine-tuning. Comparing fine-tuned LLMs against fine-tuned encoders holds fine-tuning constant and varies the fusion mechanism, but also varies the backbone architecture, so that comparison is confounded and we interpret it with care (Section~\ref{sec:findings-implications}).}

To further validate that the performance gains attributable to additional context are driven by semantic content rather than by the enlarged representation capacity introduced by vector-level concatenation, we conduct a fixed-dimension projection control experiment described in Section~\ref{sec:projection_control}. This control applies only to encoder-only models (CodeBERT, GraphCodeBERT), where representation-level vector fusion occurs explicitly. For encoder-decoder models (CodeT5, PLBART), additional context is prepended to the input token sequence before encoding; the encoder processes all tokens jointly and produces a single fixed-size representation regardless of context length, so no capacity expansion arises.

We clarify three aspects of how historical versions are handled in our version history encoding. First, not all historical commits represent meaningful changes to a method. We apply Lizard static analysis to filter version history: a historical version is included only if the method's token count changed relative to the current version, indicating a substantive modification. Commits touching only comments, whitespace, or imports without altering the method body are excluded. Second, historical versions are concatenated in most-recent-first order, so that the version immediately preceding the current state appears first within the 512-token encoder budget; the oldest versions are naturally dropped when the budget is exceeded. Across the SeSaMe dataset, 70--77\% of methods have version histories that fit within the token budget without truncation. Third, to investigate whether performance gains depend on access to many historical versions, we conduct a history depth ablation study described in Section~\ref{sec:vh_depth_ablation}.

\subsection{Evaluation Metrics.}

In this study, we select well-established metrics tailored to the nature of the downstream tasks to evaluate our approach. For Code Clone Detection and Code Classification tasks, we measure the classification performance with Precision, Recall, and F1-score.
In Code Summarisation, we recruit four widely-adopted automated metrics, including BLEU, ROUGE-L, BERTscore, and METEOR, which provide complementary views on the quality of the generated summaries.

\subsubsection{Classification Metrics:}

\paragraph{Precision, Recall, and F1-score:} These are well-established metrics measuring the performance of classification models. Precision (P) quantifies the proportion of correct positive predictions against all positive predictions made, and Recall (R) measures the ratio of actual positive instances that were correctly recognised. Meanwhile,   F1-score is particularly important for classification problems, e.g. code clone detection, where it minimises problems in imbalanced datasets by balancing precision and recall, making it a more robust metric than accuracy. F1-score ranges from 0 to 1, where 1 represents the best possible balance.

\paragraph{Macro-F1 Score:} This metric is specifically recruited in the Code Classification task to capture overall performance across multiple labels. The macro-F1 score (also ranging from 0 to 1) mitigates the bias from dominant classes by treating them all equally important; thus, it offers a more comprehensive measure of the model’s effectiveness across diverse categories.

\subsubsection{Generation Metrics:}

\paragraph{BLEU-4} (BilinguaL Evaluation Understudy)~\cite{papineni2002bleu} is a precision-based metric that measures the similarity between the generated summaries and the reference summaries by comparing n-gram matches. Thus, BLEU-4 examines up to 4-word sequences, which is suitable for evaluating code summarisation or machine translation tasks according to their implementation and explanability.

\paragraph{ROUGE-L}~\cite{lin2004rouge} is an evaluation metrics that assess text summarisation quality by focusing on the longest common subsequence between the generated summary and the reference. It's appropriate for code summarisation because it examines the structural similarity of texts, which is essential for capturing the order and meaning of code's logic and intention.

\paragraph{BERTscore} leveraging contextual embeddings from BERT, BERTscore~\cite{zhang2019bertscore} is an automated measure that evaluates semantic similarity between a generated text and a reference text. BERTScore-F1, a BERTScore variant combining precision and recall values, captures nuanced differences between the texts and is ideal for a code summarisation task where synonyms and paraphrases are common.

\paragraph{METEOR} (Metric for Evaluation of Translation with Explicit ORdering)~\cite{banerjee2005meteor} also incorporates precision and recall. This metric is particularly well-suited for code summarisation because it focuses on synonyms and stem matches, making it more robust to word choice and order variations. This implementation helps to recognise that different summaries contain similar meaning using different terms or words.

Therefore, these selected metrics provide a diverse evaluation of our approach by measuring the syntactic and semantic quality of the generated code summaries and the effectiveness of the models in classification and generation tasks.

\subsection{Human Evaluation Process}
\label{sec:human-eval}

To obtain a deeper understanding of the actual benefits of explicitly encoding additional context into deep learning models, we conducted a structured human evaluation. This manual assessment allows us to verify whether augmented models truly produce summaries that are more \emph{accurate}, \emph{adequate}, and \emph{concise} than those generated by models trained only on source code.

We pursued human judgment for two reasons. First, while automated metrics (e.g., BLEU-4, METEOR, ROUGE-L, BERTScore) are standard in code summarisation, they emphasise surface-level lexical overlap and may under-represent semantic correctness~\cite{mastropaolo2024evaluating, chen2021my, gao2023evaluating}, information adequacy, or concision. Second, recent publication~\cite{shi2022we} analyses have reported substantial noise in the CodeSearchNet dataset (e.g., HTML artefacts, partial or non-literal comments), which can distort reference-based evaluation; thus, human assessment provides a necessary reliability check on model quality.

\paragraph{Material.}
We randomly sampled 100 Java code snippets from the CodeSearchNet test split, a sample size consistent with prior work on human evaluation of code summarisation~\cite{sun2024extractive, ding2024code, gao2023code, liu2025too}. This design strikes a balance between feasibility and reliability, ensuring that the annotation process is manageable while still providing statistically meaningful insights. For each snippet, we evaluated \textbf{nine} summaries: seven produced by context-augmented CodeT5 variants, one produced by a CodeT5 baseline using source code only, and the human-written reference from the dataset. \revised{The study therefore evaluates CodeT5 and its context-augmented variants only; summaries generated by the seven LLMs in Section~\ref{sec:llm_baselines} were not included, given the annotation budget of the study, and the LLM summarisation results in Section~\ref{sec:result_rq1} rest on automatic metrics alone.} We focused on three quality dimensions commonly used in human studies: \textbf{Accuracy}, \textbf{Content Adequacy}, and \textbf{Conciseness}. Furthermore, the reliability of our evaluation is reinforced by a structured annotation protocol, including pilot phases and inter-rater agreement checks, which provide additional confidence in the robustness of the findings. In total, this evaluation required \(100 \times 9 \times 3 = \textbf{2{,}700}\) annotations.

\paragraph{Annotators and roles.}

The evaluation involved four annotators across two phases. \textit{Two primary annotators} conducted the main 100-snippet study: (i) the first author, with more than 12 years of professional experience in Java programming and Software Engineering research (University of Melbourne); and (ii) a graduate research engineer at Singapore Management University (SMU), with approximately 4 years of SE/Java experience, recruited through a referral process and interviewed specifically on Java programming expertise prior to participation. Eligibility of the second annotator was further verified through the pilot phase: both primary annotators independently labelled 10 snippets, and their results were reviewed by the two supervisors before the main study proceeded. \textit{Two senior researchers} participated as calibration reviewers in the pilot phase: (iii) the first author's PhD supervisor, an Associate Professor at the University of Melbourne with more than 15 years of SE research experience; and (iv) a co-supervisor, an Associate Professor at SMU with more than 20 years of Software Engineering research experience. All four annotators worked independently with sources blinded (model identity hidden during ranking).

\paragraph{Task and interface.}
For each code snippet and dimension, annotators were presented with the nine summaries in random order and asked to \emph{rank} them (allowing ties) from best to worst for that specific dimension. We provided written guidelines and positive/negative examples for each criterion to promote consistency.

\paragraph{Pilot phase and agreement.}
To calibrate our annotation process, all four annotators first independently labelled a pilot set of \(10\) snippets (\(10 \times 9 \times 3 = \textbf{270}\) annotations). For this initial phase with four participants, we measured inter-rater agreement using Kendall’s coefficient of concordance (\(W\)), a metric specifically designed to assess group-level agreement among three or more raters. The resulting agreement was strong, with a mean \(W\) of \(\mathbf{0.7261}\) and a median \(W\) of \(\mathbf{0.7540}\). Based on this high initial concordance, we clarified the rubric and resolved ambiguities before proceeding to the main study.

\paragraph{Main study and monitoring.}
Following the pilot, the two primary annotators completed the main study of the remaining \(\textbf{100}\) snippets. As this phase involved only two participants, we measured pairwise agreement using Kendall’s \(\boldsymbol{\tau_b}\) (tau-b). This is not a methodological shift but a direct consequence of the number of raters; when there are only two annotators, Kendall’s \(W\) mathematically reduces to Kendall’s \(\tau\) (up to a scaling constant), making \(\tau_b\) the appropriate pairwise equivalent for ordinal data with ties. The main study was conducted in two rounds to ensure continued consistency. After the first \textbf{Round 1} (\(30\) snippets), agreement was already substantial (mean \(\tau_b = 0.6347\), median \(\tau_b = 0.8636\)). After a final discussion to resolve minor divergences, the annotators completed the \textbf{remaining 70} snippets. Across the full 100-snippet main study, the final pairwise agreement was very high, with a mean \(\tau_b\) of \(\mathbf{0.8899}\) (strong positive agreement) and a median of \(\mathbf{1.0000}\) (perfect agreement).

\paragraph{Outcome.}
The evaluation phase provides a reliable ranking of \emph{generated summaries} per snippet and per dimension (accuracy, content adequacy, conciseness) for the code summarisation task. With sources blinded and rank-ordering with ties format, these human annotation enable us to: (1) quantify how much adding contextual information improves summary quality over models that rely on source code only; (2) determine whether context-augmented models are consistently better than the baseline or only under specific context types and scenarios, and why; and (3) compare single-context versus multi-context inputs to identify when multiple types of context provide additional benefit. We assess statistical significance and effect sizes using Wilcoxon signed-rank tests and Cliff’s~$\delta$, and evaluate inter-rater agreement with Kendall’s~$\tau_b$. Note that in this generative setting, multiple contextual inputs are combined into a single input to the model; hence, representation-level aggregation strategies are not evaluated here.

\subsection{Encoding and Aggregation}
\label{sec:encoding_aggregation}

\paragraph{Code Models.}
Effective \emph{aggregation} of additional context with code representations can amplify useful information or dilute important signals, depending on the task and architecture. To understand which aggregation operators work best, our evaluation is shaped by several constraints, including the input format required by each task (e.g., single vs. paired inputs), model-specific limitations like maximum token length, and the need for representation-shape compatibility across different model families. We therefore evaluate five foundational code models spanning three architectures: \textbf{ASTNN} (tree-based)~\cite{zhang2019novel,tian2022adding}, \textbf{CodeBERT} and \textbf{GraphCodeBERT} (encoder-only Transformers)~\cite{feng2020codebert,guo2020graphcodebert}, and \textbf{CodeT5} and \textbf{PLBART} (encoder--decoder Transformers)~\cite{wang2021codet5,ahmad2021unified}. The evaluation is further extended to seven code-specialised LLMs, described in Section~\ref{sec:llm_baselines}. Across these architectures, we encode four context types---the method's source code, \emph{version history}, \emph{call-graph neighbourhood} (caller/callee), and \emph{method age} (a scalar indicating how long a method has existed). For aggregation, we selected \emph{Concatenation (Concat)} and \emph{Max-Pooling} due to their broad compatibility. We also evaluate \emph{Concatenation of Absolute Differences (Diff-Concat)}, a strategy specifically applicable to the pairwise nature of our clone detection task.

\paragraph{Step 1: Encoding.}
We use each model's native encoder to convert the target method and its additional context into numerical representations. The method's source code produces a single vector (ASTNN, encoder-only Transformers) or a token sequence to the encoder (encoder--decoder models). Version history is encoded as a \emph{set of vectors} by processing each historical version of the method.
For call-graph context, we follow prior work to select and encode the \emph{longest caller} and \emph{longest callee} as two separate vectors to avoid combinatorial explosion~\cite{tian2022adding}. The \emph{longest} selection criterion (by token count) is adopted from Tian and Treude~\cite{tian2022adding} for direct comparability; the rationale is that a longer method contains more tokens and therefore may include more semantic cues about the caller's or callee's purpose. Alternative criteria, such as call frequency, would require dynamic profiling unavailable from static analysis alone. The restriction to one caller and one callee at depth-1 is a deliberate design choice driven by three practical constraints: (i) token budget --- our analysis in Section~\ref{sec:cg_error_analysis} shows that the selected callee alone has a mean of 268 tokens, and encoding multiple callers would push the majority of inputs beyond the 512-token limit; (ii) aggregation consistency --- a fixed number of context vectors ensures a clean, reproducible comparison of aggregation strategies across all five model families; and (iii) architectural interpretability --- text-based depth-1 encoding avoids introducing graph neural network components that would confound the comparison between context types and aggregation strategies. Encoding multi-hop neighbourhoods, class-level context, or module-level context is a natural extension that we identify as future work (Section~\ref{sec:limitation}).
The \emph{method age} feature is represented as a single normalised scalar (days since the first commit), derived from the underlying version-control metadata without a learned projection. After encoding, we have up to five inputs: the current-method vector, a set of history vectors, caller and callee vectors, and the method age scalar.

\textbf{Justification and encoding details for method age.}
The inclusion of method age is grounded in the hypothesis that a method's \emph{temporal maturity} is a genuinely informative prior for program comprehension. This hypothesis rests on three foundations.

\textit{(i) Established empirical evidence.} Prior SE research has shown that temporal features, including file age, are among the strongest predictors of fault-proneness and code stability~\cite{ruthruff2008predicting}, and that evolutionary aggregates from version control systems correlate with maintenance effort and code quality~\cite{thongtanunam2013mining}. These results establish that how long a method has existed is a proxy for the number of revision cycles it has survived, the degree to which its interface has been stress-tested, and the maturity of its semantics.

\textit{(ii) Task-specific reasoning.} For \emph{code clone detection}, two methods of similar age in the same codebase are likely to share not only structural similarity but an evolutionary trajectory, reinforcing the clone signal along a temporal dimension that static code structure alone cannot capture. For \emph{code classification}, a method's category tends to crystallise over time: mature methods, having survived multiple review and refactoring cycles, more cleanly represent canonical design patterns, whereas newly introduced methods may still be in transition. For \emph{code summarisation}, mature methods tend to have settled, stable semantics that are easier to describe accurately; method age provides a lightweight prior about this semantic settledness.

\textit{(iii) Complementarity with other context types.} The three context types operate on categorically different dimensions of code knowledge: version history captures \emph{what changed} (sequential semantic evolution); call graphs capture \emph{who calls and is called} (structural role); method age captures \emph{how long the method has existed} (temporal maturity and stability). These are complementary signals, \revised{and our intuition for pairing method age with version history in particular is as follows. Version history supplies the content of a method's earlier versions, but not the span of time over which those versions accumulated: the same three prior versions may represent three weeks of rapid iteration on a new method or three years of occasional maintenance on a settled one. Method age, computed from the first commit found in the same mining pass (Section~\ref{sec:context_augmentation}), supplies that span as a single scalar. We hypothesise that it helps a model weight the historical versions it is given. For an old method, the current version is likely to be the stable end point of a long evolution, and differences from earlier versions are more likely to be refinements than reversals; for a young method, the current version may still be in flux, and earlier versions may carry token patterns that the final form has not yet settled on. Age also indicates how much of a method's history the version-history context represents: in CodeSearchNet, methods younger than three months have 1.2 prior versions on average and methods older than ten years have 3.8. In this sense method age is a complementary signal rather than an independent one. It is derived from the same source as version history and is meaningful in relation to it, and for that reason we evaluate it only in combination with version history. The largest gains in configurations that include method age arise when it is combined with the other context types (e.g., PLBART +15.04\% F1 in Code Clone Detection with Version History and Method Age; CodeT5 +21.48\% macro-F1 in Code Classification with all three context types). We stress that the intuition above is a hypothesis: the present design does not isolate the contribution of method age, and a method-age-only condition is listed as an untested ablation in Section~\ref{sec:limitation}.}

\begin{figure*}[h]
  \centering
  \includegraphics[width=\linewidth]{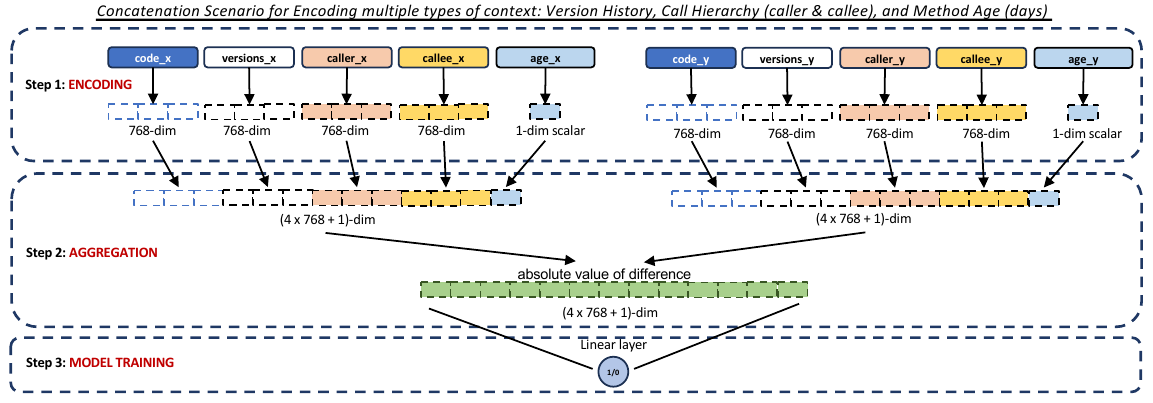}
  \caption{Concatenation Scenario.}
  \label{fig:fig_concat}
  \Description{Concatenation.}
\end{figure*}

\begin{figure*}[h]
  \centering
  \includegraphics[width=\linewidth]{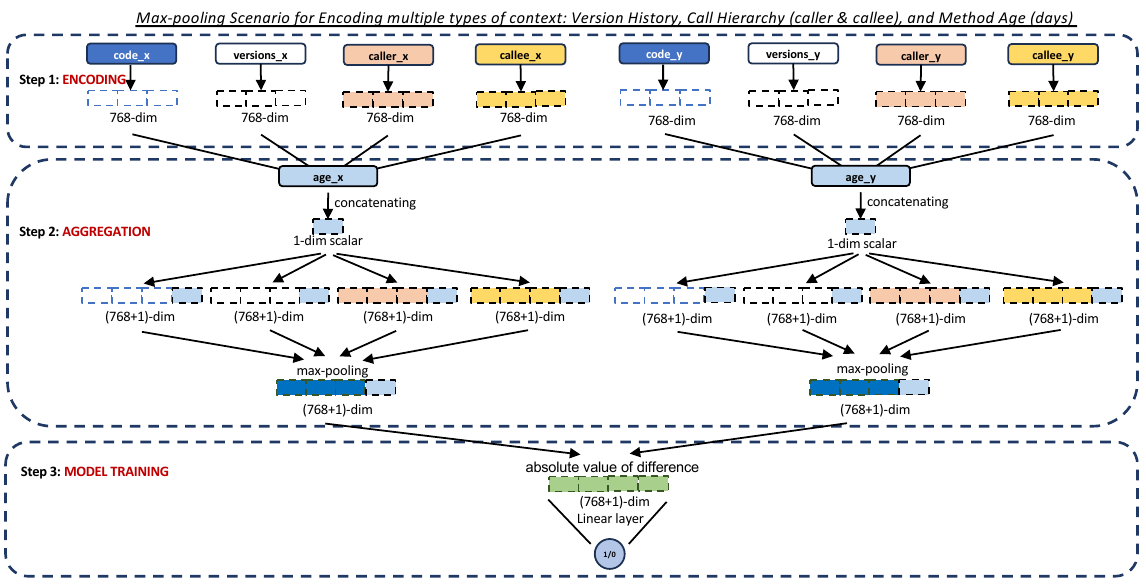}
  \caption{Max-Pooling Scenario.}
  \label{fig:fig_maxpool}
  \Description{Maxpool.}
\end{figure*}

\begin{figure*}[h]
  \centering
  \includegraphics[width=\linewidth]{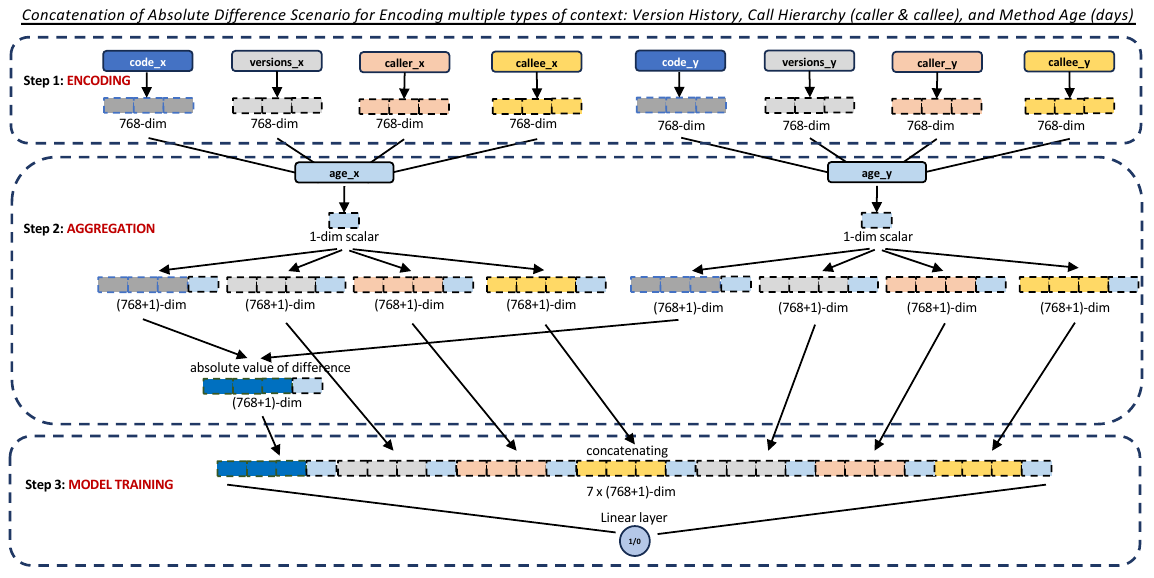}
  \caption{Concatenation of Absolute Difference Scenario.}
  \label{fig:fig_diffconcat}
  \Description{Concatenation of Absolute Difference.}
\end{figure*}

\textit{Precise encoding mechanism.} For encoder-based models (ASTNN, CodeBERT, GraphCodeBERT), the age scalar (days since first commit, normalised at the dataset level) is concatenated directly as a \textbf{1-dim feature} with the 768-dim encoder output vectors. No projection layer is applied; the task head's linear layer learns one dedicated weight for this scalar dimension alongside the weights for the 768-dim encoder streams. The exact placement of the scalar within the combined vector depends on the aggregation strategy, as shown in Figures~\ref{fig:fig_concat}, \ref{fig:fig_maxpool}, and~\ref{fig:fig_diffconcat}. In the Concatenation scenario, the scalar is appended at the end of the full concatenated vector, yielding a $(4 \times 768 + 1)$-dim input to the task head. In the Max-Pooling scenario, the scalar is appended to each individual 768-dim stream before pooling, making each stream $(768{+}1)$-dim and preserving the scalar through the element-wise maximum. In the Diff-Concat scenario, the scalar is similarly appended to each stream before the absolute difference is computed, resulting in a $7 \times (768{+}1)$-dim final representation. For encoder--decoder models (CodeT5, PLBART), method age is instead injected as a structured text tag (e.g., \texttt{[METHOD\_AGE: 365 days]}) appended to the input sequence, preserving the standard seq2seq fine-tuning objective without requiring architecture modifications.

\paragraph{Step 2: Aggregation (representation-level).}
We adopt three generic aggregation operators to combine selected vectors into a single representation:
\begin{itemize}
    \item \textbf{Concatenation (Concat):} stack vectors end-to-end (Figure~\ref{fig:fig_concat}).
    \item \textbf{Max-Pooling:} take the element-wise maximum across vectors (Figure~\ref{fig:fig_maxpool}).
    \item \textbf{Concatenation of Absolute Differences (Diff-Concat):} for a pair of methods, first compute the absolute difference between their code vectors, $\,\lvert \mathbf{c}_1 - \mathbf{c}_2 \rvert\,$, then concatenate this difference with context vectors~\cite{tian2022adding, nguyen2024encoding} (Figure~\ref{fig:fig_diffconcat}).
\end{itemize}

\textit{Task applicability.} In \textbf{Code Clone Detection} (pairwise), we use Concat, Max-Pooling, and \textbf{Diff-Concat}, because the task naturally provides two inputs and the absolute-difference signal is informative for similarity. Figures~\ref{fig:fig_concat}, \ref{fig:fig_maxpool}, and \ref{fig:fig_diffconcat} illustrate these three aggregation scenarios for the Code Clone Detection task, using the most complex combination of additional contexts. In \textbf{Code Classification} (single-input) we use Concat and Max-Pooling; Diff-Concat does not apply because there is no second method.

\paragraph{Code Summarisation (sequence generation).}
For encoder--decoder models (CodeT5, PLBART) we do \emph{not} introduce representation-level aggregation modules. Instead, we aggregate \emph{at the text level}: we concatenate the source code with selected context snippets (e.g., version excerpts, caller/callee code, and a structured text tag encoding the method age (e.g., \texttt{[METHOD\_AGE: 365 days]})) into one delimited input sequence that is fed to the model. This preserves the standard code-to-text fine-tuning objective, avoids architecture changes that would confound comparisons, and reflects common practice for injecting auxiliary signals in seq2seq setups.

\paragraph{Step 3: Task-specific training and metrics.}
After aggregation, the resulting representation feeds a linear layer with \emph{sigmoid} for clone detection and a linear layer with \emph{softmax} for code classification; summarisation uses the standard encoder--decoder stack. We report \textbf{F1-score} for Code Clone Detection, \textbf{macro-F1} for Code Classification (multi-class), and \textbf{BLEU-4}, \textbf{ROUGE-L}, \textbf{METEOR}, and \textbf{BERTScore-F1} for Code Summarisation.

\paragraph{Experimental setup.}
\emph{SeSaMe tasks.} For \textbf{Code Clone Detection}, we use the human-annotated clone pairs in SeSaMe; each pair has a binary label derived from confidence-weighted annotations, and we evaluate with F1-score. For \textbf{Code Classification}, we use the original 11 project names as labels and evaluate with \emph{macro-F1} to account for class imbalance in multi-class settings.
\emph{CodeSearchNet task.} For \textbf{Code Summarisation} on the Java subset of CodeSearchNet, we fine-tune encoder--decoder models with text-level concatenation of code and context, and evaluate with the four automatic metrics listed above, complemented by a separate human evaluation (Section~\ref{sec:human-eval}).

\paragraph{Training settings.}
To ensure fair comparisons on SeSaMe, we adopt an \textbf{80/10/10} train/\allowbreak validation/\allowbreak test split, following prior practice~\cite{tian2022adding, nguyen2024encoding}. For CodeSearchNet (Java), we \emph{reuse the official train/\allowbreak validation/\allowbreak test partitions}, consistent with prior code-summarisation work using pre-trained language models for code~\cite{lu2021codexglue}. Hyperparameters follow each model's recommended settings in recent studies~\cite{lu2021codexglue, zeng2022extensive}; we keep optimisation workflow and loss functions fixed across conditions, select checkpoints by \emph{best validation performance}, and vary only the aggregation operator and context selection per experiment.

\paragraph{Rationale.}
Concatenation, Max-Pooling, and Diff-Concat are simple, model-agnostic aggregation strategies that enable controlled ablations of which types of context are included, working uniformly across tree and Transformer encoders. For summarisation, text-level concatenation is the most comparable way to introduce additional context without changing the model's architecture or training dynamics.

\section{Experimental Results}
\label{sec:experimental_results}

This section presents the performance of 12 representative models (comprising five foundational models and seven Large Language Models) enhanced with additional context, including version history, call graphs, and method age, encoded into source code representations. By evaluating across three software engineering tasks (Code Clone Detection, Code Classification, and Code Summarisation), we examine whether the benefits of additional context scale to modern, high-capacity architectures and how they compare to standard prompting paradigms.

\subsection*{RQ1: What is the impact of explicitly encoding additional context into learned vector representations on the performance of task-specific deep learning models?}
\label{sec:result_rq1}

We compare the representative models’ performance between (i) without additional context (baseline), (ii) with version history, and (iii) with call graph using the concatenation aggregation technique.

Table~\ref{tab:result_rq1_clone_class} shows that, in the Code Clone Detection task, concatenating version history \revised{improves or leaves unchanged the F1-score of every model, with no model getting worse}. Notably, CodeT5 effectively learns contextual information from historical versions, achieving the highest F1-score at 0.9048, with an improvement of +15.92\% against the baseline (0.7805). Meanwhile, other models such as ASTNN (+3.24\%), GraphCodeBERT (+4.70\%), and CodeBERT (+3.65\%) show \revised{smaller gains that are close to the resolution of the SeSaMe test set (Section~\ref{sec:threats}); we read them as consistent in direction rather than as reliable improvements on their own}, while PLBART's performance remains unchanged. Furthermore, encoding call graph information delivers mixed results, where CodeBERT and PLBART show improvement at +10.13\% and +4.25\%, respectively. However, it produces \revised{lower F1 for} other models, such as GraphCodeBERT (-7.69\%), CodeT5 (-3.04\%), and ASTNN (-4.88\%).

\revised{For the Code Classification task (right-hand columns of Table~\ref{tab:result_rq1_clone_class}), call graph context improves macro-F1 for four of the five foundational models, most strongly for ASTNN (+25.84\%) and CodeT5 (+16.29\%), whereas version history alone gives mixed results. We note what this result does and does not show. The labels in this task are the 11 SeSaMe source projects (Section~\ref{sec:downstream_tasks}), so a gain means that additional context helps the model recover the architectural and domain family a method belongs to. This is a meaningful comprehension signal, since project identity is not present in the input: the mined context contains method bodies only, with no file paths, package declarations, or project names. It does not show that the model recovers a method's abstract functional purpose independently of its provenance. To rule out the possibility that the gains come from project-correlated surface cues in the appended context rather than from its content, we report a shuffled-context control in Section~\ref{sec:threats} (Table~\ref{tab:shuffled_context_results}).}

\begin{table}[htbp]
\centering
\scriptsize
\caption{Comprehensive performance of foundational models and LLMs (1.5B–34B) on Code Clone Detection and Code Classification. \revised{For LLMs, results are reported in zero-shot (Z) and fine-tuned (F) modes. Columns headed \% give the relative change of each context condition over the code-only baseline (W/O Ctx) of the same model and, for LLMs, the same mode; the table does not compare zero-shot against fine-tuned results.} Accuracy has been omitted for brevity.}
\label{tab:result_rq1_clone_class}
\renewcommand{\arraystretch}{1.2}
\adjustbox{max width=\textwidth, max totalheight=0.85\textheight}{%
\begin{tabular}{l|
c >{\columncolor{baselineCol}}c >{\columncolor{versionCol}}c c c >{\columncolor{callgraphCol}}c c c |
c >{\columncolor{baselineCol}}c >{\columncolor{versionCol}}c c c >{\columncolor{callgraphCol}}c c c
}
\toprule
\rowcolor{headerCol}
\multicolumn{1}{c|}{\textbf{Models}} & \multicolumn{8}{c|}{\textbf{Code Clone Detection}} & \multicolumn{8}{c}{\textbf{Code Classification}} \\
\cmidrule(lr){2-9}\cmidrule(lr){10-17}
\rowcolor{headerCol}
 & \textbf{metric} & \textbf{W/O Ctx} & \textbf{Vers.\ Hist.} & \scriptsize\% & & \textbf{CallGraph} & \scriptsize\% & & \textbf{metric} & \textbf{W/O Ctx} & \textbf{Vers.\ Hist.} & \scriptsize\% & & \textbf{CallGraph} & \scriptsize\% & \\
\midrule

\multirow{3}{*}{\textbf{CodeBERT}}
& P & 0.7778 & 0.7273 & -6.49\% & \barPercent{-6.49} & 0.7727 & -0.65\% & \barPercent{-0.65} & P & 0.7103 & 0.6971 & -1.86\% & \barPercent{-1.86} & 0.7785 & 9.59\% & \barPercent{9.59} \\
& R & 0.6667 & 0.7619 & 14.28\% & \barPercent{14.28} & 0.8095 & 21.43\% & \barPercent{21.43} & R & 0.6498 & 0.6146 & -5.42\% & \barPercent{-5.42} & 0.7114 & 9.48\% & \barPercent{9.48} \\
& F1 & 0.7180 & 0.7442 & 3.65\% & \barPercent{3.65} & 0.7907 & 10.13\% & \barPercent{10.13} & m-F1 & 0.6698 & 0.6320 & -5.65\% & \barPercent{-5.65} & 0.7345 & 9.65\% & \barPercent{9.65} \\
\midrule

\multirow{3}{*}{\textbf{GraphCodeBERT}}
& P & 0.9286 & 0.9333 & 0.51\% & \barPercent{0.51} & 0.8571 & -7.69\% & \barPercent{-7.69} & P & 0.8214 & 0.8169 & -0.55\% & \barPercent{-0.55} & 0.8584 & 4.50\% & \barPercent{4.50} \\
& R & 0.6191 & 0.6667 & 7.69\% & \barPercent{7.69} & 0.5714 & -7.69\% & \barPercent{-7.69} & R & 0.7581 & 0.7547 & -0.46\% & \barPercent{-0.46} & 0.7221 & -4.76\% & \barPercent{-4.76} \\
& F1 & 0.7429 & 0.7778 & 4.70\% & \barPercent{4.70} & 0.6857 & -7.69\% & \barPercent{-7.69} & m-F1 & 0.7661 & 0.7781 & 1.58\% & \barPercent{1.58} & 0.7618 & -0.55\% & \barPercent{-0.55} \\
\midrule

\multirow{3}{*}{\textbf{CodeT5}}
& P & 0.8000 & 0.9048 & 13.10\% & \barPercent{13.10} & 0.8750 & 9.37\% & \barPercent{9.37} & P & 0.7005 & 0.7889 & 12.62\% & \barPercent{12.62} & 0.8537 & 21.86\% & \barPercent{21.86} \\
& R & 0.7619 & 0.9048 & 18.75\% & \barPercent{18.75} & 0.6667 & -12.50\% & \barPercent{-12.50} & R & 0.6364 & 0.6713 & 5.50\% & \barPercent{5.50} & 0.7259 & 14.07\% & \barPercent{14.07} \\
& F1 & 0.7805 & 0.9048 & 15.92\% & \barPercent{15.92} & 0.7568 & -3.04\% & \barPercent{-3.04} & m-F1 & 0.6569 & 0.7094 & 8.00\% & \barPercent{8.00} & 0.7639 & 16.29\% & \barPercent{16.29} \\
\midrule

\multirow{3}{*}{\textbf{PLBART}}
& P & 0.9333 & 0.9333 & 0.00\% & \barPercent{0.00} & 0.9375 & 0.45\% & \barPercent{0.45} & P & 0.7064 & 0.7665 & 8.51\% & \barPercent{8.51} & 0.7718 & 9.26\% & \barPercent{9.26} \\
& R & 0.6667 & 0.6667 & 0.00\% & \barPercent{0.00} & 0.7143 & 7.14\% & \barPercent{7.14} & R & 0.6837 & 0.6840 & 0.04\% & \barPercent{0.04} & 0.7683 & 12.38\% & \barPercent{12.38} \\
& F1 & 0.7778 & 0.7778 & 0.00\% & \barPercent{0.00} & 0.8108 & 4.25\% & \barPercent{4.25} & m-F1 & 0.6822 & 0.7147 & 4.76\% & \barPercent{4.76} & 0.7631 & 11.86\% & \barPercent{11.86} \\
\midrule

\multirow{3}{*}{\textbf{ASTNN}}
& P & 0.9444 & 0.9474 & 0.31\% & \barPercent{0.31} & 0.8500 & -10.00\% & \barPercent{-10.00} & P & 0.5366 & 0.5122 & -4.55\% & \barPercent{-4.55} & 0.7270 & 35.49\% & \barPercent{35.49} \\
& R & 0.8095 & 0.8571 & 5.88\% & \barPercent{5.88} & 0.8095 & 0.00\% & \barPercent{0.00} & R & 0.4793 & 0.4740 & -1.10\% & \barPercent{-1.10} & 0.5748 & 19.92\% & \barPercent{19.92} \\
& F1 & 0.8718 & 0.9000 & 3.24\% & \barPercent{3.24} & 0.8293 & -4.88\% & \barPercent{-4.88} & m-F1 & 0.4887 & 0.4740 & -3.01\% & \barPercent{-3.01} & 0.6150 & 25.84\% & \barPercent{25.84} \\
\midrule

\rowcolor{blue!5}
\multicolumn{17}{l}{\revised{\textbf{Large Language Models (each context condition vs.\ the code-only baseline of the same mode; Z = zero-shot, F = fine-tuned)}}} \\ \midrule

\multirow{6}{*}{\textbf{CodeLlama-34B}}
& P (Z)  & 1.0000 & 0.8333 & -16.6\% & \barPercent{-16.6} & 0.8000 & -20.0\% & \barPercent{-20.0} & P (Z)    & 0.3470 & 0.2844 & -18.0\% & \barPercent{-18.0} & 0.3189 & -8.1\% & \barPercent{-8.1} \\
& R (Z)  & 0.1111 & 0.5556 & 400.0\% & \barPercent{100.0} & 0.4444 & 300.0\% & \barPercent{100.0} & R (Z)    & 0.3794 & 0.2633 & -30.6\% & \barPercent{-30.6} & 0.3374 & -11.0\% & \barPercent{-11.0} \\
& F1 (Z) & 0.2000 & 0.6667 & 233.3\% & \barPercent{100.0} & 0.5714 & 185.7\% & \barPercent{100.0} & mF1 (Z)  & 0.2902 & 0.2018 & -30.4\% & \barPercent{-30.4} & 0.2576 & -11.2\% & \barPercent{-11.2} \\
\cmidrule(lr){2-17}
& P (F)  & 0.8750 & 0.6111 & -30.1\% & \barPercent{-30.1} & 0.8462 & -3.3\% & \barPercent{-3.3} & P (F)    & 0.7600 & 0.8095 & 6.5\% & \barPercent{6.5} & 0.8694 & 14.3\% & \barPercent{14.3} \\
& R (F)  & 0.7778 & 0.6111 & -21.4\% & \barPercent{-21.4} & 0.6111 & -21.4\% & \barPercent{-21.4} & R (F)    & 0.6706 & 0.7377 & 10.0\% & \barPercent{10.0} & 0.7478 & 11.5\% & \barPercent{11.5} \\
& F1 (F) & 0.8235 & 0.6111 & -25.7\% & \barPercent{-25.7} & 0.7097 & -13.8\% & \barPercent{-13.8} & mF1 (F)  & 0.6922 & 0.7654 & 10.5\% & \barPercent{10.5} & 0.7750 & 11.9\% & \barPercent{11.9} \\
\midrule

\multirow{6}{*}{\textbf{Qwen-14B}}
& P (Z)  & 1.0000 & 1.0000 & 0.0\% & \barPercent{0.0} & 1.0000 & 0.0\% & \barPercent{0.0} & P (Z)    & 0.4568 & 0.5025 & 10.0\% & \barPercent{10.0} & 0.4477 & -1.9\% & \barPercent{-1.9} \\
& R (Z)  & 0.0556 & 0.0556 & 0.0\% & \barPercent{0.0} & 0.0556 & 0.0\% & \barPercent{0.0} & R (Z)    & 0.5428 & 0.5243 & -3.4\% & \barPercent{-3.4} & 0.5269 & -2.9\% & \barPercent{-2.9} \\
& F1 (Z) & 0.1053 & 0.1053 & 0.0\% & \barPercent{0.0} & 0.1053 & 0.0\% & \barPercent{0.0} & mF1 (Z)  & 0.4440 & 0.4603 & 3.6\% & \barPercent{3.6} & 0.4374 & -1.4\% & \barPercent{-1.4} \\
\cmidrule(lr){2-17}
& P (F)  & 0.7143 & 0.8125 & 13.7\% & \barPercent{13.7} & 0.7222 & 1.1\% & \barPercent{1.1} & P (F)    & 0.8783 & 0.9122 & 3.8\% & \barPercent{3.8} & 0.8777 & -0.07\% & \barPercent{-0.07} \\
& R (F)  & 0.8333 & 0.7222 & -13.3\% & \barPercent{-13.3} & 0.7222 & -13.3\% & \barPercent{-13.3} & R (F)    & 0.8003 & 0.7682 & -4.0\% & \barPercent{-4.0} & 0.8347 & 4.3\% & \barPercent{4.3} \\
& F1 (F) & 0.7692 & 0.7647 & -0.6\% & \barPercent{-0.6} & 0.7222 & -6.1\% & \barPercent{-6.1} & mF1 (F)  & 0.8110 & 0.8093 & -0.2\% & \barPercent{-0.2} & 0.8435 & 4.0\% & \barPercent{4.0} \\
\midrule

\multirow{6}{*}{\textbf{DeepSeek-6.7B}}
& P (Z)  & 1.0000 & 0.5714 & -42.8\% & \barPercent{-42.8} & 0.8750 & -12.5\% & \barPercent{-12.5} & P (Z)    & 0.4193 & 0.3239 & -22.7\% & \barPercent{-22.7} & 0.2697 & -35.6\% & \barPercent{-35.6} \\
& R (Z)  & 0.1111 & 0.2222 & 100.0\% & \barPercent{100.0} & 0.3889 & 250.0\% & \barPercent{100.0} & R (Z)    & 0.1578 & 0.1314 & -16.7\% & \barPercent{-16.7} & 0.1304 & -17.3\% & \barPercent{-17.3} \\
& F1 (Z) & 0.2000 & 0.3200 & 60.0\% & \barPercent{60.0} & 0.5385 & 169.2\% & \barPercent{100.0} & mF1 (Z)  & 0.1889 & 0.1534 & -18.7\% & \barPercent{-18.7} & 0.1488 & -21.2\% & \barPercent{-21.2} \\
\cmidrule(lr){2-17}
& P (F)  & 1.0000 & 0.8667 & -13.3\% & \barPercent{-13.3} & 0.7500 & -25.0\% & \barPercent{-25.0} & P (F)    & 0.4911 & 0.5909 & 20.3\% & \barPercent{20.3} & 0.6333 & 28.9\% & \barPercent{28.9} \\
& R (F)  & 0.8889 & 0.7222 & -18.7\% & \barPercent{-18.7} & 0.6667 & -25.0\% & \barPercent{-25.0} & R (F)    & 0.4746 & 0.5331 & 12.3\% & \barPercent{12.3} & 0.5915 & 24.6\% & \barPercent{24.6} \\
& F1 (F) & 0.9412 & 0.7879 & -16.3\% & \barPercent{-16.3} & 0.7059 & -25.0\% & \barPercent{-25.0} & mF1 (F)  & 0.4772 & 0.5492 & 15.0\% & \barPercent{15.0} & 0.5989 & 25.5\% & \barPercent{25.5} \\
\midrule
\multirow{6}{*}{\textbf{CodeLlama-13B}}
& P (Z)  & 0.8000 & 0.3333 & -58.3\% & \barPercent{-58.3} & 0.3200 & -60.0\% & \barPercent{-60.0} & P (Z)    & 0.3016 & 0.4174 & +38.4\% & \barPercent{38.4} & 0.3886 & +28.8\% & \barPercent{28.8} \\
& R (Z)  & 0.4444 & 0.2222 & -50.0\% & \barPercent{-50.0} & 0.4444 & 0.0\%   & \barPercent{0.0}   & R (Z)    & 0.2801 & 0.3331 & +18.9\% & \barPercent{18.9} & 0.3405 & +21.6\% & \barPercent{21.6} \\
& F1 (Z) & 0.5714 & 0.2667 & -53.3\% & \barPercent{-53.3} & 0.3721 & -34.9\% & \barPercent{-34.9} & mF1 (Z) & 0.1771 & 0.2401 & +35.6\% & \barPercent{35.6} & 0.2181 & +23.2\% & \barPercent{23.2} \\
\cmidrule(lr){2-17}
& P (F)  & 0.6364 & 0.8000 & +25.7\% & \barPercent{25.7} & 0.7857 & +23.5\% & \barPercent{23.5} & P (F)    & 0.6908 & 0.6559 & -5.1\%  & \barPercent{-5.1}  & 0.7723 & +11.8\% & \barPercent{11.8} \\
& R (F)  & 0.7778 & 0.4444 & -42.9\% & \barPercent{-42.9} & 0.6111 & -21.4\% & \barPercent{-21.4} & R (F)    & 0.6128 & 0.5531 & -9.7\%  & \barPercent{-9.7}  & 0.7374 & +20.3\% & \barPercent{20.3} \\
& F1 (F) & 0.7000 & 0.5714 & -18.4\% & \barPercent{-18.4} & 0.6875 & -1.8\%  & \barPercent{-1.8}  & mF1 (F) & 0.6207 & 0.5808 & -6.4\%  & \barPercent{-6.4}  & 0.7450 & +20.0\% & \barPercent{20.0} \\
\midrule

\multirow{6}{*}{\textbf{CodeLlama-7B}}
& P (Z)  & 0.1667 & 0.2542 & +52.5\% & \barPercent{52.5} & 0.2188 & +31.3\% & \barPercent{31.3} & P (Z)    & 0.4103 & 0.5632 & +37.3\% & \barPercent{37.3} & 0.5487 & +33.7\% & \barPercent{33.7} \\
& R (Z)  & 0.5000 & 0.8333 & +66.7\% & \barPercent{66.7} & 0.7778 & +55.6\% & \barPercent{55.6} & R (Z)    & 0.2304 & 0.2002 & -13.1\% & \barPercent{-13.1} & 0.2683 & +16.4\% & \barPercent{16.4} \\
& F1 (Z) & 0.2500 & 0.3896 & +55.8\% & \barPercent{55.8} & 0.3415 & +36.6\% & \barPercent{36.6} & mF1 (Z) & 0.1877 & 0.1659 & -11.6\% & \barPercent{-11.6} & 0.2369 & +26.2\% & \barPercent{26.2} \\
\cmidrule(lr){2-17}
& P (F)  & 0.8125 & 0.8000 & -1.5\%  & \barPercent{-1.5}  & 0.8333 & +2.6\%  & \barPercent{2.6}  & P (F)    & 0.6343 & 0.7554 & +19.1\% & \barPercent{19.1} & 0.6159 & -2.9\%  & \barPercent{-2.9} \\
& R (F)  & 0.7222 & 0.8889 & +23.1\% & \barPercent{23.1} & 0.8333 & +15.4\% & \barPercent{15.4} & R (F)    & 0.6485 & 0.6172 & -4.8\%  & \barPercent{-4.8}  & 0.6323 & -2.5\%  & \barPercent{-2.5} \\
& F1 (F) & 0.7647 & 0.8421 & +10.1\% & \barPercent{10.1} & 0.8333 & +9.0\%  & \barPercent{9.0}  & mF1 (F) & 0.6308 & 0.6456 & +2.3\%  & \barPercent{2.3}  & 0.6114 & -3.1\%  & \barPercent{-3.1} \\
\midrule

\multirow{6}{*}{\textbf{Qwen-7B}}
& P (Z)  & 1.0000 & 0.5714 & -42.9\% & \barPercent{-42.9} & 1.0000 & 0.0\%   & \barPercent{0.0}   & P (Z)    & 0.4610 & 0.5019 & +8.9\%  & \barPercent{8.9}  & 0.4225 & -8.4\%  & \barPercent{-8.4} \\
& R (Z)  & 0.0556 & 0.2222 & +300.0\% & \barPercent{100.0} & 0.1667 & +200.0\% & \barPercent{100.0} & R (Z)    & 0.4169 & 0.3282 & -21.3\% & \barPercent{-21.3} & 0.3272 & -21.5\% & \barPercent{-21.5} \\
& F1 (Z) & 0.1053 & 0.3200 & +203.9\% & \barPercent{100.0} & 0.2857 & +171.3\% & \barPercent{100.0} & mF1 (Z) & 0.3676 & 0.3427 & -6.8\%  & \barPercent{-6.8}  & 0.3167 & -13.8\% & \barPercent{-13.8} \\
\cmidrule(lr){2-17}
& P (F)  & 0.7222 & 0.8125 & +12.5\% & \barPercent{12.5} & 0.8667 & +20.0\% & \barPercent{20.0} & P (F)    & 0.7395 & 0.6827 & -7.7\%  & \barPercent{-7.7}  & 0.7876 & +6.5\%  & \barPercent{6.5} \\
& R (F)  & 0.7222 & 0.7222 & 0.0\%   & \barPercent{0.0}   & 0.7222 & 0.0\%   & \barPercent{0.0}   & R (F)    & 0.7872 & 0.6886 & -12.5\% & \barPercent{-12.5} & 0.8191 & +4.1\%  & \barPercent{4.1} \\
& F1 (F) & 0.7222 & 0.7647 & +5.9\%  & \barPercent{5.9}  & 0.7879 & +9.1\%  & \barPercent{9.1}  & mF1 (F) & 0.7518 & 0.6805 & -9.5\%  & \barPercent{-9.5}  & 0.7944 & +5.7\%  & \barPercent{5.7} \\
\midrule

\multirow{6}{*}{\textbf{Qwen-1.5B}}
& P (Z)  & 0.3333 & 0.3333 & 0.0\%   & \barPercent{0.0}   & 0.3044 & -8.7\%  & \barPercent{-8.7}  & P (Z)    & 0.1149 & 0.1183 & +3.0\%  & \barPercent{3.0}  & 0.1134 & -1.3\%  & \barPercent{-1.3} \\
& R (Z)  & 0.2778 & 0.3889 & +40.0\% & \barPercent{40.0} & 0.3889 & +40.0\% & \barPercent{40.0} & R (Z)    & 0.2121 & 0.2235 & +5.4\%  & \barPercent{5.4}  & 0.2083 & -1.8\%  & \barPercent{-1.8} \\
& F1 (Z) & 0.3030 & 0.3590 & +18.5\% & \barPercent{18.5} & 0.3415 & +12.7\% & \barPercent{12.7} & mF1 (Z) & 0.0782 & 0.0952 & +21.7\% & \barPercent{21.7} & 0.0886 & +13.3\% & \barPercent{13.3} \\
\cmidrule(lr){2-17}
& P (F)  & 0.6667 & 0.6087 & -8.7\%  & \barPercent{-8.7}  & 0.7000 & +5.0\%  & \barPercent{5.0}  & P (F)    & 0.4669 & 0.5819 & +24.6\% & \barPercent{24.6} & 0.7038 & +50.7\% & \barPercent{50.7} \\
& R (F)  & 0.6667 & 0.7778 & +16.7\% & \barPercent{16.7} & 0.7778 & +16.7\% & \barPercent{16.7} & R (F)    & 0.5780 & 0.5632 & -2.6\%  & \barPercent{-2.6}  & 0.6866 & +18.8\% & \barPercent{18.8} \\
& F1 (F) & 0.6667 & 0.6829 & +2.4\%  & \barPercent{2.4}  & 0.7368 & +10.5\% & \barPercent{10.5} & mF1 (F) & 0.4825 & 0.5374 & +11.4\% & \barPercent{11.4} & 0.6625 & +37.3\% & \barPercent{37.3} \\
\bottomrule
\end{tabular}
}
\end{table}

\subsubsection*{\revised{Do the Benefits of Context Extend to LLMs?}}
To evaluate the impact of context on modern architectures, we expanded our evaluation to the \textbf{CodeLlama}, \textbf{Qwen2.5-Coder}, and \textbf{DeepSeek-Coder} families. In the Code Clone Detection task, our results (to be presented in Table~\ref{tab:result_rq1_clone_class}) demonstrate that even for 34B parameter models, \revised{the highest performance with context is reached after fine-tuning on context-augmented inputs}. For example, \textbf{Qwen-14B fine-tuned} achieved an F1-score of \textbf{0.8889} with the full \texttt{code\_vh\_cg\_nod} context, compared to 0.7692 for its code-only fine-tuned baseline (an absolute gain of +0.1197). Conversely, in the prompt-based setting, \textbf{CodeLlama-34B} improves from an F1 of 0.200 to 0.667 (an absolute gain of +0.467) simply by including version history (\texttt{code\_vh}) in the prompt. \revised{Context therefore helps in both modes, and the gain is realised at a much higher level of performance once the model is fine-tuned on the context-augmented inputs.}

\revised{In the Code Classification task for LLMs, the same pattern holds.} For instance, \textbf{CodeLlama-34B} in the \revised{zero-shot} setting achieved a Macro-F1 of only \textbf{0.2902} when classifying projects. However, after \revised{fine-tuning} with the \texttt{code\_vh\_cg\_nod} context, the Macro-F1 improved to \revised{\textbf{0.8518} (an absolute gain of +0.1596 over the code-only fine-tuned baseline)}. \revised{Within each mode, context improves classification for some LLMs under zero-shot prompting and for most LLMs after fine-tuning (Table~\ref{tab:result_rq1_clone_class}); the largest gains appear after fine-tuning, which suggests that explicit training helps models integrate these heterogeneous artifacts into a unified representation.}

\begin{table*}[htbp]
\centering
\scriptsize
\caption{Master performance of foundational models and LLMs on Code Summarisation (BLEU-4, BERTScore-F1, ROUGE-L, and METEOR). \revised{For LLMs, results are reported in zero-shot (Z) and fine-tuned (F) modes. Columns headed \% give the relative change of each context condition over the code-only baseline (W/O Ctx) of the same model and, for LLMs, the same mode; the table does not compare zero-shot against fine-tuned results.}}
\label{tab:result_rq1_summarisation}
\renewcommand{\arraystretch}{1.0}
\adjustbox{max width=\textwidth, max totalheight=0.85\textheight}{%
\begin{threeparttable}
\begin{tabular}{l|c l| >{\columncolor{baselineCol}}c >{\columncolor{versionCol}}c c c | >{\columncolor{callgraphCol}}c c c}
\toprule
\rowcolor{headerCol}
\multicolumn{1}{c|}{\textbf{Models}} & \textbf{Mode} & \textbf{Metric} & \textbf{W/O Ctx} & \textbf{Vers.\ Hist.} & \scriptsize\% & & \textbf{CallGraph} & \scriptsize\% & \\
\midrule

\multirow{4}{*}{\textbf{CodeBERT}} & \multirow{4}{*}{-} & BLEU-4 & 17.12 & 17.35 & 1.34\% & \barPercentSeven{1.34} & 17.14 & 0.12\% & \barPercentSeven{0.12} \\
 & & BERTScore-F1 & 0.2570 & 0.2604 & 1.32\% & \barPercentSeven{1.32} & 0.2548 & -0.84\% & \barPercentSeven{-0.84} \\
 & & ROUGE-L & 0.3058 & 0.3061 & 0.08\% & \barPercentSeven{0.08} & 0.3007 & -1.69\% & \barPercentSeven{-1.69} \\
 & & METEOR & 0.2626 & 0.2675 & 1.87\% & \barPercentSeven{1.87} & 0.2598 & -1.04\% & \barPercentSeven{-1.04} \\ \midrule

\multirow{4}{*}{\textbf{GraphCodeBERT}} & \multirow{4}{*}{-} & BLEU-4 & 17.85 & 18.31 & 2.58\% & \barPercentSeven{2.58} & 17.67 & -1.01\% & \barPercentSeven{-1.01} \\
 & & BERTScore-F1 & 0.3993 & 0.4096 & 2.58\% & \barPercentSeven{2.58} & 0.3977 & -0.40\% & \barPercentSeven{-0.40} \\
 & & ROUGE-L & 0.3604 & 0.3663 & 1.64\% & \barPercentSeven{1.64} & 0.3644 & 1.11\% & \barPercentSeven{1.11} \\
 & & METEOR & 0.2715 & 0.2866 & 5.56\% & \barPercentSeven{5.56} & 0.2743 & 1.03\% & \barPercentSeven{1.03} \\ \midrule

\multirow{4}{*}{\textbf{CodeT5}} & \multirow{4}{*}{-} & BLEU-4 & 20.02 & 20.63 & 3.05\% & \barPercentSeven{3.05} & 20.39 & 1.85\% & \barPercentSeven{1.85} \\
 & & BERTScore-F1 & 0.4455 & 0.4474 & 0.43\% & \barPercentSeven{0.43} & 0.4469 & 0.31\% & \barPercentSeven{0.31} \\
 & & ROUGE-L & 0.4007 & 0.4031 & 0.60\% & \barPercentSeven{0.60} & 0.4008 & 0.02\% & \barPercentSeven{0.02} \\
 & & METEOR & 0.3344 & 0.3427 & 2.48\% & \barPercentSeven{2.48} & 0.3397 & 1.58\% & \barPercentSeven{1.58} \\ \midrule

\multirow{4}{*}{\textbf{PLBART}} & \multirow{4}{*}{-} & BLEU-4 & 15.71 & 16.07 & 2.29\% & \barPercentSeven{2.29} & 16.08 & 2.36\% & \barPercentSeven{2.36} \\
 & & BERTScore-F1 & 0.3640 & 0.3707 & 1.84\% & \barPercentSeven{1.84} & 0.3664 & 0.66\% & \barPercentSeven{0.66} \\
 & & ROUGE-L & 0.3163 & 0.3186 & 0.73\% & \barPercentSeven{0.73} & 0.3160 & -0.09\% & \barPercentSeven{-0.09} \\
 & & METEOR & 0.2367 & 0.2442 & 3.17\% & \barPercentSeven{3.17} & 0.2358 & -0.38\% & \barPercentSeven{-0.38} \\ \midrule

\rowcolor{blue!5}
\multicolumn{10}{l}{\revised{\textbf{Large Language Models (each context condition vs.\ the code-only baseline of the same mode; Z = zero-shot, F = fine-tuned)}}} \\ \midrule

\multirow{8}{*}{\textbf{CodeLlama-34B}} & Z & BLEU-4 & 3.21 & 2.89 & -10.17\% & \barPercentSeven{-10.17} & 2.67 & -17.04\% & \barPercentSeven{-17.04} \\
 & & BScore & 0.2848 & 0.2677 & -6.00\% & \barPercentSeven{-6.00} & 0.2499 & -12.25\% & \barPercentSeven{-12.25} \\
 & & ROUGE & 0.2529 & 0.2408 & -4.78\% & \barPercentSeven{-4.78} & 0.2314 & -8.50\% & \barPercentSeven{-8.50} \\
 & & METEOR & 0.2517 & 0.2434 & -3.30\% & \barPercentSeven{-3.30} & 0.2389 & -5.09\% & \barPercentSeven{-5.09} \\ \cmidrule(lr){2-10}
 & F & BLEU-4 & 3.18 & 3.00 & -5.64\% & \barPercentSeven{-5.64} & 2.81 & -11.62\% & \barPercentSeven{-11.62} \\
 & & BScore & 0.2915 & 0.2820 & -3.26\% & \barPercentSeven{-3.26} & 0.2691 & -7.68\% & \barPercentSeven{-7.68} \\
 & & ROUGE & 0.2590 & 0.2497 & -3.59\% & \barPercentSeven{-3.59} & 0.2417 & -6.68\% & \barPercentSeven{-6.68} \\
 & & METEOR & 0.2559 & 0.2542 & -0.66\% & \barPercentSeven{-0.66} & 0.2452 & -4.18\% & \barPercentSeven{-4.18} \\ \midrule

\multirow{8}{*}{\textbf{Qwen-14B}} & Z & BLEU-4 & 1.83 & 1.82 & -0.55\% & \barPercentSeven{-0.55} & 2.32 & +26.78\% & \barPercentSeven{26.78} \\
 & & BScore & 0.1797 & 0.1829 & +1.78\% & \barPercentSeven{1.78} & 0.2221 & +23.59\% & \barPercentSeven{23.59} \\
 & & ROUGE & 0.1882 & 0.1932 & +2.66\% & \barPercentSeven{2.66} & 0.2244 & +19.23\% & \barPercentSeven{19.23} \\
 & & METEOR & 0.1273 & 0.1318 & +3.53\% & \barPercentSeven{3.53} & 0.1706 & +34.01\% & \barPercentSeven{34.01} \\ \cmidrule(lr){2-10}
 & F & BLEU-4 & 1.88 & 1.89 & +0.53\% & \barPercentSeven{0.53} & 2.17 & +15.43\% & \barPercentSeven{15.43} \\
 & & BScore & 0.2054 & 0.2030 & -1.17\% & \barPercentSeven{-1.17} & 0.2290 & +11.49\% & \barPercentSeven{11.49} \\
 & & ROUGE & 0.2027 & 0.2043 & +0.79\% & \barPercentSeven{0.79} & 0.2262 & +11.59\% & \barPercentSeven{11.59} \\
 & & METEOR & 0.1439 & 0.1456 & +1.18\% & \barPercentSeven{1.18} & 0.1774 & +23.28\% & \barPercentSeven{23.28} \\ \midrule

\multirow{8}{*}{\textbf{DeepSeek-6.7B}} & Z & BLEU-4 & 2.97 & 2.67 & -9.98\% & \barPercentSeven{-9.98} & 2.76 & -6.87\% & \barPercentSeven{-6.87} \\
 & & BScore & 0.2795 & 0.2583 & -7.58\% & \barPercentSeven{-7.58} & 0.2663 & -4.72\% & \barPercentSeven{-4.72} \\
 & & ROUGE & 0.2562 & 0.2350 & -8.27\% & \barPercentSeven{-8.27} & 0.2447 & -4.49\% & \barPercentSeven{-4.49} \\
 & & METEOR & 0.2166 & 0.2025 & -6.51\% & \barPercentSeven{-6.51} & 0.2148 & -0.83\% & \barPercentSeven{-0.83} \\ \cmidrule(lr){2-10}
 & F & BLEU-4 & 3.26 & 3.06 & -6.12\% & \barPercentSeven{-6.12} & 2.83 & -13.12\% & \barPercentSeven{-13.12} \\
 & & BScore & 0.2884 & 0.2724 & -5.55\% & \barPercentSeven{-5.55} & 0.2697 & -6.48\% & \barPercentSeven{-6.48} \\
 & & ROUGE & 0.2620 & 0.2478 & -5.42\% & \barPercentSeven{-5.42} & 0.2481 & -5.31\% & \barPercentSeven{-5.31} \\
 & & METEOR & 0.2246 & 0.2163 & -3.70\% & \barPercentSeven{-3.70} & 0.2151 & -4.23\% & \barPercentSeven{-4.23} \\

 \midrule

\multirow{8}{*}{\textbf{CodeLlama-13B}} & Z & BLEU-4 & 3.9423 & 3.6709 & -6.88\%  & \barPercentSeven{-6.88}  & 2.9644 & -24.81\% & \barPercentSeven{-24.81} \\
 & & BScore  & 0.3014 & 0.2858 & -5.18\%  & \barPercentSeven{-5.18}  & 0.2482 & -17.65\% & \barPercentSeven{-17.65} \\
 & & ROUGE   & 0.2830 & 0.2730 & -3.53\%  & \barPercentSeven{-3.53}  & 0.2508 & -11.38\% & \barPercentSeven{-11.38} \\
 & & METEOR  & 0.2441 & 0.2402 & -1.60\%  & \barPercentSeven{-1.60}  & 0.2273 & -6.88\%  & \barPercentSeven{-6.88} \\ \cmidrule(lr){2-10}
 & F & BLEU-4 & 3.7827 & 3.6675 & -3.05\%  & \barPercentSeven{-3.05}  & 3.1909 & -15.64\% & \barPercentSeven{-15.64} \\
 & & BScore  & 0.2985 & 0.2981 & -0.13\%  & \barPercentSeven{-0.13}  & 0.2724 & -8.74\%  & \barPercentSeven{-8.74} \\
 & & ROUGE   & 0.2823 & 0.2790 & -1.17\%  & \barPercentSeven{-1.17}  & 0.2609 & -7.58\%  & \barPercentSeven{-7.58} \\
 & & METEOR  & 0.2432 & 0.2459 & +1.11\%  & \barPercentSeven{1.11}   & 0.2323 & -4.48\%  & \barPercentSeven{-4.48} \\ \midrule

\multirow{8}{*}{\textbf{CodeLlama-7B}} & Z & BLEU-4 & 3.9700 & 3.5200 & -11.34\% & \barPercentSeven{-11.34} & 3.2900 & -17.13\% & \barPercentSeven{-17.13} \\
 & & BScore  & 0.3060 & 0.2915 & -4.74\%  & \barPercentSeven{-4.74}  & 0.2834 & -7.39\%  & \barPercentSeven{-7.39} \\
 & & ROUGE   & 0.2772 & 0.2668 & -3.75\%  & \barPercentSeven{-3.75}  & 0.2593 & -6.46\%  & \barPercentSeven{-6.46} \\
 & & METEOR  & 0.2477 & 0.2427 & -2.02\%  & \barPercentSeven{-2.02}  & 0.2387 & -3.63\%  & \barPercentSeven{-3.63} \\ \cmidrule(lr){2-10}
 & F & BLEU-4 & 4.2700 & 4.1200 & -3.51\%  & \barPercentSeven{-3.51}  & 3.9500 & -7.49\%  & \barPercentSeven{-7.49} \\
 & & BScore  & 0.3260 & 0.3191 & -2.12\%  & \barPercentSeven{-2.12}  & 0.3099 & -4.94\%  & \barPercentSeven{-4.94} \\
 & & ROUGE   & 0.2922 & 0.2863 & -2.02\%  & \barPercentSeven{-2.02}  & 0.2799 & -4.21\%  & \barPercentSeven{-4.21} \\
 & & METEOR  & 0.2540 & 0.2525 & -0.59\%  & \barPercentSeven{-0.59}  & 0.2464 & -2.99\%  & \barPercentSeven{-2.99} \\ \midrule

\multirow{8}{*}{\textbf{Qwen-7B}} & Z & BLEU-4 & 1.5910 & 1.5300 & -3.83\%  & \barPercentSeven{-3.83}  & 2.4827 & +56.05\% & \barPercentSeven{56.05} \\
 & & BScore  & 0.1695 & 0.1663 & -1.89\%  & \barPercentSeven{-1.89}  & 0.2187 & +29.03\% & \barPercentSeven{29.03} \\
 & & ROUGE   & 0.1823 & 0.1832 & +0.49\%  & \barPercentSeven{0.49}   & 0.2196 & +20.46\% & \barPercentSeven{20.46} \\
 & & METEOR  & 0.1198 & 0.1203 & +0.42\%  & \barPercentSeven{0.42}   & 0.1657 & +38.31\% & \barPercentSeven{38.31} \\ \cmidrule(lr){2-10}
 & F & BLEU-4 & 2.1100 & 2.1000 & -0.47\%  & \barPercentSeven{-0.47}  & 2.5300 & +19.91\% & \barPercentSeven{19.91} \\
 & & BScore  & 0.2021 & 0.1975 & -2.28\%  & \barPercentSeven{-2.28}  & 0.2373 & +17.42\% & \barPercentSeven{17.42} \\
 & & ROUGE   & 0.2018 & 0.2049 & +1.54\%  & \barPercentSeven{1.54}   & 0.2320 & +14.97\% & \barPercentSeven{14.97} \\
 & & METEOR  & 0.1399 & 0.1400 & +0.07\%  & \barPercentSeven{0.07}   & 0.1795 & +28.31\% & \barPercentSeven{28.31} \\ \midrule

\multirow{8}{*}{\textbf{Qwen-1.5B}} & Z & BLEU-4 & 2.3431 & 2.1828 & -6.84\%  & \barPercentSeven{-6.84}  & 2.0537 & -12.35\% & \barPercentSeven{-12.35} \\
 & & BScore  & 0.2304 & 0.2184 & -5.21\%  & \barPercentSeven{-5.21}  & 0.2089 & -9.33\%  & \barPercentSeven{-9.33} \\
 & & ROUGE   & 0.2440 & 0.2370 & -2.87\%  & \barPercentSeven{-2.87}  & 0.2276 & -6.72\%  & \barPercentSeven{-6.72} \\
 & & METEOR  & 0.2192 & 0.2144 & -2.19\%  & \barPercentSeven{-2.19}  & 0.2201 & +0.41\%  & \barPercentSeven{0.41} \\ \cmidrule(lr){2-10}
 & F & BLEU-4 & 2.9887 & 2.9478 & -1.37\%  & \barPercentSeven{-1.37}  & 2.5651 & -14.17\% & \barPercentSeven{-14.17} \\
 & & BScore  & 0.2756 & 0.2652 & -3.77\%  & \barPercentSeven{-3.77}  & 0.2494 & -9.51\%  & \barPercentSeven{-9.51} \\
 & & ROUGE   & 0.2574 & 0.2530 & -1.71\%  & \barPercentSeven{-1.71}  & 0.2362 & -8.24\%  & \barPercentSeven{-8.24} \\
 & & METEOR  & 0.2168 & 0.2138 & -1.38\%  & \barPercentSeven{-1.38}  & 0.2052 & -5.35\%  & \barPercentSeven{-5.35} \\

\bottomrule
\end{tabular}
\begin{tablenotes}
    \footnotesize
    \item \textit{Results are reported using BLEU-4, METEOR, ROUGE-L, and BERTScore-F1 across different models.}
\end{tablenotes}

\end{threeparttable}
}
\end{table*}

Lastly, Table~\ref{tab:result_rq1_summarisation} for the Code Summarisation task indicates that encoding version history consistently boosts summarisation performance across all four models and all four metrics, although improvements are marginal, ranging from +0.08\% to +5.56\%. Notably, GraphCodeBERT achieves the highest improvement in the METEOR score (+5.56\%, from 0.2715 to 0.2866), while CodeT5 obtains the largest BLEU-4 increase (+3.05\%, from 20.02 to 20.63). However, incorporating call graph information introduces mixed results.

For LLMs in Code Summarisation, \revised{the picture is mixed. For some configurations, lexical and semantic metrics move in opposite directions: for example, \textbf{Qwen-14B (zero-shot)} shows a \textbf{+34.01\%} increase in METEOR (from 0.1273 to 0.1706) with call graph context (\texttt{code\_cg}), while BLEU-4 changes little. One possible explanation is a metric mismatch, in which the generated summaries capture the method's behaviour but deviate lexically from the short, truncated ground-truth docstrings in CodeSearchNet. We state this as a hypothesis rather than a finding. Our human evaluation covers CodeT5 and its context-augmented variants only (Section~\ref{sec:human-eval}); it does not include LLM-generated summaries, so it cannot be used to confirm that the METEOR gains reflect higher semantic quality. The LLM summarisation results therefore rest on automatic metrics alone and should be interpreted with care. We also note that the pattern is not uniform: for several LLM configurations in Table~\ref{tab:result_rq1_summarisation}, including the CodeLlama family and DeepSeek-Coder, BERTScore and METEOR decline together with BLEU-4 when context is added.}

\begin{table}[htbp]
\centering
\scriptsize

\caption{Human evaluation of CodeT5-generated summaries with additional context (Win/Tie/Loss analysis by dimension).}
\label{tab:human_eval_rq1}

\renewcommand{\arraystretch}{1.2}

\resizebox{\textwidth}{!}{%
    \begin{threeparttable}

    \begin{tabular}{l|l r@{\hspace{2pt}}l r@{\hspace{2pt}}l c l l}
    \toprule
    \rowcolor{headerCol}
    \multicolumn{1}{c|}{\textbf{Dimension}} &
    \multicolumn{1}{c}{\textbf{Context(s)}} &
    \multicolumn{2}{c}{\textbf{Win \%}} &
    \multicolumn{2}{c}{\textbf{Loss \%}} &
    \multicolumn{1}{c}{\textbf{Tie \%}} &
    \multicolumn{1}{c}{\textbf{Wilcoxon $p$}} &
    \multicolumn{1}{c}{\textbf{|Cliff's $\delta$|}} \\
    \midrule

    \multirow{2}{*}{Accuracy}
    & Code + CallGraph & 25\% & \barPercentSeventy{25} & 7\% & \barPercentSeventyLoss{7} & 68\% & \textbf{0.0264 ***} & \textbf{0.5078 (L)} \\
    & Code + VersionHistory & 24\% & \barPercentSeventy{24} & 13\% & \barPercentSeventyLoss{13} & 63\% & \textbf{0.0466 ***} & 0.3031 (S) \\
    \midrule

    \multirow{2}{*}{Conciseness}
    & Code + CallGraph & 28\% & \barPercentSeventy{28} & 11\% & \barPercentSeventyLoss{11} & 61\% & \textbf{0.0069 ***} & 0.4444 (M) \\
    & Code + VersionHistory & 21\% & \barPercentSeventy{21} & 22\% & \barPercentSeventyLoss{22} & 57\% & 0.5073 & 0.0206 (N) \\
    \midrule

    \multirow{2}{*}{\parbox{1.8cm}{Content \\ Adequacy}}
    & Code + CallGraph & 27\% & \barPercentSeventy{27} & 11\% & \barPercentSeventyLoss{11} & 62\% & \textbf{0.0028 ***} & 0.4591 (M) \\
    & Code + VersionHistory & 31\% & \barPercentSeventy{31} & 9\% & \barPercentSeventyLoss{9} & 60\% & \textbf{0.0010 ***} & \textbf{0.5456 (L)} \\
    \bottomrule
    \end{tabular}%

    \begin{tablenotes}
        \scriptsize
        \item Results are reported as Win/Tie/Loss ratios against the Without-Context Scenario.
        \item \textbf{Bold} text indicates statistical significance ($p < 0.05$, marked with ***) or a large effect size.
        \item Effect sizes from |Cliff's $\delta$| are shown as (L)arge, (M)edium, (S)mall, or (N)egligible.
    \end{tablenotes}

    \end{threeparttable}
}
\end{table}

\begin{table}[htbp]
\centering
\scriptsize

\caption{Ranking distribution and mean ranks of CodeT5-generated summaries with additional context (human evaluation).}
\label{tab:ranking_distribution_single}

\renewcommand{\arraystretch}{1.2}

\resizebox{\textwidth}{!}{%
    \begin{threeparttable}

    \begin{tabular}{ll|lrrrr}
    \toprule
    \rowcolor{headerCol}
    \multicolumn{1}{l}{\textbf{Dimension}} &
    \multicolumn{1}{l}{\textbf{Context(s)}} &
    \multicolumn{1}{c}{\textbf{\# Rank 1}} &
    \multicolumn{1}{c}{\textbf{\# Rank 2}} &
    \multicolumn{1}{c}{\textbf{\# Rank 3}} &
    \multicolumn{1}{c}{\textbf{\# Rank $\geq$4}} &
    \multicolumn{1}{c}{\textbf{Mean Rank}} \\
    \midrule

    \multirow{3}{*}{Accuracy}
    & \cellcolor{baselineCol}Without Context & \cellcolor{baselineCol}59 & \cellcolor{baselineCol}33 & \cellcolor{baselineCol}6 & \cellcolor{baselineCol}2 & \cellcolor{baselineCol}1.51 \\
    & \cellcolor{versionCol}Code + CallGraph & \cellcolor{versionCol}73 (\textcolor{darkgreen}{$\uparrow$}23.73\%) & \cellcolor{versionCol}22 & \cellcolor{versionCol}5 & \cellcolor{versionCol}0 & \cellcolor{versionCol}1.32 \\
    & \cellcolor{versionCol}Code + VersionHistory & \cellcolor{versionCol}67 (\textcolor{darkgreen}{$\uparrow$}13.56\%) & \cellcolor{versionCol}28 & \cellcolor{versionCol}5 & \cellcolor{versionCol}0 & \cellcolor{versionCol}1.38 \\
    \midrule

    \multirow{3}{*}{Conciseness}
    & \cellcolor{baselineCol}Without Context & \cellcolor{baselineCol}55 & \cellcolor{baselineCol}31 & \cellcolor{baselineCol}14 & \cellcolor{baselineCol}0 & \cellcolor{baselineCol}1.59 \\
    & \cellcolor{versionCol}Code + CallGraph & \cellcolor{versionCol}69 (\textcolor{darkgreen}{$\uparrow$}25.45\%) & \cellcolor{versionCol}25 & \cellcolor{versionCol}4 & \cellcolor{versionCol}2 & \cellcolor{versionCol}1.39 \\
    & \cellcolor{versionCol}Code + VersionHistory & \cellcolor{versionCol}53 (\textcolor{red}{$\downarrow$}3.64\%) & \cellcolor{versionCol}33 & \cellcolor{versionCol}13 & \cellcolor{versionCol}1 & \cellcolor{versionCol}1.62 \\
    \midrule

    \multirow{3}{*}{\parbox{1.8cm}{Content \\ Adequacy}}
    & \cellcolor{baselineCol}Without Context & \cellcolor{baselineCol}43 & \cellcolor{baselineCol}43 & \cellcolor{baselineCol}12 & \cellcolor{baselineCol}2 & \cellcolor{baselineCol}1.74 \\
    & \cellcolor{versionCol}Code + CallGraph & \cellcolor{versionCol}56 (\textcolor{darkgreen}{$\uparrow$}30.23\%) & \cellcolor{versionCol}37 & \cellcolor{versionCol}6 & \cellcolor{versionCol}1 & \cellcolor{versionCol}1.52 \\
    & \cellcolor{versionCol}Code + VersionHistory & \cellcolor{versionCol}64 (\textcolor{darkgreen}{$\uparrow$}48.84\%) & \cellcolor{versionCol}29 & \cellcolor{versionCol}7 & \cellcolor{versionCol}0 & \cellcolor{versionCol}1.43 \\
    \bottomrule
    \end{tabular}

    \begin{tablenotes}
        \scriptsize
        \item Note: Lower Mean Rank indicates better performance.
        \item Legend: \colorbox{baselineCol}{\phantom{X}} Without Context (Baseline); \colorbox{versionCol}{\phantom{X}} Single-Context Scenarios.
        \item Values in the `\# Rank 1' column are followed by the percentage change (\textcolor{darkgreen}{$\uparrow$} increase, \textcolor{red}{$\downarrow$} decrease) relative to the 'Without Context' baseline.
    \end{tablenotes}

    \end{threeparttable}
}
\end{table}

Human evaluation in Table~\ref{tab:human_eval_rq1} and Table~\ref{tab:ranking_distribution_single} further confirms that explicitly encoding additional context substantially improves summarisation quality. For \textbf{Accuracy}, context-augmented models most frequently achieve top rankings. In particular, Code+CallGraph obtains 73 Rank~1 outcomes out of 100 samples (+23.7\% over the baseline) and shows a significant win rate of 25\% with a large effect size ($p=0.026$, Cliff’s $\delta=-0.51$). Code+VersionHistory also performs strongly (67 Rank~1s, +13.6\%), although its improvements are less consistent in pairwise tests. Beyond Rank~1 counts, these models produce fewer Rank~2/3 placements compared to the baseline, suggesting more stable superiority. Tie rates remain high (e.g., 68\% for Code+CallGraph), indicating that while improvements are real, multiple models can produce comparable summaries for some code snippets. For \textbf{Conciseness}, Code+CallGraph again outperforms the baseline (69 Rank~1s, +25.5\%; win=28\%, $p=0.007$, medium effect), whereas Code+VersionHistory shows no significant advantage, reflecting that benefits for conciseness are context-dependent. Finally, for \textbf{Content Adequacy}, Code+VersionHistory displays the strongest improvements, with 64 Rank~1s (+48.8\%) and significant improvements (31\% win, $p=0.001$, Cliff’s $\delta=-0.55$, large effect), while Code+CallGraph also improves adequacy with 56 Rank~1s (+30.2\%). Together, these findings show that adding contextual signals consistently enhances performance compared to the baseline, though the degree of benefit depends on which dimension of quality is considered: call graphs are especially valuable for accuracy and conciseness, while version history provides the largest gains in content adequacy.

\begin{tcolorbox}[left=1pt, top=1pt, right=1pt, bottom=1pt]
\textbf{Summary}:
\revised{Adding additional context, such as version history or call graph information, improves classification performance for all five foundational models and, after fine-tuning, for most of the seven LLMs. For summarisation, the benefit depends on the model family: all four foundational models improve with version history, whereas among the LLMs only Qwen2.5-Coder 7B and 14B benefit from prompt-level context (mainly call graph), and the other five decline on lexical and semantic metrics alike.} \revised{For the foundational models, version history is the most reliable single signal:} \revised{it improves performance in the same direction across all tasks and all five foundational models, with the largest gains for CodeT5 in clone detection (+15.92\% F1) and for summarisation on CodeSearchNet; the smaller SeSaMe gains are close to the resolution of that test set (Section~\ref{sec:threats})}. \revised{Furthermore, on the two classification tasks every LLM reaches its highest performance with context after fine-tuning on context-augmented inputs, indicating that task-specific training helps models exploit structural and temporal software artifacts.}
\end{tcolorbox}

\textbf{RQ2: How does combining multiple types of encoded context into source code representations affect the performance of these models?}
\label{sec:result_rq2}

Table~\ref{tab:result_rq2_clone_detection} illustrates the impact of encoding multiple types of context into source code representations on Code Clone Detection performance. Generally, combining multiple context types introduces code representations with more contextual information, resulting in better performance compared to the baseline model (without-context results). However, the degree of improvement varies significantly among different models, particularly compared to single-context representations analysed in RQ1, which may be explained by the sequence and type of combined contextual information.

Specifically, combining Version History and Method Age achieves significant improvements in F1-score for PLBART (+15.04\%) and GraphCodeBERT (+13.36\%). Furthermore, GraphCodeBERT also achieves +14.42\% improvement in F1-score when incorporating all three types of context (Version History, Call Graph, and Method Age). \revised{For ASTNN, the changes under combined context (between -1.68\% and +0.72\%) are within the resolution of the SeSaMe test set (Section~\ref{sec:threats}), so we do not conclude that it either benefits or suffers. Two factors could limit any gain: ASTNN's baseline is already high (F1-score: 0.8718), and its RNN and GRU components are known to have difficulty with long-range dependencies, which may hinder its use of a larger amount of contextual input such as multiple code versions.}

\begin{table}[htbp]
\centering
\scriptsize
\caption{Model performance on Code Clone Detection with combined context types (automated evaluation). \revised{Columns headed \% give the relative change of each context condition over the code-only baseline (W/O Ctx) of the same model and, for LLMs, the same mode (Z = zero-shot, F = fine-tuned); the table does not compare zero-shot against fine-tuned results.}}
\label{tab:result_rq2_clone_detection}
\renewcommand{\arraystretch}{1.2}
\adjustbox{max width=\textwidth, max totalheight=0.85\textheight}{%
\begin{tabular}{l|
c
>{\columncolor{baselineCol}}c
>{\columncolor{versionCol}}c
c
c
>{\columncolor{callgraphCol}}c
c
c
>{\columncolor{green!20}}c
c
}
\toprule
\rowcolor{headerCol}
\multicolumn{1}{c|}{\textbf{Models}} &
\multicolumn{10}{>{\columncolor{headerCol}}c}{\textbf{Code Clone Detection}} \\
\cmidrule(l){2-11}
\rowcolor{headerCol}
\multicolumn{1}{c|}{} &
\textbf{Metric} &
\textbf{W/O Ctx} &
\textbf{\shortstack{Vers. Hist.\\+ CG}} &
\textbf{\%} &
 &
\textbf{\shortstack{Vers. Hist.\\+ M. Age}} &
\textbf{\%} &
 &
\textbf{\shortstack{Vers. Hist.\\+ CG + M. Age}} &
\textbf{\%} \\
\midrule

\textbf{CodeBERT}
& P
  & 0.7778
  & 0.7895 & 1.50\%   & \barPercent{1.50}
  & 0.7391 & -4.97\%  & \barPercent{-4.97}
  & 0.7895 & 1.50\%
\\
& R
  & 0.6667
  & 0.7143 & 7.14\%   & \barPercent{7.14}
  & 0.8095 & 21.43\%  & \barPercent{21.43}
  & 0.7143 & 7.14\%
\\
& F1
  & 0.7180
  & 0.7500 & 4.46\%   & \barPercent{4.46}
  & 0.7727 & 7.63\%   & \barPercent{7.63}
  & 0.7500 & 4.46\%
\\
\midrule

\textbf{GraphCodeBERT}
& P
  & 0.9286
  & 0.9286 & 0.00\%   & \barPercent{0.00}
  & 0.9412 & 1.36\%   & \barPercent{1.36}
  & 0.8947 & -3.64\%
\\
& R
  & 0.6191
  & 0.6191 & 0.00\%   & \barPercent{0.00}
  & 0.7619 & 23.08\%  & \barPercent{23.08}
  & 0.8095 & 30.77\%
\\
& F1
  & 0.7429
  & 0.7429 & 0.00\%   & \barPercent{0.00}
  & 0.8421 & 13.36\%  & \barPercent{13.36}
  & 0.8500 & 14.42\%
\\
\midrule

\textbf{CodeT5}
& P
  & 0.8000
  & 0.8889 & 11.11\%  & \barPercent{11.11}
  & 0.7826 & -2.17\%  & \barPercent{-2.17}
  & 0.8421 & 5.26\%
\\
& R
  & 0.7619
  & 0.7619 & 0.00\%   & \barPercent{0.00}
  & 0.8571 & 12.50\%  & \barPercent{12.50}
  & 0.7619 & 0.00\%
\\
& F1
  & 0.7805
  & 0.8205 & 5.13\%   & \barPercent{5.13}
  & 0.8182 & 4.83\%   & \barPercent{4.83}
  & 0.8000 & 2.50\%
\\
\midrule

\textbf{PLBART}
& P
  & 0.9333
  & 0.9375 & 0.45\%   & \barPercent{0.45}
  & 1.0000 & 7.14\%   & \barPercent{7.14}
  & 0.9375 & 0.45\%
\\
& R
  & 0.6667
  & 0.7143 & 7.14\%   & \barPercent{7.14}
  & 0.8095 & 21.43\%  & \barPercent{21.43}
  & 0.7143 & 7.14\%
\\
& F1
  & 0.7778
  & 0.8108 & 4.25\%   & \barPercent{4.25}
  & 0.8947 & 15.04\%  & \barPercent{15.04}
  & 0.8108 & 4.25\%
\\
\midrule

\textbf{ASTNN}
& P
  & 0.9444
  & 0.8571 & -9.24\%  & \barPercent{-9.24}
  & 0.8571 & -9.24\%  & \barPercent{-9.24}
  & 0.9000 & -4.71\%
\\
& R
  & 0.8095
  & 0.8571 & 5.88\%   & \barPercent{5.88}
  & 0.8571 & 5.88\%   & \barPercent{5.88}
  & 0.8571 & 5.88\%
\\
& F1
  & 0.8718
  & 0.8571 & -1.68\%  & \barPercent{-1.68}
  & 0.8571 & -1.68\%  & \barPercent{-1.68}
  & 0.8781 & 0.72\%
\\
\rowcolor{blue!5}
\multicolumn{11}{l}{\revised{\textbf{Large Language Models (each context condition vs.\ the code-only baseline of the same mode; Z = zero-shot, F = fine-tuned)}}} \\ \midrule

\multirow{6}{*}{\textbf{CodeLlama-13B}}
& P (Z)
  & 0.8000
  & 0.3684
  & -53.9\%
  & \barPercent{-53.9}
  & 0.5000
  & -37.5\%
  & \barPercent{-37.5}
  & 0.3333
  & -58.3\%
\\
& R (Z)
  & 0.4444
  & 0.3889
  & -12.5\%
  & \barPercent{-12.5}
  & 0.3333
  & -25.0\%
  & \barPercent{-25.0}
  & 0.1667
  & -62.5\%
\\
& F1 (Z)
  & 0.5714
  & 0.3784
  & -33.8\%
  & \barPercent{-33.8}
  & 0.4000
  & -30.0\%
  & \barPercent{-30.0}
  & 0.2222
  & -61.1\%
\\
\cmidrule(lr){2-11}
& P (F)
  & 0.6364
  & 0.5625
  & -11.6\%
  & \barPercent{-11.6}
  & 0.7000
  & +10.0\%
  & \barPercent{10.0}
  & 0.8571
  & +34.7\%
\\
& R (F)
  & 0.7778
  & 1.0000
  & +28.6\%
  & \barPercent{28.6}
  & 0.7778
  & +0.0\%
  & \barPercent{0.0}
  & 0.6667
  & -14.3\%
\\
& F1 (F)
  & 0.7000
  & 0.7200
  & +2.9\%
  & \barPercent{2.9}
  & 0.7368
  & +5.3\%
  & \barPercent{5.3}
  & 0.7500
  & +7.1\%
\\
\midrule

\multirow{6}{*}{\textbf{Qwen-14B}}
& P (Z)
  & 1.0000
  & 1.0000
  & +0.0\%
  & \barPercent{0.0}
  & 1.0000
  & +0.0\%
  & \barPercent{0.0}
  & 1.0000
  & +0.0\%
\\
& R (Z)
  & 0.0556
  & 0.0556
  & +0.0\%
  & \barPercent{0.0}
  & 0.0556
  & +0.0\%
  & \barPercent{0.0}
  & 0.0556
  & +0.0\%
\\
& F1 (Z)
  & 0.1053
  & 0.1053
  & +0.0\%
  & \barPercent{0.0}
  & 0.1053
  & +0.0\%
  & \barPercent{0.0}
  & 0.1053
  & +0.0\%
\\
\cmidrule(lr){2-11}
& P (F)
  & 0.7143
  & 0.8823
  & +23.5\%
  & \barPercent{23.5}
  & 0.7000
  & -2.0\%
  & \barPercent{-2.0}
  & 0.8889
  & +24.4\%
\\
& R (F)
  & 0.8333
  & 0.8333
  & +0.0\%
  & \barPercent{0.0}
  & 0.7778
  & -6.7\%
  & \barPercent{-6.7}
  & 0.8889
  & +6.7\%
\\
& F1 (F)
  & 0.7692
  & 0.8571
  & +11.4\%
  & \barPercent{11.4}
  & 0.7368
  & -4.2\%
  & \barPercent{-4.2}
  & 0.8889
  & +15.6\%
\\
\midrule

\multirow{6}{*}{\textbf{CodeLlama-7B}}
& P (Z)
  & 0.1667
  & 0.2885
  & +73.1\%
  & \barPercent{73.1}
  & 0.1765
  & +5.9\%
  & \barPercent{5.9}
  & 0.2000
  & +20.0\%
\\
& R (Z)
  & 0.5000
  & 0.8333
  & +66.7\%
  & \barPercent{66.7}
  & 0.3333
  & -33.3\%
  & \barPercent{-33.3}
  & 0.5000
  & +0.0\%
\\
& F1 (Z)
  & 0.2500
  & 0.4286
  & +71.4\%
  & \barPercent{71.4}
  & 0.2308
  & -7.7\%
  & \barPercent{-7.7}
  & 0.2857
  & +14.3\%
\\
\cmidrule(lr){2-11}
& P (F)
  & 0.8125
  & 0.7000
  & -13.8\%
  & \barPercent{-13.8}
  & 0.4706
  & -42.1\%
  & \barPercent{-42.1}
  & 0.4474
  & -44.9\%
\\
& R (F)
  & 0.7222
  & 0.7778
  & +7.7\%
  & \barPercent{7.7}
  & 0.8889
  & +23.1\%
  & \barPercent{23.1}
  & 0.9444
  & +30.8\%
\\
& F1 (F)
  & 0.7647
  & 0.7368
  & -3.6\%
  & \barPercent{-3.6}
  & 0.6154
  & -19.5\%
  & \barPercent{-19.5}
  & 0.6071
  & -20.6\%
\\
\midrule

\multirow{6}{*}{\textbf{Qwen-7B}}
& P (Z)
  & 1.0000
  & 0.5714
  & -42.9\%
  & \barPercent{-42.9}
  & 0.5000
  & -50.0\%
  & \barPercent{-50.0}
  & 0.5714
  & -42.9\%
\\
& R (Z)
  & 0.0556
  & 0.2222
  & +299.6\%
  & \barPercent{100.0}
  & 0.2222
  & +299.6\%
  & \barPercent{100.0}
  & 0.2222
  & +299.6\%
\\
& F1 (Z)
  & 0.1053
  & 0.3200
  & +203.9\%
  & \barPercent{100.0}
  & 0.3077
  & +192.2\%
  & \barPercent{100.0}
  & 0.3200
  & +203.9\%
\\
\cmidrule(lr){2-11}
& P (F)
  & 0.7222
  & 0.9167
  & +26.9\%
  & \barPercent{26.9}
  & 1.0000
  & +38.5\%
  & \barPercent{38.5}
  & 1.0000
  & +38.5\%
\\
& R (F)
  & 0.7222
  & 0.6111
  & -15.4\%
  & \barPercent{-15.4}
  & 0.5000
  & -30.8\%
  & \barPercent{-30.8}
  & 0.5556
  & -23.1\%
\\
& F1 (F)
  & 0.7222
  & 0.7333
  & +1.5\%
  & \barPercent{1.5}
  & 0.6667
  & -7.7\%
  & \barPercent{-7.7}
  & 0.7143
  & -1.1\%
\\
\midrule

\multirow{6}{*}{\textbf{DeepSeek-6.7B}}
& P (Z)
  & 1.0000
  & 0.4211
  & -57.9\%
  & \barPercent{-57.9}
  & 0.5500
  & -45.0\%
  & \barPercent{-45.0}
  & 0.2778
  & -72.2\%
\\
& R (Z)
  & 0.1111
  & 0.4444
  & +300.0\%
  & \barPercent{100.0}
  & 0.6111
  & +450.0\%
  & \barPercent{100.0}
  & 0.5556
  & +400.1\%
\\
& F1 (Z)
  & 0.2000
  & 0.4324
  & +116.2\%
  & \barPercent{100.0}
  & 0.5790
  & +189.5\%
  & \barPercent{100.0}
  & 0.3704
  & +85.2\%
\\
\cmidrule(lr){2-11}
& P (F)
  & 1.0000
  & 0.5625
  & -43.8\%
  & \barPercent{-43.8}
  & 0.8125
  & -18.8\%
  & \barPercent{-18.8}
  & 0.7647
  & -23.5\%
\\
& R (F)
  & 0.8889
  & 1.0000
  & +12.5\%
  & \barPercent{12.5}
  & 0.7222
  & -18.8\%
  & \barPercent{-18.8}
  & 0.7222
  & -18.8\%
\\
& F1 (F)
  & 0.9412
  & 0.7200
  & -23.5\%
  & \barPercent{-23.5}
  & 0.7647
  & -18.8\%
  & \barPercent{-18.8}
  & 0.7429
  & -21.1\%
\\
\midrule

\multirow{6}{*}{\textbf{Qwen-1.5B}}
& P (Z)
  & 0.3333
  & 0.1818
  & -45.5\%
  & \barPercent{-45.5}
  & 0.3684
  & +10.5\%
  & \barPercent{10.5}
  & 0.3333
  & +0.0\%
\\
& R (Z)
  & 0.2778
  & 0.2222
  & -20.0\%
  & \barPercent{-20.0}
  & 0.3889
  & +40.0\%
  & \barPercent{40.0}
  & 0.3333
  & +20.0\%
\\
& F1 (Z)
  & 0.3030
  & 0.2000
  & -34.0\%
  & \barPercent{-34.0}
  & 0.3784
  & +24.9\%
  & \barPercent{24.9}
  & 0.3333
  & +10.0\%
\\
\cmidrule(lr){2-11}
& P (F)
  & 0.6667
  & 0.7778
  & +16.7\%
  & \barPercent{16.7}
  & 0.6667
  & +0.0\%
  & \barPercent{0.0}
  & 0.7647
  & +14.7\%
\\
& R (F)
  & 0.6667
  & 0.7778
  & +16.7\%
  & \barPercent{16.7}
  & 0.7778
  & +16.7\%
  & \barPercent{16.7}
  & 0.7222
  & +8.3\%
\\
& F1 (F)
  & 0.6667
  & 0.7778
  & +16.7\%
  & \barPercent{16.7}
  & 0.7179
  & +7.7\%
  & \barPercent{7.7}
  & 0.7429
  & +11.4\%
\\
\midrule
\multirow{6}{*}{\textbf{CodeLlama-34B}}
& P (Z)
  & 1.0000
  & 0.8333
  & -16.7\%
  & \barPercent{-16.7}
  & 1.0000
  & +0.0\%
  & \barPercent{0.0}
  & 1.0000
  & +0.0\%
\\
& R (Z)
  & 0.1111
  & 0.5556
  & +400.0\%
  & \barPercent{400.0}
  & 0.3333
  & +200.0\%
  & \barPercent{200.0}
  & 0.1111
  & +0.0\%
\\
& F1 (Z)
  & 0.2000
  & 0.6667
  & +233.3\%
  & \barPercent{233.3}
  & 0.5000
  & +150.0\%
  & \barPercent{150.0}
  & 0.2000
  & +0.0\%
\\
\cmidrule(lr){2-11}
& P (F)
  & 0.8750
  & 1.0000
  & +14.3\%
  & \barPercent{14.3}
  & 0.8571
  & -2.0\%
  & \barPercent{-2.0}
  & 1.0000
  & +14.3\%
\\
& R (F)
  & 0.7778
  & 0.5000
  & -35.7\%
  & \barPercent{-35.7}
  & 0.6667
  & -14.3\%
  & \barPercent{-14.3}
  & 0.5556
  & -28.6\%
\\
& F1 (F)
  & 0.8235
  & 0.6667
  & -19.0\%
  & \barPercent{-19.0}
  & 0.7500
  & -8.9\%
  & \barPercent{-8.9}
  & 0.7143
  & -13.3\%
\\
\midrule
\bottomrule
\end{tabular}%
}
\end{table}

\begin{table}[htbp]
\centering
\scriptsize
\caption{Model performance on Code Classification with combined context types (automated evaluation). \revised{Columns headed \% give the relative change of each context condition over the code-only baseline (W/O Ctx) of the same model and, for LLMs, the same mode (Z = zero-shot, F = fine-tuned); the table does not compare zero-shot against fine-tuned results.}}
\label{tab:result_rq2_classification}
\renewcommand{\arraystretch}{1.2}
\adjustbox{max width=\textwidth, max totalheight=0.85\textheight}{%
\begin{tabular}{l|
c
>{\columncolor{baselineCol}}c
>{\columncolor{versionCol}}c
c
c
>{\columncolor{callgraphCol}}c
c
c
>{\columncolor{green!20}}c
c
}
\toprule
\rowcolor{headerCol}
\multicolumn{1}{c|}{\textbf{Models}} &
\multicolumn{10}{>{\columncolor{headerCol}}c}{\textbf{Code Classification}} \\
\cmidrule(l){2-11}
\rowcolor{headerCol}
\multicolumn{1}{c|}{} &
\textbf{Metric} &
\textbf{W/O Ctx} &
\textbf{\shortstack{Vers. Hist.\\+ CG}} &
\textbf{\%} &
 &
\textbf{\shortstack{Vers. Hist.\\+ M. Age}} &
\textbf{\%} &
 &
\textbf{\shortstack{Vers. Hist.\\+ CG + M. Age}} &
\textbf{\%} \\
\midrule

\textbf{CodeBERT}
& P
  & 0.7103
  & 0.7512 & 5.75\%   & \barPercentThirty{5.75}
  & 0.6881 & -3.14\%  & \barPercentThirty{-3.14}
  & 0.8504 & 19.72\%
\\
& R
  & 0.6498
  & 0.7148 & 10.01\%  & \barPercentThirty{10.01}
  & 0.6562 & 0.99\%   & \barPercentThirty{0.99}
  & 0.7522 & 15.76\%
\\
& macro-F1
  & 0.6698
  & 0.7304 & 9.04\%   & \barPercentThirty{9.04}
  & 0.6624 & -1.11\%  & \barPercentThirty{-1.11}
  & 0.7813 & 16.65\%
\\
\midrule

\textbf{GraphCodeBERT}
& P
  & 0.8214
  & 0.8168 & -0.56\%  & \barPercentThirty{-0.56}
  & 0.8567 & 4.30\%   & \barPercentThirty{4.30}
  & 0.8632 & 5.09\%
\\
& R
  & 0.7581
  & 0.7666 & 1.12\%   & \barPercentThirty{1.12}
  & 0.7591 & 0.12\%   & \barPercentThirty{0.12}
  & 0.7188 & -5.19\%
\\
& macro-F1
  & 0.7661
  & 0.7797 & 1.78\%   & \barPercentThirty{1.78}
  & 0.7835 & 2.28\%   & \barPercentThirty{2.28}
  & 0.7645 & -0.21\%
\\
\midrule

\textbf{CodeT5}
& P
  & 0.7005
  & 0.7999 & 14.18\%  & \barPercentThirty{14.18}
  & 0.7777 & 11.02\%  & \barPercentThirty{11.02}
  & 0.8419 & 20.18\%
\\
& R
  & 0.6364
  & 0.7194 & 13.05\%  & \barPercentThirty{13.05}
  & 0.6865 & 7.87\%   & \barPercentThirty{7.87}
  & 0.7889 & 23.97\%
\\
& macro-F1
  & 0.6569
  & 0.7517 & 14.44\%  & \barPercentThirty{14.44}
  & 0.7129 & 8.53\%   & \barPercentThirty{8.53}
  & 0.7980 & 21.48\%
\\
\midrule

\textbf{PLBART}
& P
  & 0.7064
  & 0.8009 & 13.38\%  & \barPercentThirty{13.38}
  & 0.7566 & 7.11\%   & \barPercentThirty{7.11}
  & 0.8581 & 21.48\%
\\
& R
  & 0.6837
  & 0.7818 & 14.35\%  & \barPercentThirty{14.35}
  & 0.7115 & 4.07\%   & \barPercentThirty{4.07}
  & 0.8062 & 17.92\%
\\
& macro-F1
  & 0.6822
  & 0.7842 & 14.95\%  & \barPercentThirty{14.95}
  & 0.7195 & 5.46\%   & \barPercentThirty{5.46}
  & 0.8191 & 20.07\%
\\
\midrule

\textbf{ASTNN}
& P
  & 0.5366
  & 0.5903 & 10.01\%  & \barPercentThirty{10.01}
  & 0.6772 & 26.21\%  & \barPercentThirty{26.21}
  & 0.6157 & 14.74\%
\\
& R
  & 0.4793
  & 0.5162 & 7.71\%   & \barPercentThirty{7.71}
  & 0.5415 & 12.99\%  & \barPercentThirty{12.99}
  & 0.5261 & 9.78\%
\\
& macro-F1
  & 0.4887
  & 0.5376 & 10.00\%  & \barPercentThirty{10.00}
  & 0.5679 & 16.20\%  & \barPercentThirty{16.20}
  & 0.5505 & 12.64\%
\\
\rowcolor{blue!5}
\multicolumn{11}{l}{\revised{\textbf{Large Language Models (each context condition vs.\ the code-only baseline of the same mode; Z = zero-shot, F = fine-tuned)}}} \\ \midrule

\multirow{6}{*}{\textbf{CodeLlama-34B}}
& P (Z)
  & 0.3470
  & 0.2598
  & -25.1\%
  & \barPercentThirty{-25.1}
  & 0.3106
  & -10.5\%
  & \barPercentThirty{-10.5}
  & 0.3045
  & -12.2\%
\\
& R (Z)
  & 0.3795
  & 0.2555
  & -32.7\%
  & \barPercentThirty{-32.7}
  & 0.3285
  & -13.4\%
  & \barPercentThirty{-13.4}
  & 0.2863
  & -24.6\%
\\
& mF1 (Z)
  & 0.2903
  & 0.1890
  & -34.9\%
  & \barPercentThirty{-34.9}
  & 0.2489
  & -14.3\%
  & \barPercentThirty{-14.3}
  & 0.2152
  & -25.8\%
\\
\cmidrule(lr){2-11}
& P (F)
  & 0.7601
  & 0.8315
  & +9.4\%
  & \barPercentThirty{9.4}
  & 0.8940
  & +17.6\%
  & \barPercentThirty{17.6}
  & 0.9174
  & +20.7\%
\\
& R (F)
  & 0.6706
  & 0.7044
  & +5.0\%
  & \barPercentThirty{5.0}
  & 0.8087
  & +20.6\%
  & \barPercentThirty{20.6}
  & 0.8164
  & +21.7\%
\\
& mF1 (F)
  & 0.6922
  & 0.7385
  & +6.7\%
  & \barPercentThirty{6.7}
  & 0.8316
  & +20.1\%
  & \barPercentThirty{20.1}
  & 0.8518
  & +23.1\%
\\
\midrule

\multirow{6}{*}{\textbf{CodeLlama-13B}}
& P (Z)
  & 0.3016
  & 0.4174
  & +38.4\%
  & \barPercentThirty{38.4}
  & 0.4377
  & +45.1\%
  & \barPercentThirty{45.1}
  & 0.4841
  & +60.5\%
\\
& R (Z)
  & 0.2801
  & 0.3464
  & +23.7\%
  & \barPercentThirty{23.7}
  & 0.3170
  & +13.2\%
  & \barPercentThirty{13.2}
  & 0.3381
  & +20.7\%
\\
& mF1 (Z)
  & 0.1771
  & 0.2499
  & +41.1\%
  & \barPercentThirty{41.1}
  & 0.2323
  & +31.2\%
  & \barPercentThirty{31.2}
  & 0.2547
  & +43.8\%
\\
\cmidrule(lr){2-11}
& P (F)
  & 0.6908
  & 0.6595
  & -4.5\%
  & \barPercentThirty{-4.5}
  & 0.6734
  & -2.5\%
  & \barPercentThirty{-2.5}
  & 0.6736
  & -2.5\%
\\
& R (F)
  & 0.6128
  & 0.5959
  & -2.8\%
  & \barPercentThirty{-2.8}
  & 0.5593
  & -8.7\%
  & \barPercentThirty{-8.7}
  & 0.6244
  & +1.9\%
\\
& mF1 (F)
  & 0.6207
  & 0.6172
  & -0.6\%
  & \barPercentThirty{-0.6}
  & 0.5818
  & -6.3\%
  & \barPercentThirty{-6.3}
  & 0.6429
  & +3.6\%
\\
\midrule

\multirow{6}{*}{\textbf{Qwen-14B}}
& P (Z)
  & 0.4568
  & 0.4813
  & +5.4\%
  & \barPercentThirty{5.4}
  & 0.5014
  & +9.8\%
  & \barPercentThirty{9.8}
  & 0.4824
  & +5.6\%
\\
& R (Z)
  & 0.5428
  & 0.5114
  & -5.8\%
  & \barPercentThirty{-5.8}
  & 0.5222
  & -3.8\%
  & \barPercentThirty{-3.8}
  & 0.5001
  & -7.9\%
\\
& mF1 (Z)
  & 0.4440
  & 0.4481
  & +0.9\%
  & \barPercentThirty{0.9}
  & 0.4635
  & +4.4\%
  & \barPercentThirty{4.4}
  & 0.4462
  & +0.5\%
\\
\cmidrule(lr){2-11}
& P (F)
  & 0.8783
  & 0.8738
  & -0.5\%
  & \barPercentThirty{-0.5}
  & 0.8717
  & -0.8\%
  & \barPercentThirty{-0.8}
  & 0.8337
  & -5.1\%
\\
& R (F)
  & 0.8003
  & 0.8128
  & +1.6\%
  & \barPercentThirty{1.6}
  & 0.8122
  & +1.5\%
  & \barPercentThirty{1.5}
  & 0.8338
  & +4.2\%
\\
& mF1 (F)
  & 0.8110
  & 0.8324
  & +2.6\%
  & \barPercentThirty{2.6}
  & 0.8292
  & +2.2\%
  & \barPercentThirty{2.2}
  & 0.8244
  & +1.7\%
\\
\midrule

\multirow{6}{*}{\textbf{CodeLlama-7B}}
& P (Z)
  & 0.4103
  & 0.5504
  & +34.1\%
  & \barPercentThirty{34.1}
  & 0.4900
  & +19.4\%
  & \barPercentThirty{19.4}
  & 0.4238
  & +3.3\%
\\
& R (Z)
  & 0.2304
  & 0.2010
  & -12.8\%
  & \barPercentThirty{-12.8}
  & 0.2490
  & +8.1\%
  & \barPercentThirty{8.1}
  & 0.2229
  & -3.3\%
\\
& mF1 (Z)
  & 0.1877
  & 0.1747
  & -6.9\%
  & \barPercentThirty{-6.9}
  & 0.2048
  & +9.1\%
  & \barPercentThirty{9.1}
  & 0.1759
  & -6.3\%
\\
\cmidrule(lr){2-11}
& P (F)
  & 0.6343
  & 0.6767
  & +6.7\%
  & \barPercentThirty{6.7}
  & 0.5592
  & -11.8\%
  & \barPercentThirty{-11.8}
  & 0.6698
  & +5.6\%
\\
& R (F)
  & 0.6485
  & 0.6374
  & -1.7\%
  & \barPercentThirty{-1.7}
  & 0.4898
  & -24.5\%
  & \barPercentThirty{-24.5}
  & 0.5995
  & -7.6\%
\\
& mF1 (F)
  & 0.6308
  & 0.6506
  & +3.1\%
  & \barPercentThirty{3.1}
  & 0.4506
  & -28.6\%
  & \barPercentThirty{-28.6}
  & 0.6106
  & -3.2\%
\\
\midrule

\multirow{6}{*}{\textbf{Qwen-7B}}
& P (Z)
  & 0.4610
  & 0.4811
  & +4.4\%
  & \barPercentThirty{4.4}
  & 0.4731
  & +2.6\%
  & \barPercentThirty{2.6}
  & 0.4373
  & -5.1\%
\\
& R (Z)
  & 0.4169
  & 0.3169
  & -24.0\%
  & \barPercentThirty{-24.0}
  & 0.3500
  & -16.0\%
  & \barPercentThirty{-16.0}
  & 0.3132
  & -24.9\%
\\
& mF1 (Z)
  & 0.3676
  & 0.3288
  & -10.6\%
  & \barPercentThirty{-10.6}
  & 0.3359
  & -8.6\%
  & \barPercentThirty{-8.6}
  & 0.3052
  & -17.0\%
\\
\cmidrule(lr){2-11}
& P (F)
  & 0.7395
  & 0.7128
  & -3.6\%
  & \barPercentThirty{-3.6}
  & 0.8375
  & +13.3\%
  & \barPercentThirty{13.3}
  & 0.7916
  & +7.0\%
\\
& R (F)
  & 0.7872
  & 0.7033
  & -10.7\%
  & \barPercentThirty{-10.7}
  & 0.8152
  & +3.6\%
  & \barPercentThirty{3.6}
  & 0.7898
  & +0.3\%
\\
& mF1 (F)
  & 0.7518
  & 0.7011
  & -6.7\%
  & \barPercentThirty{-6.7}
  & 0.8187
  & +8.9\%
  & \barPercentThirty{8.9}
  & 0.7866
  & +4.6\%
\\
\midrule

\multirow{6}{*}{\textbf{DeepSeek-6.7B}}
& P (Z)
  & 0.4193
  & 0.2322
  & -44.6\%
  & \barPercentThirty{-44.6}
  & 0.1983
  & -52.7\%
  & \barPercentThirty{-52.7}
  & 0.1271
  & -69.7\%
\\
& R (Z)
  & 0.1578
  & 0.1107
  & -29.8\%
  & \barPercentThirty{-29.8}
  & 0.1475
  & -6.5\%
  & \barPercentThirty{-6.5}
  & 0.0988
  & -37.4\%
\\
& mF1 (Z)
  & 0.1889
  & 0.1303
  & -31.0\%
  & \barPercentThirty{-31.0}
  & 0.1436
  & -24.0\%
  & \barPercentThirty{-24.0}
  & 0.0960
  & -49.2\%
\\
\cmidrule(lr){2-11}
& P (F)
  & 0.4911
  & 0.7358
  & +49.8\%
  & \barPercentThirty{49.8}
  & 0.7214
  & +46.9\%
  & \barPercentThirty{46.9}
  & 0.8236
  & +67.7\%
\\
& R (F)
  & 0.4746
  & 0.6277
  & +32.3\%
  & \barPercentThirty{32.3}
  & 0.6860
  & +44.5\%
  & \barPercentThirty{44.5}
  & 0.6655
  & +40.2\%
\\
& mF1 (F)
  & 0.4772
  & 0.6654
  & +39.4\%
  & \barPercentThirty{39.4}
  & 0.6994
  & +46.6\%
  & \barPercentThirty{46.6}
  & 0.7251
  & +51.9\%
\\
\midrule

\multirow{6}{*}{\textbf{Qwen-1.5B}}
& P (Z)
  & 0.1149
  & 0.1152
  & +0.3\%
  & \barPercentThirty{0.3}
  & 0.1148
  & -0.1\%
  & \barPercentThirty{-0.1}
  & 0.1186
  & +3.2\%
\\
& R (Z)
  & 0.2121
  & 0.1954
  & -7.9\%
  & \barPercentThirty{-7.9}
  & 0.2121
  & +0.0\%
  & \barPercentThirty{0.0}
  & 0.2083
  & -1.8\%
\\
& mF1 (Z)
  & 0.0782
  & 0.0897
  & +14.7\%
  & \barPercentThirty{14.7}
  & 0.0893
  & +14.2\%
  & \barPercentThirty{14.2}
  & 0.0936
  & +19.7\%
\\
\cmidrule(lr){2-11}
& P (F)
  & 0.4669
  & 0.5994
  & +28.4\%
  & \barPercentThirty{28.4}
  & 0.6304
  & +35.0\%
  & \barPercentThirty{35.0}
  & 0.6996
  & +49.8\%
\\
& R (F)
  & 0.5780
  & 0.5942
  & +2.8\%
  & \barPercentThirty{2.8}
  & 0.6261
  & +8.3\%
  & \barPercentThirty{8.3}
  & 0.6727
  & +16.4\%
\\
& mF1 (F)
  & 0.4825
  & 0.5793
  & +20.1\%
  & \barPercentThirty{20.1}
  & 0.6084
  & +26.1\%
  & \barPercentThirty{26.1}
  & 0.6728
  & +39.4\%
\\
\midrule
\bottomrule
\end{tabular}%
}
\end{table}

The result in the Code Classification task (Table~\ref{tab:result_rq2_classification}) presents the impact of combining multiple types of additional context into source code representations for the Code Classification task. Overall, encoding multiple context types enhances classification performance across most context combinations and models, with macro-F1 score improvements ranging from modest (+2.28\%) to substantial (+21.48\%). Notably, CodeT5 significantly benefits from combining Version History, Call Graph, and Method Age, improving its macro-F1 score by +21.48\% (from 0.6569 to 0.7980).

The four models, including CodeBERT, CodeT5, PLBART, and ASTNN, indicate varying degrees of improvement depending on the specific context combination; however, the most significant enhancements are consistently observed when all three types of context (i.e. Version History, Call Graph, and Method Age) are combined. For instance, CodeBERT (+16.65\%), PLBART (+20.07\%), and ASTNN (+12.64\%) achieve their highest improvements with this three-context combination.

\revised{GraphCodeBERT is the exception: its changes under the three combinations (+1.78\%, +2.28\%, and -0.21\%) are all within the resolution of the SeSaMe test set (Section~\ref{sec:threats}), so we do not read them as evidence for or against combined context on this model.}

\begin{table}[htbp]
\centering
\scriptsize

\caption{Automated evaluation of Code Summarisation with combined context types (BLEU-4, METEOR, ROUGE-L, and BERTScore-F1). \revised{Columns headed \% give the relative change of each context condition over the code-only baseline (W/O Ctx) of the same model and, for LLMs, the same mode (Z = zero-shot, F = fine-tuned); the table does not compare zero-shot against fine-tuned results.}}
\label{tab:result_rq2_summarisation}

\renewcommand{\arraystretch}{1.0}

\resizebox{0.9\textwidth}{!}{%
    \begin{threeparttable}

    \begin{tabular}{l >{\centering\arraybackslash\small}p{1em} l |
    >{\columncolor{baselineCol}}c
    >{\columncolor{versionCol}}c
    c
    c
    >{\columncolor{callgraphCol}}c
    c
    c
    >{\columncolor{green!20}}c
    c
    c
    >{\columncolor{blue!20}}c
    c
    c
    >{\columncolor{purple!20}}c
    c
    c
    }
    \toprule
    \rowcolor{headerCol}
    \multicolumn{2}{l}{\textbf{Models}} & \textbf{Metrics} &
    \textbf{W/O Ctx} &
    \textbf{\shortstack{Vers. Hist.\\+\\CG}} &
    \% &
    & \textbf{\shortstack{Vers. Hist.\\+\\M. Age}} &
    \% &
    & \textbf{\shortstack{Vers. Hist.\\+\\CG\\+\\M. Age}} &
    \% &
    & \textbf{\shortstack{CG\\+\\Vers. Hist.}} &
    \% &
    & \textbf{\shortstack{CG\\+\\Vers. Hist.\\+\\M. Age}} &
    \% &
    \\
    \midrule

    \multirow{4}{*}{\textbf{CodeBERT}}
    & & BLEU-4
      & 17.12 & 17.26 & 0.82\% & \barPercentSeven{0.82}
      & 17.34 & 1.29\% & \barPercentSeven{1.29}
      & 17.42 & 1.75\% & \barPercentSeven{1.75}
      & 17.23 & 0.64\% & \barPercentSeven{0.64}
      & 17.29 & 0.99\% & \barPercentSeven{0.99} \\
    & & BERTScore-F1
      & 0.2570 & 0.2537 & -1.28\% & \barPercentSeven{-1.28}
      & 0.2516 & -2.11\% & \barPercentSeven{-2.11}
      & 0.2504 & -2.56\% & \barPercentSeven{-2.56}
      & 0.2532 & -1.47\% & \barPercentSeven{-1.47}
      & 0.2523 & -1.82\% & \barPercentSeven{-1.82} \\
    & & ROUGE-L
      & 0.3058 & 0.3035 & -0.76\% & \barPercentSeven{-0.76}
      & 0.3047 & -0.38\% & \barPercentSeven{-0.38}
      & 0.3059 & 0.02\% & \barPercentSeven{0.02}
      & 0.3020 & -1.27\% & \barPercentSeven{-1.27}
      & 0.3024 & -1.11\% & \barPercentSeven{-1.11} \\
    & & METEOR
      & 0.2626 & 0.2636 & 0.40\% & \barPercentSeven{0.40}
      & 0.2610 & -0.61\% & \barPercentSeven{-0.61}
      & 0.2614 & -0.46\% & \barPercentSeven{-0.46}
      & 0.2626 & 0.00\% & \barPercentSeven{0.00}
      & 0.2614 & -0.45\% & \barPercentSeven{-0.45} \\
    \midrule

    \multirow{4}{*}{\textbf{GraphCodeBERT}}
    & & BLEU-4
      & 17.85 & 18.18 & 1.85\% & \barPercentSeven{1.85}
      & 17.74 & -0.62\% & \barPercentSeven{-0.62}
      & 18.09 & 1.34\% & \barPercentSeven{1.34}
      & 17.74 & -0.62\% & \barPercentSeven{-0.62}
      & 17.97 & 0.67\% & \barPercentSeven{0.67} \\
    & & BERTScore-F1
      & 0.3993 & 0.4045 & 1.30\% & \barPercentSeven{1.30}
      & 0.4022 & 0.73\% & \barPercentSeven{0.73}
      & 0.4072 & 1.98\% & \barPercentSeven{1.98}
      & 0.3966 & -0.68\% & \barPercentSeven{-0.68}
      & 0.4053 & 1.50\% & \barPercentSeven{1.50} \\
    & & ROUGE-L
      & 0.3604 & 0.3638 & 0.94\% & \barPercentSeven{0.94}
      & 0.3603 & -0.03\% & \barPercentSeven{-0.03}
      & 0.3639 & 0.97\% & \barPercentSeven{0.97}
      & 0.3588 & -0.44\% & \barPercentSeven{-0.44}
      & 0.3626 & 0.61\% & \barPercentSeven{0.61} \\
    & & METEOR
      & 0.2715 & 0.2846 & 4.83\% & \barPercentSeven{4.83}
      & 0.2804 & 3.28\% & \barPercentSeven{3.28}
      & 0.2781 & 2.43\% & \barPercentSeven{2.43}
      & 0.2704 & -0.41\% & \barPercentSeven{-0.41}
      & 0.2805 & 3.31\% & \barPercentSeven{3.31} \\
    \midrule

    \multirow{4}{*}{\textbf{CodeT5}}
    & & BLEU-4
      & 20.02 & 19.94 & -0.40\% & \barPercentSeven{-0.40}
      & 21.18 & 5.79\% & \barPercentSeven{5.79}
      & 20.41 & 1.95\% & \barPercentSeven{1.95}
      & 20.32 & 1.50\% & \barPercentSeven{1.50}
      & 20.57 & 2.75\% & \barPercentSeven{2.75} \\
    & & BERTScore-F1
      & 0.4455 & 0.4420 & -0.79\% & \barPercentSeven{-0.79}
      & 0.4516 & 1.37\% & \barPercentSeven{1.37}
      & 0.4434 & -0.47\% & \barPercentSeven{-0.47}
      & 0.4468 & 0.29\% & \barPercentSeven{0.29}
      & 0.4506 & 1.14\% & \barPercentSeven{1.14} \\
    & & ROUGE-L
      & 0.4007 & 0.3951 & -1.40\% & \barPercentSeven{-1.40}
      & 0.4041 & 0.85\% & \barPercentSeven{0.85}
      & 0.4015 & 0.20\% & \barPercentSeven{0.20}
      & 0.3992 & -0.37\% & \barPercentSeven{-0.37}
      & 0.4051 & 1.10\% & \barPercentSeven{1.10} \\
    & & METEOR
      & 0.3344 & 0.3360 & 0.48\% & \barPercentSeven{0.48}
      & 0.3446 & 3.05\% & \barPercentSeven{3.05}
      & 0.3390 & 1.38\% & \barPercentSeven{1.38}
      & 0.3404 & 1.79\% & \barPercentSeven{1.79}
      & 0.3460 & 3.47\% & \barPercentSeven{3.47} \\
    \midrule

    \multirow{4}{*}{\textbf{PLBART}}
    & & BLEU-4
      & 15.71 & 15.72 & 0.06\% & \barPercentSeven{0.06}
      & 16.05 & 2.16\% & \barPercentSeven{2.16}
      & 15.85 & 0.89\% & \barPercentSeven{0.89}
      & 15.72 & 0.06\% & \barPercentSeven{0.06}
      & 15.89 & 1.15\% & \barPercentSeven{1.15} \\
    & & BERTScore-F1
      & 0.3640 & 0.3632 & -0.22\% & \barPercentSeven{-0.22}
      & 0.3685 & 1.24\% & \barPercentSeven{1.24}
      & 0.3626 & -0.38\% & \barPercentSeven{-0.38}
      & 0.3559 & -2.23\% & \barPercentSeven{-2.23}
      & 0.3647 & 0.19\% & \barPercentSeven{0.19} \\
    & & ROUGE-L
      & 0.3163 & 0.3108 & -1.74\% & \barPercentSeven{-1.74}
      & 0.3178 & 0.47\% & \barPercentSeven{0.47}
      & 0.3122 & -1.30\% & \barPercentSeven{-1.30}
      & 0.3049 & -3.60\% & \barPercentSeven{-3.60}
      & 0.3144 & -0.60\% & \barPercentSeven{-0.60} \\
    & & METEOR
      & 0.2367 & 0.2338 & -1.23\% & \barPercentSeven{-1.23}
      & 0.2380 & 0.55\% & \barPercentSeven{0.55}
      & 0.2340 & -1.14\% & \barPercentSeven{-1.14}
      & 0.2344 & -0.97\% & \barPercentSeven{-0.97}
      & 0.2389 & 0.93\% & \barPercentSeven{0.93} \\
    \midrule

    \rowcolor{blue!5}
    \multicolumn{19}{l}{\revised{\textbf{Large Language Models (each context condition vs.\ the code-only baseline of the same mode; Z = zero-shot, F = fine-tuned)}}} \\ \midrule

    \multirow{8}{*}{\textbf{Qwen-1.5B}}
    & Z & BLEU-4
      & 2.3431
      & 2.0707
      & -11.6\%
      & \barPercentSeven{-11.6}
      & 2.0208
      & -13.8\%
      & \barPercentSeven{-13.8}
      & 2.0405
      & -12.9\%
      & \barPercentSeven{-12.9}
      & 2.1196
      & -9.5\%
      & \barPercentSeven{-9.5}
      & 2.0189
      & -13.8\%
      & \barPercentSeven{-13.8}
    \\
    & Z & BERTScore-F1
      & 0.2304
      & 0.2069
      & -10.2\%
      & \barPercentSeven{-10.2}
      & 0.2065
      & -10.4\%
      & \barPercentSeven{-10.4}
      & 0.2007
      & -12.9\%
      & \barPercentSeven{-12.9}
      & 0.2084
      & -9.5\%
      & \barPercentSeven{-9.5}
      & 0.2021
      & -12.3\%
      & \barPercentSeven{-12.3}
    \\
    & Z & ROUGE-L
      & 0.2440
      & 0.2278
      & -6.6\%
      & \barPercentSeven{-6.6}
      & 0.2285
      & -6.4\%
      & \barPercentSeven{-6.4}
      & 0.2220
      & -9.0\%
      & \barPercentSeven{-9.0}
      & 0.2294
      & -6.0\%
      & \barPercentSeven{-6.0}
      & 0.2234
      & -8.4\%
      & \barPercentSeven{-8.4}
    \\
    & Z & METEOR
      & 0.2192
      & 0.2180
      & -0.5\%
      & \barPercentSeven{-0.5}
      & 0.2133
      & -2.7\%
      & \barPercentSeven{-2.7}
      & 0.2165
      & -1.2\%
      & \barPercentSeven{-1.2}
      & 0.2180
      & -0.5\%
      & \barPercentSeven{-0.5}
      & 0.2168
      & -1.1\%
      & \barPercentSeven{-1.1}
    \\
    \cmidrule(lr){4-19}
    & F & BLEU-4
      & 2.9887
      & 2.8246
      & -5.5\%
      & \barPercentSeven{-5.5}
      & 2.6902
      & -10.0\%
      & \barPercentSeven{-10.0}
      & 2.8076
      & -6.1\%
      & \barPercentSeven{-6.1}
      & 2.7655
      & -7.5\%
      & \barPercentSeven{-7.5}
      & 2.6589
      & -11.0\%
      & \barPercentSeven{-11.0}
    \\
    & F & BERTScore-F1
      & 0.2756
      & 0.2525
      & -8.4\%
      & \barPercentSeven{-8.4}
      & 0.2566
      & -6.9\%
      & \barPercentSeven{-6.9}
      & 0.2494
      & -9.5\%
      & \barPercentSeven{-9.5}
      & 0.2556
      & -7.3\%
      & \barPercentSeven{-7.3}
      & 0.2502
      & -9.2\%
      & \barPercentSeven{-9.2}
    \\
    & F & ROUGE-L
      & 0.2574
      & 0.2419
      & -6.0\%
      & \barPercentSeven{-6.0}
      & 0.2459
      & -4.5\%
      & \barPercentSeven{-4.5}
      & 0.2395
      & -7.0\%
      & \barPercentSeven{-7.0}
      & 0.2460
      & -4.4\%
      & \barPercentSeven{-4.4}
      & 0.2388
      & -7.2\%
      & \barPercentSeven{-7.2}
    \\
    & F & METEOR
      & 0.2168
      & 0.2091
      & -3.6\%
      & \barPercentSeven{-3.6}
      & 0.2115
      & -2.4\%
      & \barPercentSeven{-2.4}
      & 0.2107
      & -2.8\%
      & \barPercentSeven{-2.8}
      & 0.2106
      & -2.9\%
      & \barPercentSeven{-2.9}
      & 0.2099
      & -3.2\%
      & \barPercentSeven{-3.2}
    \\
    \midrule

    \multirow{8}{*}{\textbf{DeepSeek-6.7B}}
    & Z & BLEU-4
      & 2.9683
      & 2.6071
      & -12.2\%
      & \barPercentSeven{-12.2}
      & 2.5298
      & -14.8\%
      & \barPercentSeven{-14.8}
      & 2.5426
      & -14.3\%
      & \barPercentSeven{-14.3}
      & 2.5025
      & -15.7\%
      & \barPercentSeven{-15.7}
      & 2.5738
      & -13.3\%
      & \barPercentSeven{-13.3}
    \\
    & Z & BERTScore-F1
      & 0.2795
      & 0.2597
      & -7.1\%
      & \barPercentSeven{-7.1}
      & 0.2515
      & -10.0\%
      & \barPercentSeven{-10.0}
      & 0.2560
      & -8.4\%
      & \barPercentSeven{-8.4}
      & 0.2603
      & -6.9\%
      & \barPercentSeven{-6.9}
      & 0.2549
      & -8.8\%
      & \barPercentSeven{-8.8}
    \\
    & Z & ROUGE-L
      & 0.2562
      & 0.2401
      & -6.3\%
      & \barPercentSeven{-6.3}
      & 0.2299
      & -10.3\%
      & \barPercentSeven{-10.3}
      & 0.2378
      & -7.2\%
      & \barPercentSeven{-7.2}
      & 0.2402
      & -6.2\%
      & \barPercentSeven{-6.2}
      & 0.2362
      & -7.8\%
      & \barPercentSeven{-7.8}
    \\
    & Z & METEOR
      & 0.2166
      & 0.2085
      & -3.7\%
      & \barPercentSeven{-3.7}
      & 0.2003
      & -7.5\%
      & \barPercentSeven{-7.5}
      & 0.2082
      & -3.9\%
      & \barPercentSeven{-3.9}
      & 0.2087
      & -3.6\%
      & \barPercentSeven{-3.6}
      & 0.2088
      & -3.6\%
      & \barPercentSeven{-3.6}
    \\
    \cmidrule(lr){4-19}
    & F & BLEU-4
      & 3.2595
      & 2.8592
      & -12.3\%
      & \barPercentSeven{-12.3}
      & 2.9336
      & -10.0\%
      & \barPercentSeven{-10.0}
      & 2.9109
      & -10.7\%
      & \barPercentSeven{-10.7}
      & 3.0190
      & -7.4\%
      & \barPercentSeven{-7.4}
      & 2.9614
      & -9.1\%
      & \barPercentSeven{-9.1}
    \\
    & F & BERTScore-F1
      & 0.2884
      & 0.2704
      & -6.2\%
      & \barPercentSeven{-6.2}
      & 0.2685
      & -6.9\%
      & \barPercentSeven{-6.9}
      & 0.2688
      & -6.8\%
      & \barPercentSeven{-6.8}
      & 0.2720
      & -5.7\%
      & \barPercentSeven{-5.7}
      & 0.2701
      & -6.3\%
      & \barPercentSeven{-6.3}
    \\
    & F & ROUGE-L
      & 0.2620
      & 0.2497
      & -4.7\%
      & \barPercentSeven{-4.7}
      & 0.2435
      & -7.1\%
      & \barPercentSeven{-7.1}
      & 0.2499
      & -4.6\%
      & \barPercentSeven{-4.6}
      & 0.2527
      & -3.5\%
      & \barPercentSeven{-3.5}
      & 0.2510
      & -4.2\%
      & \barPercentSeven{-4.2}
    \\
    & F & METEOR
      & 0.2246
      & 0.2160
      & -3.8\%
      & \barPercentSeven{-3.8}
      & 0.2122
      & -5.5\%
      & \barPercentSeven{-5.5}
      & 0.2160
      & -3.8\%
      & \barPercentSeven{-3.8}
      & 0.2215
      & -1.4\%
      & \barPercentSeven{-1.4}
      & 0.2167
      & -3.5\%
      & \barPercentSeven{-3.5}
    \\
    \midrule

    \multirow{8}{*}{\textbf{CodeLlama-13B}}
    & Z & BLEU-4
      & 3.9423
      & 2.9117
      & -26.1\%
      & \barPercentSeven{-26.1}
      & 3.2065
      & -18.7\%
      & \barPercentSeven{-18.7}
      & 2.5475
      & -35.4\%
      & \barPercentSeven{-35.4}
      & 3.0116
      & -23.6\%
      & \barPercentSeven{-23.6}
      & 2.7273
      & -30.8\%
      & \barPercentSeven{-30.8}
    \\
    & Z & BERTScore-F1
      & 0.3014
      & 0.2369
      & -21.4\%
      & \barPercentSeven{-21.4}
      & 0.2676
      & -11.2\%
      & \barPercentSeven{-11.2}
      & 0.2193
      & -27.2\%
      & \barPercentSeven{-27.2}
      & 0.2470
      & -18.0\%
      & \barPercentSeven{-18.0}
      & 0.2284
      & -24.2\%
      & \barPercentSeven{-24.2}
    \\
    & Z & ROUGE-L
      & 0.2830
      & 0.2445
      & -13.6\%
      & \barPercentSeven{-13.6}
      & 0.2570
      & -9.2\%
      & \barPercentSeven{-9.2}
      & 0.2323
      & -17.9\%
      & \barPercentSeven{-17.9}
      & 0.2507
      & -11.4\%
      & \barPercentSeven{-11.4}
      & 0.2366
      & -16.4\%
      & \barPercentSeven{-16.4}
    \\
    & Z & METEOR
      & 0.2441
      & 0.2266
      & -7.2\%
      & \barPercentSeven{-7.2}
      & 0.2358
      & -3.4\%
      & \barPercentSeven{-3.4}
      & 0.2224
      & -8.9\%
      & \barPercentSeven{-8.9}
      & 0.2292
      & -6.1\%
      & \barPercentSeven{-6.1}
      & 0.2238
      & -8.3\%
      & \barPercentSeven{-8.3}
    \\
    & F & BLEU-4
      & 3.7827
      & \revised{3.4431}
      & -8.98\%
      & \barPercentSeven{-8.98}
      & 3.3459
      & -11.5\%
      & \barPercentSeven{-11.5}
      & 2.84
      & -24.9\%
      & \barPercentSeven{-24.9}
      & 3.4431
      & -9.0\%
      & \barPercentSeven{-9.0}
      & 3.03
      & -19.9\%
      & \barPercentSeven{-19.9}
    \\
    & F & BERTScore-F1
      & 0.2985
      & 0.2747
      & -7.97\%
      & \barPercentSeven{-7.97}
      & 0.2845
      & -4.7\%
      & \barPercentSeven{-4.7}
      & 0.2528
      & -15.3\%
      & \barPercentSeven{-15.3}
      & 0.2747
      & -8.0\%
      & \barPercentSeven{-8.0}
      & 0.2617
      & -12.3\%
      & \barPercentSeven{-12.3}
    \\
    & F & ROUGE-L
      & 0.2823
      & 0.2664
      & -5.63\%
      & \barPercentSeven{-5.63}
      & 0.2655
      & -6.0\%
      & \barPercentSeven{-6.0}
      & 0.2467
      & -12.6\%
      & \barPercentSeven{-12.6}
      & 0.2664
      & -5.6\%
      & \barPercentSeven{-5.6}
      & 0.2542
      & -10.0\%
      & \barPercentSeven{-10.0}
    \\
    & F & METEOR
      & 0.2432
      & 0.2426
      & -0.25\%
      & \barPercentSeven{-0.25}
      & 0.2460
      & +1.2\%
      & \barPercentSeven{1.2}
      & 0.2321
      & -4.6\%
      & \barPercentSeven{-4.6}
      & 0.2426
      & -0.2\%
      & \barPercentSeven{-0.2}
      & 0.2377
      & -2.3\%
      & \barPercentSeven{-2.3}
    \\
    \midrule

    \multirow{8}{*}{\textbf{CodeLlama-34B}}
    & Z & BLEU-4
      & 3.2124
      & 2.6454
      & -17.7\%
      & \barPercentSeven{-17.7}
      & 2.7496
      & -14.4\%
      & \barPercentSeven{-14.4}
      & 2.4824
      & -22.7\%
      & \barPercentSeven{-22.7}
      & 2.6314
      & -18.1\%
      & \barPercentSeven{-18.1}
      & 2.5438
      & -20.8\%
      & \barPercentSeven{-20.8}
    \\
    & Z & BERTScore-F1
      & 0.2848
      & 0.2430
      & -14.7\%
      & \barPercentSeven{-14.7}
      & 0.2653
      & -6.8\%
      & \barPercentSeven{-6.8}
      & 0.2396
      & -15.9\%
      & \barPercentSeven{-15.9}
      & 0.2436
      & -14.5\%
      & \barPercentSeven{-14.5}
      & 0.2409
      & -15.4\%
      & \barPercentSeven{-15.4}
    \\
    & Z & ROUGE-L
      & 0.2529
      & 0.2284
      & -9.7\%
      & \barPercentSeven{-9.7}
      & 0.2359
      & -6.7\%
      & \barPercentSeven{-6.7}
      & 0.2220
      & -12.2\%
      & \barPercentSeven{-12.2}
      & 0.2269
      & -10.3\%
      & \barPercentSeven{-10.3}
      & 0.2224
      & -12.1\%
      & \barPercentSeven{-12.1}
    \\
    & Z & METEOR
      & 0.2517
      & 0.2375
      & -5.6\%
      & \barPercentSeven{-5.6}
      & 0.2420
      & -3.9\%
      & \barPercentSeven{-3.9}
      & 0.2339
      & -7.1\%
      & \barPercentSeven{-7.1}
      & 0.2366
      & -6.0\%
      & \barPercentSeven{-6.0}
      & 0.2352
      & -6.6\%
      & \barPercentSeven{-6.6}
    \\
    \cmidrule(lr){4-19}
    & F & BLEU-4
      & 3.1794
      & 2.78
      & -12.6\%
      & \barPercentSeven{-12.6}
      & 2.88
      & -9.4\%
      & \barPercentSeven{-9.4}
      & 2.63
      & -17.3\%
      & \barPercentSeven{-17.3}
      & 2.83
      & -11.0\%
      & \barPercentSeven{-11.0}
      & 2.76
      & -13.2\%
      & \barPercentSeven{-13.2}
    \\
    & F & BERTScore-F1
      & 0.2915
      & 0.2651
      & -9.1\%
      & \barPercentSeven{-9.1}
      & 0.2757
      & -5.4\%
      & \barPercentSeven{-5.4}
      & 0.2595
      & -11.0\%
      & \barPercentSeven{-11.0}
      & 0.2662
      & -8.7\%
      & \barPercentSeven{-8.7}
      & 0.2611
      & -10.4\%
      & \barPercentSeven{-10.4}
    \\
    & F & ROUGE-L
      & 0.259
      & 0.2392
      & -7.6\%
      & \barPercentSeven{-7.6}
      & 0.2445
      & -5.6\%
      & \barPercentSeven{-5.6}
      & 0.2332
      & -10.0\%
      & \barPercentSeven{-10.0}
      & 0.2409
      & -7.0\%
      & \barPercentSeven{-7.0}
      & 0.2351
      & -9.2\%
      & \barPercentSeven{-9.2}
    \\
    & F & METEOR
      & 0.2559
      & 0.2432
      & -5.0\%
      & \barPercentSeven{-5.0}
      & 0.2522
      & -1.4\%
      & \barPercentSeven{-1.4}
      & 0.2432
      & -5.0\%
      & \barPercentSeven{-5.0}
      & 0.2471
      & -3.4\%
      & \barPercentSeven{-3.4}
      & 0.246
      & -3.9\%
      & \barPercentSeven{-3.9}
    \\
    \midrule
    \multirow{8}{*}{\revised{\textbf{Qwen-14B}}}
    & \revised{Z} & \revised{BLEU-4}
      & \revised{1.83}
      & \revised{2.31}
      & \revised{+26.2\%}
      & \revised{\barPercentSeven{26.2}}
      & \revised{2.00}
      & \revised{+9.3\%}
      & \revised{\barPercentSeven{9.3}}
      & \revised{2.40}
      & \revised{+31.1\%}
      & \revised{\barPercentSeven{31.1}}
      & \revised{2.46}
      & \revised{+34.4\%}
      & \revised{\barPercentSeven{34.4}}
      & \revised{2.48}
      & \revised{+35.5\%}
      & \revised{\barPercentSeven{35.5}}
    \\
    & \revised{Z} & \revised{BERTScore-F1}
      & \revised{0.1797}
      & \revised{0.2182}
      & \revised{+21.4\%}
      & \revised{\barPercentSeven{21.4}}
      & \revised{0.1973}
      & \revised{+9.8\%}
      & \revised{\barPercentSeven{9.8}}
      & \revised{0.2257}
      & \revised{+25.6\%}
      & \revised{\barPercentSeven{25.6}}
      & \revised{0.2243}
      & \revised{+24.8\%}
      & \revised{\barPercentSeven{24.8}}
      & \revised{0.2280}
      & \revised{+26.9\%}
      & \revised{\barPercentSeven{26.9}}
    \\
    & \revised{Z} & \revised{ROUGE-L}
      & \revised{0.1882}
      & \revised{0.2267}
      & \revised{+20.5\%}
      & \revised{\barPercentSeven{20.5}}
      & \revised{0.2064}
      & \revised{+9.7\%}
      & \revised{\barPercentSeven{9.7}}
      & \revised{0.2308}
      & \revised{+22.6\%}
      & \revised{\barPercentSeven{22.6}}
      & \revised{0.2300}
      & \revised{+22.2\%}
      & \revised{\barPercentSeven{22.2}}
      & \revised{0.2339}
      & \revised{+24.3\%}
      & \revised{\barPercentSeven{24.3}}
    \\
    & \revised{Z} & \revised{METEOR}
      & \revised{0.1273}
      & \revised{0.1695}
      & \revised{+33.2\%}
      & \revised{\barPercentSeven{33.2}}
      & \revised{0.1470}
      & \revised{+15.5\%}
      & \revised{\barPercentSeven{15.5}}
      & \revised{0.1751}
      & \revised{+37.5\%}
      & \revised{\barPercentSeven{37.5}}
      & \revised{0.1746}
      & \revised{+37.2\%}
      & \revised{\barPercentSeven{37.2}}
      & \revised{0.1800}
      & \revised{+41.4\%}
      & \revised{\barPercentSeven{41.4}}
    \\
    \cmidrule(lr){4-19}
    & \revised{F} & \revised{BLEU-4}
      & \revised{1.88}
      & \revised{2.22}
      & \revised{+18.1\%}
      & \revised{\barPercentSeven{18.1}}
      & \revised{2.07}
      & \revised{+10.1\%}
      & \revised{\barPercentSeven{10.1}}
      & \revised{2.20}
      & \revised{+17.0\%}
      & \revised{\barPercentSeven{17.0}}
      & \revised{2.40}
      & \revised{+27.7\%}
      & \revised{\barPercentSeven{27.7}}
      & \revised{2.34}
      & \revised{+24.5\%}
      & \revised{\barPercentSeven{24.5}}
    \\
    & \revised{F} & \revised{BERTScore-F1}
      & \revised{0.2054}
      & \revised{0.2268}
      & \revised{+10.4\%}
      & \revised{\barPercentSeven{10.4}}
      & \revised{0.2155}
      & \revised{+4.9\%}
      & \revised{\barPercentSeven{4.9}}
      & \revised{0.2283}
      & \revised{+11.1\%}
      & \revised{\barPercentSeven{11.1}}
      & \revised{0.2359}
      & \revised{+14.8\%}
      & \revised{\barPercentSeven{14.8}}
      & \revised{0.2338}
      & \revised{+13.8\%}
      & \revised{\barPercentSeven{13.8}}
    \\
    & \revised{F} & \revised{ROUGE-L}
      & \revised{0.2027}
      & \revised{0.2271}
      & \revised{+12.0\%}
      & \revised{\barPercentSeven{12.0}}
      & \revised{0.2149}
      & \revised{+6.0\%}
      & \revised{\barPercentSeven{6.0}}
      & \revised{0.2280}
      & \revised{+12.5\%}
      & \revised{\barPercentSeven{12.5}}
      & \revised{0.2338}
      & \revised{+15.3\%}
      & \revised{\barPercentSeven{15.3}}
      & \revised{0.2314}
      & \revised{+14.2\%}
      & \revised{\barPercentSeven{14.2}}
    \\
    & \revised{F} & \revised{METEOR}
      & \revised{0.1439}
      & \revised{0.1750}
      & \revised{+21.6\%}
      & \revised{\barPercentSeven{21.6}}
      & \revised{0.1589}
      & \revised{+10.4\%}
      & \revised{\barPercentSeven{10.4}}
      & \revised{0.1785}
      & \revised{+24.0\%}
      & \revised{\barPercentSeven{24.0}}
      & \revised{0.1819}
      & \revised{+26.4\%}
      & \revised{\barPercentSeven{26.4}}
      & \revised{0.1835}
      & \revised{+27.5\%}
      & \revised{\barPercentSeven{27.5}}
    \\
    \midrule

    \multirow{8}{*}{\revised{\textbf{CodeLlama-7B}}}
    & \revised{Z} & \revised{BLEU-4}
      & \revised{3.97}
      & \revised{3.42}
      & \revised{-13.9\%}
      & \revised{\barPercentSeven{-13.9}}
      & \revised{3.39}
      & \revised{-14.6\%}
      & \revised{\barPercentSeven{-14.6}}
      & \revised{3.34}
      & \revised{-15.9\%}
      & \revised{\barPercentSeven{-15.9}}
      & \revised{3.37}
      & \revised{-15.1\%}
      & \revised{\barPercentSeven{-15.1}}
      & \revised{3.17}
      & \revised{-20.2\%}
      & \revised{\barPercentSeven{-20.2}}
    \\
    & \revised{Z} & \revised{BERTScore-F1}
      & \revised{0.3060}
      & \revised{0.2803}
      & \revised{-8.4\%}
      & \revised{\barPercentSeven{-8.4}}
      & \revised{0.2882}
      & \revised{-5.8\%}
      & \revised{\barPercentSeven{-5.8}}
      & \revised{0.2784}
      & \revised{-9.0\%}
      & \revised{\barPercentSeven{-9.0}}
      & \revised{0.2787}
      & \revised{-8.9\%}
      & \revised{\barPercentSeven{-8.9}}
      & \revised{0.2748}
      & \revised{-10.2\%}
      & \revised{\barPercentSeven{-10.2}}
    \\
    & \revised{Z} & \revised{ROUGE-L}
      & \revised{0.2772}
      & \revised{0.2606}
      & \revised{-6.0\%}
      & \revised{\barPercentSeven{-6.0}}
      & \revised{0.2622}
      & \revised{-5.4\%}
      & \revised{\barPercentSeven{-5.4}}
      & \revised{0.2578}
      & \revised{-7.0\%}
      & \revised{\barPercentSeven{-7.0}}
      & \revised{0.2595}
      & \revised{-6.4\%}
      & \revised{\barPercentSeven{-6.4}}
      & \revised{0.2543}
      & \revised{-8.3\%}
      & \revised{\barPercentSeven{-8.3}}
    \\
    & \revised{Z} & \revised{METEOR}
      & \revised{0.2477}
      & \revised{0.2413}
      & \revised{-2.6\%}
      & \revised{\barPercentSeven{-2.6}}
      & \revised{0.2433}
      & \revised{-1.8\%}
      & \revised{\barPercentSeven{-1.8}}
      & \revised{0.2402}
      & \revised{-3.0\%}
      & \revised{\barPercentSeven{-3.0}}
      & \revised{0.2411}
      & \revised{-2.7\%}
      & \revised{\barPercentSeven{-2.7}}
      & \revised{0.2405}
      & \revised{-2.9\%}
      & \revised{\barPercentSeven{-2.9}}
    \\
    \cmidrule(lr){4-19}
    & \revised{F} & \revised{BLEU-4}
      & \revised{4.27}
      & \revised{4.16}
      & \revised{-2.6\%}
      & \revised{\barPercentSeven{-2.6}}
      & \revised{4.06}
      & \revised{-4.9\%}
      & \revised{\barPercentSeven{-4.9}}
      & \revised{4.06}
      & \revised{-4.9\%}
      & \revised{\barPercentSeven{-4.9}}
      & \revised{4.05}
      & \revised{-5.2\%}
      & \revised{\barPercentSeven{-5.2}}
      & \revised{4.09}
      & \revised{-4.2\%}
      & \revised{\barPercentSeven{-4.2}}
    \\
    & \revised{F} & \revised{BERTScore-F1}
      & \revised{0.3260}
      & \revised{0.3101}
      & \revised{-4.9\%}
      & \revised{\barPercentSeven{-4.9}}
      & \revised{0.3158}
      & \revised{-3.1\%}
      & \revised{\barPercentSeven{-3.1}}
      & \revised{0.3083}
      & \revised{-5.4\%}
      & \revised{\barPercentSeven{-5.4}}
      & \revised{0.3069}
      & \revised{-5.9\%}
      & \revised{\barPercentSeven{-5.9}}
      & \revised{0.3085}
      & \revised{-5.4\%}
      & \revised{\barPercentSeven{-5.4}}
    \\
    & \revised{F} & \revised{ROUGE-L}
      & \revised{0.2922}
      & \revised{0.2835}
      & \revised{-3.0\%}
      & \revised{\barPercentSeven{-3.0}}
      & \revised{0.2845}
      & \revised{-2.6\%}
      & \revised{\barPercentSeven{-2.6}}
      & \revised{0.2812}
      & \revised{-3.8\%}
      & \revised{\barPercentSeven{-3.8}}
      & \revised{0.2825}
      & \revised{-3.3\%}
      & \revised{\barPercentSeven{-3.3}}
      & \revised{0.2821}
      & \revised{-3.5\%}
      & \revised{\barPercentSeven{-3.5}}
    \\
    & \revised{F} & \revised{METEOR}
      & \revised{0.2540}
      & \revised{0.2509}
      & \revised{-1.2\%}
      & \revised{\barPercentSeven{-1.2}}
      & \revised{0.2528}
      & \revised{-0.5\%}
      & \revised{\barPercentSeven{-0.5}}
      & \revised{0.2494}
      & \revised{-1.8\%}
      & \revised{\barPercentSeven{-1.8}}
      & \revised{0.2520}
      & \revised{-0.8\%}
      & \revised{\barPercentSeven{-0.8}}
      & \revised{0.2512}
      & \revised{-1.1\%}
      & \revised{\barPercentSeven{-1.1}}
    \\
    \midrule

    \multirow{8}{*}{\revised{\textbf{Qwen-7B}}}
    & \revised{Z} & \revised{BLEU-4}
      & \revised{1.59}
      & \revised{2.27}
      & \revised{+42.7\%}
      & \revised{\barPercentSeven{42.7}}
      & \revised{1.60}
      & \revised{+0.6\%}
      & \revised{\barPercentSeven{0.6}}
      & \revised{2.11}
      & \revised{+32.6\%}
      & \revised{\barPercentSeven{32.6}}
      & \revised{2.27}
      & \revised{+42.7\%}
      & \revised{\barPercentSeven{42.7}}
      & \revised{2.26}
      & \revised{+42.0\%}
      & \revised{\barPercentSeven{42.0}}
    \\
    & \revised{Z} & \revised{BERTScore-F1}
      & \revised{0.1695}
      & \revised{0.2095}
      & \revised{+23.6\%}
      & \revised{\barPercentSeven{23.6}}
      & \revised{0.1671}
      & \revised{-1.4\%}
      & \revised{\barPercentSeven{-1.4}}
      & \revised{0.2061}
      & \revised{+21.6\%}
      & \revised{\barPercentSeven{21.6}}
      & \revised{0.2081}
      & \revised{+22.8\%}
      & \revised{\barPercentSeven{22.8}}
      & \revised{0.2057}
      & \revised{+21.4\%}
      & \revised{\barPercentSeven{21.4}}
    \\
    & \revised{Z} & \revised{ROUGE-L}
      & \revised{0.1823}
      & \revised{0.2188}
      & \revised{+20.0\%}
      & \revised{\barPercentSeven{20.0}}
      & \revised{0.1829}
      & \revised{+0.3\%}
      & \revised{\barPercentSeven{0.3}}
      & \revised{0.2154}
      & \revised{+18.2\%}
      & \revised{\barPercentSeven{18.2}}
      & \revised{0.2180}
      & \revised{+19.6\%}
      & \revised{\barPercentSeven{19.6}}
      & \revised{0.2168}
      & \revised{+18.9\%}
      & \revised{\barPercentSeven{18.9}}
    \\
    & \revised{Z} & \revised{METEOR}
      & \revised{0.1198}
      & \revised{0.1641}
      & \revised{+37.0\%}
      & \revised{\barPercentSeven{37.0}}
      & \revised{0.1213}
      & \revised{+1.3\%}
      & \revised{\barPercentSeven{1.3}}
      & \revised{0.1600}
      & \revised{+33.6\%}
      & \revised{\barPercentSeven{33.6}}
      & \revised{0.1635}
      & \revised{+36.5\%}
      & \revised{\barPercentSeven{36.5}}
      & \revised{0.1637}
      & \revised{+36.6\%}
      & \revised{\barPercentSeven{36.6}}
    \\
    \cmidrule(lr){4-19}
    & \revised{F} & \revised{BLEU-4}
      & \revised{2.11}
      & \revised{2.48}
      & \revised{+17.5\%}
      & \revised{\barPercentSeven{17.5}}
      & \revised{1.91}
      & \revised{-9.5\%}
      & \revised{\barPercentSeven{-9.5}}
      & \revised{2.42}
      & \revised{+14.7\%}
      & \revised{\barPercentSeven{14.7}}
      & \revised{2.56}
      & \revised{+21.3\%}
      & \revised{\barPercentSeven{21.3}}
      & \revised{2.52}
      & \revised{+19.4\%}
      & \revised{\barPercentSeven{19.4}}
    \\
    & \revised{F} & \revised{BERTScore-F1}
      & \revised{0.2021}
      & \revised{0.2316}
      & \revised{+14.6\%}
      & \revised{\barPercentSeven{14.6}}
      & \revised{0.1938}
      & \revised{-4.1\%}
      & \revised{\barPercentSeven{-4.1}}
      & \revised{0.2268}
      & \revised{+12.2\%}
      & \revised{\barPercentSeven{12.2}}
      & \revised{0.2345}
      & \revised{+16.0\%}
      & \revised{\barPercentSeven{16.0}}
      & \revised{0.2291}
      & \revised{+13.4\%}
      & \revised{\barPercentSeven{13.4}}
    \\
    & \revised{F} & \revised{ROUGE-L}
      & \revised{0.2018}
      & \revised{0.2334}
      & \revised{+15.7\%}
      & \revised{\barPercentSeven{15.7}}
      & \revised{0.2009}
      & \revised{-0.4\%}
      & \revised{\barPercentSeven{-0.4}}
      & \revised{0.2278}
      & \revised{+12.9\%}
      & \revised{\barPercentSeven{12.9}}
      & \revised{0.2348}
      & \revised{+16.4\%}
      & \revised{\barPercentSeven{16.4}}
      & \revised{0.2298}
      & \revised{+13.9\%}
      & \revised{\barPercentSeven{13.9}}
    \\
    & \revised{F} & \revised{METEOR}
      & \revised{0.1399}
      & \revised{0.1790}
      & \revised{+27.9\%}
      & \revised{\barPercentSeven{27.9}}
      & \revised{0.1395}
      & \revised{-0.3\%}
      & \revised{\barPercentSeven{-0.3}}
      & \revised{0.1751}
      & \revised{+25.2\%}
      & \revised{\barPercentSeven{25.2}}
      & \revised{0.1818}
      & \revised{+29.9\%}
      & \revised{\barPercentSeven{29.9}}
      & \revised{0.1794}
      & \revised{+28.2\%}
      & \revised{\barPercentSeven{28.2}}
    \\
    \midrule
    \bottomrule
    \end{tabular}

    \begin{tablenotes}
        \small
        \item \textit{Results are reported using BLEU-4, METEOR, ROUGE-L, and BERTScore-F1 across different models.}
    \end{tablenotes}

    \end{threeparttable}
}
\end{table}

The result in Table~\ref{tab:result_rq2_summarisation} indicates the impact of encoding multiple context types into source code representations for the Code Summarisation task. Overall, none of the context combinations consistently enhance performance across all models. However, there are some notable improvements achieved in selected cases. Specifically, the combination of Version History and Method Age significantly improves CodeT5's BLEU-4 score by +5.79\% (from 20.02 to 21.18), surpassing even its best single-context performance (CodeT5 with Version History at 20.63 in RQ1).

GraphCodeBERT demonstrates notable and consistent METEOR score's improvement that shows across four out of five combinations of additional context, with enhancements ranging from +2.43\% to +4.83\%. Nevertheless, in the BERTscore-F1 and ROUGE-L metrics, there are no clear patterns of consistent improvement or degradation across different combinations and models, highlighting a more complex interaction between context types and summarisation performance.

\revised{For the seven LLMs, the combined-context results in Table~\ref{tab:result_rq2_summarisation} follow the same split by model family as the single-context results in Table~\ref{tab:result_rq1_summarisation}. Qwen2.5-Coder-7B and Qwen2.5-Coder-14B improve on nearly every metric under all five combinations, in both zero-shot and fine-tuned modes, and the combinations that include call graph context give the largest gains: Qwen-14B zero-shot reaches +35.5\% BLEU-4 and +41.4\% METEOR with \texttt{code\_cg\_vh\_nod}, above the +26.8\% and +34.0\% obtained with call graph alone. The remaining five LLMs (Qwen2.5-Coder-1.5B, the three CodeLlama sizes, and DeepSeek-Coder) decline on all four metrics under all five combinations, with a single exception (CodeLlama-13B fine-tuned, METEOR, +1.2\% with version history and method age). Combining context therefore amplifies the benefit for models that already gain from a single context type, but does not rescue models that do not; model family, rather than the choice of combination, determines whether prompt-level context helps summarisation.}

These results suggest that while encoding various types of context can selectively benefit certain models and metrics, the improvements are neither uniform nor universally positive. Therefore, careful selection and validation of context combinations are critical for effective summarisation performance enhancement.

\begin{table}[htbp]
\centering
\scriptsize

\caption{Human evaluation of CodeT5-generated summaries with combined context types (Win/Tie/Loss analysis by dimension).}
\label{tab:human_eval_new}

\renewcommand{\arraystretch}{1.2}

\resizebox{\textwidth}{!}{%
    \begin{threeparttable}

    \begin{tabular}{l|l r@{\hspace{2pt}}l r@{\hspace{2pt}}l c l l}
    \toprule
    \rowcolor{headerCol}
    \multicolumn{1}{c|}{\textbf{Dimension}} &
    \multicolumn{1}{c}{\textbf{Context(s)}} &
    \multicolumn{2}{c}{\textbf{Win \%}} &
    \multicolumn{2}{c}{\textbf{Loss \%}} &
    \multicolumn{1}{c}{\textbf{Tie \%}} &
    \multicolumn{1}{c}{\textbf{Wilcoxon $p$}} &
    \multicolumn{1}{c}{\textbf{|Cliff's $\delta$|}} \\
    \midrule

    \multirow{5}{*}{Accuracy}
    & Code + CallGraph + VersionHistory & 22\% & \barPercentSeventy{22} & 9\% & \barPercentSeventyLoss{9} & 69\% & \textbf{0.0252 ***} & 0.3913 (M) \\
    & Code + CallGraph + VersionHistory + MethodAge & 21\% & \barPercentSeventy{21} & 10\% & \barPercentSeventyLoss{10} & 69\% & 0.0817 & 0.3236 (S) \\
    & Code + VersionHistory + CallGraph & 25\% & \barPercentSeventy{25} & 11\% & \barPercentSeventyLoss{11} & 64\% & \textbf{0.0187 ***} & 0.3889 (M) \\
    & Code + VersionHistory + CallGraph + MethodAge & 24\% & \barPercentSeventy{24} & 12\% & \barPercentSeventyLoss{12} & 64\% & \textbf{0.0384 ***} & 0.3225 (S) \\
    & Code + VersionHistory + MethodAge & 14\% & \barPercentSeventy{14} & 13\% & \barPercentSeventyLoss{13} & 73\% & 0.2685 & 0.0809 (N) \\
    \midrule

    \multirow{5}{*}{Conciseness}
    & Code + CallGraph + VersionHistory & 20\% & \barPercentSeventy{20} & 13\% & \barPercentSeventyLoss{13} & 67\% & 0.0568 & 0.2332 (S) \\
    & Code + CallGraph + VersionHistory + MethodAge & 29\% & \barPercentSeventy{29} & 17\% & \barPercentSeventyLoss{17} & 54\% & \textbf{0.0341 ***} & 0.2637 (S) \\
    & Code + VersionHistory + CallGraph & 22\% & \barPercentSeventy{22} & 16\% & \barPercentSeventyLoss{16} & 62\% & 0.1733 & 0.1496 (S) \\
    & Code + VersionHistory + CallGraph + MethodAge & 32\% & \barPercentSeventy{32} & 14\% & \barPercentSeventyLoss{14} & 54\% & \textbf{0.0014 ***} & 0.4376 (M) \\
    & Code + VersionHistory + MethodAge & 19\% & \barPercentSeventy{19} & 16\% & \barPercentSeventyLoss{16} & 65\% & 0.4312 & 0.0580 (N) \\
    \midrule

    \multirow{5}{*}{\parbox{1.8cm}{Content \\ Adequacy}}
    & Code + CallGraph + VersionHistory & 24\% & \barPercentSeventy{24} & 11\% & \barPercentSeventyLoss{11} & 65\% & \textbf{0.0041 ***} & 0.4171 (M) \\
    & Code + CallGraph + VersionHistory + MethodAge & 23\% & \barPercentSeventy{23} & 16\% & \barPercentSeventyLoss{16} & 61\% & 0.1830 & 0.1742 (S) \\
    & Code + VersionHistory + CallGraph & 30\% & \barPercentSeventy{30} & 11\% & \barPercentSeventyLoss{11} & 59\% & \textbf{0.0127 ***} & 0.4253 (M) \\
    & Code + VersionHistory + CallGraph + MethodAge & 26\% & \barPercentSeventy{26} & 13\% & \barPercentSeventyLoss{13} & 61\% & \textbf{0.0373 ***} & 0.3162 (S) \\
    & Code + VersionHistory + MethodAge & 20\% & \barPercentSeventy{20} & 12\% & \barPercentSeventyLoss{12} & 68\% & 0.1272 & 0.2285 (S) \\
    \bottomrule
    \end{tabular}%

    \begin{tablenotes}
        \scriptsize
        \item Results are reported as Win/Tie/Loss ratios against the Without-Context Scenario.
        \item \textbf{Bold} text indicates statistical significance ($p < 0.05$, marked with ***) or a large effect size.
        \item Effect sizes from |Cliff's $\delta$| are shown as (L)arge, (M)edium, (S)mall, or (N)egligible.
    \end{tablenotes}

    \end{threeparttable}
}
\end{table}

\begin{table}[htbp]
\centering
\scriptsize

\caption{Ranking distribution and mean ranks of CodeT5-generated summaries with combined context types (human evaluation).}
\label{tab:ranking_distribution_all}

\renewcommand{\arraystretch}{1.2}

\resizebox{\textwidth}{!}{%
    \begin{threeparttable}

    \begin{tabular}{ll|lrrrr}
    \toprule
    \rowcolor{headerCol}
    \multicolumn{1}{l}{\textbf{Dimension}} &
    \multicolumn{1}{l}{\textbf{Context(s)}} &
    \multicolumn{1}{c}{\textbf{\# Rank 1}} &
    \multicolumn{1}{c}{\textbf{\# Rank 2}} &
    \multicolumn{1}{c}{\textbf{\# Rank 3}} &
    \multicolumn{1}{c}{\textbf{\# Rank $\geq$4}} &
    \multicolumn{1}{c}{\textbf{Mean Rank}} \\
    \midrule

    \multirow{8}{*}{Accuracy}
    & \cellcolor{baselineCol}Without Context & \cellcolor{baselineCol}59 & \cellcolor{baselineCol}33 & \cellcolor{baselineCol}6 & \cellcolor{baselineCol}2 & \cellcolor{baselineCol}1.51 \\
    & \cellcolor{versionCol}Code + CallGraph & \cellcolor{versionCol}73 (\textcolor{darkgreen}{$\uparrow$}23.73\%) & \cellcolor{versionCol}22 & \cellcolor{versionCol}5 & \cellcolor{versionCol}0 & \cellcolor{versionCol}1.32 \\
    & \cellcolor{versionCol}Code + VersionHistory & \cellcolor{versionCol}67 (\textcolor{darkgreen}{$\uparrow$}13.56\%) & \cellcolor{versionCol}28 & \cellcolor{versionCol}5 & \cellcolor{versionCol}0 & \cellcolor{versionCol}1.38 \\
    & \cellcolor{callgraphCol}Code + CallGraph + VersionHistory & \cellcolor{callgraphCol}72 (\textcolor{darkgreen}{$\uparrow$}22.03\%) & \cellcolor{callgraphCol}21 & \cellcolor{callgraphCol}6 & \cellcolor{callgraphCol}1 & \cellcolor{callgraphCol}1.36 \\
    & \cellcolor{callgraphCol}Code + VersionHistory + CallGraph & \cellcolor{callgraphCol}72 (\textcolor{darkgreen}{$\uparrow$}22.03\%) & \cellcolor{callgraphCol}22 & \cellcolor{callgraphCol}6 & \cellcolor{callgraphCol}0 & \cellcolor{callgraphCol}1.34 \\
    & \cellcolor{callgraphCol}Code + CallGraph + VersionHistory + MethodAge & \cellcolor{callgraphCol}65 (\textcolor{darkgreen}{$\uparrow$}10.17\%) & \cellcolor{callgraphCol}33 & \cellcolor{callgraphCol}2 & \cellcolor{callgraphCol}0 & \cellcolor{callgraphCol}1.37 \\
    & \cellcolor{callgraphCol}Code + VersionHistory + CallGraph + MethodAge & \cellcolor{callgraphCol}70 (\textcolor{darkgreen}{$\uparrow$}18.64\%) & \cellcolor{callgraphCol}25 & \cellcolor{callgraphCol}5 & \cellcolor{callgraphCol}0 & \cellcolor{callgraphCol}1.35 \\
    & \cellcolor{callgraphCol}Code + VersionHistory + MethodAge & \cellcolor{callgraphCol}60 (\textcolor{darkgreen}{$\uparrow$}1.69\%) & \cellcolor{callgraphCol}32 & \cellcolor{callgraphCol}6 & \cellcolor{callgraphCol}2 & \cellcolor{callgraphCol}1.50 \\
    \midrule

    \multirow{8}{*}{Conciseness}
    & \cellcolor{baselineCol}Without Context & \cellcolor{baselineCol}55 & \cellcolor{baselineCol}31 & \cellcolor{baselineCol}14 & \cellcolor{baselineCol}0 & \cellcolor{baselineCol}1.59 \\
    & \cellcolor{versionCol}Code + CallGraph & \cellcolor{versionCol}69 (\textcolor{darkgreen}{$\uparrow$}25.45\%) & \cellcolor{versionCol}25 & \cellcolor{versionCol}4 & \cellcolor{versionCol}2 & \cellcolor{versionCol}1.39 \\
    & \cellcolor{versionCol}Code + VersionHistory & \cellcolor{versionCol}53 (\textcolor{red}{$\downarrow$}3.64\%) & \cellcolor{versionCol}33 & \cellcolor{versionCol}13 & \cellcolor{versionCol}1 & \cellcolor{versionCol}1.62 \\
    & \cellcolor{callgraphCol}Code + CallGraph + VersionHistory & \cellcolor{callgraphCol}61 (\textcolor{darkgreen}{$\uparrow$}10.91\%) & \cellcolor{callgraphCol}28 & \cellcolor{callgraphCol}9 & \cellcolor{callgraphCol}2 & \cellcolor{callgraphCol}1.52 \\
    & \cellcolor{callgraphCol}Code + VersionHistory + CallGraph & \cellcolor{callgraphCol}61 (\textcolor{darkgreen}{$\uparrow$}10.91\%) & \cellcolor{callgraphCol}26 & \cellcolor{callgraphCol}11 & \cellcolor{callgraphCol}2 & \cellcolor{callgraphCol}1.54 \\
    & \cellcolor{callgraphCol}Code + CallGraph + VersionHistory + MethodAge & \cellcolor{callgraphCol}66 (\textcolor{darkgreen}{$\uparrow$}20.00\%) & \cellcolor{callgraphCol}26 & \cellcolor{callgraphCol}8 & \cellcolor{callgraphCol}0 & \cellcolor{callgraphCol}1.42 \\
    & \cellcolor{callgraphCol}Code + VersionHistory + CallGraph + MethodAge & \cellcolor{callgraphCol}73 (\textcolor{darkgreen}{$\uparrow$}32.73\%) & \cellcolor{callgraphCol}22 & \cellcolor{callgraphCol}5 & \cellcolor{callgraphCol}0 & \cellcolor{callgraphCol}1.32 \\
    & \cellcolor{callgraphCol}Code + VersionHistory + MethodAge & \cellcolor{callgraphCol}54 (\textcolor{red}{$\downarrow$}1.82\%) & \cellcolor{callgraphCol}37 & \cellcolor{callgraphCol}7 & \cellcolor{callgraphCol}2 & \cellcolor{callgraphCol}1.57 \\
    \midrule

    \multirow{8}{*}{\parbox{1.8cm}{Content \\ Adequacy}}
    & \cellcolor{baselineCol}Without Context & \cellcolor{baselineCol}43 & \cellcolor{baselineCol}43 & \cellcolor{baselineCol}12 & \cellcolor{baselineCol}2 & \cellcolor{baselineCol}1.74 \\
    & \cellcolor{versionCol}Code + CallGraph & \cellcolor{versionCol}56 (\textcolor{darkgreen}{$\uparrow$}30.23\%) & \cellcolor{versionCol}37 & \cellcolor{versionCol}6 & \cellcolor{versionCol}1 & \cellcolor{versionCol}1.52 \\
    & \cellcolor{versionCol}Code + VersionHistory & \cellcolor{versionCol}64 (\textcolor{darkgreen}{$\uparrow$}48.84\%) & \cellcolor{versionCol}29 & \cellcolor{versionCol}7 & \cellcolor{versionCol}0 & \cellcolor{versionCol}1.43 \\
    & \cellcolor{callgraphCol}Code + CallGraph + VersionHistory & \cellcolor{callgraphCol}56 (\textcolor{darkgreen}{$\uparrow$}30.23\%) & \cellcolor{callgraphCol}34 & \cellcolor{callgraphCol}9 & \cellcolor{callgraphCol}1 & \cellcolor{callgraphCol}1.55 \\
    & \cellcolor{callgraphCol}Code + VersionHistory + CallGraph & \cellcolor{callgraphCol}59 (\textcolor{darkgreen}{$\uparrow$}37.21\%) & \cellcolor{callgraphCol}33 & \cellcolor{callgraphCol}8 & \cellcolor{callgraphCol}0 & \cellcolor{callgraphCol}1.49 \\
    & \cellcolor{callgraphCol}Code + CallGraph + VersionHistory + MethodAge & \cellcolor{callgraphCol}47 (\textcolor{darkgreen}{$\uparrow$}9.30\%) & \cellcolor{callgraphCol}44 & \cellcolor{callgraphCol}7 & \cellcolor{callgraphCol}2 & \cellcolor{callgraphCol}1.65 \\
    & \cellcolor{callgraphCol}Code + VersionHistory + CallGraph + MethodAge & \cellcolor{callgraphCol}56 (\textcolor{darkgreen}{$\uparrow$}30.23\%) & \cellcolor{callgraphCol}33 & \cellcolor{callgraphCol}9 & \cellcolor{callgraphCol}2 & \cellcolor{callgraphCol}1.58 \\
    & \cellcolor{callgraphCol}Code + VersionHistory + MethodAge & \cellcolor{callgraphCol}51 (\textcolor{darkgreen}{$\uparrow$}18.60\%) & \cellcolor{callgraphCol}35 & \cellcolor{callgraphCol}12 & \cellcolor{callgraphCol}2 & \cellcolor{callgraphCol}1.65 \\
    \bottomrule
    \end{tabular}

    \begin{tablenotes}
        \footnotesize
        \item Note: Lower Mean Rank indicates better performance.
        \item Legend: \colorbox{baselineCol}{\phantom{X}} Without Context (Baseline); \colorbox{versionCol}{\phantom{X}} Single-context Scenarios; \colorbox{callgraphCol}{\phantom{X}} Multiple-types-of-context Scenarios.
        \item Values in the `\# Rank 1' column are followed by the percentage change (\textcolor{darkgreen}{$\uparrow$} increase, \textcolor{red}{$\downarrow$} decrease) relative to the 'Without Context' baseline.
    \end{tablenotes}

    \end{threeparttable}
}
\end{table}

Human evaluation in Table~\ref{tab:human_eval_new} and Table~\ref{tab:ranking_distribution_all} further refines these observations by examining Accuracy, Conciseness, and Content Adequacy. For \textit{Accuracy}, models that incorporate multiple types of context (e.g., Code + Version History + Call Graph) achieve a win rate of 25\% ($p=0.019$, Cliff’s $\delta=-0.39$, medium effect), comparable to single-context call graph models (25\%, $p=0.026$, $\delta=-0.51$, large effect). Ranking distributions (Table~\ref{tab:ranking_distribution_all}) confirm that multi-context models consistently achieve more Rank~1 outcomes (up to 72/100) and fewer Rank~2/3 results than the baseline, highlighting their robustness in producing accurate summaries.

For \textit{Conciseness}, benefits are less uniform. A plausible reason is that adding richer context encourages models to present more details, which helps \emph{accuracy} and \emph{content adequacy} but can introduce verbosity. In our study, a single structural context already improves conciseness (Code + Call Graph: win $28\%$, $p=0.007$, $\delta=-0.44$, medium), while adding more context brings only a marginal enhancement (Code + Version History + Call Graph + Method Age: win $32\%$, $p=0.001$, $\delta=-0.44$) and several other mixtures show limited or no significant gains. This pattern supports the intuition that more context is not automatically better for brevity; when conciseness is the priority, leaner structural context (e.g., call graph alone) and explicit length control are preferable to stacking multiple context types.

For \textit{Content Adequacy}, the advantage of multiple types of context is more consistent. Models such as Code + Version History + Call Graph achieve a 30\% win rate ($p=0.013$, $\delta=-0.43$, medium effect), close to the best single-context performance with version history (31\%, $\delta=-0.55$, large effect). Ranking analysis indicates that multiple-context models nearly double the proportion of Rank~1 outcomes compared to the baseline (up to +79.3\%), confirming that contextual enrichment strongly improves adequacy.

Overall, these results suggest that while encoding various context types often boosts performance beyond the baseline, they are not always superior to the strongest single-context models. The most consistent gains are observed in \textit{Accuracy} and \textit{Content Adequacy}, where multiple context types deliver both statistically significant improvements and more stable top-rank outcomes. In contrast, improvements in \textit{Conciseness} depend strongly on the specific context combination. These findings indicate that practitioners should prioritise multi-context models when accuracy and adequacy are critical, while carefully selecting context types when conciseness is the main goal.

\begin{tcolorbox}[left=1pt, top=1pt, right=1pt, bottom=1pt]
\textbf{Summary}:
Combining multiple types of additional context can further provide performance gains, though the improvements are highly dependent on the specific task, model, and context combination. For classification tasks, the benefits are often substantial, with performance gains \revised{reaching up to \textbf{+21.48\% macro-F1} in Code Classification (CodeT5 with all three context types) and \textbf{+15.04\% F1} in Code Clone Detection (PLBART with Version History and Method Age).} For the Code Summarisation task, the improvements are more selective, yet the most effective combinations still deliver notable gains (e.g., up to \textbf{+5.79\% BLEU-4}). \revised{Among the LLMs, combined context amplifies the gains for the two models that already benefit from single context (Qwen2.5-Coder 7B and 14B, up to +41.4\% METEOR) and does not help the other five, which decline under every combination.} \revised{A follow-up human evaluation of CodeT5 summaries is consistent with this trend: models augmented with multiple types of context often generate summaries that are ranked higher by human annotators for accuracy and adequacy.}

\end{tcolorbox}

\textbf{RQ3: What is the impact of different representation-level aggregation techniques on integrating source code with additional context for these models?}
\label{sec:result_rq3}

Table~\ref{tab:rq3_clone_detection} shows the influence of different aggregation techniques, including Concatenation (Concat), Max-Pooling, and Concatenation of Absolute Differences (Diff Concat), on Code Clone Detection performance across different additional context settings. Generally, no single aggregation technique consistently improves F1 scores across all models and context types. However, \textbf{Diff Concat achieves the most substantial improvements} compared to other techniques when considering identical types of context across all models. Particularly notable is CodeBERT, which shows the largest F1-score enhancements using Diff Concat in four of five context settings: Version History + CallGraph (+24.62\%), Version History + Method Age (+24.62\%), Version History alone (+20.46\%), and Version History + CallGraph + Method Age (+20.46\%).

\begin{table}[htbp]
\centering
\scriptsize

\caption{Code Clone Detection Results with 3 Different Aggregation Techniques.}
\label{tab:rq3_clone_detection}

\renewcommand{\arraystretch}{1.2}

\resizebox{\textwidth}{!}{%
\begin{threeparttable}

\begin{tabular}{ll|
c r l|
c r l|
c r l|
c r l|
c r l}
\toprule
\rowcolor{gray!20}
\multicolumn{2}{c|}{\textbf{Context}} &
\multicolumn{3}{c|}{\textbf{CodeBERT}} &
\multicolumn{3}{c|}{\textbf{GraphCodeBERT}} &
\multicolumn{3}{c|}{\textbf{CodeT5}} &
\multicolumn{3}{c|}{\textbf{PLBART}} &
\multicolumn{3}{c}{\textbf{ASTNN}} \\
\cmidrule(lr){3-5}\cmidrule(lr){6-8}\cmidrule(lr){9-11}\cmidrule(lr){12-14}\cmidrule(lr){15-17}
\rowcolor{gray!20}
\multicolumn{2}{c|}{} &
\textbf{F1} & \textbf{\%} & &
\textbf{F1} & \textbf{\%} & &
\textbf{F1} & \textbf{\%} & &
\textbf{F1} & \textbf{\%} & &
\textbf{F1} & \textbf{\%} & \\
\midrule

\multirow{1}{*}{\textbf{W/O Ctx}} &
--
& 0.7180 & &
& 0.7429 & &
& 0.7805 & &
& 0.7778 & &
& 0.8718 & &
\\
\midrule

\multirow{3}{*}{\textbf{Vers. Hist.}}
& ConCat
  & 0.7442 & 3.65\% & \barPercentThirty{3.65}
  & 0.7778 & 4.70\% & \barPercentThirty{4.70}
  & 0.9048 & 15.92\% & \barPercentThirty{15.92}
  & 0.7778 & 0.00\% & \barPercentThirty{0.00}
  & 0.9000 & 3.24\% & \barPercentThirty{3.24}
\\
& Max~Pooling
  & 0.7368 & 2.63\% & \barPercentThirty{2.63}
  & \cellcolor{yellow!30}0.8718 & \cellcolor{yellow!30}17.36\% & \cellcolor{yellow!30}\barPercentThirty{17.36}
  & 0.8781 & 12.50\% & \barPercentThirty{12.50}
  & 0.8000 & 2.86\% & \barPercentThirty{2.86}
  & 0.9000 & 3.24\% & \barPercentThirty{3.24}
\\
& Diff~Concat
  & \cellcolor{yellow!30}0.8649 & \cellcolor{yellow!30}20.46\% & \cellcolor{yellow!30}\barPercentThirty{20.46}
  & 0.8000 & 7.69\% & \barPercentThirty{7.69}
  & \cellcolor{yellow!30}0.9231 & \cellcolor{yellow!30}18.27\% & \cellcolor{yellow!30}\barPercentThirty{18.27}
  & \cellcolor{yellow!30}0.9000 & \cellcolor{yellow!30}15.71\% & \cellcolor{yellow!30}\barPercentThirty{15.71}
  & \cellcolor{yellow!30}0.9048 & \cellcolor{yellow!30}3.78\% & \cellcolor{yellow!30}\barPercentThirty{3.78}
\\
\midrule

\multirow{3}{*}{\textbf{CG}}
& ConCat
  & \cellcolor{yellow!30}0.7907 & \cellcolor{yellow!30}10.13\% & \cellcolor{yellow!30}\barPercentThirty{10.13}
  & 0.6857 & -7.69\% & \barPercentThirty{-7.69}
  & \cellcolor{yellow!30}0.7568 & \cellcolor{yellow!30}-3.04\% & \cellcolor{yellow!30}\barPercentThirty{-3.04}
  & 0.8108 & 4.25\% & \barPercentThirty{4.25}
  & 0.8293 & -4.88\% & \barPercentThirty{-4.88}
\\
& Max~Pooling
  & 0.7180 & 0.00\% & \barPercentThirty{0.00}
  & 0.7647 & 2.94\% & \barPercentThirty{2.94}
  & 0.7442 & -4.65\% & \barPercentThirty{-4.65}
  & 0.8333 & 7.14\% & \barPercentThirty{7.14}
  & \cellcolor{yellow!30}0.8718 & \cellcolor{yellow!30}0.00\% & \cellcolor{yellow!30}\barPercentThirty{0.00}
\\
& Diff~Concat
  & 0.6546 & -8.83\% & \barPercentThirty{-8.83}
  & \cellcolor{yellow!30}0.8108 & \cellcolor{yellow!30}9.15\% & \cellcolor{yellow!30}\barPercentThirty{9.15}
  & 0.6957 & -10.87\% & \barPercentThirty{-10.87}
  & \cellcolor{yellow!30}0.8781 & \cellcolor{yellow!30}12.89\% & \cellcolor{yellow!30}\barPercentThirty{12.89}
  & 0.8636 & -0.93\% & \barPercentThirty{-0.93}
\\
\midrule

\multirow{3}{*}{\textbf{\shortstack{Vers. Hist.\\+ CG}}}
& ConCat
  & 0.7500 & 4.46\% & \barPercentThirty{4.46}
  & 0.7429 & 0.00\% & \barPercentThirty{0.00}
  & 0.8205 & 5.13\% & \barPercentThirty{5.13}
  & 0.8108 & 4.25\% & \barPercentThirty{4.25}
  & 0.8571 & -1.68\% & \barPercentThirty{-1.68}
\\
& Max~Pooling
  & 0.7180 & 0.00\% & \barPercentThirty{0.00}
  & 0.7368 & -0.81\% & \barPercentThirty{-0.81}
  & 0.8500 & 8.91\% & \barPercentThirty{8.91}
  & 0.8205 & 5.49\% & \barPercentThirty{5.49}
  & 0.8205 & -5.88\% & \barPercentThirty{-5.88}
\\
& Diff~Concat
  & \cellcolor{yellow!30}0.8947 & \cellcolor{yellow!30}24.62\% & \cellcolor{yellow!30}\barPercentThirty{24.62}
  & \cellcolor{yellow!30}0.7895 & \cellcolor{yellow!30}6.27\% & \cellcolor{yellow!30}\barPercentThirty{6.27}
  & \cellcolor{yellow!30}0.8947 & \cellcolor{yellow!30}14.64\% & \cellcolor{yellow!30}\barPercentThirty{14.64}
  & \cellcolor{yellow!30}0.9000 & \cellcolor{yellow!30}15.71\% & \cellcolor{yellow!30}\barPercentThirty{15.71}
  & \cellcolor{yellow!30}0.8837 & \cellcolor{yellow!30}1.37\% & \cellcolor{yellow!30}\barPercentThirty{1.37}
\\
\midrule

\multirow{3}{*}{\textbf{\shortstack{Vers. Hist.\\+ M. Age}}}
& ConCat
  & 0.7727 & 7.63\% & \barPercentThirty{7.63}
  & \cellcolor{yellow!30}0.8421 & \cellcolor{yellow!30}13.36\% & \cellcolor{yellow!30}\barPercentThirty{13.36}
  & 0.8182 & 4.83\% & \barPercentThirty{4.83}
  & \cellcolor{yellow!30}0.8947 & \cellcolor{yellow!30}15.04\% & \cellcolor{yellow!30}\barPercentThirty{15.04}
  & 0.8571 & -1.68\% & \barPercentThirty{-1.68}
\\
& Max~Pooling
  & 0.7895 & 9.96\% & \barPercentThirty{9.96}
  & 0.7500 & 0.96\% & \barPercentThirty{0.96}
  & 0.7907 & 1.31\% & \barPercentThirty{1.31}
  & 0.8649 & 11.20\% & \barPercentThirty{11.20}
  & \cellcolor{yellow!30}0.9500 & \cellcolor{yellow!30}8.97\% & \cellcolor{yellow!30}\barPercentThirty{8.97}
\\
& Diff~Concat
  & \cellcolor{yellow!30}0.8947 & \cellcolor{yellow!30}24.62\% & \cellcolor{yellow!30}\barPercentThirty{24.62}
  & 0.7895 & 6.27\% & \barPercentThirty{6.27}
  & \cellcolor{yellow!30}0.8500 & \cellcolor{yellow!30}8.91\% & \cellcolor{yellow!30}\barPercentThirty{8.91}
  & 0.8421 & 8.27\% & \barPercentThirty{8.27}
  & 0.9268 & 6.31\% & \barPercentThirty{6.31}
\\
\midrule

\multirow{3}{*}{\textbf{\shortstack{Vers. Hist.\\+ CG + M. Age}}}
& ConCat
  & 0.7500 & 4.46\% & \barPercentThirty{4.46}
  & \cellcolor{yellow!30}0.8500 & \cellcolor{yellow!30}14.42\% & \cellcolor{yellow!30}\barPercentThirty{14.42}
  & 0.8000 & 2.50\% & \barPercentThirty{2.50}
  & 0.8108 & 4.25\% & \barPercentThirty{4.25}
  & 0.8781 & 0.72\% & \barPercentThirty{0.72}
\\
& Max~Pooling
  & 0.6342 & -11.67\% & \barPercentThirty{-11.67}
  & 0.8421 & 13.36\% & \barPercentThirty{13.36}
  & 0.8293 & 6.25\% & \barPercentThirty{6.25}
  & \cellcolor{yellow!30}0.9000 & \cellcolor{yellow!30}15.71\% & \cellcolor{yellow!30}\barPercentThirty{15.71}
  & 0.8781 & 0.72\% & \barPercentThirty{0.72}
\\
& Diff~Concat
  & \cellcolor{yellow!30}0.8649 & \cellcolor{yellow!30}20.46\% & \cellcolor{yellow!30}\barPercentThirty{20.46}
  & 0.8205 & 10.45\% & \barPercentThirty{10.45}
  & \cellcolor{yellow!30}0.8718 & \cellcolor{yellow!30}11.70\% & \cellcolor{yellow!30}\barPercentThirty{11.70}
  & \cellcolor{yellow!30}0.9000 & \cellcolor{yellow!30}15.71\% & \cellcolor{yellow!30}\barPercentThirty{15.71}
  & \cellcolor{yellow!30}0.8837 & \cellcolor{yellow!30}1.37\% & \cellcolor{yellow!30}\barPercentThirty{1.37}
\\
\bottomrule
\end{tabular}

\begin{tablenotes}
    \footnotesize
    \item The highlighted cell indicates the best-performing aggregation technique (F1-score) for within the same model and same context(s).
\end{tablenotes}

\end{threeparttable}
}
\end{table}

Further analysis of aggregation techniques within the same model and context highlights Diff Concat’s stronger performance, achieving the highest F1-score improvement in 17 out of 26 experiments (highlighted cells)\revised{, 13 of them by a margin above the 2-point resolution of the SeSaMe test set (Section~\ref{sec:threats}); the remaining four are within that resolution}. In contrast, ConCat and Max-Pooling achieve the best improvements in only 5 and 4 cases, respectively. Diff Concat also stands out in four out of five context scenarios in CodeT5 and PLBART, indicating that this approach not only improves encoder-only models (e.g., CodeBERT) but also supports encoder–decoder architectures.

Meanwhile, Max-Pooling aggregation generally produces only modest improvements and in several cases negatively impacts performance. For example, GraphCodeBERT and ASTNN experience slight F1-score reductions using Max-Pooling with context types, such as Version History + CallGraph (-0.81\%) and CallGraph (-0.93\%). In contrast, the Concat technique typically delivers small but stable gains compared to the baseline, except in notable cases such as CodeT5 (+15.92\% with Version History) and PLBART (+15.04\% with Version History and Method Age). These findings suggest that while Max-Pooling may oversimplify contextual information in binary classification settings, Concat and particularly Diff Concat are better at preserving signal differences critical for clone detection.

In conclusion, Diff Concat emerges as the most effective aggregation method in the Code Clone Detection task, significantly enhancing performance by explicitly encoding differences between two code snippets before incorporating additional context. Conversely, Max-Pooling appears less effective, likely due to oversimplification or the loss of nuanced contextual information, while Concat offers moderate but consistent improvements.

\begin{table}[htbp]
\centering
\scriptsize

\caption{Code Classification Results with 3 Different Aggregation Techniques.}
\label{tab:rq3_code_classification}

\renewcommand{\arraystretch}{1.2}

\resizebox{\textwidth}{!}{%
\begin{threeparttable}

\begin{tabular}{ll|
c r l|
c r l|
c r l|
c r l|
c r l}
\toprule
\rowcolor{gray!20}
\multicolumn{2}{c|}{\textbf{Context}} &
\multicolumn{3}{c|}{\textbf{CodeBERT}} &
\multicolumn{3}{c|}{\textbf{GraphCodeBERT}} &
\multicolumn{3}{c|}{\textbf{CodeT5}} &
\multicolumn{3}{c|}{\textbf{PLBART}} &
\multicolumn{3}{c}{\textbf{ASTNN}} \\
\cmidrule(lr){3-5}\cmidrule(lr){6-8}\cmidrule(lr){9-11}\cmidrule(lr){12-14}\cmidrule(lr){15-17}
\rowcolor{gray!20}
\multicolumn{2}{c|}{} &
\textbf{macro-F1} & \textbf{\%} & &
\textbf{macro-F1} & \textbf{\%} & &
\textbf{macro-F1} & \textbf{\%} & &
\textbf{macro-F1} & \textbf{\%} & &
\textbf{macro-F1} & \textbf{\%} & \\
\midrule

\multirow{1}{*}{\textbf{W/O Ctx}} &
--
& 0.6698 & &
& 0.7661 & &
& 0.6569 & &
& 0.6822 & &
& 0.4887 & &
\\
\midrule

\multirow{2}{*}{\textbf{Vers. Hist.}}
& Concat
  & 0.6320 & -5.65\% & \barPercentThirty{-5.65}
  & \cellcolor{yellow!30}0.7781 & \cellcolor{yellow!30}1.58\% & \cellcolor{yellow!30}\barPercentThirty{1.58}
  & 0.7094 & 8.00\% & \barPercentThirty{8.00}
  & \cellcolor{yellow!30}0.7147 & \cellcolor{yellow!30}4.76\% & \cellcolor{yellow!30}\barPercentThirty{4.76}
  & 0.4740 & -3.01\% & \barPercentThirty{-3.01}
\\
& Max~Pooling
  & \cellcolor{yellow!30}0.7584 & \cellcolor{yellow!30}13.22\% & \cellcolor{yellow!30}\barPercentThirty{13.22}
  & 0.7773 & 1.47\% & \barPercentThirty{1.47}
  & \cellcolor{yellow!30}0.7569 & \cellcolor{yellow!30}15.23\% & \cellcolor{yellow!30}\barPercentThirty{15.23}
  & 0.6907 & 1.25\% & \barPercentThirty{1.25}
  & \cellcolor{yellow!30}0.5572 & \cellcolor{yellow!30}14.01\% & \cellcolor{yellow!30}\barPercentThirty{14.01}
\\
\midrule

\multirow{2}{*}{\textbf{CG}}
& Concat
  & \cellcolor{yellow!30}0.7345 & \cellcolor{yellow!30}9.65\% & \cellcolor{yellow!30}\barPercentThirty{9.65}
  & 0.7618 & -0.55\% & \barPercentThirty{-0.55}
  & \cellcolor{yellow!30}0.7639 & \cellcolor{yellow!30}16.29\% & \cellcolor{yellow!30}\barPercentThirty{16.29}
  & \cellcolor{yellow!30}0.7631 & \cellcolor{yellow!30}11.86\% & \cellcolor{yellow!30}\barPercentThirty{11.86}
  & \cellcolor{yellow!30}0.6150 & \cellcolor{yellow!30}25.84\% & \cellcolor{yellow!30}\barPercentThirty{25.84}
\\
& Max~Pooling
  & 0.6408 & -4.33\% & \barPercentThirty{-4.33}
  & \cellcolor{yellow!30}0.8162 & \cellcolor{yellow!30}6.55\% & \cellcolor{yellow!30}\barPercentThirty{6.55}
  & 0.7031 & 7.04\% & \barPercentThirty{7.04}
  & 0.7191 & 5.41\% & \barPercentThirty{5.41}
  & 0.5451 & 11.55\% & \barPercentThirty{11.55}
\\
\midrule

\multirow{2}{*}{\textbf{\shortstack{Vers. Hist.\\+ CG}}}
& Concat
  & 0.7304 & 9.04\% & \barPercentThirty{9.04}
  & 0.7797 & 1.78\% & \barPercentThirty{1.78}
  & \cellcolor{yellow!30}0.7517 & \cellcolor{yellow!30}14.44\% & \cellcolor{yellow!30}\barPercentThirty{14.44}
  & 0.7842 & 14.95\% & \barPercentThirty{14.95}
  & \cellcolor{yellow!30}0.5376 & \cellcolor{yellow!30}10.00\% & \cellcolor{yellow!30}\barPercentThirty{10.00}
\\
& Max~Pooling
  & \cellcolor{yellow!30}0.7451 & \cellcolor{yellow!30}11.23\% & \cellcolor{yellow!30}\barPercentThirty{11.23}
  & \cellcolor{yellow!30}0.8185 & \cellcolor{yellow!30}6.84\% & \cellcolor{yellow!30}\barPercentThirty{6.84}
  & 0.7377 & 12.30\% & \barPercentThirty{12.30}
  & \cellcolor{yellow!30}0.7956 & \cellcolor{yellow!30}16.62\% & \cellcolor{yellow!30}\barPercentThirty{16.62}
  & 0.5346 & 9.39\% & \barPercentThirty{9.39}
\\
\midrule

\multirow{2}{*}{\textbf{\shortstack{Vers. Hist.\\+ M. Age}}}
& Concat
  & 0.6624 & -1.11\% & \barPercentThirty{-1.11}
  & 0.7835 & 2.28\% & \barPercentThirty{2.28}
  & \cellcolor{yellow!30}0.7129 & \cellcolor{yellow!30}8.53\% & \cellcolor{yellow!30}\barPercentThirty{8.53}
  & 0.7195 & 5.46\% & \barPercentThirty{5.46}
  & 0.5679 & 16.20\% & \barPercentThirty{16.20}
\\
& Max~Pooling
  & \cellcolor{yellow!30}0.7494 & \cellcolor{yellow!30}11.88\% & \cellcolor{yellow!30}\barPercentThirty{11.88}
  & \cellcolor{yellow!30}0.8329 & \cellcolor{yellow!30}8.72\% & \cellcolor{yellow!30}\barPercentThirty{8.72}
  & 0.6557 & -0.17\% & \barPercentThirty{-0.17}
  & \cellcolor{yellow!30}0.7246 & \cellcolor{yellow!30}6.21\% & \cellcolor{yellow!30}\barPercentThirty{6.21}
  & \cellcolor{yellow!30}0.5953 & \cellcolor{yellow!30}21.81\% & \cellcolor{yellow!30}\barPercentThirty{21.81}
\\
\midrule

\multirow{2}{*}{\textbf{\shortstack{Vers. Hist.\\+ CG + M. Age}}}
& Concat
  & \cellcolor{yellow!30}0.7813 & \cellcolor{yellow!30}16.65\% & \cellcolor{yellow!30}\barPercentThirty{16.65}
  & \cellcolor{yellow!30}0.7645 & \cellcolor{yellow!30}-0.21\% & \cellcolor{yellow!30}\barPercentThirty{-0.21}
  & \cellcolor{yellow!30}0.7980 & \cellcolor{yellow!30}21.48\% & \cellcolor{yellow!30}\barPercentThirty{21.48}
  & \cellcolor{yellow!30}0.8191 & \cellcolor{yellow!30}20.07\% & \cellcolor{yellow!30}\barPercentThirty{20.07}
  & \cellcolor{yellow!30}0.5505 & \cellcolor{yellow!30}12.64\% & \cellcolor{yellow!30}\barPercentThirty{12.64}
\\
& Max~Pooling
  & 0.7003 & 4.54\% & \barPercentThirty{4.54}
  & 0.7353 & -4.02\% & \barPercentThirty{-4.02}
  & 0.7447 & 13.37\% & \barPercentThirty{13.37}
  & 0.7707 & 12.97\% & \barPercentThirty{12.97}
  & 0.5365 & 9.78\% & \barPercentThirty{9.78}
\\
\bottomrule
\end{tabular}

\begin{tablenotes}
    \footnotesize
    \item The highlighted cell in each context group indicates the best-performing aggregation technique (macro-F1 score) for that model.
\end{tablenotes}

\end{threeparttable}
}
\end{table}

Table~\ref{tab:rq3_code_classification} illustrates the impacts of two aggregation techniques, Concat and Max-Pooling, on the performance of the Code Classification task. Generally, neither aggregation method consistently outperforms the other across all models and context types. However, Max-Pooling appears slightly more effective overall, delivering the largest macro-F1 improvements in 13 out of 25 experiments (highlighted cells), compared to ConCat’s 12 cases.

Notably, Max-Pooling achieves substantial macro-F1 improvements for CodeT5 (+15.23\% with Version History), ASTNN (+21.81\% with Version History and Method Age), and PLBART (+16.62\% with Version History and Call Graph). Conversely, the Concat aggregation technique yields significant improvements in specific multi-context scenarios, particularly for CodeT5 (+21.48\% with Version History, Call Graph, and Method Age) and ASTNN (+25.84\% with Call Graph). These results indicate that Max-Pooling is often favoured by encoder-only and encoder–decoder models when context types are moderately complex, whereas Concat may better capture richer signals when multiple types of context are combined.

Each model also demonstrates varying sensitivity to aggregation techniques. For example, CodeBERT achieves stronger results with Max-Pooling, gaining +13.22\% with Version History, while ASTNN benefits significantly from Concat when using Call Graph context (+25.84\%). This suggests that architecture plays a central role in determining which aggregation strategy is most effective.

In conclusion, while neither aggregation technique universally dominates in Code Classification, Max-Pooling generally demonstrates slightly greater consistency. Still, Concat remains highly competitive, particularly in richer multi-context settings. Therefore, careful selection and tuning of aggregation methods based on both model architecture and context type are essential to maximise classification performance.

\begin{tcolorbox}[left=1pt, top=1pt, right=1pt, bottom=1pt]
\textbf{Summary}:
The choice of aggregation technique is crucial, as no single strategy proves universally superior across all scenarios.
In \textit{Code Clone Detection} (binary classification), the Concatenation of Absolute Differences (Diff Concat) \revised{is the best strategy in most settings, and by a clear margin for encoder-only models like CodeBERT}, by explicitly encoding similarity signals.
In \textit{Code Classification} (multi-class prediction), no single technique dominates: Max-Pooling provides slightly more consistent gains for encoder-decoder models (e.g., CodeT5, PLBART), while Concat is often superior when richer context combinations are used.
ASTNN exhibits the greatest sensitivity to aggregation choice, with large improvements in certain context types (e.g., +25.84\% with Concat on Call Graph) but regressions in others.
Overall, the optimal aggregation strategy depends on the task type (binary vs multi-class), the richness of context, and the underlying model architecture.
\end{tcolorbox}

\section{Discussion}
\label{sec:discussion}

Our findings across three research questions indicate that while incorporating additional context generally improves model performance, the choice of which context to use and how to integrate it is a critical factor. In this section, we synthesise the results to discuss the specific roles of version history and call graphs (RQ1, Section~\ref{sec:result_rq1}), the impact of combining multiple types of context (RQ2, Section~\ref{sec:result_rq2}), and the sensitivity to different aggregation strategies (RQ3, Section~\ref{sec:result_rq3}). We then triangulate these automated metrics with our human evaluation to ground our findings, and conclude by outlining the practical implications of our work, its limitations, and directions for future work.

\subsection{Empirical Findings and Implications}
\label{sec:findings-implications}

\paragraph{Overview.}
Across \emph{Clone Detection}, \emph{Code Classification}, and \emph{Code Summarisation}, both our automated metrics and the human evaluation point to the same headline finding: \emph{adding context helps overall}, but \emph{which} context to add and \emph{how} to combine it depends on the \emph{task} and the \emph{model}.

\paragraph{Contextual impact: Version history as a reliable signal.}
As our primary finding for RQ1, version history \revised{improves performance in the same direction across tasks for all five foundational models} (Section~\ref{sec:result_rq1} - RQ1). It is particularly effective for clone detection and summarisation in the automated evaluations, and it aligns with human judgements on \emph{accuracy} and \emph{content adequacy} in the qualitative study. \revised{For the LLMs the picture differs: version history injected into the prompt gives little or no gain on summarisation even for the two models that benefit from context, where the useful signal is the call graph (Section~\ref{sec:result_rq1}).} In practice, version history is a strong default \revised{for representation-level encoding in foundational models} when seeking robust gains without heavy tuning.
\emph{Implications for researchers:} use version history as the default ablation baseline; report per-task and per-metric gains; study selective version-history sampling to mitigate token-length constraints.
\emph{Implications for practitioners:} prefer version history when the budget is tight; expect steady lifts in clone detection and in summarisation accuracy/adequacy.

\paragraph{Mixed results for call graphs: task/model dependence.}
Call-graph context delivers substantial improvements for code classification but shows mixed outcomes for clone detection and summarisation as detailed in section~\ref{sec:result_rq1} - RQ1 (a mechanistic analysis of why this occurs is provided in Section~\ref{sec:cg_error_analysis}). \revised{For the two LLMs whose summarisation improves with prompt-level context (Qwen2.5-Coder 7B and 14B), call graph is the context that drives the gain, and combinations containing it give the largest improvements (Section~\ref{sec:result_rq2}).} In human evaluations, summary outputs from augmented models with call hierarchy context often improve both \emph{accuracy} and \emph{conciseness} in specific setups; however, call graphs are not universally superior across all models or items.
\emph{Implications for researchers:} stratify analyses by task and backbone; measure dimension-specific effects (accuracy/adequacy vs.\ conciseness).
\emph{Implications for practitioners:} add call graph for classification or when targeting accuracy/conciseness in summaries; do not assume universal gains.

\paragraph{Combined multiple types of context: Version history, call graph, and method age.}
Encoding multiple context types can selectively improve performance (RQ2, Section~\ref{sec:result_rq2}). For classification tasks, combining multiple types of context generally \emph{amplifies} improvements, suggesting complementary signals between historical, structural, and temporal cues. For summarisation, gains are \emph{selective}: the best combinations depend on the \emph{target dimension} and the \emph{model}. \revised{Among the LLMs, combining context types amplifies the benefit for Qwen2.5-Coder 7B and 14B but does not help the five models that do not benefit from any single context type; whether prompt-level context helps summarisation is decided by model family before the choice of combination matters.} Concretely, when prioritising \emph{accuracy} and \emph{content adequacy}, pairing version history with call graph and/or method age data tends to be beneficial; when prioritising \emph{conciseness}, lighter combinations (e.g., call graph alone or call graph with minimal additions) can be as effective as richer multi-context settings. These patterns emphasise the need to \emph{tailor} context choices to the task and evaluation dimension rather than expecting a single combination to dominate everywhere.
\emph{Implications for researchers:} compare single vs. multi-context under token-length constraints; report effect sizes and interaction effects.
\emph{Implications for practitioners:} for classification, prefer multiple context combination (\emph{version history + call graph} $\pm$ \emph{method age}); for summarisation, choose by goal: accuracy/adequacy $\rightarrow$ version history + call graph or version history + method age; conciseness $\rightarrow$ call-graph–heavy or light setups.

\paragraph{Human evaluation provides practical validation.}
The blinded-source and rank-ordering with ties study confirms the automated trends \revised{for CodeT5}: context-augmented summaries are preferred in terms of \emph{accuracy} and \emph{content adequacy}, while gains in \emph{conciseness} depend on the specific context and model. \revised{This validation applies to the CodeT5 results only; the LLM summarisation results were not part of the human study and rest on automatic metrics.} High tie rates on some items further indicate overlap in performance and the importance of matching context choice to the intended objective.
\emph{Implications for researchers:} pair automated metrics with blinded rank-based human judgements; report effect sizes, not only $p$-values.
\emph{Implications for practitioners:} pilot small human reviews to validate dimension-level gains before wider rollout.

\paragraph{The critical role of aggregation.}
Finally, our analysis for RQ3 established that performance is highly sensitive to the choice of representation-level aggregation (Section~\ref{sec:result_rq3}). Diff-Concat is consistently the strongest strategy for the binary classification task of clone detection. For multi-class classification, however, no single technique dominates, with the optimal choice depending on the model and the richness of the context. \emph{Implications for both:} treat aggregation (Diff-Concat, Concat, Max-Pooling) as a tunable design choice; select per task and model rather than fixing one operator.

\paragraph{\revised{Fine-tuning, fusion mechanism, and the use of context by LLMs.}}
\revised{Our evaluation of seven LLMs (CodeLlama, Qwen2.5-Coder, and DeepSeek-Coder) involves two variables, whether the model is fine-tuned and how context is fused, and we draw two separate conclusions from it (Section~\ref{sec:llm_baselines}).}

\revised{The first concerns fine-tuning and is a within-model comparison. Zero-shot prompting can benefit from context (for example, CodeLlama-34B improves from an F1 of 0.200 to 0.667 on clone detection when version history is included in the prompt), but the same model reaches its highest performance with context after fine-tuning on context-augmented inputs: Qwen2.5-Coder-14B rises from an F1 of 0.1053 zero-shot to 0.8889 fine-tuned on clone detection with the full context, and CodeLlama-34B from a macro-F1 of 0.2152 to 0.8518 on classification under the same full context. When structural and temporal artefacts are present during training, the model learns to align code semantics with their historical and dependency context; when the same information is appended only at inference time, it must infer that mapping without supervision. This is the finding our evidence supports cleanly, and it holds across all seven LLMs on the two classification tasks.}

\revised{The second concerns the fusion mechanism and is confounded. Comparing fine-tuned LLMs with prompt-level context against fine-tuned encoders with representation-level fusion holds fine-tuning constant, but no single backbone is evaluated under both mechanisms, so architecture and mechanism vary together. On clone detection the two are close (Qwen2.5-Coder-14B 0.8889; CodeT5 0.9048 with concatenation). On code classification the fine-tuned LLMs reach the highest macro-F1 in the study (CodeLlama-34B 0.8518; PLBART 0.8191 and CodeT5 0.7980 for the encoders, Table~\ref{tab:result_rq2_classification}). We therefore do not claim that representation-level fusion is the better mechanism. What the encoder results establish is that simple, learned aggregation over a 512-token budget already captures a large part of the available signal, and that the choice of aggregation operator matters (RQ3). Whether a learned fusion head would improve a decoder-only LLM further is an open question that we identify as future work (Section~\ref{sec:limitation}).}

\subsubsection{Implications for Researchers}
\label{sec:contribution-researchers}

\paragraph{Advancement in Code Representation}
This study advances neural code representation by integrating historical (version history) and structural (call graph) information directly into learned vector encodings, moving beyond code-only inputs. By treating these artifacts as first-class signals during fine-tuning, we capture evolutionary cues (what changed, how often, and in what patterns) and dependency cues (who calls whom and in which context), enabling models to encode semantics that are difficult to infer from a static snapshot alone.

\paragraph{Empirical Evidence}
Converging evidence from automated metrics across classification and generation, together with human preferences in a blinded-source and rank-ordering evaluation, indicates that context-enhanced representations improve model effectiveness \revised{for the five foundational models, and for a subset of the LLMs on summarisation; the evidence does not support a general claim across all model families}. For code summarisation in particular, the dimension-level analysis is instructive: accuracy and content adequacy benefit most reliably from context, while conciseness depends on the specific context and model. These results collectively strengthen the case that representation-level context encoding is a principled and practical avenue for improving program comprehension tasks.

\paragraph{Methodological Suggestions}
Our results suggest actionable guidance for research practice. As a default, consider encoding version history due to its broad reliability across tasks and models; add call graphs when the objective prioritises accuracy and conciseness in summaries or when tackling classification tasks that benefit from structural signals. For aggregation strategies, prefer Diff-Concat for clone detection, and empirically compare Concat versus Max-Pooling for classification, tuning per model and dataset. We also recommend reporting both automated metrics and human judgements, using rank-based analyses and effect sizes to characterise practical impact beyond statistical significance and to illuminate which evaluation dimensions (accuracy, adequacy, conciseness) benefit from which additional context.

\subsubsection{Implications for Practitioners}
\label{sec:contribution-practitioners}

\paragraph{Enhancements to Program-Comprehension Tools}
Context-enriched code representations can strengthen developer-facing tooling across the lifecycle. For clone detection, they reduce false positives and negatives by incorporating evolutionary and structural cues. For code summarisation, their information is more precise, complete, and descriptive, which improves readability and onboarding. These benefits can support IDE features (e.g., inline summaries, similarity hints), CI/CD checks (e.g., detection of duplicate or fractured logic), and code review assistants that prioritise high-impact changes.

\paragraph{Practical Recommendations}
The empirical results suggest a simple, actionable playbook for integrating context into production workflows:
\begin{itemize}
  \item \emph{Default to version history:} adopt version history as the baseline contextual signal due to its reliability across tasks and models.
  \item \emph{Add call graph selectively:} include the call graph for classification tasks or when targeting higher accuracy and conciseness in summaries.
  \item \emph{Aggregation techniques by task:} for clone detection, try \emph{Diff-Concat} first; for classification, compare \emph{Concat} versus \emph{Max-Pooling} under your data distribution and model.
  \item \emph{Summarisation setup:} consider version history combined with a lightweight signal for recency or stability (e.g., method age) and run a small human evaluation before organisation-wide rollout.
  \item \revised{\emph{Validate per model family before injecting context into an LLM prompt:} for summarisation, prompt-level context helped only Qwen2.5-Coder 7B and 14B in our study and lowered every metric for CodeLlama, DeepSeek-Coder, and Qwen2.5-Coder-1.5B, zero-shot and fine-tuned alike. Where representation-level encoding in a foundational model is an option, it gives larger and more reliable gains on classification-style tasks at bounded inference cost (Section~\ref{sec:limitation}).}
  \item \revised{\emph{For retrieval-augmented and agentic pipelines, consider historical context alongside the structural context already retrieved:} coding agents already follow dependency structure and retrieve related files~\cite{bairi2024codeplan,xia2025agentless}, which corresponds to the structural context studied here. Version history is not typically part of the context these systems retrieve, and it was the most reliable signal at the method level in our study; it can be mined at repository scale from version control at low cost (Section~\ref{sec:limitation}). Our results do not evaluate such pipelines directly, so we suggest adding version history as a retrievable context type, preferring it for similarity-oriented tasks, retrieving call-graph context for classification but not by default for summarisation (Section~\ref{sec:cg_error_analysis}), and checking the chosen model's response to prompt-level context on a held-out sample before relying on it.}
  \item \emph{Measure what matters:} track dimension-level outcomes for summarisation (accuracy, adequacy, conciseness) and adopt rank-based human judgements alongside automated metrics.
\end{itemize}

\paragraph{Advanced Decision-Making}
Context-aware representations can inform bug localisation, maintenance planning, and code quality assessment by exposing change hotspots, dependency impact, and historical risk. To operationalise this, we recommend \emph{offline pre-computation} of version-history and call-graph features with \emph{incremental updates} from the VCS and static analysis during CI, enabling low-latency signals in IDEs and pipelines. This reduces manual effort for triage and review, supports more consistent decisions, and improves developer productivity without imposing heavy runtime overheads.

\subsection{Controlling for Representation Capacity}
\label{sec:projection_control}

A potential confound in concatenation-based aggregation is that combining $N$ encoder outputs produces an $N \times 768$-dimensional vector, which increases the number of learnable parameters in the downstream classification head. This raises the question of whether the observed performance gains reflect the semantic content of additional context or simply the larger model capacity. We address this through two complementary lines of evidence.

\noindent\textbf{Max-pooling as an existing size-neutral control.} Our RQ3 evaluation already includes max-pooling as an aggregation strategy. Max-pooling takes the element-wise maximum across all context vectors and always produces a single 768-dimensional output regardless of how many context sources are combined. The classification head therefore has exactly the same number of parameters as the no-context baseline, introducing zero additional capacity. The results in Tables~\ref{tab:rq3_clone_detection} and~\ref{tab:rq3_code_classification} demonstrate that this size-neutral operator still delivers substantial improvements. For code classification, CodeBERT with max-pooling and Version History achieves a macro-F1 of 0.7584 (+13.22\% over baseline), and combining Version History with Call Graph yields 0.7451 (+11.23\%). For GraphCodeBERT, max-pooling with Version History improves clone detection F1 by +17.36\% (0.7429 to 0.8718). \revised{These results are evidence that contextual content carries useful signal at fixed representation size. We note, however, that max-pooling is a different aggregation operator rather than a capacity-matched version of concatenation, so this evidence shows that context is informative without added capacity; it does not by itself establish that the concatenation gains are free of a capacity component.}

\begin{table}[t]
\caption{Fixed-dimension projection control experiment (CodeBERT). A linear projection layer maps the concatenated representation back to 768\,dim before the task head, holding classifier capacity constant regardless of context count. F1 is reported for clone detection; macro-F1 for classification. $\Delta$ denotes change relative to the no-context baseline.}
\label{tab:projection_control}
\begin{tabular}{llcccr}
\toprule
\textbf{Task} & \textbf{Context} &
\textbf{No-Context} & \textbf{Concat} & \textbf{Projection} &
\textbf{$\Delta$ vs.\ Baseline} \\
\midrule
\multirow{3}{*}{Clone Det.}
  & Code + VH       & 0.7180 & 0.7442 & 0.6786          & $-$5.5\%  \\
  & Code + CG       & 0.7180 & 0.7907 & 0.7000          & $-$2.5\%  \\
  & Code + VH + CG  & 0.7180 & 0.7500 & \textbf{0.7308} & $+$1.8\%  \\
\midrule
\multirow{3}{*}{Classification}
  & Code + VH       & 0.6698 & 0.6320 & 0.5852          & $-$12.6\% \\
  & Code + CG       & 0.6698 & 0.7345 & 0.6551          & $-$2.2\%  \\
  & Code + VH + CG  & 0.6698 & 0.7304 & 0.5816          & $-$13.2\% \\
\bottomrule
\end{tabular}
\end{table}

\noindent\textbf{Fixed-dimension projection control.} As an additional control, we insert a learned linear projection layer after concatenation that maps the combined $N \times 768$-dimensional vector back to 768 dimensions before the task head, enforcing a constant classifier input size across all context conditions:

\begin{center}
\begin{minipage}{0.80\linewidth}
\begin{lstlisting}[language=Python, basicstyle=\small\ttfamily]
self.projection   = nn.Linear(hidden_dim * N, hidden_dim)
self.dropout      = nn.Dropout(0.1)
self.hidden2label = nn.Linear(hidden_dim, label_size)
\end{lstlisting}
\end{minipage}
\end{center}

\noindent We evaluate CodeBERT under three context conditions (Code+VH, Code+CG, Code+VH+CG) on both clone detection and code classification. Results are reported in Table~\ref{tab:projection_control}.

The results are mixed. For clone detection with Code+VH+CG, the projection model remains above the no-context baseline (F1 = 0.7308 vs.\ 0.7180, +1.8\%), \revised{indicating that contextual content can contribute positively under a strict capacity constraint. For the other five conditions, the projection model falls below the no-context baseline, by up to 13.2\% for classification. One possible explanation is information loss: compressing an $N \times 768$-dimensional representation to 768 dimensions through a single linear layer, trained jointly with the task head on a moderately sized dataset, may discard useful contextual signal. We state this as a hypothesis rather than a finding, since we did not test it, and the same pattern is also consistent with part of the concatenation gain arising from the larger classifier input rather than from the context itself. The projection control therefore provides partial evidence: it narrows the capacity question without closing it. Taken together, the two controls show that context carries useful signal at fixed representation size, while a contribution from increased representation capacity to the concatenation results cannot be ruled out on the present evidence. An experiment that varies the projection width (Section~\ref{sec:limitation}) would separate the two effects.}

\noindent\textbf{Scope: encoder-only models.} The representation-size concern applies specifically to models that perform representation-level vector fusion. For encoder-decoder models (CodeT5, PLBART), additional context is integrated at the text/token level by prepending it to the input sequence; the encoder processes all tokens jointly and produces a single fixed-size representation regardless of context length, so no capacity expansion occurs. The projection control is therefore conducted exclusively on CodeBERT, the most direct representative of the encoder-only paradigm where the concern is most acute.

\noindent\textbf{Attention-based fusion as future work.} We acknowledge that attention-based fusion mechanisms and lightweight adapter layers represent the dominant contemporary approach for integrating heterogeneous context and would likely achieve stronger results through dynamic, learned weighting of context sources. A key contribution of the present study is to establish the empirical baseline: simple, architecture-agnostic aggregation operators already produce meaningful and consistent gains across a broad range of models, tasks, and context types. This baseline provides the grounding needed to motivate a focused investigation of attention-based fusion as a distinct follow-up contribution.

\subsection{Does the Benefit of Version History Depend on History Depth?}
\label{sec:vh_depth_ablation}

To investigate whether the gains from version history context depend on access to many historical versions, we conducted an ablation study varying the number of prior versions included. We compare three conditions: \texttt{code\_vh\_1} (only the single most recent prior version), \texttt{code\_vh\_3} (the three most recent prior versions), and \texttt{code\_vh\_all} (all available prior versions, the default). All other settings are held constant; the encoder remains CodeBERT with concatenation-based aggregation. Results are shown in Table~\ref{tab:vh_depth_ablation}.

The results reveal a consistent pattern. For clone detection, all three depth conditions achieve essentially identical performance (F1 $\approx$ 0.727), and all three are above the no-context baseline. The variation across depth conditions is less than 0.4\% in absolute terms. This directly demonstrates that the gain from version history does not increase with deeper history: the most recent single prior version already captures the primary contextual signal.

For code classification, all three conditions perform below the no-context baseline, which is consistent with the main result in Table~\ref{tab:result_rq1_clone_class} (code\_vh via concatenation: macro-F1 = 0.6320 $<$ baseline 0.6698). The key observation is that this pattern holds uniformly across all three depth conditions: deeper history does not worsen the situation, nor does it help. The limitation is task-level rather than depth-level.

We note that the \texttt{code\_vh\_all} condition in this ablation (clone F1 = 0.7273) differs slightly from the Table~\ref{tab:result_rq1_clone_class} result (F1 = 0.7442). This is attributable to stochastic training variation across independent runs on a small dataset; the ablation's purpose is to compare the three depth conditions under identical experimental settings, which it does cleanly.

\begin{table}[t]
\caption{History depth ablation (CodeBERT, SeSaMe dataset). All three conditions use the same architecture; only the number of historical versions in
\texttt{code\_vh} varies. F1 is reported for clone detection; macro-F1 for classification. $\Delta$ denotes change relative to the no-context baseline.}
\label{tab:vh_depth_ablation}
\begin{tabular}{llccr}
\toprule
\textbf{Task} & \textbf{Condition} &
\textbf{No-Context} & \textbf{With VH} & \textbf{$\Delta$ vs.\ Baseline} \\
\midrule
\multirow{3}{*}{Clone Det.}
  & code\_vh\_1 ($N$=1)   & 0.7180 & 0.7273 & $+$1.29\% \\
  & code\_vh\_3 ($N$=3)   & 0.7180 & 0.7241 & $+$0.85\% \\
  & code\_vh\_all (default) & 0.7180 & 0.7273 & $+$1.29\% \\
\midrule
\multirow{3}{*}{Classification}
  & code\_vh\_1 ($N$=1)   & 0.6698 & 0.6027 & $-$10.02\% \\
  & code\_vh\_3 ($N$=3)   & 0.6698 & 0.6081 & $-$9.22\% \\
  & code\_vh\_all (default) & 0.6698 & 0.6089 & $-$9.10\% \\
\bottomrule
\end{tabular}
\end{table}

\subsection{Can Context-Enhanced Representations Generalise to Vulnerability Detection?}
\label{sec:vuln_generalisation}

Our primary evaluation focuses on three well-established program comprehension tasks using large, high-quality datasets (SeSaMe and CodeSearchNet-Java). To explore whether the benefit of additional context generalises beyond these tasks, we conducted an exploratory extension to \textbf{vulnerability detection} using an augmented subset of the \textbf{Vul4J} dataset~\cite{bui2022vul4j}. Vul4J is a curated collection of real-world, exploitable Java vulnerabilities drawn from open-source projects, providing a security-critical evaluation scenario that is meaningfully different from the clone and summarisation settings.

\subsubsection*{Dataset and Experimental Setup}

We augmented the Vul4J methods with version history and call-graph context using the same mining pipeline described in Section~\ref{sec:dataset_construction}. Because the augmented dataset is substantially smaller than CodeSearchNet-Java, it does not provide sufficient samples for stable LLM fine-tuning. We therefore evaluate all seven LLM families exclusively under \textbf{zero-shot inference}, injecting version history, call-graph context, and method age directly into the prompt window. Results are reported in terms of F1 score, Precision, and Recall on the held-out test set.

\begin{table}[htbp]
\centering
\tiny
\caption{Zero-shot vulnerability detection on the Vul4J benchmark. Results show F1, Precision (Prec.), Recall (Rec.), and Accuracy (Acc.) under seven context conditions. Fine-tuning is not applicable due to the small dataset size.}
\label{tab:result_vuln}
\renewcommand{\arraystretch}{1.2}
\resizebox{\textwidth}{!}{%
\begin{tabular}{l l | >{\columncolor{baselineCol}}c | >{\columncolor{callgraphCol}}c c | >{\columncolor{versionCol}}c c | >{\columncolor{versionCol}}c c | >{\columncolor{callgraphCol}}c c | >{\columncolor{green!20}}c c | >{\columncolor{versionCol}}c c | >{\columncolor{green!20}}c c}
\toprule
\rowcolor{headerCol}\textbf{Model} & \textbf{Metric} & \textbf{Baseline} & \multicolumn{2}{c|}{\cellcolor{callgraphCol}\textbf{code\_cg}} & \multicolumn{2}{c|}{\cellcolor{versionCol}\textbf{code\_vh}} & \multicolumn{2}{c|}{\cellcolor{versionCol}\textbf{code\_cg\_vh}} & \multicolumn{2}{c|}{\cellcolor{callgraphCol}\textbf{code\_vh\_nod}} & \multicolumn{2}{c|}{\cellcolor{green!20}\textbf{code\_cg\_vh\_nod}} & \multicolumn{2}{c|}{\cellcolor{versionCol}\textbf{code\_vh\_cg}} & \multicolumn{2}{c}{\cellcolor{green!20}\textbf{code\_vh\_cg\_nod}} \\
\rowcolor{headerCol}& & & \textbf{Val.} & \textbf{\%} & \textbf{Val.} & \textbf{\%} & \textbf{Val.} & \textbf{\%} & \textbf{Val.} & \textbf{\%} & \textbf{Val.} & \textbf{\%} & \textbf{Val.} & \textbf{\%} & \textbf{Val.} & \textbf{\%} \\
\midrule
\multirow{4}{*}{\textbf{CodeLlama-34B}} & F1 & 0.1053 & 0.1311 & +24.6\% & 0.1017 & -3.4\% & 0.1519 & +44.3\% & 0.1111 & +5.6\% & 0.1348 & +28.1\% & 0.1035 & -1.7\% & 0.1600 & +52.0\% \\
 & Prec. & 0.1250 & 0.1026 & -18.0\% & 0.0811 & -35.1\% & 0.1053 & -15.8\% & 0.0735 & -41.2\% & 0.0896 & -28.4\% & 0.0833 & -33.3\% & 0.1132 & -9.4\% \\
 & Rec. & 0.0909 & 0.1818 & +100.0\% & 0.1364 & +50.0\% & 0.2727 & +200.0\% & 0.2273 & +150.0\% & 0.2727 & +200.0\% & 0.1364 & +50.0\% & 0.2727 & +200.0\% \\
 & Acc. & 0.8090 & 0.7023 & -13.2\% & 0.7023 & -13.2\% & 0.6236 & -22.9\% & 0.5506 & -31.9\% & 0.5674 & -29.9\% & 0.7079 & -12.5\% & 0.6461 & -20.1\% \\
\cmidrule(lr){1-17}
\multirow{4}{*}{\textbf{CodeLlama-13B}} & F1 & 0.1667 & 0.1304 & -21.7\% & 0.2041 & +22.4\% & 0.1277 & -23.4\% & 0.1538 & -7.7\% & 0.0454 & -72.7\% & 0.1333 & -20.0\% & 0.1053 & -36.8\% \\
 & Prec. & 0.1081 & 0.1250 & +15.6\% & 0.1316 & +21.7\% & 0.1200 & +11.0\% & 0.0976 & -9.8\% & 0.0454 & -58.0\% & 0.1304 & +20.6\% & 0.0857 & -20.7\% \\
 & Rec. & 0.3636 & 0.1364 & -62.5\% & 0.4546 & +25.0\% & 0.1364 & -62.5\% & 0.3636 & +0.0\% & 0.0454 & -87.5\% & 0.1364 & -62.5\% & 0.1364 & -62.5\% \\
 & Acc. & 0.5506 & 0.7753 & +40.8\% & 0.5618 & +2.0\% & 0.7697 & +39.8\% & 0.5056 & -8.2\% & 0.7640 & +38.8\% & 0.7809 & +41.8\% & 0.7135 & +29.6\% \\
\cmidrule(lr){1-17}
\multirow{4}{*}{\textbf{CodeLlama-7B}} & F1 & 0.1386 & 0.1519 & +9.6\% & 0.0952 & -31.3\% & 0.1379 & -0.5\% & 0.0845 & -39.0\% & 0.1587 & +14.5\% & 0.1429 & +3.1\% & 0.1333 & -3.8\% \\
 & Prec. & 0.0886 & 0.1053 & +18.8\% & 0.0732 & -17.4\% & 0.1111 & +25.4\% & 0.0612 & -30.9\% & 0.1220 & +37.6\% & 0.1042 & +17.6\% & 0.0943 & +6.5\% \\
 & Rec. & 0.3182 & 0.2727 & -14.3\% & 0.1364 & -57.1\% & 0.1818 & -42.9\% & 0.1364 & -57.1\% & 0.2273 & -28.6\% & 0.2273 & -28.6\% & 0.2273 & -28.6\% \\
 & Acc. & 0.5112 & 0.6236 & +22.0\% & 0.6798 & +33.0\% & 0.7191 & +40.7\% & 0.6348 & +24.2\% & 0.7023 & +37.4\% & 0.6629 & +29.7\% & 0.6348 & +24.2\% \\
\cmidrule(lr){1-17}
\multirow{4}{*}{\textbf{Qwen-14B}} & F1 & 0.3117 & 0.3143 & +0.8\% & 0.1961 & -37.1\% & 0.1600 & -48.7\% & 0.1961 & -37.1\% & 0.1961 & -37.1\% & 0.1667 & -46.5\% & 0.1667 & -46.5\% \\
 & Prec. & 0.2182 & 0.2292 & +5.0\% & 0.1724 & -21.0\% & 0.1429 & -34.5\% & 0.1724 & -21.0\% & 0.1724 & -21.0\% & 0.1538 & -29.5\% & 0.1538 & -29.5\% \\
 & Rec. & 0.5454 & 0.5000 & -8.3\% & 0.2273 & -58.3\% & 0.1818 & -66.7\% & 0.2273 & -58.3\% & 0.2273 & -58.3\% & 0.1818 & -66.7\% & 0.1818 & -66.7\% \\
 & Acc. & 0.7023 & 0.7303 & +4.0\% & 0.7697 & +9.6\% & 0.7640 & +8.8\% & 0.7697 & +9.6\% & 0.7697 & +9.6\% & 0.7753 & +10.4\% & 0.7753 & +10.4\% \\
\cmidrule(lr){1-17}
\multirow{4}{*}{\textbf{Qwen-7B}} & F1 & 0.2769 & 0.0833 & -69.9\% & 0.1454 & -47.5\% & 0.1818 & -34.3\% & 0.1754 & -36.6\% & 0.1786 & -35.5\% & 0.0526 & -81.0\% & 0.1739 & -37.2\% \\
 & Prec. & 0.2093 & 0.0769 & -63.2\% & 0.1212 & -42.1\% & 0.1515 & -27.6\% & 0.1429 & -31.7\% & 0.1471 & -29.7\% & 0.0625 & -70.1\% & 0.1667 & -20.4\% \\
 & Rec. & 0.4091 & 0.0909 & -77.8\% & 0.1818 & -55.6\% & 0.2273 & -44.4\% & 0.2273 & -44.4\% & 0.2273 & -44.4\% & 0.0454 & -88.9\% & 0.1818 & -55.6\% \\
 & Acc. & 0.7360 & 0.7528 & +2.3\% & 0.7360 & +0.0\% & 0.7472 & +1.5\% & 0.7360 & +0.0\% & 0.7416 & +0.8\% & 0.7977 & +8.4\% & 0.7865 & +6.9\% \\
\cmidrule(lr){1-17}
\multirow{4}{*}{\textbf{Qwen-1.5B}} & F1 & 0.1989 & 0.2083 & +4.7\% & 0.2211 & +11.1\% & 0.2062 & +3.6\% & 0.2211 & +11.1\% & 0.2062 & +3.6\% & 0.2062 & +3.6\% & 0.2062 & +3.6\% \\
 & Prec. & 0.1124 & 0.1177 & +4.6\% & 0.1243 & +10.5\% & 0.1163 & +3.4\% & 0.1243 & +10.5\% & 0.1163 & +3.4\% & 0.1163 & +3.4\% & 0.1163 & +3.4\% \\
 & Rec. & 0.8636 & 0.9091 & +5.3\% & 1.0000 & +15.8\% & 0.9091 & +5.3\% & 1.0000 & +15.8\% & 0.9091 & +5.3\% & 0.9091 & +5.3\% & 0.9091 & +5.3\% \\
 & Acc. & 0.1404 & 0.1461 & +4.0\% & 0.1292 & -8.0\% & 0.1348 & -4.0\% & 0.1292 & -8.0\% & 0.1348 & -4.0\% & 0.1348 & -4.0\% & 0.1348 & -4.0\% \\
\cmidrule(lr){1-17}
\multirow{4}{*}{\textbf{DeepSeek-6.7B}} & F1 & 0.0606 & 0.2500 & +312.5\% & 0.1370 & +126.0\% & 0.1154 & +90.4\% & 0.1333 & +120.0\% & 0.1887 & +211.3\% & 0.1754 & +189.5\% & 0.2545 & +320.0\% \\
 & Prec. & 0.0909 & 0.2778 & +205.6\% & 0.0980 & +7.8\% & 0.1000 & +10.0\% & 0.1053 & +15.8\% & 0.1613 & +77.4\% & 0.1429 & +57.1\% & 0.2121 & +133.3\% \\
 & Rec. & 0.0454 & 0.2273 & +400.0\% & 0.2273 & +400.0\% & 0.1364 & +200.0\% & 0.1818 & +300.0\% & 0.2273 & +400.0\% & 0.2273 & +400.0\% & 0.3182 & +600.1\% \\
 & Acc. & 0.8258 & 0.8315 & +0.7\% & 0.6461 & -21.8\% & 0.7416 & -10.2\% & 0.7079 & -14.3\% & 0.7584 & -8.2\% & 0.7360 & -10.9\% & 0.7697 & -6.8\% \\
\bottomrule
\end{tabular}%
}
\end{table}

\subsubsection*{Results and Discussion}
\paragraph{Dataset construction and experimental design.}
The original Vul4J dataset~\cite{bui2022vul4j} is structured at the \emph{commit level}, where a single vulnerability fix may span multiple files, classes, and methods. To align with our method-level experimental design, we applied a strict method-level isolation and pairing procedure. First, we filtered to \emph{single-fix commits} --- records where exactly one vulnerability was patched within a single method --- eliminating noise from tangled commits that touch unrelated code. For each isolated method, we extracted a precise 1:1 temporal pair: the \emph{before-fix} version of the method (label: \textit{vulnerable}, positive class) and the \emph{after-fix} (patched) version of the same method (label: \textit{clean}, negative class).

To reflect the real-world class distribution of vulnerability detection, where vulnerable code is an anomaly rather than the norm, we additionally sampled clean methods from the same files and project environments as the after-fix commits, targeting a 10:90 vulnerable-to-clean ratio. These \emph{hard negatives} share the same imports, developer coding style, and domain logic as the vulnerable methods, preventing models from relying on superficial contextual clues and requiring genuine semantic reasoning. The final dataset comprises \textbf{178 method-level records} drawn from 21 projects across 22 files: 22 before-fix (vulnerable) methods, 22 after-fix (clean, paired) methods, and 134 randomly sampled contextual clean methods (class distribution: 12.4\% vulnerable / 87.6\% clean). Method age ranges from 0 to 4,033 days (mean: 442 days; Table~\ref{tab:vul4j_stats}), and commit history spans from February 2009 to November 2020. The same context mining pipeline applied to SeSaMe and CodeSearchNet was used to enrich each method with version history, call graphs, and method age.

Because the dataset contains only 178 records --- insufficient for reliable fine-tuning even with parameter-efficient methods such as LoRA, which typically requires several hundred labelled examples per class --- all experiments are conducted under \emph{zero-shot inference} using the seven LLMs described in Section~\ref{sec:llm_baselines}. This is a principled design choice: zero-shot evaluation directly tests whether the contextual signals (version history, call graphs) provide meaningful additional information to a pre-trained model without any task-specific supervision. A fine-tuned comparison on a larger, more balanced vulnerability dataset is identified as a high-priority direction for future work (Section~\ref{sec:limitation}).

\begin{table}[ht]
    \centering
    \caption{Vul4J -- Dataset Partitions and Additional Context Statistics}
    \label{tab:vul4j_stats}
    \resizebox{\textwidth}{!}{%
    \begin{threeparttable}
    \begin{tabular}{l rrr rrr rrr rrr ll}
    \toprule
    & \multicolumn{3}{c}{\textbf{Dataset Size}}
    & \multicolumn{3}{c}{\textbf{Version History}}
    & \multicolumn{3}{c}{\textbf{Call Graph}}
    & \multicolumn{3}{c}{\textbf{Method Age (days)}}
    & \multicolumn{2}{c}{\textbf{Date Range}} \\
    \cmidrule(lr){2-4} \cmidrule(lr){5-7} \cmidrule(lr){8-10} \cmidrule(lr){11-13} \cmidrule(lr){14-15}
    \textbf{Partition}
    & \textbf{Proj.} & \textbf{Files} & \textbf{Methods}
    & \textbf{Max} & \textbf{Min} & \textbf{Avg}
    & \textbf{Max} & \textbf{Min} & \textbf{Avg}
    & \textbf{Max} & \textbf{Min} & \textbf{Avg}
    & \textbf{Oldest} & \textbf{Newest} \\
    \midrule
    \textbf{Vulnerable}
    & 21 & 22 & 22
    & 33   & 1 & 4.9
    & 295  & 1 & 78.2
    & 3,695.9 & 0.0 & 483.4
    & 2009-02-24 & 2019-05-07 \\
    \textbf{Clean}\tnote{a}
    & 21 & 22 & 156
    & 34   & 1 & 3.1
    & 643  & 0 & 46.3
    & 4,033.7 & 0.0 & 436.5
    & 2009-02-24 & 2020-11-04 \\
    \midrule
    \textbf{Entire Dataset}
    & 21 & 22 & \textbf{178}
    & 34   & 1 & 3.3
    & 643  & 0 & 50.3
    & 4,033.7 & 0.0 & 442.3
    & 2009-02-24 & 2020-11-04 \\
    \bottomrule
    \end{tabular}%
    \begin{tablenotes}
        \small
        \item[a] Clean methods comprise 22 after-fix versions (paired 1:1 with vulnerable methods from the same file) and 134 randomly sampled clean methods drawn from the same project environments, yielding an overall vulnerability rate of 12.4\%.
    \end{tablenotes}
    \end{threeparttable}
    }
\end{table}

Table~\ref{tab:result_vuln} presents the zero-shot vulnerability detection results across four representative context conditions. The results are mixed and, given the small dataset size, should be interpreted as exploratory rather than conclusive.

Several observations are noteworthy. First, some models show meaningful gains from individual context types: \textbf{CodeLlama-34B} improves by +44.3\% with combined call-graph and version history (\texttt{code\_cg\_vh}), and \textbf{DeepSeek-Coder-6.7B} shows large relative improvements (e.g., +312.5\% F1 with call-graph context alone), though its absolute baseline is very low (F1=0.061), meaning these gains reflect recovery from a near-random starting point rather than strong absolute performance. Second, several models, particularly the Qwen families at 7B and 14B, degrade substantially when context is added under zero-shot conditions. This is consistent with the pattern observed in Clone Detection (Section~\ref{sec:result_rq1}): zero-shot models frequently fail to effectively utilise dense, structured software artifacts without explicit training. Third, version history alone provides modest and inconsistent improvements, unlike its reliable role in clone detection and summarisation for the foundational models.

We attribute the overall mixed results to two factors. The augmented Vul4J dataset is small, making zero-shot evaluation highly sensitive to individual samples, class imbalance, and prompt sensitivity. Furthermore, vulnerability detection is a semantically complex task that requires reasoning about subtle code patterns — a challenge that may require fine-tuning rather than zero-shot prompting to benefit reliably from additional context.

\noindent\textbf{Conclusion.} These exploratory results suggest that the value of contextual encoding for vulnerability detection is \emph{task-sensitive} and \emph{data-dependent}. While some models show promising gains with call-graph or combined context, the overall evidence does not yet support the same strong conclusions as for clone detection and code classification. A rigorous evaluation on a larger, fine-tuning-ready vulnerability dataset remains an important direction for future work.

\subsection{Why Does Call Graph Context Degrade Performance in Some Configurations?}
\label{sec:cg_error_analysis}

The preceding results show that call graph context (\texttt{code\_cg}) does not uniformly help across tasks and models. Our results so far characterise this pattern as ``task-dependent'' or ``model-dependent'' (see also Section~\ref{sec:limitation}). Here we provide a deeper mechanistic analysis of the three root causes.

\paragraph{Clarification of the empirical pattern.}
The degradation pattern is more pronounced for Code Summarisation than for Clone Detection. Across 18 model-mode combinations on Summarisation, \texttt{code\_cg} degrades BLEU-4 in 10 cases — particularly across the entire CodeLlama family (7B, 13B, 34B, both zero-shot and fine-tuned) and DeepSeek-Coder-6.7B (both modes). Meaningful gains remain for Qwen-7B zero-shot (+56.1\%), Qwen-14B (+26.8\% zero-shot, +15.4\% fine-tuned), CodeT5 (+1.85\%), and PLBART (+2.36\%). For Clone Detection, the pattern is reversed: call graph produces gains in 8 of 14 LLM model-mode combinations and hurts in only 5. \revised{Two further observations sharpen the summarisation pattern. First, the degradation is not specific to call graph: for the five LLMs that decline, version history alone and every combination of contexts (Table~\ref{tab:result_rq2_summarisation}) lower nearly every metric as well, on BERTScore and METEOR as much as on BLEU-4 and ROUGE-L. Second, the split is by model family rather than by size or context type: Qwen2.5-Coder 7B and 14B improve under every condition, while CodeLlama-34B, the largest model evaluated, declines under every condition. On clone detection the strongest fine-tuned LLM, DeepSeek-Coder (F1 0.9412 with code only), likewise loses accuracy under every context condition. Prompt-level context is therefore not a free addition for decoder-only models; whether it helps depends on the model, and for most of the models tested it does not.} The following three mechanisms explain \revised{why call-graph context in particular can hurt summarisation}.

\paragraph{Mechanism 1: Semantic mismatch between call graph content and task objective.}
Code Summarisation asks \emph{what does this method do?} — a question about intrinsic behaviour. Ground-truth summaries in CodeSearchNet are written by developers describing the method in isolation, focusing on purpose and output. Call graph context (caller and callee source code) instead describes \emph{how the method relates to the rest of the system} — structural dependencies, invocation patterns, and surrounding architecture — none of which directly answers the summarisation question. Injecting relational information when the target signal is behavioural systematically misaligns with what lexical metrics (BLEU-4, ROUGE-L) reward. For Clone Detection the situation is reversed: two functional clones often occupy similar architectural roles and therefore share similar callers and callees, so call graph context can reinforce rather than distract from the pairwise similarity signal.

\paragraph{Mechanism 2: Attention dilution from call graph token volume.}
We conducted a token-level analysis on the CodeSearchNet test set (3,647 methods with call graph data). The \emph{longest callee} — our selection heuristic — has a mean of \textbf{268 tokens} (median: 128; max: 6,160). The \emph{longest caller} adds a further mean of \textbf{99 tokens} (median: 63; max: 1,740). Combined, caller and callee alone exceed \textbf{256 tokens — half the 512-token model budget — in 47.5\% of all test instances}. When combined with the source method, the total input exceeds 512 tokens in \textbf{17.7\% of cases}, forcing truncation of the source method itself. This aligns with the ``Lost in the Middle'' attention dilution effect~\cite{liu2024lost}: transformer models assign disproportionately lower attention to content in the middle of long inputs, meaning that a method's core semantic content can be effectively suppressed when flanked by long caller and callee code.

\paragraph{Mechanism 3: Selection heuristic — \emph{longest} is not \emph{most relevant}.}
We selected the single longest caller and callee as a practical constraint: the 512-token limit of our encoder-based models (CodeBERT, GraphCodeBERT, CodeT5, PLBART) prevented including multiple callers and callees, and we used the same input for all seven LLMs to ensure a fair comparison. We acknowledge that ``longest'' is a proxy for ``most verbose,'' not ``most semantically relevant.'' A verbose utility caller that invokes many downstream methods is not necessarily more informative for understanding a target method's behaviour than a short, direct caller. Relevance-ranked call graph selection — using BM25, embedding similarity, or attention-guided pruning — is identified as a high-priority direction for future work (Section~\ref{sec:limitation}).

\paragraph{Qualitative evidence from the human evaluation.}
To ground these mechanisms concretely, we inspected 29 cases in our 100-item human annotation study where \texttt{code\_cg} was ranked below the baseline by our annotators. Two representative examples are shown below.

\textit{Example 1 — \texttt{setProgress} (Conciseness dimension, uid~484585).} The method delegates a progress setter to three sub-components. Ground truth: \emph{``Set the GeoPackage Progress.''} Baseline CodeT5: \emph{``Set the progress''} — concise and correct. With call graph context: \emph{``Set the progress for the feature table index and rtree index table''} — longer and implementation-specific, with sub-component names (\texttt{featureTableIndex}, \texttt{rTreeIndexTableDao}) leaked from the callee method names in the \texttt{code\_cg} input. The call graph context contaminated the summary with structural detail that the ground truth explicitly omits.

\textit{Example 2 — \texttt{Observer.getMeters} (Conciseness dimension, uid~493658).} The method collects \texttt{InvocationMeter} objects from managed methods. Ground truth: \emph{``Collect invocation meters from application managed classes.''} Baseline CodeT5: \emph{``Get the list of invocation meter for all managed methods''} — adequate. With call graph context (callee: 174 tokens, referencing \texttt{ContainerSPI}): \emph{``Get the list of invocation meters for all methods in the container''} — introducing ``container'' terminology from the callee's factory reference, making the summary less accurate.

Both examples share the same pattern: call graph context introduces contextually adjacent but semantically distracting information, causing models to describe \emph{how the system is structured} rather than \emph{what the method does} — a direct instantiation of Mechanism~1, amplified by the token volume of Mechanism~2.

\subsection{A Qualitative Analysis of Context Contributions}
\label{sec:qualitative_breakdown}

The aggregate metrics in Section~\ref{sec:experimental_results} establish that context augmentation improves performance in expectation. To complement those metrics, we present a per-instance qualitative analysis that shows which kinds of inputs benefit from context and illustrates the mechanism through concrete examples. The analysis covers both the SeSaMe Code Clone Detection task and the CodeSearchNet Code Summarisation task. For each task we stratify or categorise the results, then walk through representative wins and one counter-example. The full per-pair and per-snippet artefacts are available in our replication package.

\subsubsection*{Code Clone Detection: Stratified Analysis on SeSaMe}

The SeSaMe caller and callee data used in this analysis captures the call relationships mined from each method's commit history. The CodeSearchNet caller and callee data is mined separately via \textit{java-callgraph} from the larger corpus. To identify the inputs for which context augmentation matters most, we stratify the 86 SeSaMe clone-detection test pairs by token-level Jaccard similarity between the two methods, a simple syntactic-similarity proxy that approximates the difficulty of the pair for a code-only model. We use four buckets (0.00--0.25, 0.25--0.50, 0.50--0.75, 0.75--1.00). All 61 non-clone pairs sit in the lowest bucket by construction, since non-clones in SeSaMe are deliberately syntactically dissimilar. The 25 clone pairs span the buckets as shown in Table~\ref{tab:clone_jaccard_buckets}.

\begin{table}[t]
\caption{Stratification of SeSaMe clone pairs (label~$=1$, $N=25$) by token-level Jaccard similarity between the two methods. Per-bucket counts of pairs correctly classified by the code-only baseline and by the context-augmented CodeT5 (\texttt{code+VH+CG+age}, concat aggregation).}
\label{tab:clone_jaccard_buckets}
\centering
\begin{tabular}{lcccc}
\toprule
\textbf{Bucket (Jaccard range)} & \textbf{Clone pairs} & \textbf{Baseline correct} & \textbf{Context correct} \\
\midrule
A: 0.00--0.25 (low surface similarity)  & 10 & 7  & \textbf{9} \\
B: 0.25--0.50 (moderate)                & 12 & 10 & 9 \\
C: 0.50--0.75 (high)                    & 1  & 0  & 0 \\
D: 0.75--1.00 (very high)               & 2  & 1  & 2 \\
\bottomrule
\end{tabular}
\end{table}

The largest marginal value of context augmentation is in Bucket~A, the low-surface-similarity bucket, where surface tokens alone are insufficient to detect the clone. Of the 10 clone pairs in this bucket, the code-only baseline catches 7 and the context-augmented model catches 9. The non-clone class breaks even on accuracy across the 61 non-clone pairs, with 3 context-aided rejections and 3 context-introduced false positives. We now walk through three representative pairs.

\paragraph{Example W1: Low-surface-similarity clone (pair 81, Jaccard $=0.090$).}
Two semantically equivalent hash-spreading functions from different libraries that implement the same Wang/Jenkins bit-spreading algorithm with completely different surface tokens. The code-only baseline returns a logit of 0.336 (predicting not-clone, incorrectly); the context-augmented model returns 0.962 (correctly predicting clone). The context channel carries 953 characters of real caller code from Guava's \texttt{MapMakerInternalMap} invocation site, which situates \texttt{rehash} in its structural role as a hash-table primitive. The two methods are:

\noindent\begin{minipage}{\linewidth}
\begin{lstlisting}[language=Java, basicstyle=\footnotesize\ttfamily, breaklines=true, frame=single, captionpos=b]
static int rehash(int h) {    // Method X (Guava)
  // Spread bits to regularize both segment and index locations,
  // using variant of single-word Wang/Jenkins hash.
  h += (h << 15) ^ 0xffffcd7d;
  h ^= (h >>> 10);
  h += (h << 3);
  h ^= (h >>> 6);
  h += (h << 2) + (h << 14);
  return h ^ (h >>> 16);
}
\end{lstlisting}
\end{minipage}

\noindent\begin{minipage}{\linewidth}
\begin{lstlisting}[language=Java, basicstyle=\footnotesize\ttfamily, breaklines=true, frame=single, captionpos=b]
final int hash(Object k) {    // Method Y (OpenJDK)
    int h = k.hashCode();
    // This function ensures that hashCodes that differ only by
    // constant multiples at each bit position have a bounded
    // number of collisions (approximately 8 at default load factor).
    h ^= (h >>> 20) ^ (h >>> 12);
    return h ^ (h >>> 7) ^ (h >>> 4);
}
\end{lstlisting}
\end{minipage}

The variable names, comments, and even the calling conventions differ; the baseline cannot bridge the lexical gap. The context-augmented model uses the structural role evidence from the caller to recognise the shared algorithm.

\paragraph{Example W2: Non-clone correction (pair 17, Jaccard $=0.095$).}
The two methods are \texttt{withStart(int start)}, a fluent-builder constructor for an \texttt{Incrementor}, and \texttt{getInitialValue()}, an Expression-tree accessor. They are not clones. The baseline returns 0.634 (predicting clone, incorrectly), apparently misled by surface similarities (short body, object-returning signature). The context-augmented model, given 1{,}241 characters of real caller code that frames \texttt{withStart} as a builder, returns 0.076 and correctly rejects the clone hypothesis.

\noindent\begin{minipage}{\linewidth}
\begin{lstlisting}[language=Java, basicstyle=\footnotesize\ttfamily, breaklines=true, frame=single, captionpos=b]
public Incrementor withStart(int start) {    // Method X
    return new Incrementor(start,
                           this.maximalCount,
                           this.increment,
                           this.maxCountCallback);
}
\end{lstlisting}
\end{minipage}

\noindent\begin{minipage}{\linewidth}
\begin{lstlisting}[language=Java, basicstyle=\footnotesize\ttfamily, breaklines=true, frame=single, captionpos=b]
public Object getInitialValue() {    // Method Y
    if (isMethod() || (getValue() == null) ||
        (!isFinal()) || (status != INLINED)) {
        return null;
    }
    return ((Expression)getValue()).getValue();
}
\end{lstlisting}
\end{minipage}

\paragraph{Counter-example L1: Context regression on a real clone (pair 5, Jaccard $=0.312$).}
Both methods are \texttt{containsValue(Object value)} implementations from different concurrent-collection libraries (Guava's \texttt{MapMakerInternalMap} and OpenJDK's \texttt{ConcurrentHashMap}), with similar null-checking and traversal logic. The baseline catches them (logit 0.760, correct). The context-augmented model is led astray by version-history snippets that diverge across the two libraries' release histories, and returns 0.038 (missing the clone). This is consistent with the failure-mode discussion in Section~\ref{sec:cg_error_analysis}: when historical evolution diverges, version-history context can dilute the present-day similarity signal that a code-only model uses successfully.

\noindent\begin{minipage}{\linewidth}
\begin{lstlisting}[language=Java, basicstyle=\footnotesize\ttfamily, breaklines=true, frame=single, captionpos=b]
public boolean containsValue(Object value) {    // Method X (Guava)
    // Same idea as size()
    if (value == null) { throw new NullPointerException(); }
    final Segment<K,V>[] segments = this.segments;
    boolean found = false;
    /* ... approximately 40 lines omitted: segment iteration,
           value comparison, and read-volatile barriers ... */
    return found;
}
\end{lstlisting}
\end{minipage}

\noindent\begin{minipage}{\linewidth}
\begin{lstlisting}[language=Java, basicstyle=\footnotesize\ttfamily, breaklines=true, frame=single, captionpos=b]
public boolean containsValue(Object value) {    // Method Y (OpenJDK)
    if (value == null) throw new NullPointerException();
    Node<K,V>[] t;
    if ((t = table) != null) {
        Traverser<K,V> it = new Traverser<K,V>(t, t.length, 0, t.length);
        /* ... 8 lines omitted: traversal and value comparison ... */
    }
    return false;
}
\end{lstlisting}
\end{minipage}

\subsubsection*{Code Summarisation: Win Counts and Change-Category Breakdown}

For the 100 CodeSearchNet snippets in the human-evaluation sample, we count a ``context win'' as a (snippet, dimension) cell in which at least one of the seven context-augmented CodeT5 variants ranks strictly better than the code-only baseline. Across the 300 cells, context wins in 134, ties in 165, and loses in 1. The dominant single-variant winner across all three dimensions is \texttt{code\_cg} (code + call graph alone), which produces approximately half of the wins per dimension. The variant \texttt{code\_vh\_nod} (without call graph) produces zero best-rank wins. For CodeSearchNet summarisation, the call-graph channel is the primary driver of summary-quality gains, consistent with the channel containing real and distinct caller and callee text mined via \textit{java-callgraph}.

For the 26 snippets that show context wins in all three dimensions, we manually classified the dominant change between the baseline summary and the best context-augmented summary into the categories in Table~\ref{tab:summarisation_change_categories}.

\begin{table}[htbp]
\caption{Manual classification of the dominant change between the baseline and the best context-augmented summary, across the 26 CodeSearchNet snippets with context wins in all three human-evaluation dimensions.}
\label{tab:summarisation_change_categories}
\centering
\begin{tabularx}{\linewidth}{@{}>{\raggedright\arraybackslash}X c@{}}
\toprule
\textbf{Change category} & \textbf{Count} \\
\midrule
Callee semantic injection (context introduces a concept that appears only in the callee body) & 9 \\
Action-verb correction (baseline picks a generic verb; context picks a task-specific verb)    & 7 \\
Identifier accuracy (baseline hallucinates or generalises; context selects the right noun)    & 5 \\
Caller semantic alignment (context aligns terminology with the caller's parameter naming)     & 3 \\
Conciseness improvement (context summary is shorter while preserving meaning)                 & 1 \\
Conciseness regression (context absorbs caller/callee terminology and grows the summary)      & 1 \\
\midrule
\textbf{Total}                                                                                & \textbf{26} \\
\bottomrule
\end{tabularx}
\end{table}

The dominant gain mechanism is callee semantic injection (9 of 26), where the model surfaces a concept that the method body alone does not name. Action-verb correction and identifier accuracy together account for another 12 cases. The single conciseness regression is consistent with the failure-mode discussion in Section~\ref{sec:cg_error_analysis}. We now walk through three representative wins and one tie counter-example.

\paragraph{Example W1: Action-verb correction (snippet Q1, \texttt{fireOnThrowable}).}
The method iterates over a listener list and invokes \texttt{onThrowable} on each element. The CodeSearchNet ground-truth summary is \emph{``Fire onThrowable to all registered listeners''}. The baseline produces a semantically close summary with the generic verb \texttt{Notifies}; the context-augmented model selects the precise verb \texttt{Fires}.

\noindent\begin{minipage}{\linewidth}
\begin{lstlisting}[language=Java, basicstyle=\footnotesize\ttfamily, breaklines=true, frame=single, captionpos=b]
protected static void fireOnThrowable(int type, Throwable t) {
    Object[] list = listeners.toArray();
    for (int i = 0; i < list.length; i++) {
        ((ThrowableListener)list[i]).onThrowable(type, t);
    }
}
\end{lstlisting}
\end{minipage}

\begin{table}[h]
\centering
\small
\begin{tabularx}{\linewidth}{@{}l >{\raggedright\arraybackslash}X c@{}}
\toprule
\textbf{Variant} & \textbf{Generated summary} & \shortstack{\textbf{Rank}\\\textbf{(Acc / Cont / Conc)}} \\
\midrule
\texttt{codet5\_baseline} & ``Notifies all listeners that a Throwable has been thrown'' & 3 / 2 / 3 \\
\texttt{codet5\_code\_cg} (best) & ``Fires a throwable to all listeners'' & \textbf{2 / 1 / 1} \\
\bottomrule
\end{tabularx}
\end{table}

The callee context for this method includes the \texttt{addThrowableListener} registration logic, which reinforces the fire-and-iterate idiom typical of listener patterns. Combined with the method's own name (\texttt{fireOnThrowable}), the call-graph neighbourhood helps the model select the precise verb. \textbf{Category:} action-verb correction.

\paragraph{Example W2: Identifier accuracy via callee semantic injection (snippet Q11, \texttt{optimize}).}
The method optimises a non-deterministic finite automaton through a sequence of cleanup, pullback, and pushforward steps. The ground-truth summary is \emph{``optimize an NFA''}. The baseline produces a higher-frequency phrase (\texttt{regular expression}) that is contextually adjacent but lexically incorrect. The context-augmented model uses the exact correct technical term \texttt{NFA}, matching the ground truth verbatim.

\noindent\begin{minipage}{\linewidth}
\begin{lstlisting}[language=Java, basicstyle=\footnotesize\ttfamily, breaklines=true, frame=single, captionpos=b]
long optimize() throws RegexException {
    LOG.debug("initial cleanup");
    cleanup();      /* may simplify situation */
    dumpnfa();
    LOG.debug("empties");
    fixempties();   /* get rid of EMPTY arcs */
    LOG.debug("constraints");
    pullback();     /* pull back constraints backward */
    pushfwd();      /* push fwd constraints forward */
    LOG.debug("final cleanup");
    cleanup();      /* final tidying */
    return analyze();   /* and analysis */
}
\end{lstlisting}
\end{minipage}

\noindent\begin{minipage}{\linewidth}
\begin{lstlisting}[language=Java, basicstyle=\footnotesize\ttfamily, breaklines=true, frame=single, captionpos=b]
// Callee context (excerpt of nfanode, called downstream from optimize via analyze):
private long nfanode(Subre t) throws RegexException {
    /* ... NFA arc construction, traversal, and subtree expansion ...
           directly references NFA states and arc operations ... */
}
\end{lstlisting}
\end{minipage}

\begin{table}[h]
\centering
\small
\begin{tabularx}{\linewidth}{@{}l >{\raggedright\arraybackslash}X c@{}}
\toprule
\textbf{Variant} & \textbf{Generated summary} & \shortstack{\textbf{Rank}\\\textbf{(Acc / Cont / Conc)}} \\
\midrule
\texttt{codet5\_baseline} & ``Optimize the regular expression'' & 4 / 3 / 3 \\
\texttt{codet5\_code\_cg} (best) & ``Optimize the NFA'' & \textbf{2 / 1 / 1} \\
\bottomrule
\end{tabularx}
\end{table}

{\emergencystretch=1em
The caller context (the \texttt{pullback()} method, 1{,}102 characters) and the callee context (the \texttt{nfanode(\allowbreak{}Subre t)} method, 541 characters) both reference NFA construction and arc manipulation, giving the model repeated direct evidence for the technical term. \textbf{Category:} identifier accuracy via callee semantic injection.\par}

\paragraph{Example W3: Callee semantic injection (snippet Q61, \texttt{asyncDatastore}).}
The method returns either a \texttt{CachingAsyncDatastore} or a plain \texttt{AsyncDatastore} depending on a conditional check. The ground-truth summary is \emph{``Might produce a caching version if caching is enabled''}. The baseline captures the wrapping pattern but omits the central caching semantic; the context-augmented model injects the term \texttt{caching} and aligns with the ground truth.

\noindent\begin{minipage}{\linewidth}
\begin{lstlisting}[language=Java, basicstyle=\footnotesize\ttfamily, breaklines=true, frame=single, captionpos=b]
public AsyncDatastore asyncDatastore(final boolean enableGlobalCache) {
    if (this.entityMemcache != null && enableGlobalCache &&
        this.registrar.isCacheEnabled())
        return new CachingAsyncDatastore(asyncDatastore(), this.entityMemcache);
    else
        return asyncDatastore();
}
\end{lstlisting}
\end{minipage}

\noindent\begin{minipage}{\linewidth}
\begin{lstlisting}[language=Java, basicstyle=\footnotesize\ttfamily, breaklines=true, frame=single, captionpos=b]
// Caller (isCacheEnabled):
public boolean isCacheEnabled() { return this.cacheEnabled; }
 
// Callee (zero-argument asyncDatastore):
public AsyncDatastoreReaderWriter asyncDatastore() {
    return factory.asyncDatastore(ofy.getOptions().isCache());
}
\end{lstlisting}
\end{minipage}

\begin{table}[h]
\centering
\small
\begin{tabularx}{\linewidth}{@{}l >{\raggedright\arraybackslash}X c@{}}
\toprule
\textbf{Variant} & \textbf{Generated summary} & \shortstack{\textbf{Rank}\\\textbf{(Acc / Cont / Conc)}} \\
\midrule
\texttt{codet5\_baseline} & ``Returns an AsyncDatastore that wraps the underlying Datastore'' & 3 / 3 / 3 \\
\texttt{codet5\_code\_cg\_vh} (best) & ``Creates an async datastore with caching enabled or disabled'' & \textbf{1 / 1 / 2} \\
\bottomrule
\end{tabularx}
\end{table}

Both the caller \texttt{isCacheEnabled()} and the callee \texttt{asyncDatastore()} reference cache state directly (\texttt{cacheEnabled}, \texttt{isCache}), surfacing the term \texttt{caching} to the model. \textbf{Category:} callee semantic injection.

\paragraph{Counter-example L1: A tie case where context did not help (snippet Q74, \texttt{lookupTableRef}).}
The method looks up a table-reference descriptor and throws an exception if not found. The ground-truth summary is \emph{``Lookups for table reference and throws an exception if table reference not found''}. All eight summary variants (baseline plus seven context-augmented) produce summaries of the form ``Look(s) up the entity descriptor for the (given) table reference,'' correctly capturing the lookup action but omitting the exception-throwing branch. The annotator ranked every variant at the top of each dimension.

\noindent\begin{minipage}{\linewidth}
\begin{lstlisting}[language=Java, basicstyle=\footnotesize\ttfamily, breaklines=true, frame=single, captionpos=b]
protected DbEntityDescriptor lookupTableRef(final String tableRef) {
    DbEntityDescriptor ded = getTableDescriptor(tableRef);
    if (ded == null) {
        throw new DbSqlBuilderException(
            "Table reference not used in this query: " + tableRef);
    }
    return ded;
}
\end{lstlisting}
\end{minipage}

This counter-example is informative because the missing semantic (the exception throw) lives inside the method body itself, not in the caller (\texttt{getTableDescriptor}, 200 characters) or the callee (a one-line forwarder to a two-argument overload). Context augmentation cannot help here because the discriminating information is intra-method. This complements the win examples above, which all illustrate context surfacing information that the method body alone does not name.

\subsubsection*{Cross-Task Synthesis}

Taking the two tasks together, the qualitative evidence supports the model-and-task-dependent narrative already established in Section~\ref{sec:findings-implications}. For SeSaMe clone detection, the dominant beneficiary of context augmentation is the low-surface-similarity bucket where surface tokens alone fail; the principal contributing channels are version history and method age, with caller and callee text contributing in cases where the mined neighbourhood is informative. For CodeSearchNet summarisation, the dominant beneficiary is the class of methods whose central semantic appears only in the caller or callee rather than in the method body, and the call-graph channel is the primary contributor. The counter-examples reinforce two limits already discussed in the paper: version-history context can dilute present-day similarity signals when libraries evolve divergently (Section~\ref{sec:cg_error_analysis}); and context cannot supply information that is missing from the method body but absent from its neighbourhood. The replication package contains the per-pair predictions and per-snippet ranking data used to generate this analysis.

\section{Threats to Validity and Limitations}

\subsection{Threats to Validity}
\label{sec:threats}

We now discuss the threats to validity, limitations, and mitigation strategies of our context-augmented code representation study.

\paragraph{Internal validity}
We designed the pipeline so that observed effects stem from our interventions (context encoding and aggregation) rather than artifacts of data collection or training. Repository attrition (unavailable repositories, orphan commits, and build failures) was handled with uniform, pre-declared rules. We preserved the original train/\allowbreak validation/\allowbreak test partitions from prior work and performed method extraction, deduplication, and leakage checks to avoid split contamination at the clone-pair level, following the partitioning protocol of prior work ~\cite{tian2022adding, nguyen2024encoding}. A key implementation threat lies in the truncation limits of pre-trained Transformer models. To mitigate this threat, we use widely adopted architectures with a reasonable context window (512 tokens). With this limit, concatenating source code with historical versions and other types of context can exceed the window. We further minimise the impact by ordering versions from most recent to oldest and concatenating exhaustively until the limit is reached, so that the most informative recent changes are consistently retained across all settings.
We quantify this exposure precisely for CodeSearchNet (Code Summarisation): for version history context alone (code\_vh), approximately 23--30\% of training records and 19--25\% of test records exceed the 512-token limit across our four encoder models; for call graph context alone (code\_cg), approximately 32--36\% of training and 27--39\% of test records are affected; and for the combined scenario (code\_cg\_vh), approximately 51--63\% of training and 46--59\% of test records exceed the limit. These figures reflect the additional token load of encoding caller and callee source code alongside the method and its version history. The truncation is principled: versions are retained most-recent-first, so the most current and typically most informative state of the method is always preserved within the budget, while only older, more distant versions are dropped. We ensure that comparisons use the same truncation policy and maintain hyperparameters/budgets matched with the fixed seeds provided in the baseline works. In summary, we believe that the residual risk from truncation and training variance is limited and does not overturn the direction of our findings.

\paragraph{Construct validity}
Automated metrics (BLEU-4, ROUGE-L, METEOR, and BERTScore-F1) only partially capture the perceived quality of summaries. \revised{BLEU-4 and ROUGE-L measure surface n-gram overlap with a single reference; METEOR and BERTScore are partially semantic, but they measure similarity to the reference rather than correctness of the summary. For code summarisation specifically, Roy et al.~\cite{roy2021reassessing} found that these metrics agree only weakly with human judgement and that small metric differences are not perceived by humans as quality differences, and Mastropaolo et al.~\cite{mastropaolo2024evaluating} reported weak correlation between existing metrics, including BERTScore, and human ratings of code summaries.} To mitigate this, we complemented automated evaluation with a blinded human study using rank-order-with-ties, preceded by clear guidelines and a pilot to calibrate raters, and we report agreement statistics. \revised{This mitigation applies to the CodeT5 summarisation results only: the human study covers 100 snippets and one model family (Section~\ref{sec:human-eval}). The LLM summarisation results in Tables~\ref{tab:result_rq1_summarisation} and~\ref{tab:result_rq2_summarisation} rest on automatic metrics alone and carry the full limitation; the mixed pattern discussed in Section~\ref{sec:result_rq1}, where lexical and semantic metrics sometimes move in opposite directions, should be read in that light. The reference-quality issues documented for CodeSearchNet (Section~\ref{sec:human-eval}) compound this for LLM output, since LLM summaries tend to be fuller natural-language sentences that diverge from the terse reference style regardless of their correctness.} Because our annotation task is ordinal rank-ordering (ranking nine summaries from best to worst with ties allowed), we use Kendall's coefficient of concordance (\(W\)) for the four-rater pilot phase and Kendall's \(\tau_b\) for the two-rater main study --- the methodologically appropriate metrics for ordinal data with ties. Agreement was strong in the pilot (mean \(W = 0.7261\), median \(W = 0.7540\)) and very high in the main study (mean \(\tau_b = 0.8899\), median \(\tau_b = 1.0000\)); full details are reported in Section~\ref{sec:human-eval}. The human assessment focuses on a representative model and a stratified sample of items to align with practical judgments on accuracy, adequacy, and conciseness. Our operationalisation of “additional context” targets three representative sources, i.e. version history (code hosting platform), call graphs (static analysis), and method age (a simple temporal signal), acknowledging that other artifacts (e.g., issues/PRs, review comments, runtime traces) are promising but left for future work.

\paragraph{External validity}
Our results are derived from Java open-source projects (SeSaMe, CodeSearchNet, and Vul4J). This choice enables scalable mining of version history and call graphs but may limit generality to other languages, build systems, or proprietary settings. Call-graph extraction process depends on the static analysis tool and its configuration. We evaluate twelve models (five foundational code models spanning tree-based and pre-trained Transformer architectures, and seven code-specialised LLMs) but do not claim exhaustiveness. \revised{All tasks are evaluated at the method level. Modern code LLMs and coding agents are increasingly used in repository-level and task-oriented settings, where context is retrieved dynamically and combined with planning and tool use. Our findings indicate which context types carry useful signal at the method level and are, we hypothesise, relevant to the context-selection decisions such systems make, but we have not measured the size or the direction of the effect at repository level (Section~\ref{sec:limitation}).}
\paragraph{Call-graph selection bias.}
Call-graph extraction required the successful compilation of each repository from its pinned commit. Of the CodeSearchNet repositories, 39.2\% could not be compiled, primarily due to missing dependencies, Java version incompatibilities, or deprecated build scripts; the retained 63,259 methods represent approximately 35\% of the original pool. This introduces a selection effect: the retained subset consists of projects whose dependency declarations remained resolvable at the time of our experiment, which may correlate mildly with dependency stability or project maintenance discipline. \revised{To test whether the retained subset differs systematically from the discarded one, we compared the two groups directly (Table~\ref{tab:retained_vs_discarded}). Of the 4,001 repositories in the mining pipeline, 3,729 had a Maven or Gradle build file and could be cloned; 2,269 built successfully (retained) and 1,460 failed (discarded), giving the 39.2\% failure rate above. The remaining 272 repositories, which had no build file or were no longer available on GitHub, are excluded from the comparison since neither case is a build failure. Because version history mining did not require compilation, method-level properties are available for both groups: 79,730 methods in retained repositories and 68,109 in discarded ones. We report medians with interquartile ranges and Cliff's $\delta$ as the effect size; with tens of thousands of methods per group every difference is statistically detectable, so effect size is the informative quantity. The two groups are indistinguishable on every size and history dimension measured. Methods per repository, method length, number of prior versions, and method age all show negligible effect sizes ($|\delta| \leq 0.07$), and discarded repositories are marginally larger rather than smaller. On the dimensions we can measure, the retained subset is therefore representative of the discarded one. The one clear difference is the build system: 72.8\% of Maven repositories built successfully compared with 20.9\% of Gradle repositories. This is a tooling difference rather than a code-quality difference. Gradle builds at an old pinned commit depend on a wrapper version and plugin set that has frequently drifted out of the ecosystem, whereas Maven builds are more often recoverable from central repositories. The discarded set includes some of the largest and most actively maintained projects in the corpus, such as \texttt{apache/flink}, \texttt{deeplearning4j}, \texttt{Alluxio}, \texttt{apache/groovy}, and \texttt{spring-projects/spring-boot}. On this evidence, build failures are predominantly environmental rather than indicators of project complexity or code quality. Two caveats remain. Characteristics we did not measure, such as test coverage or dependency-management discipline, may still differ between the groups. In addition, because Gradle projects fail far more often, the call-graph subset over-represents Maven projects, and we state this as part of the scope of the call-graph findings.} The primary findings of this paper, which concern version history context and are derived from approximately 97\% of the full method pool (no compilation required), are unaffected by this selection. For the call-graph results specifically, the scope of generalisation is: projects where static call-graph extraction from source is feasible, a constraint shared by all static analysis tools in the literature~\cite{yamaguchi2014modeling, jasz2019systematic, gousios2019javacallgraph, guo2020graphcodebert}. Future work could broaden this scope through hermetic Docker-based build environments with pinned JDK images, bytecode-level call-graph extraction (which bypasses source compilation entirely), or dynamic call graphs derived from test-suite execution.

\begin{table}[t]
\ifrevisedhighlight\color{revisioncolor}\fi
\caption{Retained-versus-discarded comparison for call-graph mining on CodeSearchNet-Java. Retained: 2,269 repositories (79,730 methods) that built successfully from their pinned commit. Discarded: 1,460 repositories (68,109 methods) with a Maven or Gradle build file that failed to build. Values are medians with interquartile ranges in brackets; Cliff's $\delta$ is the effect size (negligible below 0.147).}
\label{tab:retained_vs_discarded}
\centering
\begin{tabular}{lccc}
\toprule
\textbf{Variable} & \textbf{Retained} & \textbf{Discarded} & \textbf{Cliff's $\delta$} \\
\midrule
Methods per repository       & 9 [3, 26]        & 11 [4, 35]       & 0.07 (negligible) \\
Method length (code tokens)  & 66 [43, 116]     & 70 [44, 124]     & 0.04 (negligible) \\
Prior versions per method    & 1 [1, 2]         & 1 [1, 2]         & 0.03 (negligible) \\
Method age (days)            & 903 [358, 1,548] & 794 [390, 1,416] & 0.03 (negligible) \\
\midrule
\textbf{Build system} & \textbf{Retained ($n$)} & \textbf{Discarded ($n$)} & \textbf{Retention rate} \\
\midrule
Maven  & 2,090 & 781 & 72.8\% \\
Gradle & 179   & 679 & 20.9\% \\
\bottomrule
\end{tabular}
\end{table}

\paragraph{Label leakage from contextual signals.}
An important question is whether the gains attributed to additional context could arise from project-correlated surface cues in the appended payload (e.g., project names, characteristic comments, or naming conventions) rather than from genuine semantic content. Our extraction pipeline already excludes the most direct carriers of such cues: PyDriller and \textit{java-callgraph}~\cite{gousios2019javacallgraph} return method bodies only, not file headers, package declarations, import blocks, directory paths, or filenames. To test the remaining channels empirically, we ran an across-project shuffled-context control on Code Classification, the task in which the leakage concern is most acute because the labels are project names. For each test method with project label $P_i$, we replaced its entire version history and call-graph payload with the corresponding payload from a uniformly randomly chosen donor method whose label is different from $P_i$, leaving the method's own source code and label unchanged. We focused the control on the \texttt{code+VH+CG+age} configuration, which is the configuration with the largest macro-F1 gain over the baseline for both tested models (CodeBERT $+$16.65\%, CodeT5 $+$21.48\%) and the one that exposes the maximum surface area for any latent leakage signal. Smaller context subsets are strict subsets of this configuration and therefore cannot exhibit more leakage than what is tested here. We selected CodeBERT and CodeT5 as representatives of the two Transformer architectural families used in the paper.

Table~\ref{tab:shuffled_context_results} summarises the result across two random seeds (42 and 1234). For CodeBERT, the original-context configuration improves macro-F1 over the baseline by $+$16.65\% (from 0.6698 to 0.7813), whereas the shuffled-context configuration moves macro-F1 in the opposite direction by an average of $-$28.29\% across the two seeds (mean macro-F1 = 0.4803, below the baseline). For CodeT5, the original context improves macro-F1 by $+$21.48\% (from 0.6569 to 0.7980), while the shuffled context yields a mean macro-F1 of 0.6555, statistically indistinguishable from the baseline ($-$0.21\%); the two seeds straddle the baseline by $-$2.09\% and $+$1.66\%, respectively, well within typical fine-tuning seed variance. These results indicate that the original-context gains are not driven by project-correlated surface cues: if such cues drove the gain, the shuffled context, which is drawn from a different project and therefore still carries its own project-correlated cues, would continue to produce gains. We also verified that the shuffled context is not systematically shorter than the original (Table~\ref{tab:shuffled_context_length_stats}); the median context-token length and interquartile range are comparable between conditions across all three splits, so the observed performance collapse cannot be attributed to a length deficit. This control isolates project-correlated cues carried by the appended payload but does not test intra-method comments or identifier-naming conventions inside the target method itself; we discuss this as a future-work item in Section~\ref{sec:limitation}.

\begin{table}[t]
\caption{Across-project shuffled-context control experiment on Code Classification (SeSaMe). For each test method with project label $P_i$, the version history and call-graph payload is replaced with that of a uniformly randomly chosen donor with a different label; the method's own source code and label are preserved. We report two random seeds (42 and 1234) and their mean. P and R denote macro-averaged precision and recall; macro-F1 is the primary metric for this multi-class task. $\Delta$ denotes percentage change in macro-F1 relative to the no-context baseline.}
\label{tab:shuffled_context_results}
\centering
\begin{tabular}{llcccr}
\toprule
\textbf{Model} & \textbf{Condition} &
\textbf{P} & \textbf{R} & \textbf{macro-F1} &
\textbf{$\Delta$ vs.\ Baseline} \\
\midrule
\multirow{5}{*}{CodeBERT}
  & Baseline (no context)            & 0.7103 & 0.6498 & 0.6698          & (reference) \\
  & Original context                 & 0.8504 & 0.7522 & 0.7813          & $+$16.65\%  \\
  & Shuffled context, seed 42        & 0.5739 & 0.5352 & 0.5411          & $-$19.21\%  \\
  & Shuffled context, seed 1234      & 0.4580 & 0.4360 & 0.4195          & $-$37.37\%  \\
  & Shuffled context, mean of seeds  & 0.5160 & 0.4856 & \textbf{0.4803} & \textbf{$-$28.29\%} \\
\midrule
\multirow{5}{*}{CodeT5}
  & Baseline (no context)            & 0.7005 & 0.6364 & 0.6569          & (reference) \\
  & Original context                 & 0.8419 & 0.7889 & 0.7980          & $+$21.48\%  \\
  & Shuffled context, seed 42        & 0.6715 & 0.6341 & 0.6432          & $-$2.09\%   \\
  & Shuffled context, seed 1234      & 0.7049 & 0.6576 & 0.6678          & $+$1.66\%   \\
  & Shuffled context, mean of seeds  & 0.6882 & 0.6459 & \textbf{0.6555} & \textbf{$-$0.21\%} \\
\bottomrule
\end{tabular}
\end{table}

\begin{table}[t]
\caption{Context-token-length distribution for the original and shuffled conditions of the leakage control experiment (Table~\ref{tab:shuffled_context_results}), by split. The shuffled context is drawn from a different project per row by uniform random sampling. Medians and interquartile ranges are comparable between conditions across all three splits, confirming that no systematic length confound was introduced by shuffling.}
\label{tab:shuffled_context_length_stats}
\centering
\begin{tabular}{llrrr}
\toprule
\textbf{Split} & \textbf{Condition} &
\textbf{Median} & \textbf{p25} & \textbf{p75} \\
\midrule
\multirow{2}{*}{Train}
  & Original   & 108 & 54   & 259 \\
  & Shuffled   & 111 & 54   & 291 \\
\midrule
\multirow{2}{*}{Dev}
  & Original   & 102 & 53   & 255 \\
  & Shuffled   & 102 & 59   & 317 \\
\midrule
\multirow{2}{*}{Test}
  & Original   & 103 & 60   & 239 \\
  & Shuffled   & 108 & 65   & 338 \\
\bottomrule
\end{tabular}
\end{table}

\paragraph{Conclusion validity}
Multiple models, context types, and aggregation techniques increase the risk of spurious findings. We therefore use paired non-parametric tests and report effect sizes, and we emphasise consistent directional trends rather than isolated p-values. High tie rates in the human ranking indicate overlapping performance on some items; we interpret gains conservatively. \revised{The SeSaMe test split is small: 86 clone pairs, of which 25 are positive, and 170 methods across 11 project classes (Table~\ref{tab:sesame_stats}). A single changed prediction therefore moves clone-detection recall by 4 percentage points and F1 by roughly 2, and macro-F1 for classification is similarly sensitive because several classes have few test methods. Independent runs of the same configuration differ by a comparable amount: 1.7 F1 points in the history-depth ablation (Section~\ref{sec:vh_depth_ablation}) and about 2\% across seeds in the shuffled-context control (Table~\ref{tab:shuffled_context_results}). We treat roughly 2 F1 points as the resolution of the SeSaMe results. Differences below that level are described as directional rather than as improvements, and the conclusions we draw rest on gains that are consistent in direction across models and on the larger effects (above 10\%), not on individual single-digit differences. The CodeSearchNet results (63,259 methods) are not subject to this limitation. Seeds and splits are controlled for comparability; full multi-seed replication of the SeSaMe tables would further reduce uncertainty and is left for future work.} Overall, mitigations are in place, and residual threats are limited and transparent.

\subsection{Limitations and Future Work}
\label{sec:limitation}

\paragraph{Model families and scope.}
In this work, we focus on \emph{code representation models}, which are designed to encode source code into dense vector representations. We extended the evaluation to include seven code-specialised large language models (CodeLlama, Qwen2.5-Coder, and DeepSeek-Coder) under both prompt-based context injection and fine-tuned settings (Section~\ref{sec:result_rq1}). \revised{Each LLM reaches its highest performance with context after fine-tuning on context-augmented inputs, and fine-tuned LLMs with prompt-level context are competitive with fine-tuned encoders on the classification tasks. However, no single backbone is evaluated under both prompt-level and representation-level fusion, so the comparison between the two fusion mechanisms is confounded with architecture family. A controlled comparison would attach a representation-level fusion head to a decoder-only LLM and compare it against the same model with prompt-level context; we regard this as future work alongside the attention-based fusion study noted above.} \revised{For summarisation, prompt-level context helps only two of the seven LLMs (Qwen2.5-Coder 7B and 14B) and lowers every metric for the other five, whereas representation-level encoding improves all four foundational models; the two paradigms are therefore complementary rather than directly substitutable, and the LLM result should not be generalised beyond the model families tested.} While Large Language Models (LLMs) have demonstrated remarkable capabilities in code generation and understanding, their decoder-only Transformer architectures are not inherently optimised for producing high-quality code representations~\cite{llm2vec}. Embedding-tuned LLMs and more sophisticated retrieval-augmented pipelines remain promising directions for future work.

\paragraph{\revised{Representation capacity under concatenation.}}
\revised{The two controls in Section~\ref{sec:projection_control} narrow but do not close the question of whether part of the concatenation gain in encoder-only models arises from the larger classifier input rather than from the contextual content. The experiment that would separate the two effects is simple to state: hold the context fixed and vary the width of the projection layer (for example 768, 1,536, and 3,072 dimensions) so that the cost of the bottleneck can be measured directly against the contribution of the context. We regard this as a natural companion to the attention-based fusion study identified in Section~\ref{sec:projection_control}.}

\paragraph{Context coverage (breadth of artifacts).}
We study three representative context types: version history (from VCS), call graphs (from static analysis), and method age (a simple temporal signal). \revised{Method age is derived from the version-history mining pass and is evaluated only in combination with version history, so its independent contribution is not isolated in this study, and the intuition given in Section~\ref{sec:encoding_aggregation} for why it should help remains a hypothesis; a method-age-only condition is a straightforward ablation that we have not run.} Other artifacts, such as issue and pull-request discussions, code review comments, API documentation, commit messages, runtime traces, or test coverage, are out of scope for this paper. Our choice reflects scalability and availability across large corpora rather than an assertion of completeness. A fruitful next step is to incorporate textual artifacts (e.g., issues/PRs) and dynamic signals (e.g., traces), and to systematically compare \emph{structural} (CG) versus \emph{textual} (issues/docs) context, including their interactions.

\paragraph{Token-budget constraints and selective context encoding.}
The 512-token limit of our encoder-based models means that approximately one in four records sees truncated version history and roughly one in two records sees truncated combined context (version history plus call graphs) in CodeSearchNet. While the most-recent-first ordering ensures the most recent context, which is often informative, is retained, this is a heuristic rather than a learned selection. Three directions address this limitation. First, \emph{salience-aware version selection}: rather than selecting by recency, versions could be ranked by semantic distance from the current method (e.g., using BM25 or embedding cosine similarity) and only the most distinct and informative versions admitted to the budget. Second, \emph{near-duplicate deduplication}: many consecutive versions differ only in whitespace, formatting, or minor variable renames; removing these before encoding would free budget for more semantically distinct historical states. Third, \emph{long-context architectures}: models such as Longformer~\cite{beltagy2020longformer} or BigBird~\cite{zaheer2020bigbird} extend the effective context window to 4,096--16,384 tokens, enabling full version histories and complete call graph neighbourhoods without truncation. We regard resolving the tension between token budgets and context richness as a high-priority direction for future work.

\paragraph{Call-graph context depth and selection strategy.}
The current study encodes one caller and one callee at depth-1 in the call graph, selected by token length following Tian and Treude~\cite{tian2022adding}. This is a practical starting point constrained by the 512-token model budget: encoding even two callers would push the majority of inputs past the token limit given that the selected callee alone has a mean of 268 tokens. Richer call-graph context is a meaningful and high-priority extension. Three concrete directions are: (i) \emph{multi-caller aggregation}, encoding all direct callers via a set encoder (e.g., mean pooling or a learned attention over caller embeddings) once token-budget constraints are relaxed; (ii) \emph{multi-hop traversal}, propagating context from callers-of-callers and callees-of-callees using graph neural network layers or graph attention, which would situate a method within a broader architectural context; and (iii) \emph{class-level and module-level context}, encoding sibling methods in the same class, class-level docstrings, or package-level summaries as additional signals. All three directions depend on resolving the token-budget bottleneck, for example through long-context architectures (Longformer, BigBird) or hierarchical encoding strategies that compress richer neighborhoods into fixed-size representations before fusion.

\paragraph{Generalisation to other SE tasks.}
We evaluate three canonical program-comprehension tasks (i.e. clone detection, code classification, and code summarisation) to study how representation-level context encoding affects both classification and generation. Other downstream tasks (e.g., bug localisation, defect prediction, code review assistance, comment generation, commit message generation, or test generation) remain untested. We view these tasks as natural extensions: the same mining pipeline can be reused, and the learned context encoders can be applied or adapted with task-specific heads. Exploring these tasks will help determine which context types (historical, structural, textual, and temporal) deliver the largest practical gains under different operational constraints.

\paragraph{Scope relative to repository-level benchmarks.}
Recent evaluation benchmarks in AI for Software Engineering, such as SWE-bench~\cite{jimenez2024swebench} and DevEval~\cite{li2024deveval}, target repository-level tasks (issue resolution, multi-file patch generation, and end-to-end code generation) that require reasoning across entire codebases and agentic tool use. \revised{Our study is method-level and does not evaluate these settings. The scope was chosen because method-level tasks isolate the contribution of contextual encoding from retrieval quality, prompt construction, and multi-step planning, all of which vary in repository-level pipelines, and they remain the primary testbed for encoder-based models in the SE community~\cite{wan2022they,guo2022unixcoder}. We nonetheless see a direct connection to repository-level work. Coding agents already operate on repository-level context: they retrieve files, follow dependency structure, and localise edits across the codebase. CodePlan~\cite{bairi2024codeplan} plans multi-file edits from an incremental dependency analysis, and Agentless~\cite{xia2025agentless} shows that a simple hierarchical localisation over repository structure outperforms more elaborate agent loops. Context selection is therefore one important component of repository-level use. The context types studied here go beyond the current snapshot of the repository that these systems consume: version history and method age describe how a method has evolved, and the call graph describes its structural role. At the method level, version history was the most reliable signal for the foundational models and call-graph context drove the summarisation gains for the LLMs that benefit from context at all.}

\revised{To the question of how far these findings carry, we can say the following. For structural context, our result is consistent with what current agents already exploit. Historical context is not typically part of the context retrieved by the systems we discuss, and our result suggests it is worth adding; both context types can be mined at repository scale from version control and static analysis, as we did for 175,240 methods. Agents also use LLMs through inference-only prompting, and our zero-shot results show that context still helps in this setting, although not uniformly across tasks and models: version history improves zero-shot clone detection for five of the seven LLMs (Table~\ref{tab:result_rq1_clone_class}), and call-graph context improves zero-shot summarisation for Qwen2.5-Coder 7B and 14B (Table~\ref{tab:result_rq1_summarisation}); exploiting the full history of a repository would in addition require models whose context windows can hold it. We did not evaluate an agent directly: the method-level design isolates the effect of context from retrieval quality, prompt construction, and multi-step planning, which vary across agent pipelines, and repository-level history and multi-file call graphs exceed the context budget of the models studied here (Section~\ref{sec:cg_error_analysis} shows that the selection and fusion problem grows with the amount of context even at the method level). The size of the benefit at repository level therefore remains a hypothesis; it will depend on retrieval quality, on the token budget, and on the model family (Section~\ref{sec:cg_error_analysis}). The experiment that would test it is to add version history and call-graph context to an agent's retrieval and evaluate on a repository-level benchmark, where call-graph context could span multiple files and version history could encode multi-commit fix patterns. We regard this as the natural next step for this work.} As a first step toward generalisation beyond program comprehension, we conducted a vulnerability detection experiment (Section~\ref{sec:vuln_generalisation}) on a real-world dataset (Vul4J), providing initial evidence that context-enriched representations transfer to a security-critical task.

\paragraph{Computational overhead.}
\label{sec:overhead}
We clarify that our framework introduces two categorically distinct costs that must be assessed separately: a one-time offline preprocessing cost and a bounded per-instance inference cost.

\textit{Preprocessing (offline, one-time).} All context — version history, call graphs, and method age — is extracted from source repositories and stored as pre-computed features before any model is trained or queried. At inference time no extraction occurs; the model receives a pre-formatted concatenated string. This is analogous to how GraphCodeBERT~\cite{guo2020graphcodebert} requires pre-extracted data flow graphs as a dataset construction step. We executed the mining pipeline on five virtual servers (4 vCPU, 16 GB RAM each; the University of Melbourne's Research Computing Platform). Version history was mined per-method by traversing each repository's git log from the pinned commit hash back to the method's first appearance; call graphs were produced by static analysis after compiling each repository with Maven or Gradle; method age was derived at zero additional cost from the first-commit timestamp already retrieved during version history mining. Table~\ref{tab:preprocessing} summarises the measured and estimated wall-clock times.

\begin{table}[ht]
\centering
\caption{Preprocessing pipeline wall-clock time (5 servers, 4 vCPU / 16 GB RAM each). CodeSearchNet figures are measured; SeSaMe and Vul4J are estimated using the same per-repository and per-method rates.}
\label{tab:preprocessing}
\resizebox{0.75\textwidth}{!}{%
\begin{tabular}{lrrr}
\toprule
\textbf{Step} & \textbf{CodeSearchNet} & \textbf{SeSaMe} & \textbf{Vul4J} \\
\midrule
Repo clone         & 40.6 h & 0.46 h & 0.70 h \\
Compile (Maven/Gradle) & 5.5 h  & 0.37 h & 0.35 h \\
Call graph extraction  & 4.1 h  & 0.05 h & 0.06 h \\
Version history mining & 8.0 h  & 0.46 h & 0.03 h \\
Method age             & \multicolumn{3}{c}{$\approx$0 (derived from VH first-commit timestamp)} \\
\midrule
\textbf{Total}         & \textbf{$\sim$58 h} & \textbf{$\sim$1.3 h} & \textbf{$\sim$1.1 h} \\
\bottomrule
\end{tabular}%
}
\end{table}

The pipeline was executed against all 4,001 CodeSearchNet repositories identified in the original corpus, requiring approximately 58 hours of wall-clock time. The resulting usable subsets differ by context type, as detailed in Section~\ref{sec:context_augmentation}: version history mining does not require a compilable project state and retained 175,240 methods ($\approx$97\% of the original corpus); call graph extraction requires a successful build, and after removing repositories unavailable on GitHub, build failures (39.2\% of repositories could not be compiled), and methods with no identifiable caller or callee, 63,259 methods were retained ($\approx$35\%). SeSaMe (11 repositories, 1,679 methods) and Vul4J (21 repositories, 178 methods) each required approximately one hour. The preprocessing cost reflects the full pipeline attempt --- including repositories that ultimately failed to build --- and is a one-time cost reused across all model families and aggregation strategies, amortising to approximately 1.2 seconds per method across both context types.

\textit{Inference overhead (online).} All aggregation strategies (Concatenation, Max-Pooling, Diff-Concat) use a \emph{single} model forward pass over the concatenated input — there is no additional cost from running separate encoders per context source. We verified the runtime overhead using WandB logs across all seven LLMs and four tasks (NVIDIA A100 GPUs, Spartan HPC, Unimelb). For zero-shot inference over the full test set, overhead relative to the code-only baseline is minimal: $+$5.7\% for Vulnerability Detection (1.8 min~$\to$~1.9 min), $\approx$0\% for Clone Detection (1.4 min~$\to$~1.3 min), and $+$37.9\% for Code Classification (2.5 min~$\to$~3.5 min). For fine-tuned models, overhead is proportional to the additional tokens per training step: $+$29.4\% for Code Summarisation (30.0 h~$\to$~38.9 h), $+$76.4\% for Code Classification (120.5 min~$\to$~212.5 min), and $+$164\% for Clone Detection (76.4 min~$\to$~202.1 min, driven by version history roughly doubling the average input length). Fine-tuning overhead is a one-time training cost; once the model is trained it serves predictions at standard inference speed, with per-query latency determined solely by the bounded token overhead of the concatenated context.

\paragraph{Identifier-masked control for in-method leakage channels.}
The shuffled-context control reported in Table~\ref{tab:shuffled_context_results} isolates leakage carried by the \emph{appended} context (version history, callers, callees). It does not test other potential leakage channels that reside inside the \emph{target} method itself, such as intra-method comments referencing class hierarchies, project-specific docstring conventions, or identifier-naming patterns. A natural complementary control is to mask these in-method signals (for example, by replacing project-specific identifiers and inline-comment tokens with placeholders) and re-evaluate model performance. If gains persist under masking, the model is exploiting structural and semantic features beyond surface identifiers; if gains disappear, those identifiers carried a non-trivial share of the signal. Designing this control requires careful choice of the masking vocabulary so that legitimate semantic content (e.g., method names that carry meaning) is not stripped along with project-specific terminology. We regard this as a high-priority follow-up to the present study.

\paragraph{\revised{Evaluation of LLM-generated summaries.}}
\revised{The LLM summarisation results in this study are evaluated with reference-based automatic metrics only, which, as discussed in Section~\ref{sec:threats}, agree only weakly with human judgement for code summaries~\cite{roy2021reassessing, mastropaolo2024evaluating}. Extending the human evaluation to LLM output, or adopting evaluation methods that do not depend on a single lexical reference, such as the reference-free SIDE metric~\cite{mastropaolo2024evaluating} or LLM-as-judge protocols calibrated against human ratings, would allow the semantic quality of context-augmented LLM summaries to be assessed directly rather than inferred from metric disagreement.}

\section{Conclusion}

This paper extended our prior MSR 2024 study~\cite{nguyen2024encoding} by systematically investigating \emph{representation-level} encoding of additional software artifacts, including version history and call graphs, into neural code representations. Across three datasets (SeSaMe, CodeSearchNet, and Vul4J), twelve models (five foundational code models spanning tree-based and pre-trained Transformer architectures, and seven code-specialised LLMs), and four downstream tasks (clone detection, code classification, code summarisation, and vulnerability detection), we showed that context-enriched representations deliver \revised{consistent, practically meaningful gains for the five foundational models on all three program comprehension tasks, and for the LLMs on the classification tasks after fine-tuning; on summarisation, prompt-level context helps two of the seven LLMs and hurts the rest,} \revised{with human judgements corroborating the automated metrics for CodeT5 summarisation}. \revised{The expanded LLM evaluation further shows that LLMs reach their highest performance with context after fine-tuning on context-augmented inputs; whether representation-level fusion is a better mechanism than prompt-level fusion at equal training remains open, since the two are confounded with backbone architecture in our design.} Our results address the three research questions: contextual signals are beneficial overall; the benefits of each context and the optimal way to combine them are task- and model-dependent; and aggregation choices at the representation level significantly impact outcomes.

\paragraph{Additional Context as Software Engineering Artifacts.}
Our findings confirm that encoding software-development artifacts into code representations improves downstream performance. In particular, \emph{version history} emerges as a reliable signal across tasks \revised{for the foundational models}, while \emph{call graphs} provide substantial but model- and task-specific advantages (notably for classification\revised{, for accuracy/conciseness in summarisation, and as the context that drives the summarisation gains of the two LLMs that benefit from prompt-level context}). These artifacts come in multiple forms—natural language (e.g., commit messages), graphs (call hierarchy), timestamps (commit date-time), and numeric signals (e.g., method age/versions)—and can be mined at scale from version-control and static-analysis pipelines. Mining, however, introduces challenges: inconsistent availability across projects, imbalanced artifact distributions (e.g., many methods with one–two versions but some with hundreds), and potential noise from near-duplicate or minor edits. We mitigated these issues with uniform mining rules, split preservation, and consistent truncation policies, yet the results highlight an agenda for \emph{selective} context encoding (e.g., salience-aware version selection, de-duplication of minor edits, or learned pruning). Overall, the evidence supports using version history as a default augmentation, adding call graphs when task demands align, and tailoring context choice to evaluation dimensions (accuracy, adequacy, conciseness) in generative settings.

\paragraph{Models and Aggregation.}
Representation-level aggregation significantly impacts how effectively models utilise context. Diff-Concat is consistently strong for clone detection, whereas Concat vs.\ Max-Pooling trade places in classification depending on model family and context richness; no single method dominates everywhere. These outcomes interact with architectural constraints: tree/RNN-style models (e.g., ASTNN) struggle with long-range dependencies, and encoder-style Transformers (e.g., CodeBERT/GraphCodeBERT) are limited by input windows that can truncate concatenated context. Two complementary directions follow. First, \emph{better aggregation}—from hierarchical or cross-attentional fusion, gating/weighting per context, to domain-specific pooling—can allow models to focus on informative history while suppressing noise. Second, \emph{better architectures}, such as Graph Transformers, hierarchical encoders, or long-context LMs, can alleviate length constraints and capture multi-hop structure more naturally. Together, these paths suggest practical recipes: pick aggregation by task (e.g., Diff-Concat for similarity-oriented classification), choose context types by objective (version history as the default; call graphs when structure matters), and consider longer-context or hierarchical encoders when histories are substantial.

In summary, this work advances an actionable understanding of \emph{when} and \emph{how} to encode software engineering artifacts to strengthen neural code representations, and it recommends practical next steps toward selective context use, richer artifacts, and learned aggregation mechanisms that generalise across models and tasks.

\textbf{Data Availability.}
The replication package is hosted across two platforms. The Figshare archive~\cite{dataset} contains: (i) the augmented datasets (SeSaMe, CodeSearchNet-Java, and Vul4J), each enriched with mined version history, call-graph context, and method age; (ii) all model training and evaluation scripts for ASTNN, CodeBERT, GraphCodeBERT, CodeT5, PLBART, and all seven LLM families across all tasks and context combinations; (iii) the human evaluation annotation data for code summarisation; and (iv) partial fine-tuned model weights for the best-performing configuration per task. The code repository, archived on Zenodo~\cite{replication}, contains the mining scripts and a step-by-step README describing how to reproduce the full context-mining pipeline from a clean project repository.

\begin{acks}
We thank the anonymous reviewers and editors for their constructive comments and feedback.
We are grateful to Jai Lal Lulla (Singapore Management University) for valuable contributions to the human evaluation phase of this project.
This research was supported by The University of Melbourne’s Research Computing Services and the Petascale Campus Initiative.
The icons in Figure~\ref{fig:fig_framework} were designed by Magnific, Vectors Tank, and Afian Rochmah Afif and obtained from \url{https://www.flaticon.com}.

\end{acks}

\bibliographystyle{ACM-Reference-Format}
\bibliography{bib/refs}


\begin{thebibliography}{79}

%%% the Web address.

\ifx \showCODEN    \undefined \def \showCODEN     #1{\unskip}     \fi
\ifx \showISBNx    \undefined \def \showISBNx     #1{\unskip}     \fi
\ifx \showISBNxiii \undefined \def \showISBNxiii  #1{\unskip}     \fi
\ifx \showISSN     \undefined \def \showISSN      #1{\unskip}     \fi
\ifx \showLCCN     \undefined \def \showLCCN      #1{\unskip}     \fi
\ifx \shownote     \undefined \def \shownote      #1{#1}          \fi
\ifx \showarticletitle \undefined \def \showarticletitle #1{#1}   \fi
\ifx \showURL      \undefined \def \showURL       {\relax}        \fi
% invisible to TeX
\providecommand\bibfield[2]{#2}
\providecommand\bibinfo[2]{#2}
\providecommand\natexlab[1]{#1}
\providecommand\showeprint[2][]{arXiv:#2}

\bibitem[Ahmad et~al\mbox{.}(2021)]%
        {ahmad2021unified}
\bibfield{author}{\bibinfo{person}{Wasi Ahmad}, \bibinfo{person}{Saikat Chakraborty}, \bibinfo{person}{Baishakhi Ray}, {and} \bibinfo{person}{Kai-Wei Chang}.} \bibinfo{year}{2021}\natexlab{}.
\newblock \showarticletitle{Unified Pre-training for Program Understanding and Generation}. In \bibinfo{booktitle}{\emph{Proceedings of the 2021 Conference of the North American Chapter of the Association for Computational Linguistics: Human Language Technologies}}, \bibfield{editor}{\bibinfo{person}{Kristina Toutanova}, \bibinfo{person}{Anna Rumshisky}, \bibinfo{person}{Luke Zettlemoyer}, \bibinfo{person}{Dilek Hakkani-Tur}, \bibinfo{person}{Iz~Beltagy}, \bibinfo{person}{Steven Bethard}, \bibinfo{person}{Ryan Cotterell}, \bibinfo{person}{Tanmoy Chakraborty}, {and} \bibinfo{person}{Yichao Zhou}} (Eds.). \bibinfo{publisher}{Association for Computational Linguistics}, \bibinfo{address}{Online}, \bibinfo{pages}{2655--2668}.
\newblock
\href{https://doi.org/10.18653/v1/2021.naacl-main.211}{doi:\nolinkurl{10.18653/v1/2021.naacl-main.211}}


\bibitem[Alon et~al\mbox{.}(2019)]%
        {alon2019code2vec}
\bibfield{author}{\bibinfo{person}{Uri Alon}, \bibinfo{person}{Meital Zilberstein}, \bibinfo{person}{Omer Levy}, {and} \bibinfo{person}{Eran Yahav}.} \bibinfo{year}{2019}\natexlab{}.
\newblock \showarticletitle{code2vec: Learning distributed representations of code}.
\newblock \bibinfo{journal}{\emph{Proceedings of the ACM on Programming Languages}} \bibinfo{volume}{3}, \bibinfo{number}{POPL} (\bibinfo{year}{2019}), \bibinfo{pages}{1--29}.
\newblock


\bibitem[Bairi et~al\mbox{.}(2024)]%
        {bairi2024codeplan}
\bibfield{author}{\bibinfo{person}{Ramakrishna Bairi}, \bibinfo{person}{Atharv Sonwane}, \bibinfo{person}{Aditya Kanade}, \bibinfo{person}{Vageesh D.~C.}, \bibinfo{person}{Arun Iyer}, \bibinfo{person}{Suresh Parthasarathy}, \bibinfo{person}{Sriram Rajamani}, \bibinfo{person}{B. Ashok}, {and} \bibinfo{person}{Shashank Shet}.} \bibinfo{year}{2024}\natexlab{}.
\newblock \showarticletitle{CodePlan: Repository-Level Coding using LLMs and Planning}.
\newblock \bibinfo{journal}{\emph{Proceedings of the ACM on Software Engineering}} \bibinfo{volume}{1}, \bibinfo{number}{FSE} (\bibinfo{year}{2024}), \bibinfo{pages}{Article 31}.
\newblock
\href{https://doi.org/10.1145/3643757}{doi:\nolinkurl{10.1145/3643757}}


\bibitem[Banerjee and Lavie(2005)]%
        {banerjee2005meteor}
\bibfield{author}{\bibinfo{person}{Satanjeev Banerjee} {and} \bibinfo{person}{Alon Lavie}.} \bibinfo{year}{2005}\natexlab{}.
\newblock \showarticletitle{{METEOR}: An Automatic Metric for {MT} Evaluation with Improved Correlation with Human Judgments}. In \bibinfo{booktitle}{\emph{Proceedings of the {ACL} Workshop on Intrinsic and Extrinsic Evaluation Measures for Machine Translation and/or Summarization}}. \bibinfo{publisher}{Association for Computational Linguistics}, \bibinfo{address}{Ann Arbor, Michigan}, \bibinfo{pages}{65--72}.
\newblock


\bibitem[Bano et~al\mbox{.}(2024)]%
        {bano2024large}
\bibfield{author}{\bibinfo{person}{Muneera Bano}, \bibinfo{person}{Rashina Hoda}, \bibinfo{person}{Didar Zowghi}, {and} \bibinfo{person}{Christoph Treude}.} \bibinfo{year}{2024}\natexlab{}.
\newblock \showarticletitle{Large language models for qualitative research in software engineering: exploring opportunities and challenges}.
\newblock \bibinfo{journal}{\emph{Automated Software Engineering}} \bibinfo{volume}{31}, \bibinfo{number}{1} (\bibinfo{year}{2024}), \bibinfo{pages}{8}.
\newblock


\bibitem[BehnamGhader et~al\mbox{.}(2024)]%
        {llm2vec}
\bibfield{author}{\bibinfo{person}{Parishad BehnamGhader}, \bibinfo{person}{Vaibhav Adlakha}, \bibinfo{person}{Marius Mosbach}, \bibinfo{person}{Dzmitry Bahdanau}, \bibinfo{person}{Nicolas Chapados}, {and} \bibinfo{person}{Siva Reddy}.} \bibinfo{year}{2024}\natexlab{}.
\newblock \showarticletitle{{LLM2V}ec: Large Language Models Are Secretly Powerful Text Encoders}. In \bibinfo{booktitle}{\emph{First Conference on Language Modeling}}.
\newblock
\urldef\tempurl%
\url{https://openreview.net/forum?id=IW1PR7vEBf}
\showURL{%
\tempurl}


\bibitem[Beltagy et~al\mbox{.}(2020)]%
        {beltagy2020longformer}
\bibfield{author}{\bibinfo{person}{Iz Beltagy}, \bibinfo{person}{Matthew~E Peters}, {and} \bibinfo{person}{Arman Cohan}.} \bibinfo{year}{2020}\natexlab{}.
\newblock \showarticletitle{Longformer: The long-document transformer}.
\newblock \bibinfo{journal}{\emph{arXiv preprint arXiv:2004.05150}} (\bibinfo{year}{2020}).
\newblock


\bibitem[Ben-Nun et~al\mbox{.}(2018)]%
        {ben2018neural}
\bibfield{author}{\bibinfo{person}{Tal Ben-Nun}, \bibinfo{person}{Alice~Shoshana Jakobovits}, {and} \bibinfo{person}{Torsten Hoefler}.} \bibinfo{year}{2018}\natexlab{}.
\newblock \showarticletitle{Neural code comprehension: A learnable representation of code semantics}.
\newblock \bibinfo{journal}{\emph{Advances in neural information processing systems}}  \bibinfo{volume}{31} (\bibinfo{year}{2018}), \bibinfo{pages}{3585--3597}.
\newblock


\bibitem[Bui et~al\mbox{.}(2022)]%
        {bui2022vul4j}
\bibfield{author}{\bibinfo{person}{Quang-Cuong Bui}, \bibinfo{person}{Riccardo Scandariato}, {and} \bibinfo{person}{Nicol\'{a}s E.~D\'{\i}az Ferreyra}.} \bibinfo{year}{2022}\natexlab{}.
\newblock \showarticletitle{Vul4J: a dataset of reproducible Java vulnerabilities geared towards the study of program repair techniques}. In \bibinfo{booktitle}{\emph{Proceedings of the 19th International Conference on Mining Software Repositories}} (Pittsburgh, Pennsylvania) \emph{(\bibinfo{series}{MSR '22})}. \bibinfo{publisher}{Association for Computing Machinery}, \bibinfo{address}{New York, NY, USA}, \bibinfo{pages}{464–468}.
\newblock
\showISBNx{9781450393034}
\href{https://doi.org/10.1145/3524842.3528482}{doi:\nolinkurl{10.1145/3524842.3528482}}


\bibitem[Chen et~al\mbox{.}(2021)]%
        {chen2021my}
\bibfield{author}{\bibinfo{person}{Qiuyuan Chen}, \bibinfo{person}{Xin Xia}, \bibinfo{person}{Han Hu}, \bibinfo{person}{David Lo}, {and} \bibinfo{person}{Shanping Li}.} \bibinfo{year}{2021}\natexlab{}.
\newblock \showarticletitle{Why my code summarization model does not work: Code comment improvement with category prediction}.
\newblock \bibinfo{journal}{\emph{ACM Transactions on Software Engineering and Methodology (TOSEM)}} \bibinfo{volume}{30}, \bibinfo{number}{2} (\bibinfo{year}{2021}), \bibinfo{pages}{1--29}.
\newblock


\bibitem[Ding et~al\mbox{.}(2024)]%
        {ding2024code}
\bibfield{author}{\bibinfo{person}{Xi Ding}, \bibinfo{person}{Rui Peng}, \bibinfo{person}{Xiangping Chen}, \bibinfo{person}{Yuan Huang}, \bibinfo{person}{Jing Bian}, {and} \bibinfo{person}{Zibin Zheng}.} \bibinfo{year}{2024}\natexlab{}.
\newblock \showarticletitle{Do code summarization models process too much information? function signature may be all that is needed}.
\newblock \bibinfo{journal}{\emph{ACM Transactions on Software Engineering and Methodology}} \bibinfo{volume}{33}, \bibinfo{number}{6} (\bibinfo{year}{2024}), \bibinfo{pages}{1--35}.
\newblock


\bibitem[Feng et~al\mbox{.}(2020)]%
        {feng2020codebert}
\bibfield{author}{\bibinfo{person}{Zhangyin Feng}, \bibinfo{person}{Daya Guo}, \bibinfo{person}{Duyu Tang}, \bibinfo{person}{Nan Duan}, \bibinfo{person}{Xiaocheng Feng}, \bibinfo{person}{Ming Gong}, \bibinfo{person}{Linjun Shou}, \bibinfo{person}{Bing Qin}, \bibinfo{person}{Ting Liu}, \bibinfo{person}{Daxin Jiang}, {and} \bibinfo{person}{Ming Zhou}.} \bibinfo{year}{2020}\natexlab{}.
\newblock \showarticletitle{{C}ode{BERT}: A Pre-Trained Model for Programming and Natural Languages}. In \bibinfo{booktitle}{\emph{Findings of the Association for Computational Linguistics: EMNLP 2020}}, \bibfield{editor}{\bibinfo{person}{Trevor Cohn}, \bibinfo{person}{Yulan He}, {and} \bibinfo{person}{Yang Liu}} (Eds.). \bibinfo{publisher}{Association for Computational Linguistics}, \bibinfo{address}{Online}, \bibinfo{pages}{1536--1547}.
\newblock
\href{https://doi.org/10.18653/v1/2020.findings-emnlp.139}{doi:\nolinkurl{10.18653/v1/2020.findings-emnlp.139}}


\bibitem[Gao et~al\mbox{.}(2023b)]%
        {gao2023evaluating}
\bibfield{author}{\bibinfo{person}{Haoyu Gao}, \bibinfo{person}{Christoph Treude}, {and} \bibinfo{person}{Mansooreh Zahedi}.} \bibinfo{year}{2023}\natexlab{b}.
\newblock \showarticletitle{Evaluating transfer learning for simplifying github readmes}. In \bibinfo{booktitle}{\emph{Proceedings of the 31st ACM Joint European Software Engineering Conference and Symposium on the Foundations of Software Engineering}}. \bibinfo{pages}{1548--1560}.
\newblock


\bibitem[Gao et~al\mbox{.}(2023a)]%
        {gao2023code}
\bibfield{author}{\bibinfo{person}{Shuzheng Gao}, \bibinfo{person}{Cuiyun Gao}, \bibinfo{person}{Yulan He}, \bibinfo{person}{Jichuan Zeng}, \bibinfo{person}{Lunyiu Nie}, \bibinfo{person}{Xin Xia}, {and} \bibinfo{person}{Michael Lyu}.} \bibinfo{year}{2023}\natexlab{a}.
\newblock \showarticletitle{Code structure--guided transformer for source code summarization}.
\newblock \bibinfo{journal}{\emph{ACM Transactions on Software Engineering and Methodology}} \bibinfo{volume}{32}, \bibinfo{number}{1} (\bibinfo{year}{2023}), \bibinfo{pages}{1--32}.
\newblock


\bibitem[GitClear(2024)]%
        {gitclear2023CodingCopilot}
\bibfield{author}{\bibinfo{person}{GitClear}.} \bibinfo{year}{2024}\natexlab{}.
\newblock \bibinfo{title}{{C}oding on {C}opilot: 2023 {D}ata {S}uggests {D}ownward {P}ressure on {C}ode {Q}uality (incl 2024 projections) - {G}it{C}lear --- gitclear.com}.
\newblock \bibinfo{howpublished}{\url{https://www.gitclear.com/coding_on_copilot_data_shows_ais_downward_pressure_on_code_quality}}.
\newblock
\newblock
\shownote{[Accessed 02-02-2024]}.


\bibitem[Gold et~al\mbox{.}(2004)]%
        {gold2004understanding}
\bibfield{author}{\bibinfo{person}{Nicolas Gold}, \bibinfo{person}{Andrew Mohan}, \bibinfo{person}{Claire Knight}, {and} \bibinfo{person}{Malcolm Munro}.} \bibinfo{year}{2004}\natexlab{}.
\newblock \showarticletitle{Understanding service-oriented software}.
\newblock \bibinfo{journal}{\emph{IEEE software}} \bibinfo{volume}{21}, \bibinfo{number}{2} (\bibinfo{year}{2004}), \bibinfo{pages}{71--77}.
\newblock


\bibitem[Gousios(2019)]%
        {gousios2019javacallgraph}
\bibfield{author}{\bibinfo{person}{Georgios Gousios}.} \bibinfo{year}{2019}\natexlab{}.
\newblock \bibinfo{title}{{Java Call Graph}}.
\newblock \bibinfo{howpublished}{\url{https://github.com/gousiosg/java-callgraph}}.
\newblock
\newblock
\shownote{GitHub repository}.


\bibitem[Guo et~al\mbox{.}(2022)]%
        {guo2022unixcoder}
\bibfield{author}{\bibinfo{person}{Daya Guo}, \bibinfo{person}{Shuai Lu}, \bibinfo{person}{Nan Duan}, \bibinfo{person}{Yanlin Wang}, \bibinfo{person}{Ming Zhou}, {and} \bibinfo{person}{Jian Yin}.} \bibinfo{year}{2022}\natexlab{}.
\newblock \showarticletitle{Unixcoder: Unified cross-modal pre-training for code representation}. In \bibinfo{booktitle}{\emph{Proceedings of the 60th Annual Meeting of the Association for Computational Linguistics (Volume 1: Long Papers)}}. \bibinfo{pages}{7212--7225}.
\newblock


\bibitem[Guo et~al\mbox{.}(2021)]%
        {guo2020graphcodebert}
\bibfield{author}{\bibinfo{person}{Daya Guo}, \bibinfo{person}{Shuo Ren}, \bibinfo{person}{Shuai Lu}, \bibinfo{person}{Zhangyin Feng}, \bibinfo{person}{Duyu Tang}, \bibinfo{person}{Shujie LIU}, \bibinfo{person}{Long Zhou}, \bibinfo{person}{Nan Duan}, \bibinfo{person}{Alexey Svyatkovskiy}, \bibinfo{person}{Shengyu Fu}, \bibinfo{person}{Michele Tufano}, \bibinfo{person}{Shao~Kun Deng}, \bibinfo{person}{Colin Clement}, \bibinfo{person}{Dawn Drain}, \bibinfo{person}{Neel Sundaresan}, \bibinfo{person}{Jian Yin}, \bibinfo{person}{Daxin Jiang}, {and} \bibinfo{person}{Ming Zhou}.} \bibinfo{year}{2021}\natexlab{}.
\newblock \showarticletitle{GraphCode{\{}BERT{\}}: Pre-training Code Representations with Data Flow}. In \bibinfo{booktitle}{\emph{International Conference on Learning Representations}}.
\newblock
\urldef\tempurl%
\url{https://openreview.net/forum?id=jLoC4ez43PZ}
\showURL{%
\tempurl}


\bibitem[Guo et~al\mbox{.}(2024)]%
        {guo2024deepseek}
\bibfield{author}{\bibinfo{person}{Daya Guo}, \bibinfo{person}{Qihao Zhu}, \bibinfo{person}{Dejian Yang}, \bibinfo{person}{Zhenda Xie}, \bibinfo{person}{Kai Dong}, \bibinfo{person}{Wentao Zhang}, \bibinfo{person}{Guanting Chen}, \bibinfo{person}{Xiao Bi}, \bibinfo{person}{Yifan Wu}, \bibinfo{person}{YK Li}, {et~al\mbox{.}}} \bibinfo{year}{2024}\natexlab{}.
\newblock \showarticletitle{DeepSeek-Coder: when the large language model meets programming--the rise of code intelligence}.
\newblock \bibinfo{journal}{\emph{arXiv preprint arXiv:2401.14196}} (\bibinfo{year}{2024}).
\newblock


\bibitem[Ho et~al\mbox{.}(2025a)]%
        {ho2025ensesmells}
\bibfield{author}{\bibinfo{person}{Anh Ho}, \bibinfo{person}{Anh~MT Bui}, \bibinfo{person}{Phuong~T Nguyen}, \bibinfo{person}{Amleto Di~Salle}, {and} \bibinfo{person}{Bach Le}.} \bibinfo{year}{2025}\natexlab{a}.
\newblock \showarticletitle{EnseSmells: Deep ensemble and programming language models for automated code smells detection}.
\newblock \bibinfo{journal}{\emph{Journal of Systems and Software}}  \bibinfo{volume}{224} (\bibinfo{year}{2025}), \bibinfo{pages}{112375}.
\newblock
\href{https://doi.org/10.1016/j.jss.2025.112375}{doi:\nolinkurl{10.1016/j.jss.2025.112375}}


\bibitem[Ho et~al\mbox{.}(2025b)]%
        {ho2025empirical}
\bibfield{author}{\bibinfo{person}{Anh Ho}, \bibinfo{person}{Thanh Le-Cong}, \bibinfo{person}{Bach Le}, {and} \bibinfo{person}{Christine Rizkallah}.} \bibinfo{year}{2025}\natexlab{b}.
\newblock \showarticletitle{From Empirical Evaluation to Context-Aware Enhancement: Repairing Regression Errors with LLMs}.
\newblock \bibinfo{journal}{\emph{arXiv preprint arXiv:2506.13182}} (\bibinfo{year}{2025}).
\newblock


\bibitem[Hoang et~al\mbox{.}(2020)]%
        {hoang2020cc2vec}
\bibfield{author}{\bibinfo{person}{Thong Hoang}, \bibinfo{person}{Hong~Jin Kang}, \bibinfo{person}{David Lo}, {and} \bibinfo{person}{Julia Lawall}.} \bibinfo{year}{2020}\natexlab{}.
\newblock \showarticletitle{CC2Vec: distributed representations of code changes}. In \bibinfo{booktitle}{\emph{Proceedings of the ACM/IEEE 42nd International Conference on Software Engineering}} (Seoul, South Korea) \emph{(\bibinfo{series}{ICSE '20})}. \bibinfo{publisher}{Association for Computing Machinery}, \bibinfo{address}{New York, NY, USA}, \bibinfo{pages}{518–529}.
\newblock
\showISBNx{9781450371216}
\href{https://doi.org/10.1145/3377811.3380361}{doi:\nolinkurl{10.1145/3377811.3380361}}


\bibitem[Hu et~al\mbox{.}(2022)]%
        {hu2022lora}
\bibfield{author}{\bibinfo{person}{Edward~J Hu}, \bibinfo{person}{Yelong Shen}, \bibinfo{person}{Phillip Wallis}, \bibinfo{person}{Zeyuan Allen-Zhu}, \bibinfo{person}{Yuanzhi Li}, \bibinfo{person}{Shean Wang}, \bibinfo{person}{Liang Wang}, \bibinfo{person}{Weizhu Chen}, {et~al\mbox{.}}} \bibinfo{year}{2022}\natexlab{}.
\newblock \showarticletitle{Lora: Low-rank adaptation of large language models.}
\newblock \bibinfo{journal}{\emph{Iclr}} \bibinfo{volume}{1}, \bibinfo{number}{2} (\bibinfo{year}{2022}), \bibinfo{pages}{3}.
\newblock


\bibitem[Hui et~al\mbox{.}(2024)]%
        {hui2024qwen2}
\bibfield{author}{\bibinfo{person}{Binyuan Hui}, \bibinfo{person}{Jian Yang}, \bibinfo{person}{Zeyu Cui}, \bibinfo{person}{Jiaxi Yang}, \bibinfo{person}{Dayiheng Liu}, \bibinfo{person}{Lei Zhang}, \bibinfo{person}{Tianyu Liu}, \bibinfo{person}{Jiajun Zhang}, \bibinfo{person}{Bowen Yu}, \bibinfo{person}{Keming Lu}, {et~al\mbox{.}}} \bibinfo{year}{2024}\natexlab{}.
\newblock \showarticletitle{Qwen2. 5-coder technical report}.
\newblock \bibinfo{journal}{\emph{arXiv preprint arXiv:2409.12186}} (\bibinfo{year}{2024}).
\newblock


\bibitem[Husain et~al\mbox{.}(2019)]%
        {husain2019codesearchnet}
\bibfield{author}{\bibinfo{person}{Hamel Husain}, \bibinfo{person}{Ho-Hsiang Wu}, \bibinfo{person}{Tiferet Gazit}, \bibinfo{person}{Miltiadis Allamanis}, {and} \bibinfo{person}{Marc Brockschmidt}.} \bibinfo{year}{2019}\natexlab{}.
\newblock \showarticletitle{Codesearchnet challenge: Evaluating the state of semantic code search}.
\newblock \bibinfo{journal}{\emph{arXiv preprint arXiv:1909.09436}} (\bibinfo{year}{2019}).
\newblock


\bibitem[J{\'a}sz et~al\mbox{.}(2019)]%
        {jasz2019systematic}
\bibfield{author}{\bibinfo{person}{Judit J{\'a}sz}, \bibinfo{person}{Istv{\'a}n Siket}, \bibinfo{person}{Edit Peng{\H{o}}}, \bibinfo{person}{Zolt{\'a}n S{\'a}godi}, {and} \bibinfo{person}{Rudolf Ferenc}.} \bibinfo{year}{2019}\natexlab{}.
\newblock \showarticletitle{Systematic Comparison of Six Open-source {Java} Call Graph Construction Tools}. In \bibinfo{booktitle}{\emph{Proceedings of the 14th International Conference on Software Technologies (ICSOFT 2019)}}. \bibinfo{publisher}{SciTePress}, \bibinfo{pages}{117--128}.
\newblock
\href{https://doi.org/10.5220/0007929201170128}{doi:\nolinkurl{10.5220/0007929201170128}}


\bibitem[Jiang et~al\mbox{.}(2022)]%
        {jiang2022hierarchical}
\bibfield{author}{\bibinfo{person}{Yuan Jiang}, \bibinfo{person}{Xiaohong Su}, \bibinfo{person}{Christoph Treude}, {and} \bibinfo{person}{Tiantian Wang}.} \bibinfo{year}{2022}\natexlab{}.
\newblock \showarticletitle{Hierarchical semantic-aware neural code representation}.
\newblock \bibinfo{journal}{\emph{Journal of Systems and Software}}  \bibinfo{volume}{191} (\bibinfo{year}{2022}), \bibinfo{pages}{111355}.
\newblock


\bibitem[Jimenez et~al\mbox{.}(2024)]%
        {jimenez2024swebench}
\bibfield{author}{\bibinfo{person}{Carlos~E Jimenez}, \bibinfo{person}{John Yang}, \bibinfo{person}{Alexander Wettig}, \bibinfo{person}{Shunyu Yao}, \bibinfo{person}{Kexin Pei}, \bibinfo{person}{Ofir Press}, {and} \bibinfo{person}{Karthik Narasimhan}.} \bibinfo{year}{2024}\natexlab{}.
\newblock \showarticletitle{Swe-bench: Can language models resolve real-world github issues?}. In \bibinfo{booktitle}{\emph{International Conference on Learning Representations}}, Vol.~\bibinfo{volume}{2024}. \bibinfo{pages}{54107--54157}.
\newblock


\bibitem[Kamp et~al\mbox{.}(2019)]%
        {kamp2019sesame}
\bibfield{author}{\bibinfo{person}{Marius Kamp}, \bibinfo{person}{Patrick Kreutzer}, {and} \bibinfo{person}{Michael Philippsen}.} \bibinfo{year}{2019}\natexlab{}.
\newblock \showarticletitle{SeSaMe: a data set of semantically similar Java methods}. In \bibinfo{booktitle}{\emph{Proceedings of the 16th International Conference on Mining Software Repositories}} \emph{(\bibinfo{series}{MSR '19})}. \bibinfo{publisher}{IEEE Press}, \bibinfo{address}{Montreal, Quebec, Canada}, \bibinfo{pages}{529–533}.
\newblock
\href{https://doi.org/10.1109/MSR.2019.00079}{doi:\nolinkurl{10.1109/MSR.2019.00079}}


\bibitem[Kanade et~al\mbox{.}(2020)]%
        {kanade2020learning}
\bibfield{author}{\bibinfo{person}{Aditya Kanade}, \bibinfo{person}{Petros Maniatis}, \bibinfo{person}{Gogul Balakrishnan}, {and} \bibinfo{person}{Kensen Shi}.} \bibinfo{year}{2020}\natexlab{}.
\newblock \showarticletitle{Learning and evaluating contextual embedding of source code}. In \bibinfo{booktitle}{\emph{International conference on machine learning}}. PMLR, \bibinfo{pages}{5110--5121}.
\newblock


\bibitem[Krinke and Ragkhitwetsagul(2022)]%
        {krinke2022bigclonebench}
\bibfield{author}{\bibinfo{person}{Jens Krinke} {and} \bibinfo{person}{Chaiyong Ragkhitwetsagul}.} \bibinfo{year}{2022}\natexlab{}.
\newblock \showarticletitle{Bigclonebench considered harmful for machine learning}. In \bibinfo{booktitle}{\emph{2022 IEEE 16th International Workshop on Software Clones (IWSC)}}. IEEE, \bibinfo{pages}{1--7}.
\newblock


\bibitem[Kulkarni and Varma(2014)]%
        {kulkarni2014supporting}
\bibfield{author}{\bibinfo{person}{Naveen Kulkarni} {and} \bibinfo{person}{Vasudeva Varma}.} \bibinfo{year}{2014}\natexlab{}.
\newblock \showarticletitle{Supporting comprehension of unfamiliar programs by modeling an expert's perception}. In \bibinfo{booktitle}{\emph{Proceedings of the 3rd International Workshop on Realizing Artificial Intelligence Synergies in Software Engineering}} (Hyderabad, India) \emph{(\bibinfo{series}{RAISE 2014})}. \bibinfo{publisher}{Association for Computing Machinery}, \bibinfo{address}{New York, NY, USA}, \bibinfo{pages}{19–24}.
\newblock
\showISBNx{9781450328463}
\href{https://doi.org/10.1145/2593801.2593805}{doi:\nolinkurl{10.1145/2593801.2593805}}


\bibitem[Le-Cong et~al\mbox{.}(2022)]%
        {le2022autopruner}
\bibfield{author}{\bibinfo{person}{Thanh Le-Cong}, \bibinfo{person}{Hong~Jin Kang}, \bibinfo{person}{Truong~Giang Nguyen}, \bibinfo{person}{Stefanus~Agus Haryono}, \bibinfo{person}{David Lo}, \bibinfo{person}{Xuan-Bach~D Le}, {and} \bibinfo{person}{Quyet~Thang Huynh}.} \bibinfo{year}{2022}\natexlab{}.
\newblock \showarticletitle{Autopruner: transformer-based call graph pruning}. In \bibinfo{booktitle}{\emph{Proceedings of the 30th ACM Joint European Software Engineering Conference and Symposium on the Foundations of Software Engineering}}. \bibinfo{pages}{520--532}.
\newblock


\bibitem[Le-Cong et~al\mbox{.}(2025)]%
        {lecongetal2025llms}
\bibfield{author}{\bibinfo{person}{Thanh Le-Cong}, \bibinfo{person}{Bach Le}, {and} \bibinfo{person}{Toby Murray}.} \bibinfo{year}{2025}\natexlab{}.
\newblock \showarticletitle{Can {LLM}s Reason About Program Semantics? A Comprehensive Evaluation of {LLM}s on Formal Specification Inference}. In \bibinfo{booktitle}{\emph{Proceedings of the 63rd Annual Meeting of the Association for Computational Linguistics (Volume 1: Long Papers)}}, \bibfield{editor}{\bibinfo{person}{Wanxiang Che}, \bibinfo{person}{Joyce Nabende}, \bibinfo{person}{Ekaterina Shutova}, {and} \bibinfo{person}{Mohammad~Taher Pilehvar}} (Eds.). \bibinfo{publisher}{Association for Computational Linguistics}, \bibinfo{address}{Vienna, Austria}, \bibinfo{pages}{21991--22014}.
\newblock
\showISBNx{979-8-89176-251-0}
\href{https://doi.org/10.18653/v1/2025.acl-long.1068}{doi:\nolinkurl{10.18653/v1/2025.acl-long.1068}}


\bibitem[LeClair et~al\mbox{.}(2020)]%
        {leclair2020improved}
\bibfield{author}{\bibinfo{person}{Alexander LeClair}, \bibinfo{person}{Sakib Haque}, \bibinfo{person}{Lingfei Wu}, {and} \bibinfo{person}{Collin McMillan}.} \bibinfo{year}{2020}\natexlab{}.
\newblock \showarticletitle{Improved code summarization via a graph neural network}. In \bibinfo{booktitle}{\emph{Proceedings of the 28th international conference on program comprehension}}. \bibinfo{pages}{184--195}.
\newblock


\bibitem[Li et~al\mbox{.}(2024)]%
        {li2024deveval}
\bibfield{author}{\bibinfo{person}{Jia Li}, \bibinfo{person}{Ge Li}, \bibinfo{person}{Yunfei Zhao}, \bibinfo{person}{Yongmin Li}, \bibinfo{person}{Huanyu Liu}, \bibinfo{person}{Hao Zhu}, \bibinfo{person}{Lecheng Wang}, \bibinfo{person}{Kaibo Liu}, \bibinfo{person}{Zheng Fang}, \bibinfo{person}{Lanshen Wang}, {et~al\mbox{.}}} \bibinfo{year}{2024}\natexlab{}.
\newblock \showarticletitle{Deveval: A manually-annotated code generation benchmark aligned with real-world code repositories}. In \bibinfo{booktitle}{\emph{Findings of the Association for Computational Linguistics: ACL 2024}}. \bibinfo{pages}{3603--3614}.
\newblock


\bibitem[Lin(2004)]%
        {lin2004rouge}
\bibfield{author}{\bibinfo{person}{Chin-Yew Lin}.} \bibinfo{year}{2004}\natexlab{}.
\newblock \showarticletitle{{ROUGE}: A Package for Automatic Evaluation of Summaries}. In \bibinfo{booktitle}{\emph{Text Summarization Branches Out}}. \bibinfo{publisher}{Association for Computational Linguistics}, \bibinfo{address}{Barcelona, Spain}, \bibinfo{pages}{74--81}.
\newblock


\bibitem[Liu et~al\mbox{.}(2025)]%
        {liu2025too}
\bibfield{author}{\bibinfo{person}{Chunhua Liu}, \bibinfo{person}{Hong~Yi Lin}, {and} \bibinfo{person}{Patanamon Thongtanunam}.} \bibinfo{year}{2025}\natexlab{}.
\newblock \showarticletitle{Too noisy to learn: Enhancing data quality for code review comment generation}. In \bibinfo{booktitle}{\emph{2025 IEEE/ACM 22nd International Conference on Mining Software Repositories (MSR)}}. IEEE, \bibinfo{pages}{236--248}.
\newblock


\bibitem[Liu et~al\mbox{.}(2024)]%
        {liu2024lost}
\bibfield{author}{\bibinfo{person}{Nelson~F. Liu}, \bibinfo{person}{Kevin Lin}, \bibinfo{person}{John Hewitt}, \bibinfo{person}{Ashwin Paranjape}, \bibinfo{person}{Michele Bevilacqua}, \bibinfo{person}{Fabio Petroni}, {and} \bibinfo{person}{Percy Liang}.} \bibinfo{year}{2024}\natexlab{}.
\newblock \showarticletitle{Lost in the Middle: How Language Models Use Long Contexts}.
\newblock \bibinfo{journal}{\emph{Transactions of the Association for Computational Linguistics}}  \bibinfo{volume}{12} (\bibinfo{year}{2024}), \bibinfo{pages}{157--173}.
\newblock
\href{https://doi.org/10.1162/tacl_a_00638}{doi:\nolinkurl{10.1162/tacl_a_00638}}


\bibitem[Long et~al\mbox{.}(2022)]%
        {long2022multi}
\bibfield{author}{\bibinfo{person}{Ting Long}, \bibinfo{person}{Yutong Xie}, \bibinfo{person}{Xianyu Chen}, \bibinfo{person}{Weinan Zhang}, \bibinfo{person}{Qinxiang Cao}, {and} \bibinfo{person}{Yong Yu}.} \bibinfo{year}{2022}\natexlab{}.
\newblock \showarticletitle{Multi-View Graph Representation for Programming Language Processing: An Investigation into Algorithm Detection}.
\newblock \bibinfo{journal}{\emph{Proceedings of the AAAI Conference on Artificial Intelligence}} \bibinfo{volume}{36}, \bibinfo{number}{5} (\bibinfo{date}{Jun.} \bibinfo{year}{2022}), \bibinfo{pages}{5792--5799}.
\newblock
\href{https://doi.org/10.1609/aaai.v36i5.20522}{doi:\nolinkurl{10.1609/aaai.v36i5.20522}}


\bibitem[Lu et~al\mbox{.}(2021)]%
        {lu2021codexglue}
\bibfield{author}{\bibinfo{person}{Shuai Lu}, \bibinfo{person}{Daya Guo}, \bibinfo{person}{Shuo Ren}, \bibinfo{person}{Junjie Huang}, \bibinfo{person}{Alexey Svyatkovskiy}, \bibinfo{person}{Ambrosio Blanco}, \bibinfo{person}{Colin Clement}, \bibinfo{person}{Dawn Drain}, \bibinfo{person}{Daxin Jiang}, \bibinfo{person}{Duyu Tang}, \bibinfo{person}{Ge Li}, \bibinfo{person}{Lidong Zhou}, \bibinfo{person}{Linjun Shou}, \bibinfo{person}{Long Zhou}, \bibinfo{person}{Michele Tufano}, \bibinfo{person}{MING GONG}, \bibinfo{person}{Ming Zhou}, \bibinfo{person}{Nan Duan}, \bibinfo{person}{Neel Sundaresan}, \bibinfo{person}{Shao~Kun Deng}, \bibinfo{person}{Shengyu Fu}, {and} \bibinfo{person}{Shujie LIU}.} \bibinfo{year}{2021}\natexlab{}.
\newblock \showarticletitle{CodeXGLUE: A Machine Learning Benchmark Dataset for Code Understanding and Generation}. In \bibinfo{booktitle}{\emph{Proceedings of the Neural Information Processing Systems Track on Datasets and Benchmarks}}, \bibfield{editor}{\bibinfo{person}{J.~Vanschoren} {and} \bibinfo{person}{S.~Yeung}} (Eds.), Vol.~\bibinfo{volume}{1}.
\newblock
\urldef\tempurl%
\url{https://datasets-benchmarks-proceedings.neurips.cc/paper_files/paper/2021/file/c16a5320fa475530d9583c34fd356ef5-Paper-round1.pdf}
\showURL{%
\tempurl}


\bibitem[Maalej et~al\mbox{.}(2014)]%
        {maalej2014comprehension}
\bibfield{author}{\bibinfo{person}{Walid Maalej}, \bibinfo{person}{Rebecca Tiarks}, \bibinfo{person}{Tobias Roehm}, {and} \bibinfo{person}{Rainer Koschke}.} \bibinfo{year}{2014}\natexlab{}.
\newblock \showarticletitle{On the comprehension of program comprehension}.
\newblock \bibinfo{journal}{\emph{ACM Transactions on Software Engineering and Methodology (TOSEM)}} \bibinfo{volume}{23}, \bibinfo{number}{4} (\bibinfo{year}{2014}), \bibinfo{pages}{1--37}.
\newblock


\bibitem[Maletic and Marcus(2001)]%
        {maletic2001supporting}
\bibfield{author}{\bibinfo{person}{Jonathan~I. Maletic} {and} \bibinfo{person}{Andrian Marcus}.} \bibinfo{year}{2001}\natexlab{}.
\newblock \showarticletitle{Supporting program comprehension using semantic and structural information}. In \bibinfo{booktitle}{\emph{Proceedings of the 23rd International Conference on Software Engineering}} (Toronto, Ontario, Canada) \emph{(\bibinfo{series}{ICSE '01})}. \bibinfo{publisher}{IEEE Computer Society}, \bibinfo{address}{USA}, \bibinfo{pages}{103–112}.
\newblock
\showISBNx{0769510507}


\bibitem[Mastropaolo et~al\mbox{.}(2024)]%
        {mastropaolo2024evaluating}
\bibfield{author}{\bibinfo{person}{Antonio Mastropaolo}, \bibinfo{person}{Matteo Ciniselli}, \bibinfo{person}{Massimiliano Di~Penta}, {and} \bibinfo{person}{Gabriele Bavota}.} \bibinfo{year}{2024}\natexlab{}.
\newblock \showarticletitle{Evaluating code summarization techniques: A new metric and an empirical characterization}. In \bibinfo{booktitle}{\emph{Proceedings of the IEEE/ACM 46th International Conference on Software Engineering}}. \bibinfo{pages}{1--13}.
\newblock


\bibitem[Murphy(2019)]%
        {murphy2019beyond}
\bibfield{author}{\bibinfo{person}{Gail~C Murphy}.} \bibinfo{year}{2019}\natexlab{}.
\newblock \showarticletitle{Beyond Integrated Development Environments: Adding Context to Software Development}. In \bibinfo{booktitle}{\emph{2019 IEEE/ACM 41st International Conference on Software Engineering: New Ideas and Emerging Results (ICSE-NIER)}}. \bibinfo{pages}{73--76}.
\newblock
\href{https://doi.org/10.1109/ICSE-NIER.2019.00027}{doi:\nolinkurl{10.1109/ICSE-NIER.2019.00027}}


\bibitem[Nguyen et~al\mbox{.}(2024)]%
        {nguyen2024encoding}
\bibfield{author}{\bibinfo{person}{Huy Nguyen}, \bibinfo{person}{Christoph Treude}, {and} \bibinfo{person}{Patanamon Thongtanunam}.} \bibinfo{year}{2024}\natexlab{}.
\newblock \showarticletitle{Encoding version history context for better code representation}. In \bibinfo{booktitle}{\emph{Proceedings of the 21st International Conference on Mining Software Repositories}}. \bibinfo{pages}{631--636}.
\newblock


\bibitem[Nguyen et~al\mbox{.}(2026a)]%
        {dataset}
\bibfield{author}{\bibinfo{person}{Huy Nguyen}, \bibinfo{person}{Christoph Treude}, {and} \bibinfo{person}{Patanamon Thongtanunam}.} \bibinfo{year}{2026}\natexlab{a}.
\newblock \bibinfo{title}{Replication Package and Datasets for Journal Paper ``Enhancing Neural Code Representation with Additional Context''}.
\newblock
\href{https://doi.org/10.6084/m9.figshare.30296542}{doi:\nolinkurl{10.6084/m9.figshare.30296542}}


\bibitem[Nguyen et~al\mbox{.}(2026b)]%
        {replication}
\bibfield{author}{\bibinfo{person}{Huy Nguyen}, \bibinfo{person}{Christoph Treude}, {and} \bibinfo{person}{Patanamon Thongtanunam}.} \bibinfo{year}{2026}\natexlab{b}.
\newblock \bibinfo{title}{Replication Package for ``Enhancing Neural Code Representation with Additional Context''}.
\newblock
\href{https://doi.org/10.5281/zenodo.23365805}{doi:\nolinkurl{10.5281/zenodo.23365805}}


\bibitem[Nguyen et~al\mbox{.}(2023)]%
        {nguyen2023multi}
\bibfield{author}{\bibinfo{person}{Truong~Giang Nguyen}, \bibinfo{person}{Thanh Le-Cong}, \bibinfo{person}{Hong~Jin Kang}, \bibinfo{person}{Ratnadira Widyasari}, \bibinfo{person}{Chengran Yang}, \bibinfo{person}{Zhipeng Zhao}, \bibinfo{person}{Bowen Xu}, \bibinfo{person}{Jiayuan Zhou}, \bibinfo{person}{Xin Xia}, \bibinfo{person}{Ahmed~E. Hassan}, \bibinfo{person}{Xuan-Bach~D. Le}, {and} \bibinfo{person}{David Lo}.} \bibinfo{year}{2023}\natexlab{}.
\newblock \showarticletitle{Multi-Granularity Detector for Vulnerability Fixes}.
\newblock \bibinfo{journal}{\emph{IEEE Transactions on Software Engineering}} \bibinfo{volume}{49}, \bibinfo{number}{8} (\bibinfo{year}{2023}), \bibinfo{pages}{4035--4057}.
\newblock
\href{https://doi.org/10.1109/TSE.2023.3281275}{doi:\nolinkurl{10.1109/TSE.2023.3281275}}


\bibitem[Pan et~al\mbox{.}(2024)]%
        {pan2024assessing}
\bibfield{author}{\bibinfo{person}{Wei~Hung Pan}, \bibinfo{person}{Ming~Jie Chok}, \bibinfo{person}{Jonathan Leong~Shan Wong}, \bibinfo{person}{Yung~Xin Shin}, \bibinfo{person}{Yeong~Shian Poon}, \bibinfo{person}{Zhou Yang}, \bibinfo{person}{Chun~Yong Chong}, \bibinfo{person}{David Lo}, {and} \bibinfo{person}{Mei~Kuan Lim}.} \bibinfo{year}{2024}\natexlab{}.
\newblock \showarticletitle{Assessing AI Detectors in Identifying AI-Generated Code: Implications for Education}. In \bibinfo{booktitle}{\emph{2024 IEEE/ACM 46th International Conference on Software Engineering: Software Engineering Education and Training (ICSE-SEET)}}. IEEE, \bibinfo{publisher}{{IEEE} / {ACM}}, \bibinfo{address}{Lisbon, Portugal}, \bibinfo{pages}{11--22}.
\newblock


\bibitem[Papineni et~al\mbox{.}(2002)]%
        {papineni2002bleu}
\bibfield{author}{\bibinfo{person}{Kishore Papineni}, \bibinfo{person}{Salim Roukos}, \bibinfo{person}{Todd Ward}, {and} \bibinfo{person}{Wei-Jing Zhu}.} \bibinfo{year}{2002}\natexlab{}.
\newblock \showarticletitle{Bleu: a method for automatic evaluation of machine translation}. In \bibinfo{booktitle}{\emph{Proceedings of the 40th annual meeting of the Association for Computational Linguistics}}. \bibinfo{pages}{311--318}.
\newblock


\bibitem[Roy et~al\mbox{.}(2021)]%
        {roy2021reassessing}
\bibfield{author}{\bibinfo{person}{Devjeet Roy}, \bibinfo{person}{Sarah Fakhoury}, {and} \bibinfo{person}{Venera Arnaoudova}.} \bibinfo{year}{2021}\natexlab{}.
\newblock \showarticletitle{Reassessing automatic evaluation metrics for code summarization tasks}. In \bibinfo{booktitle}{\emph{Proceedings of the 29th ACM Joint Meeting on European Software Engineering Conference and Symposium on the Foundations of Software Engineering (ESEC/FSE)}}. \bibinfo{publisher}{Association for Computing Machinery}, \bibinfo{pages}{1105--1116}.
\newblock
\href{https://doi.org/10.1145/3468264.3468588}{doi:\nolinkurl{10.1145/3468264.3468588}}


\bibitem[Roziere et~al\mbox{.}(2023)]%
        {roziere2023code}
\bibfield{author}{\bibinfo{person}{Baptiste Roziere}, \bibinfo{person}{Jonas Gehring}, \bibinfo{person}{Fabian Gloeckle}, \bibinfo{person}{Sten Sootla}, \bibinfo{person}{Itai Gat}, \bibinfo{person}{Xiaoqing~Ellen Tan}, \bibinfo{person}{Yossi Adi}, \bibinfo{person}{Jingyu Liu}, \bibinfo{person}{Romain Sauvestre}, \bibinfo{person}{Tal Remez}, {et~al\mbox{.}}} \bibinfo{year}{2023}\natexlab{}.
\newblock \showarticletitle{Code llama: Open foundation models for code}.
\newblock \bibinfo{journal}{\emph{arXiv preprint arXiv:2308.12950}} (\bibinfo{year}{2023}).
\newblock


\bibitem[Ruthruff et~al\mbox{.}(2008)]%
        {ruthruff2008predicting}
\bibfield{author}{\bibinfo{person}{Joseph~R Ruthruff}, \bibinfo{person}{John Penix}, \bibinfo{person}{J~David Morgenthaler}, \bibinfo{person}{Sebastian Elbaum}, {and} \bibinfo{person}{Gregg Rothermel}.} \bibinfo{year}{2008}\natexlab{}.
\newblock \showarticletitle{Predicting accurate and actionable static analysis warnings: an experimental approach}. In \bibinfo{booktitle}{\emph{Proceedings of the 30th international conference on Software engineering}}. \bibinfo{pages}{341--350}.
\newblock


\bibitem[Samoaa et~al\mbox{.}(2022)]%
        {samoaa2022systematic}
\bibfield{author}{\bibinfo{person}{Hazem~Peter Samoaa}, \bibinfo{person}{Firas Bayram}, \bibinfo{person}{Pasquale Salza}, {and} \bibinfo{person}{Philipp Leitner}.} \bibinfo{year}{2022}\natexlab{}.
\newblock \showarticletitle{A systematic mapping study of source code representation for deep learning in software engineering}.
\newblock \bibinfo{journal}{\emph{IET Software}} \bibinfo{volume}{16}, \bibinfo{number}{4} (\bibinfo{year}{2022}), \bibinfo{pages}{351--385}.
\newblock


\bibitem[Shi et~al\mbox{.}(2022)]%
        {shi2022we}
\bibfield{author}{\bibinfo{person}{Lin Shi}, \bibinfo{person}{Fangwen Mu}, \bibinfo{person}{Xiao Chen}, \bibinfo{person}{Song Wang}, \bibinfo{person}{Junjie Wang}, \bibinfo{person}{Ye Yang}, \bibinfo{person}{Ge Li}, \bibinfo{person}{Xin Xia}, {and} \bibinfo{person}{Qing Wang}.} \bibinfo{year}{2022}\natexlab{}.
\newblock \showarticletitle{Are we building on the rock? on the importance of data preprocessing for code summarization}. In \bibinfo{booktitle}{\emph{Proceedings of the 30th ACM Joint European Software Engineering Conference and Symposium on the Foundations of Software Engineering}}. \bibinfo{pages}{107--119}.
\newblock


\bibitem[Shido et~al\mbox{.}(2019)]%
        {shido2019automatic}
\bibfield{author}{\bibinfo{person}{Yusuke Shido}, \bibinfo{person}{Yasuaki Kobayashi}, \bibinfo{person}{Akihiro Yamamoto}, \bibinfo{person}{Atsushi Miyamoto}, {and} \bibinfo{person}{Tadayuki Matsumura}.} \bibinfo{year}{2019}\natexlab{}.
\newblock \showarticletitle{Automatic source code summarization with extended tree-lstm}. In \bibinfo{booktitle}{\emph{2019 International Joint Conference on Neural Networks (IJCNN)}}. IEEE, \bibinfo{pages}{1--8}.
\newblock


\bibitem[Sites(2021)]%
        {sites2021understanding}
\bibfield{author}{\bibinfo{person}{Richard Sites}.} \bibinfo{year}{2021}\natexlab{}.
\newblock \bibinfo{booktitle}{\emph{Understanding Software Dynamics}}.
\newblock \bibinfo{publisher}{Addison Wesley}, \bibinfo{address}{Boston, USA}.
\newblock


\bibitem[Spadini et~al\mbox{.}(2018)]%
        {spadini2018pydriller}
\bibfield{author}{\bibinfo{person}{Davide Spadini}, \bibinfo{person}{Maur\'{\i}cio Aniche}, {and} \bibinfo{person}{Alberto Bacchelli}.} \bibinfo{year}{2018}\natexlab{}.
\newblock \showarticletitle{PyDriller: Python framework for mining software repositories}. In \bibinfo{booktitle}{\emph{Proceedings of the 2018 26th ACM Joint Meeting on European Software Engineering Conference and Symposium on the Foundations of Software Engineering}} (Lake Buena Vista, FL, USA) \emph{(\bibinfo{series}{ESEC/FSE 2018})}. \bibinfo{publisher}{Association for Computing Machinery}, \bibinfo{address}{New York, NY, USA}, \bibinfo{pages}{908–911}.
\newblock
\showISBNx{9781450355735}
\href{https://doi.org/10.1145/3236024.3264598}{doi:\nolinkurl{10.1145/3236024.3264598}}


\bibitem[Sun et~al\mbox{.}(2024)]%
        {sun2024extractive}
\bibfield{author}{\bibinfo{person}{Weisong Sun}, \bibinfo{person}{Chunrong Fang}, \bibinfo{person}{Yuchen Chen}, \bibinfo{person}{Quanjun Zhang}, \bibinfo{person}{Guanhong Tao}, \bibinfo{person}{Yudu You}, \bibinfo{person}{Tingxu Han}, \bibinfo{person}{Yifei Ge}, \bibinfo{person}{Yuling Hu}, \bibinfo{person}{Bin Luo}, {et~al\mbox{.}}} \bibinfo{year}{2024}\natexlab{}.
\newblock \showarticletitle{An extractive-and-abstractive framework for source code summarization}.
\newblock \bibinfo{journal}{\emph{ACM Transactions on Software Engineering and Methodology}} \bibinfo{volume}{33}, \bibinfo{number}{3} (\bibinfo{year}{2024}), \bibinfo{pages}{1--39}.
\newblock


\bibitem[Svajlenko and Roy(2015)]%
        {svajlenko2015evaluating}
\bibfield{author}{\bibinfo{person}{Jeffrey Svajlenko} {and} \bibinfo{person}{Chanchal~K Roy}.} \bibinfo{year}{2015}\natexlab{}.
\newblock \showarticletitle{Evaluating clone detection tools with bigclonebench}. In \bibinfo{booktitle}{\emph{2015 IEEE international conference on software maintenance and evolution (ICSME)}}. IEEE, \bibinfo{pages}{131--140}.
\newblock


\bibitem[Thongtanunam et~al\mbox{.}(2013)]%
        {thongtanunam2013mining}
\bibfield{author}{\bibinfo{person}{Patanamon Thongtanunam}, \bibinfo{person}{Raula~G Kula}, \bibinfo{person}{Ana~EC Cruz}, \bibinfo{person}{Norihiro Yoshida}, \bibinfo{person}{Kohei Ichikawa}, {and} \bibinfo{person}{Hajimu Iida}.} \bibinfo{year}{2013}\natexlab{}.
\newblock \showarticletitle{Mining history of gamification towards finding expertise in question and answering communities: experience and practice with Stack Exchange}.
\newblock \bibinfo{journal}{\emph{The Review of Socionetwork Strategies}}  \bibinfo{volume}{7} (\bibinfo{year}{2013}), \bibinfo{pages}{115--130}.
\newblock


\bibitem[Tian and Treude(2022)]%
        {tian2022adding}
\bibfield{author}{\bibinfo{person}{Fuwei Tian} {and} \bibinfo{person}{Christoph Treude}.} \bibinfo{year}{2022}\natexlab{}.
\newblock \showarticletitle{Adding Context to Source Code Representations for Deep Learning}. In \bibinfo{booktitle}{\emph{2022 IEEE International Conference on Software Maintenance and Evolution (ICSME)}}. \bibinfo{publisher}{{IEEE} / {ACM}}, \bibinfo{address}{Limassol, Cyprus}, \bibinfo{pages}{374--378}.
\newblock
\href{https://doi.org/10.1109/ICSME55016.2022.00042}{doi:\nolinkurl{10.1109/ICSME55016.2022.00042}}


\bibitem[Treude and Storey(2025)]%
        {treude2025generative}
\bibfield{author}{\bibinfo{person}{Christoph Treude} {and} \bibinfo{person}{Margaret-Anne Storey}.} \bibinfo{year}{2025}\natexlab{}.
\newblock \showarticletitle{Generative AI and Empirical Software Engineering: A Paradigm Shift}. In \bibinfo{booktitle}{\emph{2025 2nd IEEE/ACM International Conference on AI-powered Software (AIware)}}. \bibinfo{publisher}{IEEE}, \bibinfo{address}{Seoul, Republic of Korea}, \bibinfo{pages}{233--239}.
\newblock
\href{https://doi.org/10.1109/AIware69974.2025.00033}{doi:\nolinkurl{10.1109/AIware69974.2025.00033}}


\bibitem[Vaithilingam et~al\mbox{.}(2022)]%
        {vaithilingam2022expectation}
\bibfield{author}{\bibinfo{person}{Priyan Vaithilingam}, \bibinfo{person}{Tianyi Zhang}, {and} \bibinfo{person}{Elena~L Glassman}.} \bibinfo{year}{2022}\natexlab{}.
\newblock \showarticletitle{Expectation vs. experience: Evaluating the usability of code generation tools powered by large language models}. In \bibinfo{booktitle}{\emph{Chi conference on human factors in computing systems extended abstracts}}. \bibinfo{pages}{1--7}.
\newblock


\bibitem[Wan et~al\mbox{.}(2022)]%
        {wan2022they}
\bibfield{author}{\bibinfo{person}{Yao Wan}, \bibinfo{person}{Wei Zhao}, \bibinfo{person}{Hongyu Zhang}, \bibinfo{person}{Yulei Sui}, \bibinfo{person}{Guandong Xu}, {and} \bibinfo{person}{Hai Jin}.} \bibinfo{year}{2022}\natexlab{}.
\newblock \showarticletitle{What do they capture? a structural analysis of pre-trained language models for source code}. In \bibinfo{booktitle}{\emph{Proceedings of the 44th international conference on software engineering}}. \bibinfo{pages}{2377--2388}.
\newblock


\bibitem[Wang and Lo(2014)]%
        {wang2014version}
\bibfield{author}{\bibinfo{person}{Shaowei Wang} {and} \bibinfo{person}{David Lo}.} \bibinfo{year}{2014}\natexlab{}.
\newblock \showarticletitle{Version history, similar report, and structure: putting them together for improved bug localization}. In \bibinfo{booktitle}{\emph{Proceedings of the 22nd International Conference on Program Comprehension}} (Hyderabad, India) \emph{(\bibinfo{series}{ICPC 2014})}. \bibinfo{publisher}{Association for Computing Machinery}, \bibinfo{address}{New York, NY, USA}, \bibinfo{pages}{53–63}.
\newblock
\showISBNx{9781450328791}
\href{https://doi.org/10.1145/2597008.2597148}{doi:\nolinkurl{10.1145/2597008.2597148}}


\bibitem[Wang et~al\mbox{.}(2021)]%
        {wang2021codet5}
\bibfield{author}{\bibinfo{person}{Yue Wang}, \bibinfo{person}{Weishi Wang}, \bibinfo{person}{Shafiq Joty}, {and} \bibinfo{person}{Steven~C.H. Hoi}.} \bibinfo{year}{2021}\natexlab{}.
\newblock \showarticletitle{{C}ode{T}5: Identifier-aware Unified Pre-trained Encoder-Decoder Models for Code Understanding and Generation}. In \bibinfo{booktitle}{\emph{Proceedings of the 2021 Conference on Empirical Methods in Natural Language Processing}}, \bibfield{editor}{\bibinfo{person}{Marie-Francine Moens}, \bibinfo{person}{Xuanjing Huang}, \bibinfo{person}{Lucia Specia}, {and} \bibinfo{person}{Scott Wen-tau Yih}} (Eds.). \bibinfo{publisher}{Association for Computational Linguistics}, \bibinfo{address}{Online and Punta Cana, Dominican Republic}, \bibinfo{pages}{8696--8708}.
\newblock
\href{https://doi.org/10.18653/v1/2021.emnlp-main.685}{doi:\nolinkurl{10.18653/v1/2021.emnlp-main.685}}


\bibitem[Wang et~al\mbox{.}(2023)]%
        {wang2023comparison}
\bibfield{author}{\bibinfo{person}{Yuekun Wang}, \bibinfo{person}{Yuhang Ye}, \bibinfo{person}{Yueming Wu}, \bibinfo{person}{Weiwei Zhang}, \bibinfo{person}{Yinxing Xue}, {and} \bibinfo{person}{Yang Liu}.} \bibinfo{year}{2023}\natexlab{}.
\newblock \showarticletitle{Comparison and Evaluation of Clone Detection Techniques with Different Code Representations}. In \bibinfo{booktitle}{\emph{Proceedings of the 45th International Conference on Software Engineering}} \emph{(\bibinfo{series}{ICSE '23})}. \bibinfo{publisher}{IEEE Press}, \bibinfo{address}{Melbourne, Victoria, Australia}, \bibinfo{pages}{332–344}.
\newblock
\showISBNx{9781665457019}
\href{https://doi.org/10.1109/ICSE48619.2023.00039}{doi:\nolinkurl{10.1109/ICSE48619.2023.00039}}


\bibitem[Widyasari et~al\mbox{.}(2023)]%
        {widyasari2023topic}
\bibfield{author}{\bibinfo{person}{Ratnadira Widyasari}, \bibinfo{person}{Zhipeng Zhao}, \bibinfo{person}{Thanh Le~Cong}, \bibinfo{person}{Hong~Jin Kang}, {and} \bibinfo{person}{David Lo}.} \bibinfo{year}{2023}\natexlab{}.
\newblock \showarticletitle{Topic Recommendation for GitHub Repositories: How Far Can Extreme Multi-Label Learning Go?}. In \bibinfo{booktitle}{\emph{2023 IEEE International Conference on Software Analysis, Evolution and Reengineering (SANER)}}. IEEE, \bibinfo{pages}{167--178}.
\newblock


\bibitem[Xia et~al\mbox{.}(2025)]%
        {xia2025agentless}
\bibfield{author}{\bibinfo{person}{Chunqiu~Steven Xia}, \bibinfo{person}{Yinlin Deng}, \bibinfo{person}{Soren Dunn}, {and} \bibinfo{person}{Lingming Zhang}.} \bibinfo{year}{2025}\natexlab{}.
\newblock \showarticletitle{Demystifying LLM-based Software Engineering Agents}.
\newblock \bibinfo{journal}{\emph{Proceedings of the ACM on Software Engineering}} \bibinfo{volume}{2}, \bibinfo{number}{FSE}, Article \bibinfo{articleno}{FSE037} (\bibinfo{year}{2025}), \bibinfo{numpages}{24}~pages.
\newblock
\href{https://doi.org/10.1145/3715754}{doi:\nolinkurl{10.1145/3715754}}


\bibitem[Yamaguchi et~al\mbox{.}(2014)]%
        {yamaguchi2014modeling}
\bibfield{author}{\bibinfo{person}{Fabian Yamaguchi}, \bibinfo{person}{Nico Golde}, \bibinfo{person}{Daniel Arp}, {and} \bibinfo{person}{Konrad Rieck}.} \bibinfo{year}{2014}\natexlab{}.
\newblock \showarticletitle{Modeling and discovering vulnerabilities with code property graphs}. In \bibinfo{booktitle}{\emph{2014 IEEE symposium on security and privacy}}. IEEE, \bibinfo{pages}{590--604}.
\newblock


\bibitem[Yang et~al\mbox{.}(2021)]%
        {yang2021learning}
\bibfield{author}{\bibinfo{person}{Xueqi Yang}, \bibinfo{person}{Jianfeng Chen}, \bibinfo{person}{Rahul Yedida}, \bibinfo{person}{Zhe Yu}, {and} \bibinfo{person}{Tim Menzies}.} \bibinfo{year}{2021}\natexlab{}.
\newblock \showarticletitle{Learning to recognize actionable static code warnings (is intrinsically easy)}.
\newblock \bibinfo{journal}{\emph{Empirical Software Engineering}} \bibinfo{volume}{26}, \bibinfo{number}{3} (\bibinfo{year}{2021}), \bibinfo{pages}{56}.
\newblock


\bibitem[Zaheer et~al\mbox{.}(2020)]%
        {zaheer2020bigbird}
\bibfield{author}{\bibinfo{person}{Manzil Zaheer}, \bibinfo{person}{Guru Guruganesh}, \bibinfo{person}{Kumar~Avinava Dubey}, \bibinfo{person}{Joshua Ainslie}, \bibinfo{person}{Chris Alberti}, \bibinfo{person}{Santiago Ontanon}, \bibinfo{person}{Philip Pham}, \bibinfo{person}{Anirudh Ravula}, \bibinfo{person}{Qifan Wang}, \bibinfo{person}{Li Yang}, {et~al\mbox{.}}} \bibinfo{year}{2020}\natexlab{}.
\newblock \showarticletitle{Big bird: Transformers for longer sequences}.
\newblock \bibinfo{journal}{\emph{Advances in neural information processing systems}}  \bibinfo{volume}{33} (\bibinfo{year}{2020}), \bibinfo{pages}{17283--17297}.
\newblock


\bibitem[Zeng et~al\mbox{.}(2022)]%
        {zeng2022extensive}
\bibfield{author}{\bibinfo{person}{Zhengran Zeng}, \bibinfo{person}{Hanzhuo Tan}, \bibinfo{person}{Haotian Zhang}, \bibinfo{person}{Jing Li}, \bibinfo{person}{Yuqun Zhang}, {and} \bibinfo{person}{Lingming Zhang}.} \bibinfo{year}{2022}\natexlab{}.
\newblock \showarticletitle{An extensive study on pre-trained models for program understanding and generation}. In \bibinfo{booktitle}{\emph{Proceedings of the 31st ACM SIGSOFT International Symposium on Software Testing and Analysis}} (South Korea) \emph{(\bibinfo{series}{ISSTA 2022})}. \bibinfo{publisher}{Association for Computing Machinery}, \bibinfo{address}{New York, NY, USA}, \bibinfo{pages}{39–51}.
\newblock
\showISBNx{9781450393799}
\href{https://doi.org/10.1145/3533767.3534390}{doi:\nolinkurl{10.1145/3533767.3534390}}


\bibitem[Zhang et~al\mbox{.}(2019)]%
        {zhang2019novel}
\bibfield{author}{\bibinfo{person}{Jian Zhang}, \bibinfo{person}{Xu Wang}, \bibinfo{person}{Hongyu Zhang}, \bibinfo{person}{Hailong Sun}, \bibinfo{person}{Kaixuan Wang}, {and} \bibinfo{person}{Xudong Liu}.} \bibinfo{year}{2019}\natexlab{}.
\newblock \showarticletitle{A novel neural source code representation based on abstract syntax tree}. In \bibinfo{booktitle}{\emph{Proceedings of the 41st International Conference on Software Engineering, {ICSE} 2019, Montreal, QC, Canada, May 25-31, 2019}}, \bibfield{editor}{\bibinfo{person}{Joanne~M. Atlee}, \bibinfo{person}{Tevfik Bultan}, {and} \bibinfo{person}{Jon Whittle}} (Eds.). \bibinfo{publisher}{{IEEE} / {ACM}}, \bibinfo{address}{Montreal, QC, Canada}, \bibinfo{pages}{783--794}.
\newblock
\href{https://doi.org/10.1109/ICSE.2019.00086}{doi:\nolinkurl{10.1109/ICSE.2019.00086}}


\bibitem[Zhang* et~al\mbox{.}(2020)]%
        {zhang2019bertscore}
\bibfield{author}{\bibinfo{person}{Tianyi Zhang*}, \bibinfo{person}{Varsha Kishore*}, \bibinfo{person}{Felix Wu*}, \bibinfo{person}{Kilian~Q. Weinberger}, {and} \bibinfo{person}{Yoav Artzi}.} \bibinfo{year}{2020}\natexlab{}.
\newblock \showarticletitle{BERTScore: Evaluating Text Generation with BERT}. In \bibinfo{booktitle}{\emph{International Conference on Learning Representations}}.
\newblock
\urldef\tempurl%
\url{https://openreview.net/forum?id=SkeHuCVFDr}
\showURL{%
\tempurl}


\bibitem[Zhou et~al\mbox{.}(2019)]%
        {zhou2019devign}
\bibfield{author}{\bibinfo{person}{Yaqin Zhou}, \bibinfo{person}{Shangqing Liu}, \bibinfo{person}{Jingkai Siow}, \bibinfo{person}{Xiaoning Du}, {and} \bibinfo{person}{Yang Liu}.} \bibinfo{year}{2019}\natexlab{}.
\newblock \showarticletitle{Devign: Effective vulnerability identification by learning comprehensive program semantics via graph neural networks}.
\newblock \bibinfo{journal}{\emph{Advances in neural information processing systems}}  \bibinfo{volume}{32} (\bibinfo{year}{2019}), \bibinfo{pages}{10197--10207}.
\newblock


\end{thebibliography}

\end{document}